\documentclass[onecolumn,aps,prX,a4paper,11pt,superscriptaddress,amsmath,amssymb]{revtex4}

\usepackage{graphicx}
\usepackage{epsfig}
\usepackage{verbatim}
\usepackage{subfiles}
\usepackage{dcolumn}
\usepackage{bm}
\usepackage{placeins} 
\usepackage{subfigure}
\usepackage{enumerate}

\usepackage{chapterbib}

\def\dst{\displaystyle}

\newcommand{\refp}[1]{(\ref{#1})}
\newcommand{\fref}[1]{Fig.~\ref{#1}}
\newcommand{\tref}[1]{Table~\ref{#1}}
\newcommand{\eref}[1]{Eq.~(\ref{#1})}
\newcommand{\cref}[1]{Chapter~\ref{#1}}
\newcommand{\sref}[1]{Section~\ref{#1}}

\renewcommand\Re{\operatorname{Re}}
\renewcommand\Im{\operatorname{Im}}
\newcommand{\ccc}[1]{} 

\newcommand{\abs}[1]{\ensuremath{\left| {#1} \right|}}
\newcommand{\ampl}[2][]{\ensuremath{F_{\mathrm{#1}}^{\mathrm{#2}}(s,t)}}

\newcommand{\dcss}[3][]{
\ifthenelse{\equal{#1}{}}
{\ensuremath{      \frac{\text{d}\sigma^{#2}{#3}}{\text{d}t}}}
{\ensuremath{\left.\frac{\text{d}\sigma^{#2}{#3}}{\text{d}t}\right|_{#1}}}
}
\newcommand{\rate}[2]{\ensuremath{N^{\mathrm{#1}}_{\mathrm{#2}}}}
\newcommand{\difrate}[2]{\ensuremath{\frac{\text{d}\rate{#1}{#2}}{\text{d}t}}}
\def\Journal#1#2#3#4{{#1} {\bf #2}, #3 (#4)}

\def\NIMA{{\em Nucl. Instrum. Methods} A}

\def\be{\begin{equation}}
\def\ee{\end{equation}}
\def\bea{\begin{eqnarray}}
\def\eea{\end{eqnarray}}

\begin{document}
\titlepage                                                    
\begin{flushright}                                                    
\today \\                                                    
\end{flushright}

\title{{\Large Proceedings of the Workshop} \\ 
{\huge FORWARD PHYSICS AT THE LHC} \\ 
{\Large La Biodola, Elba Island, Italy, May 27-29, 2010}}
\author{F. Ferro}\altaffiliation{Editors}\affiliation{INFN Genova,  
  Via Dodecaneso 33, Genova, Italy, I-16146}
\author{S. Lami}\altaffiliation{Editors}\affiliation{INFN Pisa,  
  Largo Pontecorvo 3, Pisa, Italy, I-56127}
\author{A. Achilli}
\affiliation{Physics Department and INFN, University of Perugia, Perugia I06123, Italy}
\author{O. Adriani}
\affiliation{%
University of Florence and INFN Sezione di Firenze
}%
\author{A. Alkin}\affiliation{
Bogolyubov Institute for Theoretical Physics,Metrologichna 14b, Kiev, Ukraine, UA-03680}

\author{S. Diglio}
\affiliation{ LAL, Univ Paris-Sud, CNRS/IN2P3, Orsay, France }
\author{M. Gallinaro}
\affiliation{The Rockefeller University (USA)}\affiliation{LIP Lisbon, Portugal}
\author{R. M. Godbole}
\affiliation{
Centre for High Energy Physics, Indian Institute of Science, Bangalore, 560012, India }
\author{K. Goulianos}
\affiliation{The Rockefeller University (USA)}
\author{A. Grau}
\affiliation{Departamento de Fisica Teorica y del Cosmos, Universidad de Granada, Spain}
\author{J. Ka\v{s}par},
\affiliation{CERN, Geneva, Switzerland}
\author{V. Kundr\'{a}t}
\affiliation{Institute of Physics,
AS~CR, v.v.i., 182 21 Praha 8, Czech Republic}
\author{M. Lokaj\'{\i}\v{c}ek}
\affiliation{Institute of Physics,
AS~CR, v.v.i., 182 21 Praha 8, Czech Republic}
\author{E. Martynov}\affiliation{
Bogolyubov Institute for Theoretical Physics,Metrologichna 14b, Kiev, Ukraine, UA-03680}
\author{K. \"Osterberg }
\affiliation{Helsinki Institute of Physics and Department of Physics, University of Helsinki, P.O. Box 64, FI-00014 Helsinki, Finland }
\author{G. Pancheri}
\affiliation{INFN Frascati National Laboratories, Va E. Fermi 40, I00044 Frascati, Italy}
\author{J. Proch\'{a}zka}
\affiliation{CERN, Geneva, Switzerland}
\author{C. Royon}
\affiliation{IRFU/Service de physique des particules, CEA/Saclay, 91191 Gif-sur-Yvette cedex, France}
\author{M. Schmelling}
\affiliation{MPI for Nuclear Physics, Saupfercheckweg 1, D-69117 Heidelberg, Germany}
\author{O. Shekhovtsova}
\affiliation{INFN Frascati National Laboratories, Va E. Fermi 40, I00044 Frascati, Italy}
\author{Y. N. Srivastava}
\affiliation{Physics Department and INFN, University of Perugia, Perugia I06123, Italy}

\author{D. Volyanskyy}
\affiliation{%
Deutsche Elektronen-Synchrotron DESY \\
Notkestrasse 85, 22607 Hamburg, Germany
}

%
\maketitle
\newpage
\clearpage

\setcounter{tocdepth}{-1}
\tableofcontents
\newpage
\clearpage

\setcounter{affil}{0}

\title
{\Large Integral and derivative dispersion relations \\ for $pp$ and
$\bar pp$ amplitudes}
\thanks{Based on the talk given by E. Martynov at the Workshop ``Forward Physics at LHC'', Elba, May 27-29, 2010.}

\author{A. Alkin}
\author{E. Martynov}%
\affiliation{
Bogolyubov Institute for Theoretical Physics,\\ 
Metrologichna 14b, Kiev, Ukraine, UA-03680}


\begin{abstract}
\noindent The methods of integral dispersion relations (IDR) and derivative dispersion relations (DDR) are applied for analysis of the data on $pp$ and $\bar pp$ total cross sections and ratios of real to imaginary part of elastic forward scattering amplitude. The models of pomeron behaving as triple, double and simple pole (with intercept larger than one) in the angular momentum plane are considered. Predictions of the models are given for the TOTEM measurements at 7 and 14 TeV.
%
\end{abstract}

\pacs{Valid PACS appear here}
\maketitle

\addcontentsline{toc}{part}{Integral and derivative dispersion relations for $pp$ and
$\bar pp$ amplitudes - {\it A.Alkin, E.Martinov}}

\section{Introduction}
At the  LHC energies  the TOTEM measurements of proton-proton total cross-section and near forward elastic proton scattering are in fact the first ``measurement of pure pomeron only''. Many estimates made in the various models of elastic scattering amplitude show that $f$- and $\omega$-reggeon contributions are negligible at LHC energies and small momenta transferred. The forward real and imaginary parts of a hadron elastic scattering amplitudes are not independent because an amplitude obeys the integral dispersion relation which is originated from their analyticity. Therefore it would be interest and of importance to investigate a goodness of   the experimental data fit by the IDR method. However to reach this we need the reliable model for imaginary part of scattering amplitude (or for total cross section)  in whole kinematical region starting from the threshold. There are a lot of the models describing the high energy data on $\sigma_{t}$ and ratio $\rho$ of real to imaginary part of forward scattering amplitude. But they cannot be directly extrapolated to low energies in order to explore the IDR.

In this paper we suggest and realize the procedure which allow us to use the IDR for analysis of any high energy model of the forward $pp$ $\bar pp$ scattering amplitude. Concretely we have considered and analyzed three pomeron models and predict the values of cross sections and ratios of real to imaginary parts of forward amplitude an LHC energies 7 TeV and 14 TeV. The procedure can be easily extended for other amplitudes, such as $\pi^{\pm}p$ and $K^{\pm}p$ ones.

\section{
Generalities and definitions}

{\bf Amplitude.} $S$-matrix theory postulates that the amplitude of any hadronic process is an analytic function of
invariant kinematic variables. For the $ab\to cd$ processes under interest, i.e. $p(\bar p)p\to p(\bar p) p$, they are 
\begin{equation}
s=(p_{a}+p_{b})^{2},\quad
t=(p_{a}-p_{c})^{2},\quad
u=(p_{a}-p_{d})^{2},\qquad 
s+t+u =m_{a}^{2}+m_{b}^{2}+m_{c}^{2}+m_{d}^{2}=4m_{p}^{2}.
\end{equation}

{\bf Crossing-symmetry.} Crossing-symmetry means that processes $p+p \longrightarrow p+p$ ($s$-channel) and $\bar p+p \longrightarrow \bar p+p$
 ($t,u$-channels)  are described by the limiting values of  unique analytic function $A(s,t,u)$ but taken in different regions of the variables $s,t$ and $u$.
As only two of three variables $s,t,u$ are independent in what follows we write down often $A(s,t)$ instead of $A(s,t,u)$.

\begin{figure}[h]
\begin{center}
\includegraphics[scale=0.4]{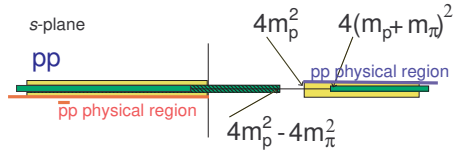}
\vspace*{-.4cm}
\caption{\label{fig:singular}Structure of singularities of $pp$ and $\bar pp$ elastic scattering amplitudes}
\end{center}
\end{figure}

{\bf Structure of singularities.} Singularities of $pp$ and $\bar pp$ elastic scattering amplitudes at $t=0$ are shown in Fig.\ref{fig:singular}. They are: i)the branch points at $s\geq 4m_{p}^{2}$ corresponding to the threshold energies of elastic and inelastic processes,\\ ii)branch points generated by the thresholds in $u$-channel  at $s\leq 0$, iii)nonphysical  (for elastic $\bar pp$ scattering) branch points generated by $u$-channel states (at $4m_{\pi}^{2}\leq u \leq 4m_{p}^{2}$ ). Thus for amplitude we have the right-hand and the left-hand cuts.

The physical amplitudes of $pp$ and $\bar pp$ elastic scattering are determined at the upper side of the cut from $4m_{p}^{2}$ to $+\infty$, precisely,
$A_{pp}(s,t)=\lim\limits_{\varepsilon\to 0}A(s+i\varepsilon,t,u )\equiv A_{-}(s,t,u)$ at $s>4m_{p}^{2}$,  $A_{\bar pp}(u,t)=\lim\limits_{\varepsilon\to 0}A(s,t,u +i\varepsilon)\equiv A_{+}(s,t,u)$ at $u>4m_{p}^{2}$.  One can derive from the definition  that
\begin{equation}\label{eq: pbarp-ampl}
A_{\bar pp}(u,t)=\lim\limits_{\varepsilon\to 0}A(s-i\varepsilon,t,u )\quad \mbox{at}\quad s+t<0.
\end{equation}

{\bf Optical theorem.} For $pp$ and $\bar pp$ it is read as
\begin{eqnarray}\label{eq:stot}
\sigma_{tot}^{\bar pp}(s)\equiv \sigma_{+}=\frac{1}{2m_{p}p}ImA_{\bar pp}(s,0)=
\frac{1}{2q_{s}\sqrt{s}}\, ImA_{+}(s,0),\\
\sigma_{tot}^{pp}(s)\equiv \sigma_{-}=\frac{1}{2m_{p}p}ImA_{\bar pp}(s,0)= \frac{1}{2q_{s}\sqrt{s}}\, ImA_{-}(s,0)
\end{eqnarray}
where $p$ is a momentum in the laboratory system, $q_{s}=\frac{1}{2}\sqrt{s-4m_{p}^{2}}$ and $A_{\pm}(s,0)=A^{\bar pp}_{pp}(s.0)$.

{\bf Polynomial behaviour.}
It is well known  that the scattering amplitude may rise at high values of $|s|$ not faster than a power of $s$, i.e. a such $N$ exists that at $|s|\to \infty$ and $t_{0}<t\leq 0$
\begin{equation}\label{eq: polybound}
|A(s,t)|<|s|^{N}.
\end{equation}

{\bf High-energy bounds for cross-sections.}
Total hadron cross sections behave at asymptotical energies in accordance with the well known Froissart-Martin-{\L}ukaszuk bound  
\begin{equation}\label{eq: FMLb}
\sigma_{t}(s)<\frac{\pi}{m_{\pi}^{2}}\ln^{2} (s/s_{0})\quad  \mbox{at}\quad  s\to \infty, \quad s_{0}\sim 1 \mbox{GeV}^{2}.
\end{equation}
The last inequality means that $|A(s,0)/s^{2}|\to 0\quad  {\rm at}\quad |s|\to \infty$.

All these properties of elastic scattering amplitude are of importance at deriving the integral dispersion relations.

\section{
 Integral Dispersion Relation (IDR)}

As analytic function of variable $s$ the amplitude $A(s,t)$ (in what follows we consider forward scattering amplitude, $t=0$) must satisfy the dispersion relation which can be derived from the Cauchy theorem for analytic function .

Because of asymptotic behaviour of the hadronic amplitudes ($|A(s,0)/s^{2}|\to 0$  but $|A(s,0)/s|\nrightarrow 0$ at $|s|\to \infty$) it is more convenient to apply Cauchy theorem to the function $A(s,0)/((s-s_{0})(s-s_{1}))$ rather than directly to amplitude. Generally, the points $s_{0}$ and $s_{1}$ are arbitrary ones but usually they are chosen at $s_{0}=s_{1}=2m_{p}^{2}$.

Deforming integration contour $C$ in the Cauchy relation for amplitude (more details can be found in the books \cite{Collins})  and neglecting the integral over circle with infinite radius (because  $|A(s,0)/s^{2}|\to 0$ at $|s|\to \infty$) one can write
\begin{eqnarray}\label{eq:dispers III}
A(s,0)=&A(s_{0},0)+(s-s_{0})A'(s_{0},0)+
\frac{(s-s_{0})^{2}}{\pi}\bigg [\int\limits_{4m_{p}^{2}}^{\infty}
\frac{D_{s}(s',0)}{(s'-s_{0})^{2}(s'-s)}ds'
+\int\limits_{-\infty}^{0}
\frac{D_{s}(s',0)}{(s'-s_{0})^{2}(s'-s)}ds'\bigg  ]\\
=&A(s_{0},0)+(s-s_{0})A'(s_{0},0)+
\frac{(s-s_{0})^{2}}{\pi}\bigg [\int\limits_{4m_{p}^{2}}^{\infty}
\frac{D_{s}(s',0)}{(s'-s_{0})^{2}(s'-s)}ds'
+\int\limits_{4m_{p}^{2}}^{\infty}
\frac{D_{u}(u',0)}{(u'-u_{0})^{2}(u'-u)}du'\bigg ] \nonumber
\end{eqnarray}
where
\begin{equation}\label{eq: discontin}
D_{s}(s,t)=\frac{1}{2i}[A(s+i\epsilon,t,u)-A(s-i\epsilon,t,u)],\qquad
D_{u}(u,t)=\frac{1}{2i}[A(s,t,u+i\epsilon)-A(s,t,u-i\epsilon)].
\end{equation}
After some simple transformations one can obtain the  standard  form of integral dispersion relations written  in the laboratory system  ($s=2m_{p}(E+m_{p}), u=2m_{p}(-E+m_{p})$, point $s_{0}$ corresponds to $E_{0}=0$ ).
\begin{equation}\label{eq:dispers IV}
\rho_{\pm}\sigma_{\pm}=\frac{A_{\pm}(2m_{p}^{2},0)}{2m_{p}p}+\frac{E\,
A_{\pm}'(2m_{p}^{2},0)}{p} +\frac{E^{2}}{\pi p}{\rm
P}\int\limits_{m_{p}}^{\infty}\left
[\frac{\sigma_{\pm}}{E'^{2}(E'-E)}+\frac{\sigma_{\mp}}{E'^{2}(E'+E)}\right
]p'\, dE'
\end{equation}
where $A'(z,0)=(d/dz)A(z,0)$.

We would like to note that  if the difference of cross sections $\Delta \sigma=\sigma_{+}-\sigma_{-}$ vanishes with the energy, i.e. if odderon does not contribute asymptotically to the forward $pp$ and $\bar pp$ amplitudes (it was confirmed by the COMPETE analysis \cite{COMPETE} that there are no indications for any visible odderon contribution to $\sigma_{tot}(s)$ and $\rho(s)$), then it can be proved that the integral dispersion relation Eq.~(\ref{eq:dispers IV}) with two subtractions is reduced to the dispersion relation  with one subtraction. Thus, if $\Delta \sigma=\sigma_{+}-\sigma_{-}$ at $s\to \infty$ the dispersion relations for $\bar pp$ and $pp$ amplitudes in terms of the cross sections and parameters $\rho$ can be written in the following form \cite{Soding}
\begin{equation}\label{eq:final dr}
\rho_{\pm}\, \sigma_{\pm} =\dst  \frac{B}{p} + \frac{E}{\pi p}{\rm
P}\int\limits_{m_{p}}^{\infty}\left
[\frac{\sigma_{\pm}}{E'(E'-E)}-\frac{\sigma_{\mp}}{E'(E'+E)}\right ]p'\,
dE'
\end{equation}
where $B$ is a constant to be determined within a some specific model or {\it e.g.} from the experimental data fit.

 Eq. (\ref{eq:final dr}) does not contain a contribution of the nonphysical cut $(0,4(m_{p}^{2}-m_{\pi}^{2}))$, however, as pointed out in \cite{Lengyels} at high energies it is small (less a few percents). Therefore, applying IDR for data description at $\sqrt{s}>$ 5GeV we suppose that the subtraction constant $B$ treated as free parameter efficiently mimics also a small nonphysical contribution.

\section{
 Derivative dispersion relations (DDR).}

The starting point to obtain the derivative dispersion relations is the dispersion integrals with two subtractions for an even
 $A^{(+)}$ and odd $A^{(-)}$ amplitudes
\begin{equation}\label{eq:signamp}
A^{(\pm)}(s,t)=\frac{1}{2}[A_{+}(s,t)\pm A_{-}(s,t)]
\end{equation}
It was proved in \cite{DDR0} that at high energy the following formal relations are valid
{\small
\begin{eqnarray}\label{eq:ddrasymp}
Re A^{(+)}(E,0)-B\approx  &&E\tan\left (\frac{\pi}{2}E\frac{d}{dE}\right )
ImA^{(+)}(E,0)/E=E\tan\left (\frac{\pi}{2}E\frac{d}{dE}\right
)\frac{2m_{p}p_{\, lab}}{E}\sigma_{t}^{+},\\
Re A^{(-)}(E,0)-C\frac{E}{m_{p}}\approx && E^{2}\tan\left (\frac{\pi}{2}E\frac{d}{dE}\right )
ImA^{(-)}(E,0)/E^{2}=E^{2}\tan\left (\frac{\pi}{2}E\frac{d}{dE}\right
)\frac{2m_{p}p_{\, lab}}{E^{2}}\sigma_{t}^{+} \nonumber
\end{eqnarray}
}
where $B$ and $C$ are  originated from subtraction constants in IDR.  At high energy the approximation $\tan\left ((\pi /2)Ed/dE\right )\approx (\pi /2)Ed/dE$ can be used.

In the papers \cite{DDRcor} all corrections to the asymptotic expressions Eqs. (\ref{eq:ddrasymp}) were derived in the form of $m_{p}/E$ power series.

\section{
Description of the $\sigma_{tot}$ and $\rho$ data at $\sqrt{s}>$ 5 GeV  in IDR and DDR}

If  the imaginary part of scattering amplitude (or total cross section) is known there are two ways to calculate its real part or ratio $\rho$.

1). $ReA(s,0)$ is calculated through IDR, however in this case the imaginary part of amplitude (or total cross section) must be known in whole kinematical region, just from the threshold.

2). $ReA(s,0)$ is calculated through DDR,  this method is good for high enough energy, but for low energy it is necessary to know many derivatives of scattering amplitude in order to determine corrections.

We compare the both methods fitting the data \cite{data} on $\sigma_{tot}$ and $\rho$ at energy $\sqrt{s}>$ 5GeV in three models for pomeron. 

\medskip
{\bf Low energy cross sections in IDR.}
As it was mentioned above the imaginary part of  the amplitude under consideration must be known in whole kinematic region
for a given process starting from the threshold. In our opinion the best way to solve the problem phenomenologically is the following.
The cross-sections at low energies  (153 $pp$ points and 385 $\bar pp$ points at $\sqrt{s}<5$ GeV) are parameterized as much well as possible.  The obtained values of the integral can be used with the various high energy models of the total cross sections, because they do not affect  high energy results.

The parametrization was taken as follows.
For $pp$ cross-section:
\begin{equation}\label{eq:siglpp}
\sigma^{pp}_{tot}(p)=\left \{
\begin{array}{llll}
c_{p1}+g_{p1}pe^{-p/p_{p1}}, &&p&<p_{1},\\
c_{p2}+g_{p2}(1+g_{p3}e^{p/p_{p2}})/(1+g_{p4}e^{p/p_{p3}}), \qquad
&p_{1}<&p&<p_{2},\\
c_{p3}+g_{p5}s^{\nu_{p}}+g_{p6}s^{\mu_{p}}, &p_{2}<&p&<p_{m}.\\
\end{array}
\right .
\end{equation}
For $\bar pp$ cross-section:
\begin{equation}\label{eq:siglapp}
\sigma^{\bar pp}_{tot}(p)=\left \{
\begin{array}{llll}
c_{a1}+g_{a1}e^{-p/p_{a1}}+g_{a2}e^{-p/p_{a2}}, \qquad \qquad  \qquad
&&p<p_{3},\\
c_{a2}+g_{a3}s^{nu_{a}}+g_{a4}s^{mu_{a}}, &p_{3}<&p<p_{m}&.
\end{array}
\right .
\end{equation}
 In the above expressions $p$ is a momentum in the laboratory system. The value of $p_{m}$ is the momentum at $s_{min}$. Parameters $c_{pk}$,
$c_{ak}$ and some ones of $g$ are not free, they are determined from the constraints $\sigma_{tot}(p_{lk}-0)=\sigma_{tot}(p_{lk}+0)$.
The values of the rest free parameters are determined by a data fit at $p<p_{m}$ under the following constraints. The values of the cross sections $\sigma(p_{m})$ calculated in accordance with the Eqs. (\ref{eq:siglpp}) and (\ref{eq:siglapp}) must be equal correspondingly to $\sigma_{tot}^{pp}(s_{min})$ and $\sigma_{tot}^{\bar pp}(s_{min})$ given by  fitting the specified model for pomeron at high energies $s>s_{min}$. One can be ascertained that such a tuning does not lead to a big variation of high energy parameters.
\begin{figure}[h]
\includegraphics[scale=0.45]{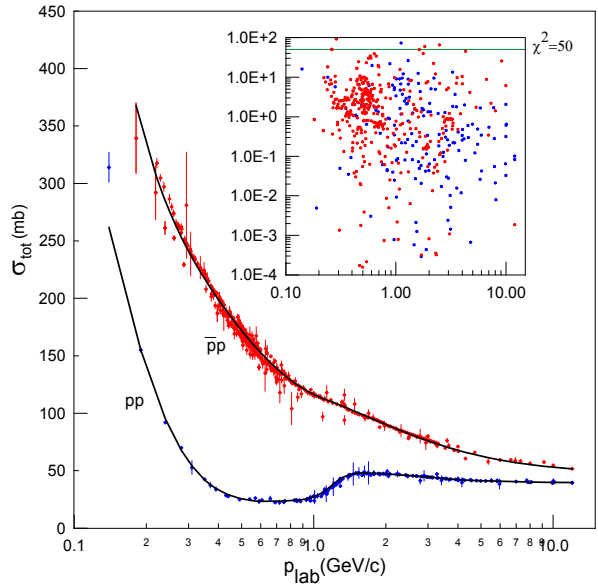}
\caption{Description of the $pp$ and $\bar pp$ cross sections at low energies, $\sqrt{s}<$ 5 GeV. The values of $\chi^{2}$ are shown for each data point.}
\label{fig:lowensig}
\end{figure}

As one can see from the Fig. \ref{fig:lowensig}, a quality of the data is far to be good, a data spread  is wide. There are a few points highly deviating from the main bulk of the data, each of them gives  $\chi^{2}>$ 50. If  the contribution of only these points to the total $\chi^{2}$ is not accounted  (but points are  not excluded from the fit) we obtain $\chi^{2}/N\approx 1.5$ for the parametrization (\ref{eq:siglpp},\ref{eq:siglapp}), so a quality of the fit is quite good.

Having a parametrization of the cross sections at low energy we are able to apply IDR for various high energy models. We perform an overall fit in the three steps.

{\small
{\bf The first step.} The chosen model for high energy cross-sections is fitted to the data on the cross sections only (without $\rho$ data) at $s>s_{min}$.

{\bf The second step.} The obtained "high-energy" parameters are fixed. The "low-energy" parameters
from Eqs. (\ref{eq:siglpp},\ref{eq:siglapp}) are determined by the fit at $s<s_{min}$,  provided the $\sigma_{pp}^{\bar pp}(s_{min})$ are given by the first step.

{\bf The third step.}  The subtraction constant $B_{+}$ is determined by the fit at
$s>s_{min}$ with all other high energy parameters being fixed (alternatively all high energy parameters are free, the results below are given for this case).
}

This procedure allow us to calculate the ratios $\rho_{pp}$ and $\rho_{\bar pp}$ at all energies above threshold. The results  are
given below for the specified high energy models.

\medskip
{\bf High energy. Pomeron models.}
We  consider three models leading to the different asymptotic behaviour of the total cross sections. For each model we
investigate how  the integral and derivative dispersion relations work. We start from the explicit parametrization of the total $pp$ and $\bar
pp$ cross-section. Then to find the ratios of the real to imaginary part we apply the IDR as it was described above and  compare the results for ratios calculated as well making use the DDR.

The  models under consideration are parameterized as the contributions of Pomeron, crossing even and crossing odd reggeons
\begin{equation}\label{sigmod}
\sigma_{pp}^{\bar pp}=\frac{1}{2m_{p}p}\left ({\cal P}(E)+R_{+}(E)\pm R_{-}(E)\right),
\end{equation}
where ${\cal P}(E)$ is a pomeron contribution while $R_{+}, R_{-}$ mimic the contributions of all crossing-even ($f, a_{2}, ...$ ) and crossing-odd terms ($\omega, \rho, ...$ ). Actually  $R_{+}$ and $R_{-}$ are very close   to $f$- and $\omega$-reggeons which are the most important  at the intermediate  energies.
\begin{equation}\label{reggeon}
R_{\pm}(E)=g_{\pm}z_{t}^{\alpha_{\pm}(0)-1}
\end{equation}
where $z_{t}=|\cos\vartheta_{t}|=|1+2s/(t-4m_{p}^{2})|(=E/m_{p} \quad \mbox{at} \quad t=0)$ is the ``natural'' Regge variable with $\vartheta_{t}$ being the scattering angle in $t$-channel \cite{Collins}.

The well known parametrization  with the replacement $E/m_{p}\to -is/s_{0}, \quad s_{0}=$ 1GeV$^{2}$ and with asymptotic form of the optical theorem, $\sigma=(1/s)Im A(s,0)$  (it is a standard Regge-type parametrization which simultaneously gives the both imaginary and real part of amplitude, we call it $``-is''$ parametrization) was considered  as well for a comparison.

\medskip
{\bf a. Simple Pomeron model (Modified Donnachie-Landshoff model \cite{DL}.}
$$
{\cal P}_{S}(E)=z_{t}\left \{g_{0}+g_{1}z_{t}^{\alpha_{\cal
P}(0)-1}\right \}, \qquad  \alpha_{\cal P}(0)=1+\varepsilon >1.
$$
The standard Regge amplitudes ( "$-is$" rule) are
\begin{equation}\label{simpleS}
(s_{0}/s)A^{\bar pp}_{pp}(s,0)=ig_{0}+ig_{1}(-is/s_{0})^{\alpha_{\cal
P}(0)-1}+ig_{+}(-is/s_{0})^{\alpha_{+}(0)-1}\pm
g_{-}(-is/s_{0})^{\alpha_{-}(0)-1}.
\end{equation}

{\bf b. Dipole Pomeron model}
$$
{\cal P}_{D}(E)=z_{t}\left \{g_{0}+g_{1}\ln z_{t}\right \}.
$$
\begin{equation}\label{doubleS}
(s_{0}/s)A^{\bar pp}_{pp}(s,0)=ig_{0}+ig_{1}\ln(-is/s_{0})
+ig_{+}(-is/s_{0})^{\alpha_{+}(0)-1}\pm g_{-}(-is/s_{0})^{\alpha_{-}(0)-1}.
\end{equation}

{\bf c. Triple Pomeron model}
$$
{\cal P}_{T}(E)=z_{t}\left \{g_{0}+g_{1}\ln z_{t}+g_{2}\ln^{2}z_{t}\right \}.
$$
\begin{eqnarray}\label{tripoleS}
(s_{0}/s)A^{\bar pp}_{pp}(s,0)&&=ig_{0}+ig_{1}\ln(-is/s_{0})+ig_{2}\ln^{2}(-is/s_{0})
+ig_{+}(-is/s_{0})^{\alpha_{+}(0)-1}\\ \nonumber
&&\pm g_{-}(-is/s_{0})^{\alpha_{-}(0)-1}.
\end{eqnarray}

\newpage
\section{ Results of the fit}

\begin{figure}[!h]
\begin{center}
\includegraphics[scale=0.4]{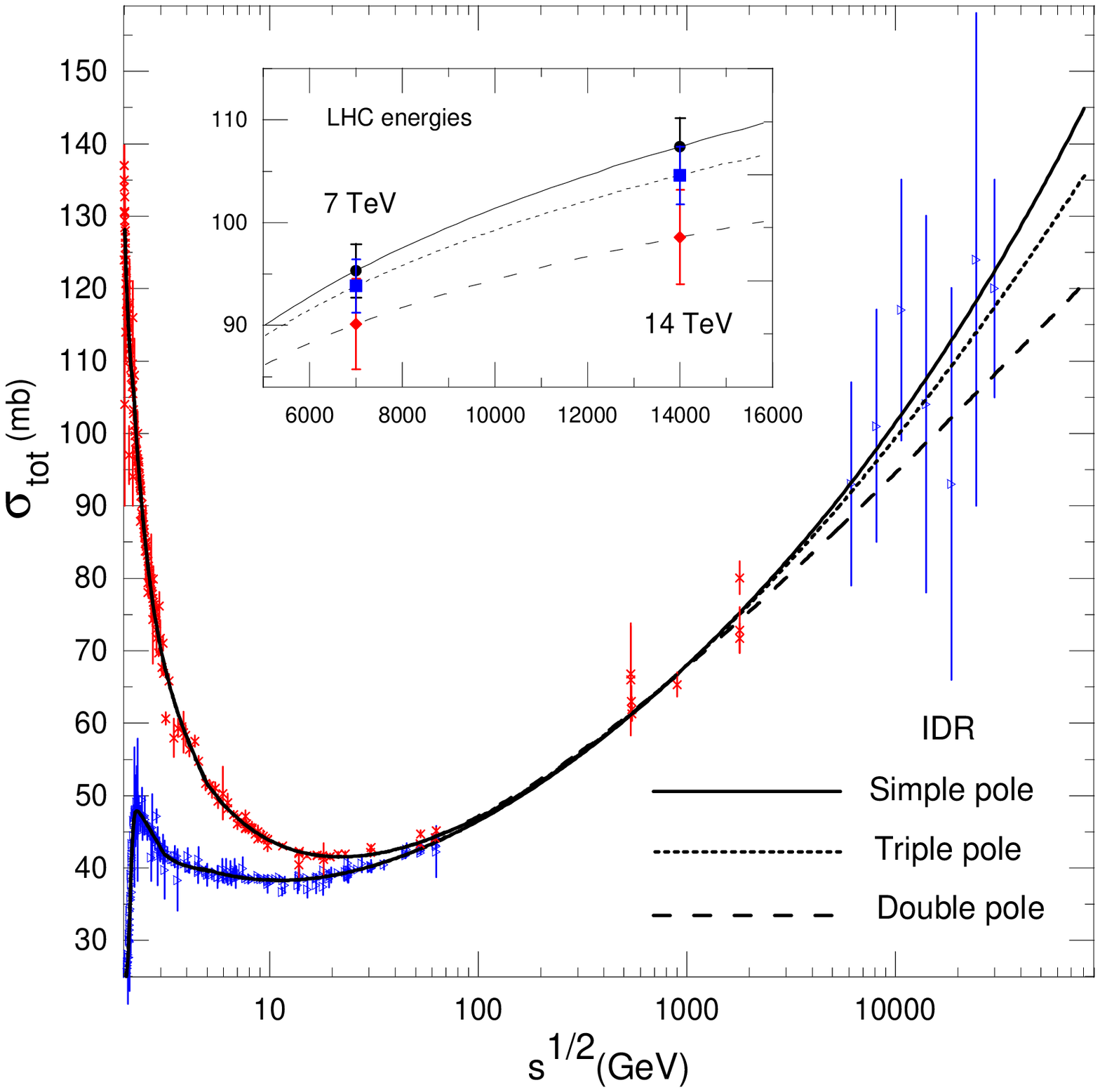}
\caption{Total $pp$ and $\bar pp$ cross sections in the simple, double and triple pomeron models obtained within IDR method}
\label{fig:sigma-idr}
\end{center}
\end{figure}

\begin{figure}[!h]
\begin{minipage}{6.cm}
\begin{center}
\includegraphics[scale=0.36]{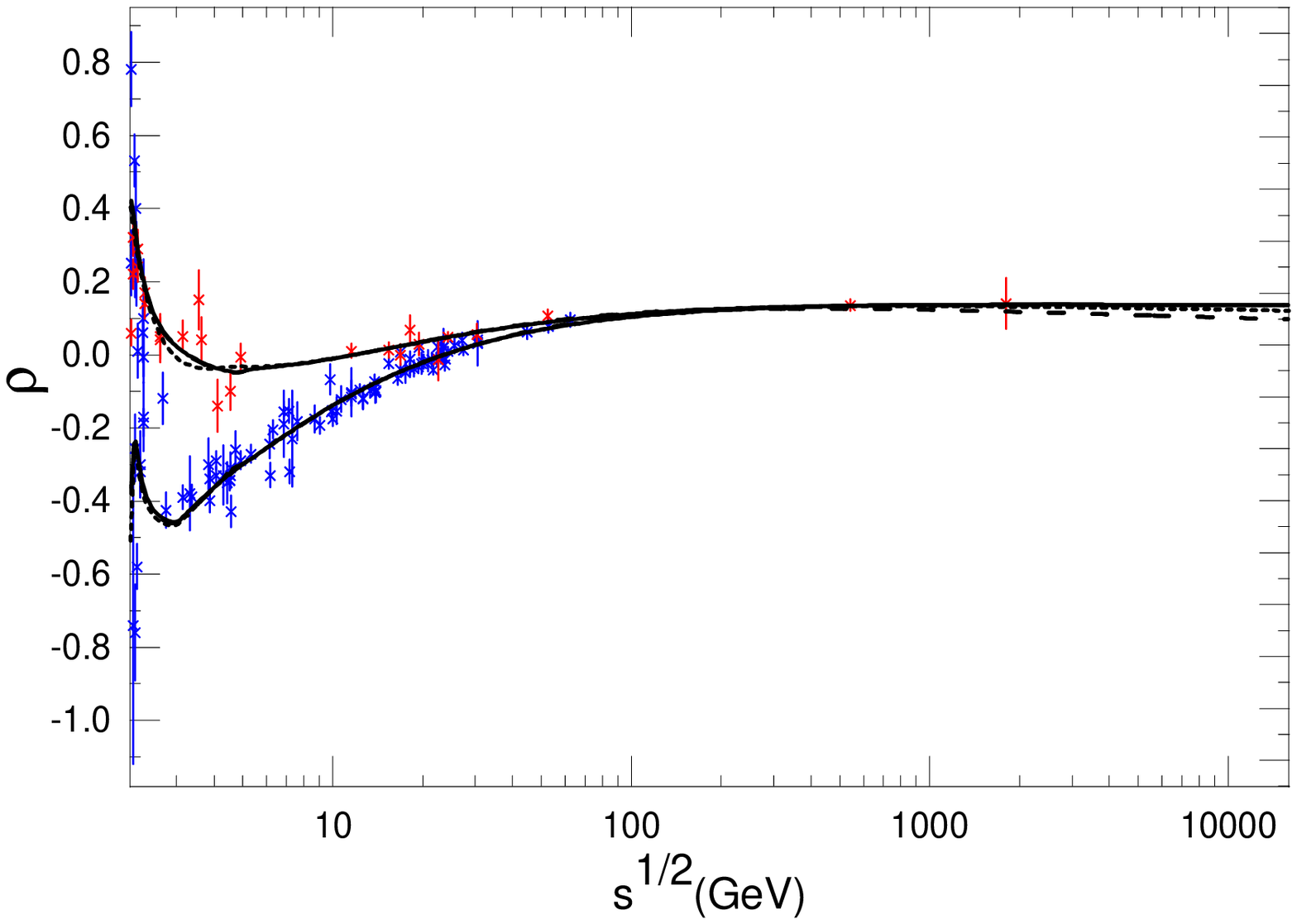}
\caption{The ratios $\rho$ for $pp$ and $\bar pp$ in three pomeron models at all energies.}
\label{fig:rho-all}
\end{center}
\end{minipage}
\hspace{2.cm}
\begin{minipage}{7.cm}
\begin{center}
\includegraphics[scale=0.36]{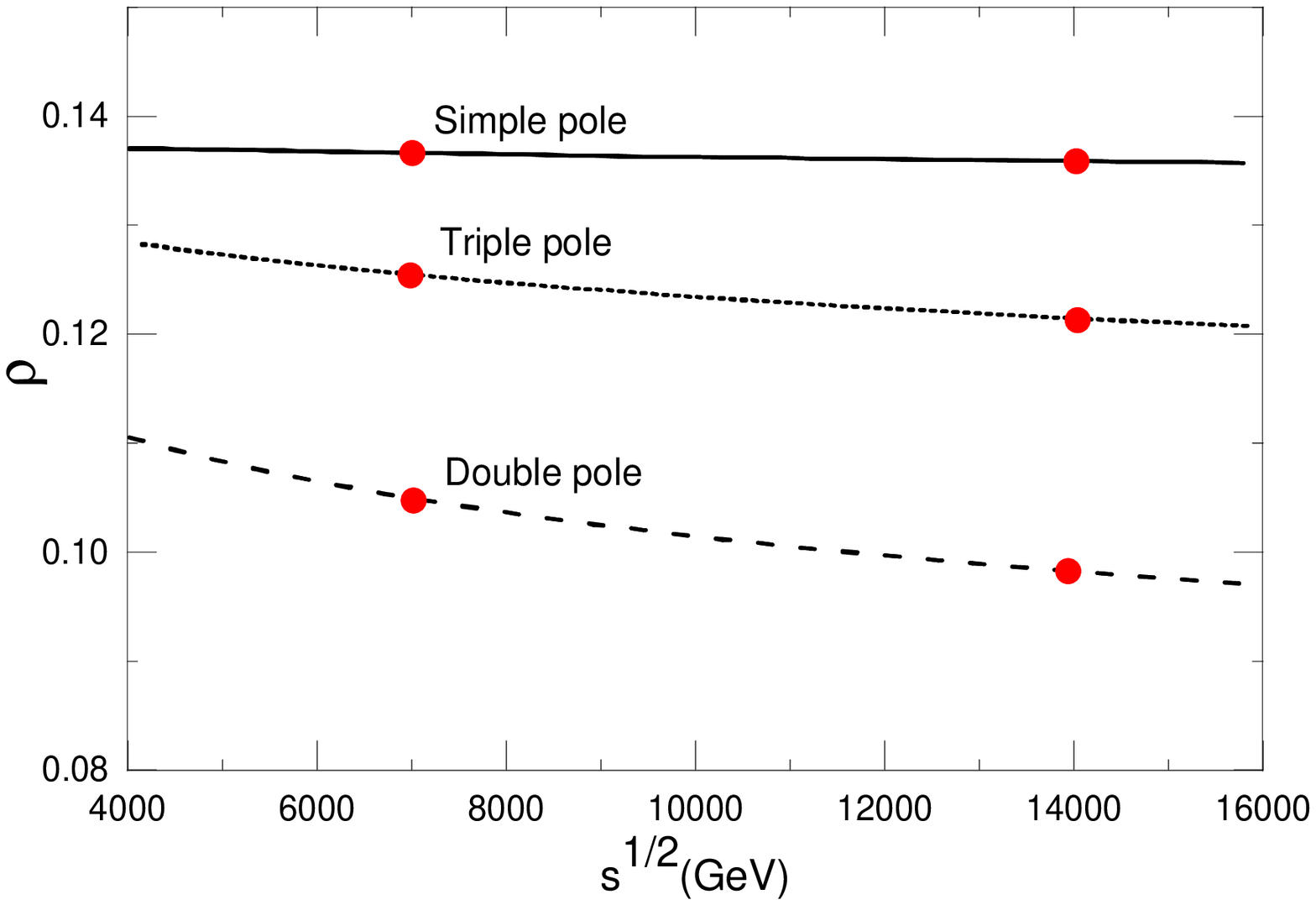}
\caption{The ratio $\rho_{pp}$ at LHC energies, predictions of three pomeron models }
\label{fig:rho-high}
\end{center}
\end{minipage}
\end{figure}

Omitting  details of the fits (postponing them for a forthcoming  paper)  we concentrate  on the main results and conclusions. The quality of the
fits is presented in the Table, where $\chi^{2}/dof=\chi^{2}/(N_{exp. points}-N_{parameters})$, and illustrated by Figs. \ref{fig:sigma-idr} and \ref{fig:rho-all}.  

As one can see from the Table, all models give good descriptions of the data on $\sigma_{tot}$ and $\rho$. Evidently, the fit
 by IDR method  is preferable. The data on $\sigma$ are described with $\chi^{2}/dof \approx 0.91$, while the data on $\rho$ are described less well, with a $\chi^{2}/N_{p}\approx 1.5$. In our opinion it is occurred because of the bad quality of the $\rho$ data.
\begin{table*}
\caption{\label{tab:globres}The values of the $pp$ and $\bar pp$ total cross sections and ratios of the real to imaginary part of forward amplitude at the LHC energies obtained in three pomeron models by IDR, DDR and ``$-is$ methods. The $\chi^{2}/dof$, where ''dof`` means degrees of freedom (number of data minus number of  free model parameters }
\begin{ruledtabular}
\begin{tabular}{cccccccccc}
&\multicolumn{9}{c}{Pomeron model}\\
 & \multicolumn{3}{c}{Simple pole} &  \multicolumn{3}{c}{Double pole} &  \multicolumn{3}{c}{Triple pole}\\
  \hline
 Fit procedure &  IDR & DDR&  $-is$ &IDR & DDR&  $-is$  &  IDR & DDR&  $-is$ \\
  \hline
$\sigma_{pp}$(mb) at 7 TeV& 95.3 &96.3 &91.2 &90.1 &90.1 & 90.5& 93.8 &93.5 &91.0 \\
$\rho_{pp}$ at 7 TeV& 0.137& 0.142&0.109 & 0.105& 0.105&0.105 & 0.125&0.137 &0.107 \\
  \hline
$\sigma_{pp}$(mb) at 14 TeV& 107.4 &109.2 & 100.2& 98.6 &98.6 & 99.1& 104.6 &104.2 &100.0 \\
$\rho_{pp}$ at 14 TeV&0.136 & 0.143&0.103 & 0.098&0.098 &0.098 & 0.121 &0.133 & 0.101\\
  \hline
 $\chi^{2}/dof$ & 1.104 &  1.101  & 1.137 &1.108 &  1.109 &1.146 & 1.103 & 1.093 &1.137 \\
\end{tabular}
\end{ruledtabular}
\end{table*}

\begin{figure}[h]
\begin{center}
\includegraphics[scale=0.35]{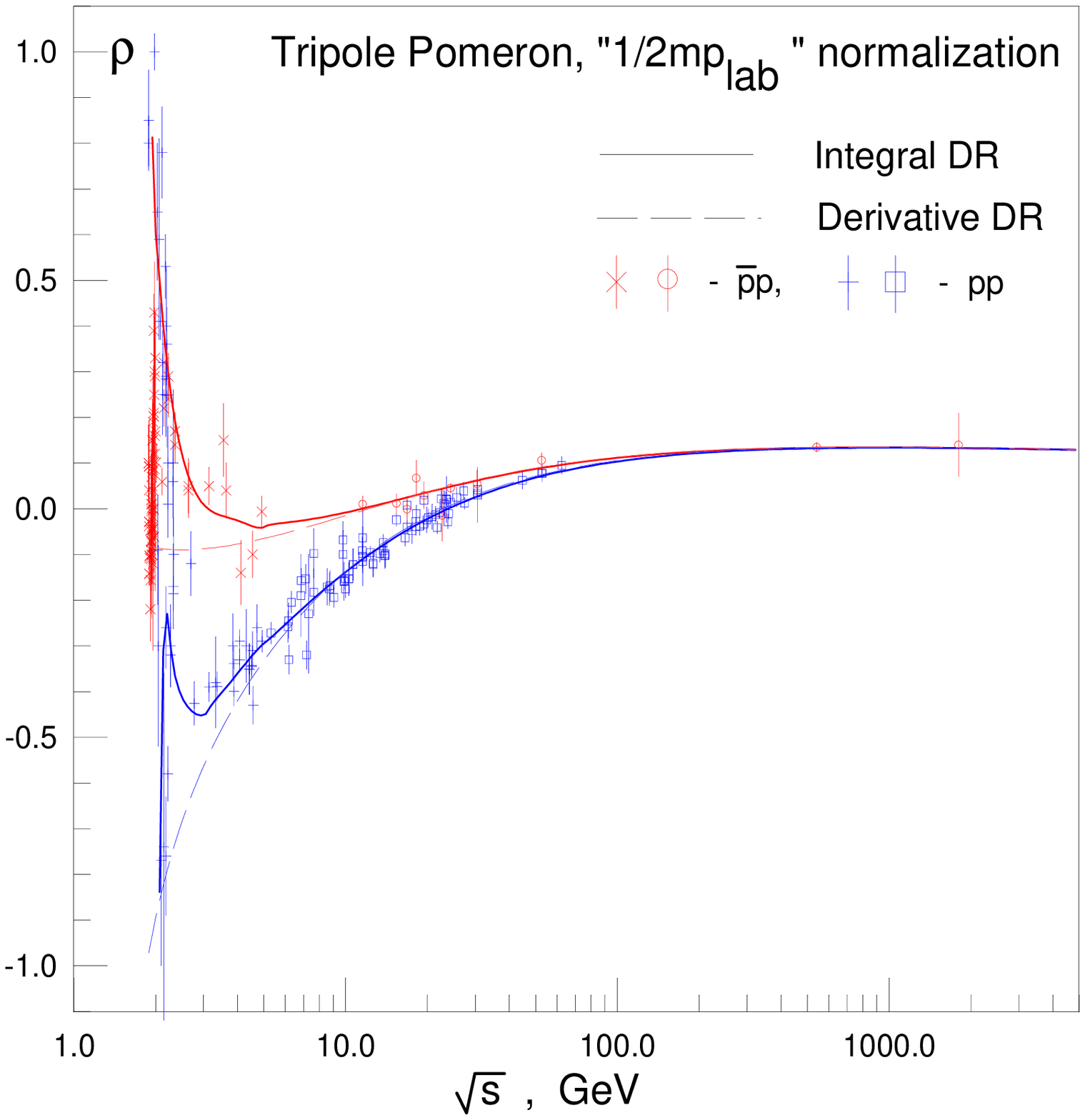}
\caption{\label{fig:rho-ddr-idr} Comparison of IDR and DDR methods applied to the triple pomeron model}
\end{center}
\end{figure}

We would like to emphasize  the fact that the values of $\rho$ calculated using DDR are deviated from those calculated by IDR even at
$\sqrt{s} \lesssim 7 -8$ GeV (see Fig. \ref{fig:rho-ddr-idr}). It means that in order to have more correct values of the $\rho$ at low energies, one must use the
IDR rather than explicit analytical expressions from the asymptotic DDR.

Some difference in predictions of COMPETE \cite{COMPETE} for $\sigma_{tot}$ and $\rho$ and those obtained there is explained mainly by different sets of data. We considered only $pp$ and $\bar pp$ data while in \cite{COMPETE} the data on $\pi p, Kp$  and on other hadrons were used. It would be interesting to extend IDR method for all available data.

Concluding we would like to emphasize  that, firstly, the IDR method allows to describe the data in a wide interval of energy. Secondly, three considered models having a quite good analytical properties predict the values of the total cross sections and ratios in not large intervals, at $\sqrt{s}=$ 14 GeV $\sigma_{pp}$ is in interval 98 - 108 mb. Nevertheless, if TOTEM measure the total cross section with accuracy about 1$\%$ one of the models can be preferred.

\newpage
\clearpage
\setcounter{affil}{0}
\setcounter{section}{0}
\setcounter{figure}{0}
\setcounter{table}{0}
\setcounter{equation}{0}

\hyphenation{PYTHIA DD SD ND}

\title{Diffractive cross sections and event final states at the LHC}

\author{Konstantin Goulianos}

\affiliation{The Rockefeller University}


\begin{abstract}
We discuss a phenomenological model that describes results on diffractive $pp$ and $\bar pp$ cross sections and event final states up to the Fermilab Tevatron energy of $\sqrt s =1.96$ TeV and use it to make predictions for Large Hadron Collider (LHC) energies up to $\sqrt s=14$~TeV and asymptotically as $\sqrt s\rightarrow\infty$. 
 The model is anchored in a saturation effect observed in single diffraction dissociation that explains quantitatively the factorization breaking observed in soft and hard $pp$ and $\bar pp$ diffractive processes and in diffractive photoproduction and low $Q^2$ deep inelastic scattering.    
\end{abstract}
\maketitle
\addcontentsline{toc}{part}{Diffractive cross sections and event final states at the LHC - {\it K.Goulianos}}
\section{Introduction}

As we entered a new energy frontier at the Large Hadron Collider (LHC) with data collected at $\sqrt s=900$~GeV, 2360~GeV, and 7~TeV from Fall 2009 to Spring 2010, it became painfully clear that the Monte Carlo (MC) simulations designed to represent the collective knowledge of the field on diffractive cross sections and event final states did not meet the challenge presented to them in this new higher energy environment. The most commonly used event generators, {\sc pythia}~\cite{PYTHIA} and {\sc phojet}~\cite{PHOJET}, were found to disagree not only with the data but also with each other. The latter clearly meant that the two simulations could not both be right. Therefore, an update of the MCs was urgently needed. Because of the importance of Minimum-Bias (MB) MC simulations in estimating trigger rates, backgrounds, and the machine luminosity at the LHC, a ``{\sc diffraction}'' workshop was organized at CERN on 7 May 2010~\cite{D-day} that brought experimentalists and theorists together to exchange ideas with the goal of producing a reliable MC generator for the LHC. This paper is based on a talk I presented at that meeting and an expanded version presented at this workshop.  
  
Diffraction dissociation in $pp$/$\bar pp$ interactions may be defined by the signature of one or more ``large'' and characteristically not exponentially suppressed~\cite{Bj} rapidity gaps (regions of rapidity devoid of particles)~\cite{rapidity} in the final state. The rapidity gap is presumed to be due to the exchange of a strongly-interacting color singlet quark/gluon combination with the quantum numbers of the vacuum, traditionally referred to as ``Pomeron''~($\pom$). Diffractive processes are classified as single diffraction (SD), double Pomeron exchange (DPE), also referred to as central dissociation (CD), and double diffraction (DD). In $\bar pp$ SD$_{\bar p}$~(SD$_p$), the $p(\bar{p})$ dissociates while the $\bar{p}(p)$ remains intact escaping the collision with momentum close to that of the original beam momentum and separated from the $p$~($\bar p$) dissociation products by a {\em forward} gap; in DPE both the $\bar p$ and the $p$ escape, resulting in {\em two forward gaps}; and in DD a {\em central} gap is formed while both the $p$ and $\bar p$ dissociate. 
The above basic diffractive processes are listed below, along with two additional two 2-gap processes which are combinations of SD and DD and are indicated as SDD:

\begin{figure}[!tb]
\centerline{\psfig{figure=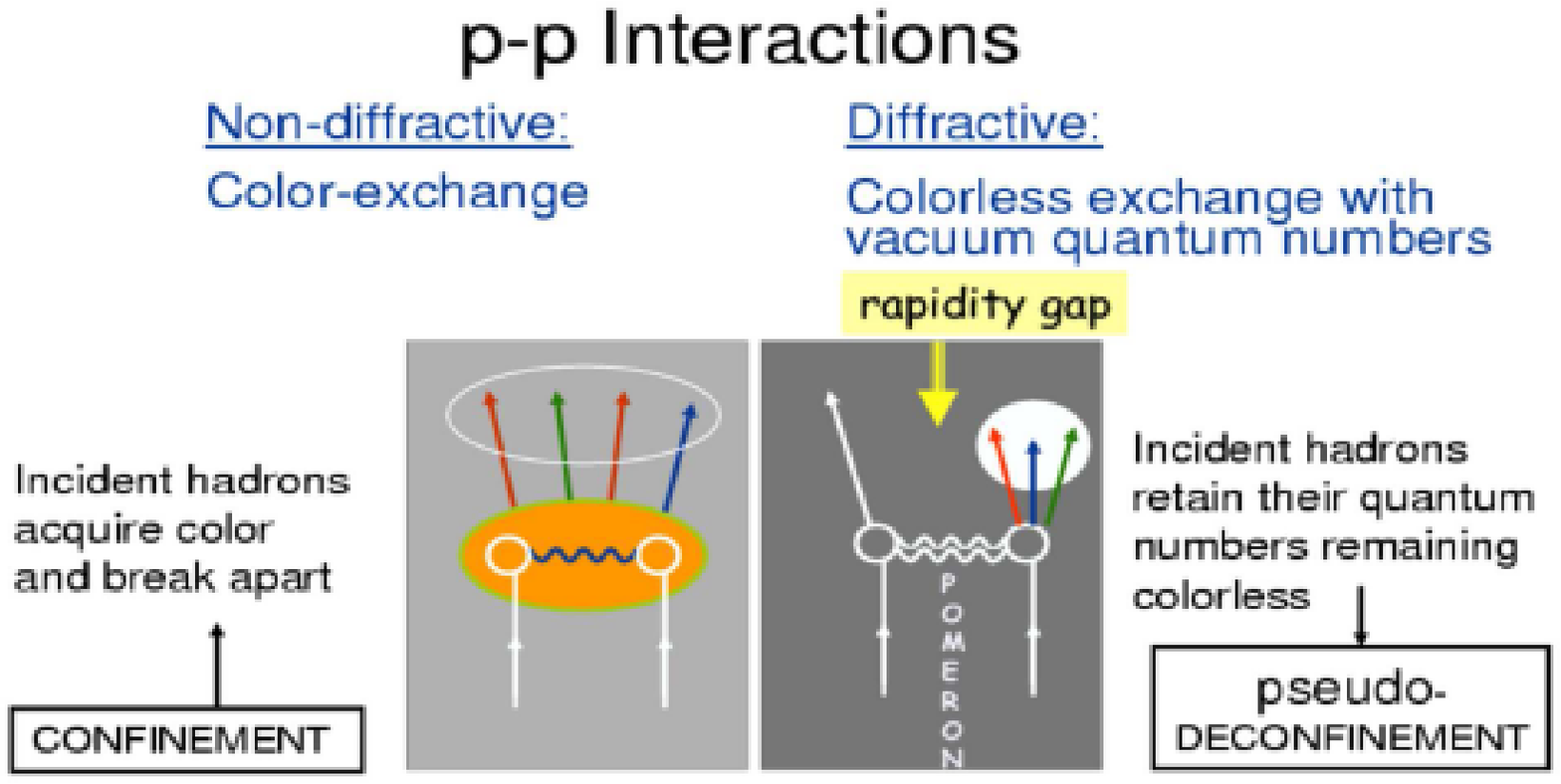,width=0.5\textwidth}}    
\caption{Non-diffractive and diffractive $pp$ interactions.}
\label{fig:rapgaps}
\end{figure}

\begin{table}[h]
\label{cross_sections}
\begin{center}
\caption{Diffractive cross sections.}
\begin{tabular}{ll}\hline\hline
\underline{acronym}&\underline{basic diffractive processes}\\
{\bf SD}$_{\bar p}$&$\bar{p}p\rightarrow \bar{p}+{\rm gap}+[p\rightarrow X_p]$,\\
{\bf SD}$_p$&$\bar{p}p\rightarrow [\bar p\rightarrow X_{\bar p}]+{\rm gap}+p,$\\
{\bf DD}&$\bar{p}p\rightarrow [\bar p\rightarrow X_{\bar p}]+{\rm gap}+[p\rightarrow X_p],$\\
{\bf DPE}&$\bar{p}p\rightarrow \bar{p}+{\rm gap}+X_c+{\rm gap}+p$,\\
&\underline{2-gap combinations of SD and DD}\\
{\bf SDD}$_{\bar p}$&$\bar{p}p\rightarrow \bar{p}+{\rm gap}+X_c+{\rm gap}+[p\rightarrow X_p]$,\\
{\bf SDD}$_p$&$\bar{p}p\rightarrow[\bar{p}\rightarrow X_{\bar p}]{\rm gap}+X_c+{\rm gap}+p$.\\
&\\
 \hline\hline
\end{tabular}
\end{center}
\end{table}

\noindent Here, $X_{\bar p}$, $X_p$ and  $X_c$ represent clusters of particles in rapidity regions not occupied by the gap(s). The 2-gap processes are examples of {\em multi-gap} diffraction, a term coined by this author to represent events with multiple diffractive rapidity gaps. A special case of DPE  is {\em  exclusive production,} where a particle state is centrally produced, as for example a dijet system or a $Z$ boson.
 
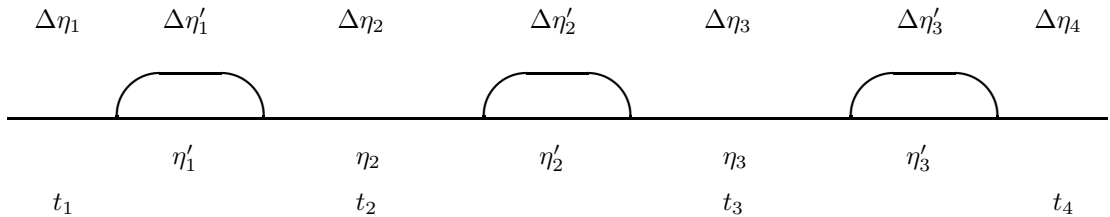
\begin{figure*}[t]
\unitlength 0.95in
\thicklines
\begin{center}
\begin{picture}(6,1)(1,0)
\put(1,0){\line(2,0){6}}
\multiput(2,0)(2,0){3}{\oval(0.8,0.5)[t]}
\put(1.9,-0.25){$\eta_1'$}
\put(2.9,-0.25){$\eta_2$}
\put(3.9,-0.25){$\eta_2'$}
\put(4.9,-0.25){$\eta_3$}
\put(5.9,-0.25){$\eta_3'$}
\put(2.8,0.5){$\Delta \eta_2$}
\put(4.8,0.5){$\Delta \eta_3$}
\put(1.25,-0.5){$t_1$}
\put(2.9,-0.5){$t_2$}
\put(4.9,-0.5){$t_3$}
\put(6.7,-0.5){$t_4$}
\put(1.15,0.5){$\Delta \eta_1$}
\put(1.85,0.5){$\Delta \eta'_1$}
\put(3.85,0.5){$\Delta \eta'_2$}
\put(5.85,0.5){$\Delta \eta'_3$}
\put(6.6,0.5){$\Delta \eta_4$}
\end{picture}
\end{center}
\vspace{3em}
\caption{Average multiplicity $dN/d\eta$ (vertical axis) vs. $\eta$ (horizontal axis) for a process with four rapidity gaps, $\Delta \eta_{i}(i=1-4)$.} 
\label{fig:soft}
\end{figure*}

Below, in Sec.~\ref{strategy} ({\sc strategy}) we outline the method we follow to implement an algorithm for a MC simulation, in Sec.~\ref{references} ({\sc cross sections and final states}) we present excerpts from previous papers on total and differential diffractive cross sections and final states, and in Sec.~\ref{conclusions} we conclude.

\section{Strategy}\label{strategy}
A phenomenology that is used to make predictions for the LHC and beyond should be based on cross sections and final states that incorporate the current knowledge in the field molded into a form that can be extrapolated to higher energies. The issues to be addressed is how to take into account saturation effects that suppress cross sections, and what formulas to use for event final state multiplicity, pseudorapidity, and transverse energy ($E_T$)~\cite{missingET} distributions.  In addition, the structure of an algorithm for implementing this knowledge into a MC simulation should also be addressed. The algorithm must be robust against changes in the collision energy, so that it may be equally well applied to simulate collisions at fixed target energies as well as at the higher energies of $pp$ and $\bar pp$ colliders and in astrophysics. In this section, we outline a strategy that addresses these issues.\\

The following input is used for cross sections and final states:

\begin{enumerate}[(i)]
\item $d^2\sigma/d\xi dt$ of the diffractive processes listed in Table~\ref{cross_sections} from the {\sc renorm} model~\cite{R};
\item $\sigma_t(s)$ from {\sc superball model}~\cite{superball};
\item optical theorem $\rightarrow\;Im\,f_{el}(t=0)$ (imaginary part of the forward scattering amplitude); 
\item dispersion relations $\rightarrow\;Re\,f_{el}(t=0)$, using low energy cross sections from global fit~\cite{global};
\item final states: use ``nesting'' to describe gap processes, where a {\em nest} is defined as a region of $\Delta\eta$ where there is particle production, in contrast to a {\em gap} region where there are no particles~\cite{{corfu01},{multigap}}.
\end{enumerate}

 Figure~\ref{fig:soft} shows a schematic $\eta$ topology of an event with four rapidity gaps and three nests of final state particles. The cross section for this configuration is presented in Sec.~\ref{x-sections}

We propose the following algorithm for generating final states:
\begin{itemize}
\item start with a $pp\rightarrow X$ inelastic collision at $\sqrt s$;
\item decide whether the collision is ND or diffractive based on the expected cross sections; if ND, use the ND final state expected at $\sqrt s$; if diffractive, select SD$_{\bar p}$, SD$_p$, DD, or DPE based on probabilities scaled to the corresponding cross sections;
\item for each diffractive event, check whether the region of $\eta$ where particles are produced, $\Delta\eta'$,  is large enough to accommodate additional diffractive rapidity gaps: if {\em yes}, decide whether or not the event will have other gaps within this region, again using probabilities scaled to the cross sections, and branch off accordingly;
\item continue this process until the region $\Delta\eta'$ is too small to accommodate another diffractive gap. 
\end{itemize}

It is important to note that in our definition of a ND collision there are no diffractive gaps whatsoever in the final state of the event. In this respect, this definition differs from those of ``inclusive'' or ``non-single-diffractive'' definitions of ND events used in the literature.     

\section{Cross sections and final states}\label{references} 

In this section,  we discuss briefly the diffraction dissociation and total cross sections using information and/or excerpts from Refs.~\cite{{R},{lathuile04}}.
  
\subsection{Diffractive cross sections}\label{x-sections}
In Ref.~\cite{pomQCD}, the following expression is obtained for the SD cross section [quoting]:

\begin{quote}
\begin{eqnarray}
\frac{d^2\sigma_{sd}(s,\Delta\eta,t)}{dt\,d\Delta\eta}=
\frac{1}{N_{gap}(s)}\times \nonumber\\
\underbrace{C_{gap}\cdot F_p^2(t)\left\{e^{\textstyle (\epsilon+\alpha' \,t)\Delta\eta}\right\}^2}_{\textstyle P_{gap}(\Delta\eta,t)}
 \cdot \;\kappa \cdot \left[\sigma_\circ\,e^{\textstyle \epsilon\Delta\eta'}\right],
\label{eq:diffPM}
\end{eqnarray}
where:\\
(i) the factor in square brackets represents the cross section due to the wee partons in the $\eta$-region of particle production $\Delta\eta'$;\\
(ii)  $\Delta\eta=\ln s$-$\Delta\eta'$ is the rapidity gap;\\
(iii)  $\kappa$ is a QCD color factor selecting color-singlet $gg$ or $q\bar q$ exchanges to form the rapidity gap;\\ 
(iv) $P_{gap}(\Delta\eta,t)$ is a gap probability factor representing the elastic scattering between the dissociated proton (cluster of dissociation particles) and the surviving proton;\\ 
(v)  $N_{gap}(s)$ is the integral of the gap probability distribution over all phase-space in $t$ and $\Delta\eta$;\\ 
(vi)  $F_p(t)$ in $P_{gap}(\Delta\eta,t)$ is the proton form factor $F_p(t)=e^{\displaystyle b_\circ t}$ ... 
; and\\
(vii) $C_{gap}$ is a normalization constant, whose value is rendered irrelevant by the renormalization division by $N_{gap}(s)$. ...\\
By a change of variables from $\Delta\eta$ to $M^2$ using $\Delta\eta'=\ln M^2$ and $\Delta\eta=\ln s-\ln M^2$, Eq.~(\ref{eq:diffPM}) takes the form:
\begin{eqnarray}
\frac{d^2\sigma (s,M^2,t)}{dM^2 dt}=
\left[\frac{\sigma_\circ}{16\pi}\sigma_\circ^{\pom p}\right]
\,\frac{s^{\displaystyle 2\epsilon}}{N(s)}
\;\frac{1}{\left(M^2\right)^{\displaystyle 1+\epsilon}}\;e^{\displaystyle b\,t}\nonumber\\
\;\;\stackrel{\displaystyle s\rightarrow \infty}{\Rightarrow}\;\;
\left[2\alpha' \,e^{\frac{\displaystyle\epsilon\,b_0}{\displaystyle\alpha' }}
\sigma_\circ^{\pom p}\right]
\frac{\ln s^{\displaystyle 2\epsilon}}{\left(M^2\right)^{\displaystyle 1+\epsilon}}\;e^{\displaystyle b\,t},
\label{eq:diffM2}
\end{eqnarray}
where $b=b_0+2\alpha'\ln\frac{\displaystyle s}{M^2}$ [$b$ is the slope of the diffractive $t$-distribution].
Integrating this expression over $M^2$ and $t$ yields the total single diffractive cross section,
\begin{equation}
\sigma_{sd}\stackrel{\displaystyle s\rightarrow \infty}{\rightarrow} 
2\,\sigma_\circ^{\pom p}\;
\exp\left[{\frac{\epsilon\,b_0}{2\alpha '}}\right]\mbox{= constant}\equiv\sigma^{\infty}_{sd}.
\label{eq:sigma_not}
\end{equation}
 
The remarkable property that the total single diffractive cross section becomes constant as $s\rightarrow \infty$ is a direct consequence of the coherence condition required for the recoil proton to escape the interaction intact. This condition selects one out of several available wee partons to provide a color-shield to the exchange and enable the formation of a diffractive rapidity gap.
\end{quote}

Details are presented in Ref.~\cite{pomQCD}, where this formulation of the cross section  is used to derive the ratio of the intercept to the slope of the Pomeron trajectory. Good agreement with the ratio extracted from measurements is obtained, providing support for the renormalization approach used in the phenomenology.  

A similar expression may be use for DD, DPE, and multigap processes, as discussed in Refs.~\cite{{corfu01},{multigap}}. For example, the differential cross section for the process displayed in Fif.~\ref{fig:soft} is derived in Ref.~\cite{multigap} as [quoting]:

\begin{eqnarray}
\frac{d^{10}\sigma^D}{\Pi_{i=1}^{10}dV_i}=
N^{-1}_{gap}\; 
\underbrace{
F^2_p(t_1)F^2_p(t_4)
\Pi_{i=1}^4\left\{e^{[\epsilon+\alpha't_i]\Delta\eta_i}\right\}^2
}_{\hbox{gap probability}}\;\nonumber\\
\times
\kappa^4\left[\sigma_0\,e^{\epsilon\sum_{i=1}^3\Delta\eta'_i}\right],
\label{eq:diffPM4}
\end{eqnarray}

\begin{quote}

\noindent where the term in square brackets is the $pp$  
total cross section at the reduced $s$-value, defined 
through $\ln (s'/s_0)=\sum_i\Delta \eta_i'$, 
$\kappa$ (one for each gap) is the QCD color factor for gap formation, 
the gap probability is the amplitude squared for elastic scattering 
between two diffractive clusters or between a diffractive cluster and a 
surviving proton with form factor $F^2_p(t)$,
and $N_{gap}$ is the (re)normalization factor defined as 
the gap probability integrated 
over all 10 independent variables $t_i$, $\eta_i$, $\eta'_i$, and 
$\Delta\eta\equiv \sum_{i=1}^4\Delta\eta_i$. 

The renormalization 
factor $N_{gap}$ is a function of $s$ only.
The color factors are $c_g=(N_c^2-1)^{-1}$ 
and $c_q=1/N_c$ for gluon and quark color-singlet exchange, respectively.  
Since the reduced energy cross section is properly normalized, the 
gap probability is (re)normalized to unity. The quark to gluon 
fraction, and thereby the Pomeron intercept parameter $\epsilon$ 
may be obtained from the inclusive parton distribution 
functions (PDFs)~\cite{lathuile04}. Thus, normalized differential multigap cross 
sections at $t=0$ may be fully derived from inclusive PDFs and QCD color 
factors without any free parameters.

The exponential dependence of the cross section on $\Delta\eta_i$ leads to 
a renormalization factor $\sim s^{2\epsilon}$ independent of the number of 
gaps in the process. This remarkable property of the renormalization model, 
which was confirmed in two-gap to one-gap cross section ratios measured by 
the CDF Collaboration (see Ref.~\cite{lathuile04}), suggests that multigap
diffraction can be used as a tool for exploring the QCD aspects of 
diffraction in an environment free of rapidity gap suppression effects.
The LHC with its large rapidity coverage provides the ideal arena 
for such studies.  
\end{quote}

\subsection{The total cross section} 
In Ref.~\cite{superball}, an analytic expression is obtained for the total cross section using a parton model approach and exploiting a saturation effect observed in the SD cross section. The abstract of Ref.~\cite{superball} reads [quoting]:
\begin{quote}
  The single-diffractive and total $pp$ cross sections at the LHC are predicted in a phenomenological approach that obeys all unitarity constraints. The approach is based on the renormalization model of diffraction and a saturated Froissart bound for the total cross section yielding $\sigma_t=(\pi/s_o)\cdot \ln^2(s/s_F)$ for $s>s_F$, where the parameters $s_o$ and $s_F$ are experimentally determined from the $\sqrt s$-dependence of the single-diffractive cross section.  
\end{quote}

\begin{figure}
\vspace*{-5em}
\centerline{\psfig{figure=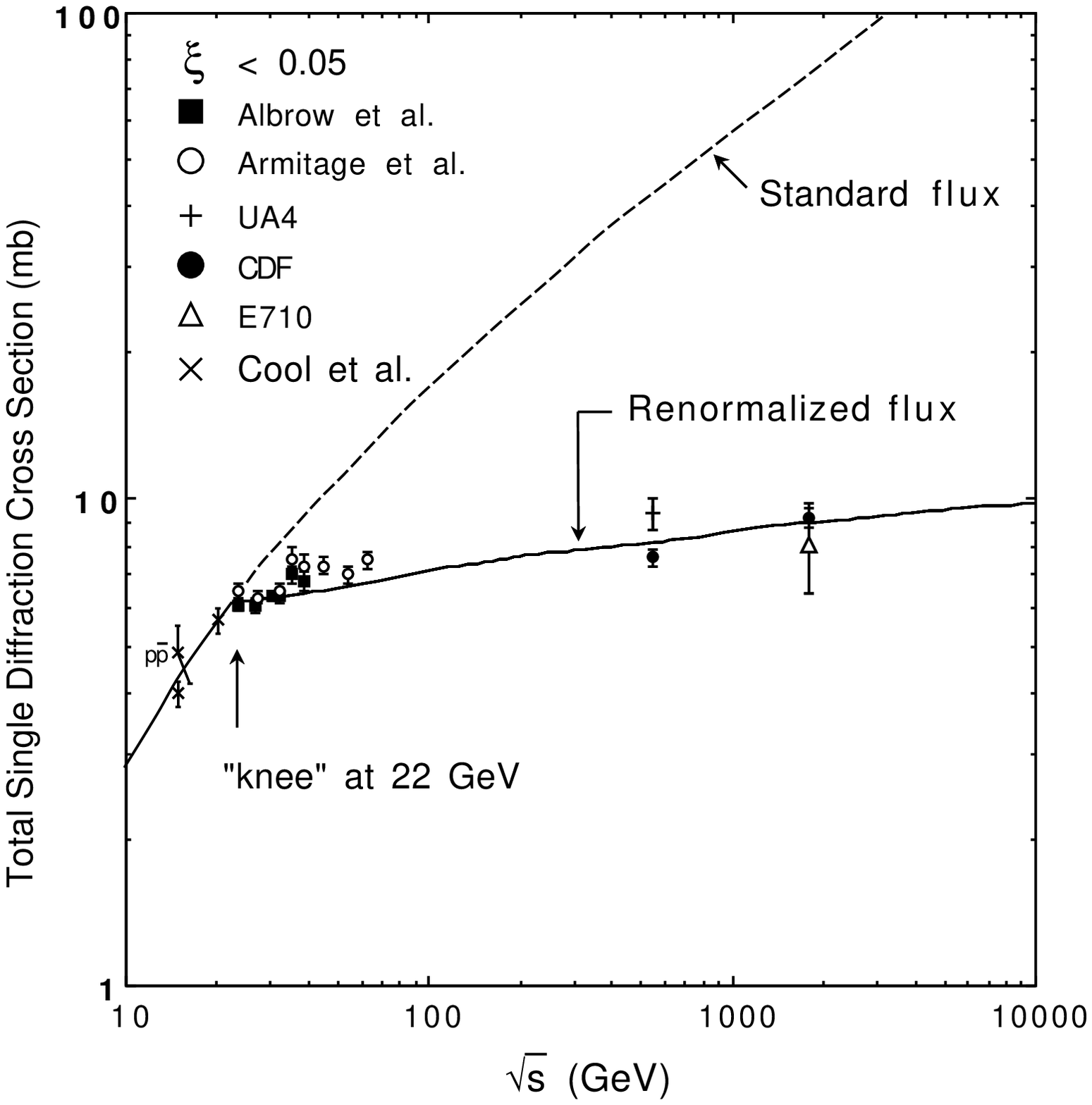,width=0.6\textwidth}}
\vspace*{-10em}  
\caption{Total \protect{$pp/\bar pp$}
single-diffraction dissociation cross section data (sum of both $\bar p$ and $p$ dissociation) for \protect{$\xi<0.05$} compared with predictions 
based on the standard and the renormalized Pomeron flux (from Ref.~\cite{R}).}
\label{fig:tot_sd}
\end{figure}

The following strategy is used in Ref.~\cite{superball} [quoting]:
 \begin{quote}

\addtolength{\itemsep}{-0.5em}
\begin{itemize}       
\item Use the Froissart formula as a {\em saturated}\, cross section rather than as a bound above $s_F$:\\ 

{\Large {$\sigma_t(s>s_F)=\sigma_t(s_F)+ \frac{\pi}{m^2}\cdot \ln^2\frac{s}{s_F}$}}\\

\item This formula should be valid above the {\em knee} in $\sigma_{sd}$ vs. $\sqrt s$ at $\sqrt s_F=22$~GeV (Fig.~\ref{fig:tot_sd}) and therefore valid at $\sqrt s=1800$~GeV.
\item Use $m^2=s_o$  in the Froissart formula multiplied by 1/0.389 to convert it to mb$^{-1}$.
\item Note that contributions from Reggeon exchanges  at $\sqrt s=1800$~GeV are negligible, as can be verified from the global fit of Ref.~\cite{global}.
\item Obtain the total cross section at the LHC:\\

{{
$\sigma_t^{\rm LHC}=\sigma_t^{\rm CDF}+
{\dfrac{\pi}{s_o}}\cdot
\left(\ln^2 \dfrac{s^{\rm LHC}}{s_F}-\ln^2 \dfrac{s^{\rm CDF}}{s_F}\right)$}
}\\
\end{itemize}

For a numerical evaluation of $\sigma^{LHC}$ we use as input the CDF cross section at $\sqrt s=1800$~GeV, $\sigma_t^{CDF}=80.03\pm 2.24$~mb, the Froissart saturation energy $\sqrt s_F=22$~GeV, and the parameter $s_o$. \\
...The resulting prediction for the total cross section at the LHC at $\sqrt s=14\;{\rm TeV}$ is:
$$\sigma^{LHC}_{14\,{\rm TeV}}= (80\pm3)+(29\pm 12)=109\pm12\mbox{\;\; mb}.$$
\end{quote}

\noindent For $\sqrt s=7$~TeV, the predicted cross section is: 
\begin{equation*}
\sigma_{7\;\rm{TeV}}^{\rm LHC}=98\pm 8\;{\rm mb}\left[{\rm at}\;\sqrt s=7\;{\rm TeV}\right],
\end{equation*}
The result for $\sqrt s=14$~TeV is in good agreement with $\sigma_t^{\rm CMG}=114\pm 5\;\rm{mb}$ obtained by the global fit of Ref.~\cite{global}, where the uncertainty was estimated from $\delta\epsilon$ and the  $s^\epsilon$ dependence from which the value of the parameter $s_o$ was obtained. 

\section{Conclusions}\label{conclusions}

We briefly discuss a phenomenological model that describes available results on diffractive $pp$ and $\bar pp$ cross sections and event final states up to the Fermilab Tevatron energy of $\sqrt s =1.96$~TeV  and refer the reader to previous publications for further details. We also outline a procedure to be used to implement the predictions of the model into a Monte Carlo simulation that is robust against changes in the collision energy, so that it may be equally well applied to simulate collisions at fixed target energies as well as at the higher energies of the Tevatron, the LHC, and beyond. The model is anchored in a saturation effect observed in single diffraction dissociation that explains quantitatively the factorization breaking observed in soft and hard $pp$ and $\bar pp$ diffractive processes and in diffractive photoproduction and low $Q^2$ deep inelastic scattering.    

\section{Acknowledgments}
I would like to thank my colleagues at The Rockefeller University and my collaborators at the Collider Detector at Fermilab for  providing the interactive environment in which this work was made possible, and the organizers of the Workshop for putting together a comprehensive program of presentations at the threshold in time of the opening of a new era in particle physics at the Large Hadron Collider.  
 


\newpage
\clearpage
\setcounter{affil}{0}
\setcounter{section}{0}
\setcounter{figure}{0}
\setcounter{table}{0}
\setcounter{equation}{0}

\begin{titlepage}
\title{\bf{Problems of phenomenological description of elastic
pp scattering at the LHC; predictions of contemporary models}}
\author{Vojt\v{e}ch Kundr\'{a}t\footnote{kundrat@fzu.cz},
Milo\v{s} Lokaj\'{\i}\v{c}ek\footnote{lokaj@fzu.cz}}
\affiliation{Institute of Physics,
AS~CR, v.v.i., 182 21 Praha 8, Czech Republic}
\author{Jan Ka\v{s}par\footnote{Jan.Kaspar@cern.ch},
Ji\v{r}\'{i} Proch\'{a}zka\footnote{Jiri.Prochazka@cern.ch}
\footnote{both on leave of absence from Institute of Physics,
AS~CR, v.v.i., 182 21 Praha 8, Czech Republic}}
\affiliation{CERN, Geneva, Switzerland}
\begin{abstract}
The standard description of common influence of 
both the Coulomb and hadronic elastic scattering
in the proton - proton elastic collisions at high 
energies with the help of West and Yennie complete
amplitude is shown to be theoretically inconsistent. 
The approach being based on the eikonal model amplitude 
removes these troubles. The preference of its application 
to the analysis of experimental data and in obtaining the 
predictions of contemporary models for proton - proton 
high energy elastic hadronic scattering are discussed.

\addcontentsline{toc}{part}{Problems of phenomenological description of elastic
pp scattering at the LHC; predictions of contemporary models - {\it V.Kundr\'{a}t, M.Lokaj\'{\i}\v{c}ek, J.Ka\v{s}par, J.Proch\'{a}zka}}
\section{Introduction}
\label{sec1}

Differential cross section of elastic scattering of charged nucleons 
at high energies can be defined as
\vspace*{-0.3cm}
\begin{equation}
\vspace*{-0.1cm}
{{ d \sigma_{el} } \over {dt}} =  {{\pi} \over {s p^2}} 
|F^{C+N}(s,t)|^2.
\label{ds1}
\end{equation}
Here $F^{C+N}(s,t)$ represents the complete elastic scattering 
amplitude which has been decomposed according to Bethe \cite{beth} 
into the sum of Coulomb component $F^C(s,t)$ known from QED and 
the elastic hadronic component $F^N(s,t)$ bound mutually by 
a relative phase $\alpha\Psi(s,t)$:
\vspace*{-0.3cm}
\begin{equation}
\vspace*{-0.2cm}
F^{C+N}(s,t) =
e^{i\alpha \Psi(s,t)} F^{C}(s,t)+ F^{N}(s,t);
\label{to1}
\end{equation}
$s$ is the square of the energy in the center-of-momentum system, 
$p$ is the momentum of incident nucleon in this system 
and $\alpha=1/137.036$ is the fine structure constant.

The complete (simplified) elastic scattering amplitude $F^{C+N}(s,t)$ has 
been proposed by West and Yennie \cite{west} (for details see, e.g., \cite{kunx}) as
\vspace*{-0.3cm}
\begin{equation}
\vspace*{-0.2cm}
F^{C+N}(s,t) =
 {\alpha s \over t} f_1(t)f_2(t)e^{i\alpha \Psi(s,t)}+
{\sigma_{tot}(s) \over {4\pi}} p\sqrt {s} 
\left ( \rho(s)+i\right ) e^{B(s)t/2}.
\label{wy1}
\end{equation}
Formula (\ref{wy1}) is valid provided the hadronic elastic 
amplitude (the second term on its right 
\end{abstract}
\maketitle

\end{titlepage}
\noindent 
hand side) has a constant
diffractive slope $B$ together with constant quantity $\rho$
(the ratio of the real to imaginary parts of hadronic amplitude)
in the whole kinematically allowed region 
of $t$. Together with the total cross section $\sigma_{tot}$ they 
may depend only on the energy. The two dipole form factors $f_1(t)$, 
$f_2(t)$ describe the electromagnetic structure of nucleons.

The $t$ dependence of the relative phase $\alpha \Psi(s,t)$ 
in Eq.~(\ref{wy1}) for $pp$ scattering has been estimated by 
West and Yennie \cite{west} to be
\vspace*{-0.3cm}
\begin{equation}
\vspace*{-0.2cm}
 \alpha \Psi(s,t) = - \; \alpha (\ln (-B(s)t/2) + \gamma )
 \label{wy2}
\end{equation}
where $\gamma =0.577215$ is the Euler constant.

In earlier analyses the data sets have been usually divided into 
two regions: {\mbox{$|t| \lesssim 0.01$}} GeV$^2$ and higher $|t|$
\cite{bloc}. It has been assumed that the Coulomb and hadronic 
interactions interfered in the first region while only the hadronic 
interactions described with the help of phenomenological elastic 
hadronic amplitude $F^N(s,t)$ (having usually more complicated
$t$ dependence than the hadronic amplitude in Eq. (\ref{wy1})). Thus
two very diverse formulas for the description of elastic differential
cross section in the two different regions of $t$ have been used.

It might seem to be more suitable to use instead of Eq. (\ref{wy2})
the integral formula \cite{kun12,kun7} 
\vspace*{-0.2cm}
\begin{equation}
\vspace*{-0.3cm}
\Psi_{WY}(s,t)= - \; \ln{-s\over t} + \int_{-4p^2}^{0}{dt'\over |t'-t|}   
\left[1-{F^N(s,t')\over F^N(s,t)}\right],
\label{wy0}
\end{equation}
derived earlier by West and Yennie \cite{west}. 
However, it may be hardly possible if the relative phase
$\alpha \Psi_{WY}(s,t)$ should be real. In such a case the
phase $\zeta^{N}(s,t)$ (defined by 
\mbox{${F^{N}(s,t) = i |F^{N}(s,t)| e^ {-i \zeta^{N} (s,t)} }$}) - or
$\rho(s,t)$ - must be $t$ independent, which is in 
disagreement with experimental data.

Therefore, the decisive preference should be given to a more 
suitable approach based on the eikonal model formulated within 
an impact parameter representation of elastic amplitudes. 
\vspace*{-0.3cm}
\section{Impact parameter representation of scattering amplitudes and mean values of impact parameter}
\label{sec2}

It has been shown in papers of Adachi et al. \cite{adac}  
that the scattering amplitude $F^N(s,t)$
may be related to the eikonal $\delta(s,b)$
by the Fourier-Bessel (FB) transformation
\vspace*{-0.2cm}
\begin{equation}
\vspace*{-0.2cm}
F(s,q^2=-t)= {s\over {4 \pi i}} \int\limits_{\Omega_b}d^2b
e^{i\vec{q}\vec{b}} \left[e^{2i\delta^N(s,b)}-1\right],
\label{eik1} 
\end{equation}
where $\Omega_b$ is the two-dimensional Euclidean space of 
the impact parameter $\vec b$.

If formula (\ref{eik1}) is to be applied at finite energies 
some problems appear as the amplitude $F^N(s,t)$ is defined 
in finite region of $t$ only. However, mathematically consistent
definition of FB transformation requires also the existence of inverse
transformation. Thus the amplitude should be defined also in its 
unphysical region \cite{adac}. The uniqueness of that problem
has been established by Islam \cite{isla,islb} by continuing
analytically the elastic hadronic amplitude $F^N(s,t)$ from
the physical to the unphysical regions.

Then the elastic hadronic amplitude in the impact parameter space
consists of two terms
\vspace*{-0.3cm}
\begin{equation}
\vspace*{-0.2cm}
h_{el}(s,b) \;=\; h_1(s,b)\;+\;h_2(s,b) 
=\int \limits_{t_{min}}^{0} dt \; F^N(s,t)\; J_{0}(b\sqrt{-t}) +
\int \limits_{-\infty}^{t_{min}} dt \; F^N(s,t)\; J_{0}(b\sqrt{-t}); 
\label{imp1}
\end{equation}
a similar relation holds also for the representation of inelastic
overlap function $g_{inel}(s,b)$ \cite{hove} in the impact parameter space.
Then the unitarity equation in the impact parameter space can be 
approximated as
\vspace*{-0.3cm}
\begin{equation}
\vspace*{-0.2cm}
\Im h_{1}(s,b) \;=\; |h_1(s,b)|^2 \;+\; g_1(s,b).
\label{imp2}
\end{equation}
And the total cross section, integrated elastic and inelastic cross 
sections may be obtained also as  
\vspace*{-0.3cm}
\begin{equation}
\vspace*{-0.2cm}
\sigma_{tot}(s) = 
8 \pi \int \limits_{0}^{\infty} b db \; \Im h_{1}(s, b),\;\;\;
\sigma_{el}(s) = 
8 \pi \int \limits_{0}^{\infty} b db \; |h_{1}(s,b)|^2,\;\;\;
\sigma_{inel}(s) = 
8 \pi \int \limits_{0}^{\infty} b db \; g_{1}(s,b).
\label{imp3}
\end{equation}

The functions $\Im h_{1}(s,b)$ and $|h_{1}(s,b)|^2$ represent then two main 
impact parameter profiles (for total and elastic processes) and describe the 
intensity of interactions between two colliding particles in the dependence on 
their mutual impact parameter. 

Expressing this amplitude with the help of modulus  
$|F^{N}(s,t)|$ and of phase $\zeta^{N}(s,t)$ it is 
possible to write for the mean-square value of elastic 
impact parameter - see Ref. \cite{kunx}
\begin{eqnarray}
\!\!\!\!\!\!\!\!\!\!\!{\langle } b^2(s){\rangle }_{el}  \; &=& \;
4 \; {{\int \limits_{t_{min}}^{0}\!\! dt \;|t| \;
\Bigl({d \over{dt}} |F^{N}(s,t)|\Bigr)^2}
\over { \int \limits_{t_{min}}^{0}\!\! dt \; |F^{N}(s,t)|^2}} +
4 \; {{\int \limits_{t_{min}}^{0}\!\! dt \;|t| \;
|F^{N}(s,t)|^2 \Bigl( {d \over {dt}} \zeta^{N}(s,t)\Bigr)^2} 
\over { \int \limits_{t_{min}}^{0} \!\!dt \; |F^{N}(s,t)|^2}} 
\nonumber  \\
&\equiv &  \;
 {\langle } b^2(s){\rangle }_{mod} + {\langle } b^2(s){\rangle }_{ph}.  
 \label{mv2}  
\end{eqnarray}
Introducing further the diffractive slope as 
\begin{equation}
B(s,t)= {d\over {dt}}\left[\ln {d \sigma^{N}\over {dt}}\right] =
{2\over |F^{N}(s,t)|}{d\over {dt}}|F^{N}(s,t)|,
\label{sl1}
\end{equation}
one can derive under the validity of optical theorem for the 
total mean-square value \cite{kunx} 
\begin{equation}
{\langle } b^2(s){\rangle }_{tot}\;\; = 
{ { {8 \pi \int \limits_{0}}^{\infty} b \;db \; b^2 \Im h_{1}(s,b)}
\over {\sigma_{tot}(s)}} = 2 B(s,0).  
\label{mv3}
\end{equation}
It is then also possible to write for the inelastic mean-square 
value \cite{kunx}
\begin{equation}
{\langle } b^2(s){\rangle }_{tot}\;\; = \; 
{{\sigma_{el}(s)} \over {\sigma_{tot}(s)}} {\langle } b^2(s){\rangle }_{el}
 \; + \; {{\sigma_{inel}(s)} \over {\sigma_{tot}(s)}} 
 {\langle } b^2(s){\rangle }_{inel}.
\label{mv4}
\end{equation}
\section{Eikonal model and influence of Coulomb scattering}
\label{sec3}
The complete elastic eikonal equals the sum of 
the Coulomb and the hadronic eikonals  and the 
Eq.~(\ref{eik1}) can be rewritten as \cite{fran1}
\vspace*{-0.3cm}
\begin{equation}
\vspace*{-0.2cm}
F^{C+N}(s,t) = F^C(s,t) + F^N(s,t) + {i\over {\pi s}}
\int\limits_{\Omega_{q'}} d^2q'
F^{C}(s,{q'^2})F^{N}(s,[\vec{q} - \vec{q'}]^2). 
\label{eik3}
\end{equation}
Eq.~(\ref{eik3}) containing the convolution integral
between Coulomb and hadronic amplitudes differs
significantly from Bethe's Eq.~(\ref{to1}). In its final 
form (valid at any $s$ and $t$) it equals \cite{kun8}
\vspace*{-0.3cm}
\begin{equation}
\vspace*{-0.2cm}
F^{C+N}(s,t) = + \;{\alpha s\over t}f_1(t)f_2(t) +
F^{N}(s,t)\left [1 \; - \; i\alpha G(s,t) \right ],  
\label{kl1}
\end{equation}
where
\vspace*{-0.3cm}
\begin{equation}
\vspace*{-0.2cm}
G(s,t) = \int\limits_{-4p^2}^0
dt'\left\{ \ln \left( {t'\over t} \right )
{d \over{dt'}}
\left[f_1(t')f_2(t')\right]
+ {1\over {2\pi}}\left [{F^{N}(s,t')\over F^{N}(s,t)}-1\right]
I(t,t')\right\},
\label{kl2}
\end{equation}
and
\vspace*{-0.3cm}
\begin{equation}
\vspace*{-0.2cm}
I(t,t')=\int\limits_0^{2\pi}d{\Phi^{\prime \prime}}
{f_1(t^{\prime \prime})f_2(t^{\prime \prime})\over t^{\prime \prime}}, \;\;
t^{\prime \prime}=t+t'+2\sqrt{tt'}\cos{\Phi}^{\prime \prime}.
\label{kl3}
\end{equation}
Instead of the $t$ independent quantities
$B$ and $\rho$, it is now necessary to consider
corresponding $t$ dependent quantities $B(s,t)$ 
defined by Eq.~(\ref{sl1}), and $\rho(s,t)$ and 
$\sigma_{tot}(s)$ as
\begin{equation}
\rho (s,t) = {{\Re F^{N}(s,t)} \over {\Im F^{N}(s,t)}},
\hspace*{1cm}
\sigma_{tot} (s) = {{4 \pi}\over {p \sqrt{s}}} \Im F^{N}(s,t=0).
\label{ro1}
\end{equation}
The form factors $f_1(t)$ and $f_2(t)$ reflect the 
electromagnetic structure of colliding nucleons; owing 
to the integration over all kinematically allowed region
of $t$ in Eq.~(\ref{kl3}) their actual $t$ parameterization 
has been taken from Ref. \cite{bork}:  
\begin{equation}
f_{j}(t) = \sum_{k=1}^4 {g_k\over{w_k-t}},
\hspace*{0.5cm}j=1,2
\label{bo1}
\end{equation}
where the values of the parameters $g_k$ and $w_k$ 
are to be taken from the quoted paper; for futher
details see Ref. \cite{kun8}).

When the Coulomb part in formula (\ref{kl1}) is taken as 
known the complete amplitude depends in principle on hadronic 
amplitude $F^N(s,t)$ only. Thus Eq.~(\ref{kl1}) may be
used in two complementary ways: either one may test the 
predictions of different models of high-energy elastic 
hadronic scattering given be corresponding hadronic 
amplitudes $F^{N}(s,t)$, or one may resolve phenomenological 
$t$ dependence of elastic hadronic amplitude $F^{N}(s,t)$ 
at a given $s$ (and for {\it{all measured  t  values}}), by 
fitting experimental elastic differential cross section data
with the help of Eq.~(\ref{ds1}) and Eqs.~(\ref{kl1})-(\ref{kl3}). 

\section{Eikonal model and assumptions of West and Yennie}
\label{sec4}
The given eikonal model approach has been originally applied to the $pp$
elastic scattering at energy of 53 GeV in Ref. \cite{kun8} (data taken from
Ref. \cite{byst}). The modulus and the phase of the elastic hadronic 
amplitude $F^N(s,t)$ have been parameterized as
\begin{equation}
|F^{N}(s,t)| = (a_1+a_2t)e^{b_1t+b_2t^2+b_3t^3} +
 (c_1+c_2t)e^{d_1t+d_2t^2+d_3t^3}\;, \;\;
 \zeta^{N} (s,t) = \arctan {\rho_0  \over {1 - { | {{t/t_{di\!f\!f}} |}} }}.
\label{wy10}
\end{equation}
The complete elastic scattering amplitude has been then established with
the help of Eqs.~(\ref{kl1})-(\ref{kl3}). 

In the following we shall demonstrate the fits \cite{proc} when some quantities 
have been limited according the crucial conditions imposed in the approach of West
and Yennie. The results are shown in Fig. 1 where the fits of differential
cross section under different assumptions are given. The value of corresponding
$\chi^2$ distribution calculated in the whole measured interval of $t$ has been 
always minimized. First, the contribution of the mere Coulomb interaction 
corresponds to the dashed and dotted line. Second, the contribution of the pure 
hadronic interaction and corresponding to $\rho = const.$ and $B=const.$ 
is represented by dotted line. When the Coulomb scattering is added 
to this case the graph with a dashed line is obtained. And finally,
when the interference between the Coulomb and hadronic scattering with
$\rho = const.$ and $B(t)$ (parameterized according to Eqs.~(\ref{wy10}))
is demonstrated by the full line roughly copying the data. Compared 
with the case from Ref. \cite{kun8} where both the parameters $\rho(t)$ 
and $B(t)$ have been fitted, it gives only a little bit higher value of $\chi^2$. 
It may be mainly the constancy of $B(s,t)$ that must be denoted as fully
unacceptable; even if the constancy of $\rho$ must be denoted as unacceptable, too. 

\section{Model predictions for $pp$ elastic scattering at the LHC}
\label{sec5}

The eikonal approach has been made use of in deriving some predictions
for the region of LHC energy. We have made use of the results of four 
models, proposed by Bourrely, Soffer and Wu \cite{bou2}, 
Petrov, Predazzi and Prokhudin \cite{petr2}, 
Block, Gregores, Halzen and Pancheri \cite{blo2}
and Islam, Luddy and Prokhudin \cite{isla5}, in which
the elastic hadronic amplitudes at individual lower
energies have been mutually correlated. The eikonal approach has been 
applied to predict the complete elastic $pp$ amplitude at 14 TeV,
as the West and Yennie approach must be regarded as irregular 
(see Eqs.~(\ref{wy1}) and (\ref{wy2})). Some of these results 
have been included already in TOTEM collective papers \cite{tot1,tot2}
and in \cite{kasp}.

\begin{figure}[ht!]
\centerline{
\begin{tabular}{cc}
\includegraphics[width=8.5cm,keepaspectratio]{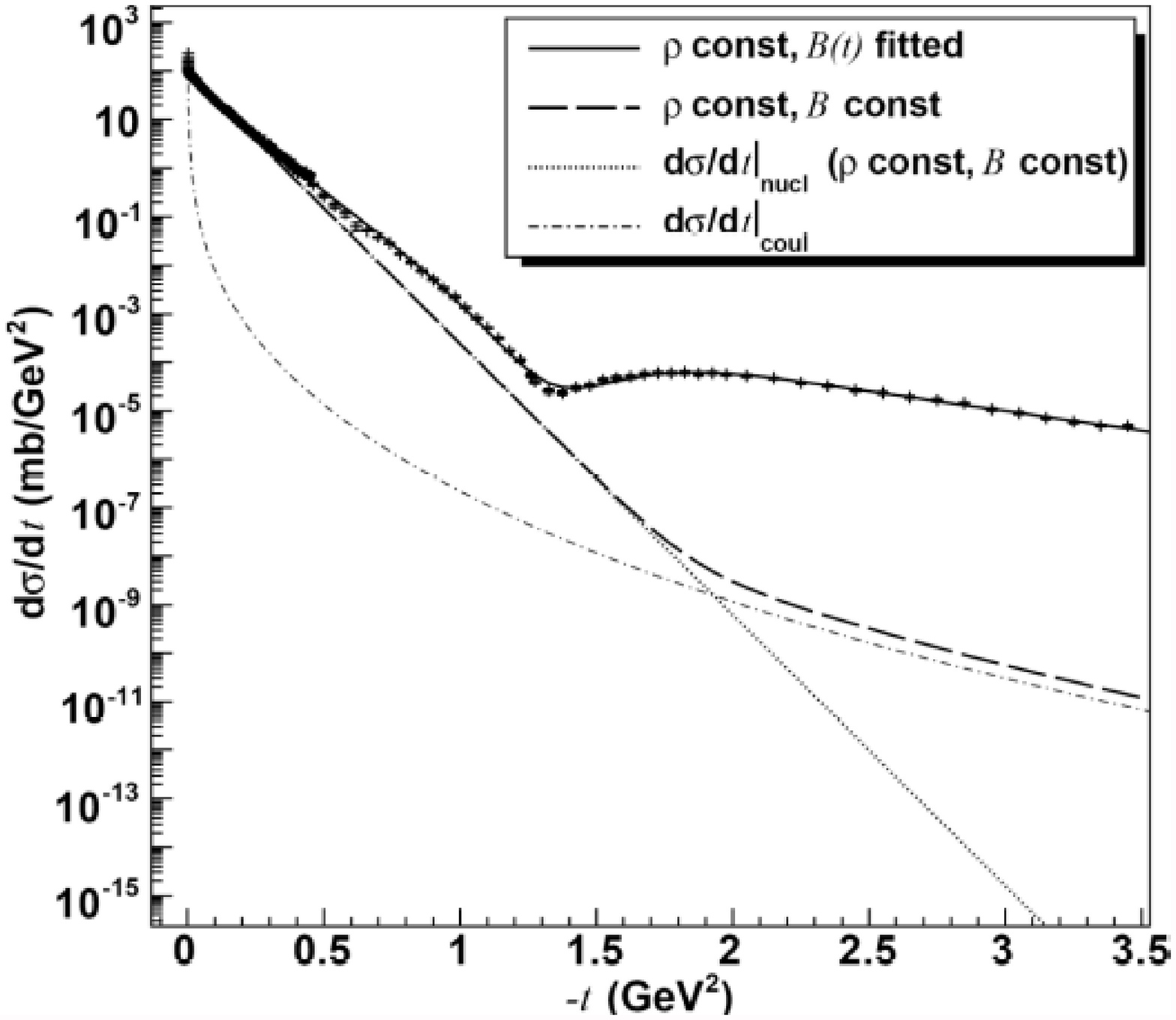} &
\includegraphics{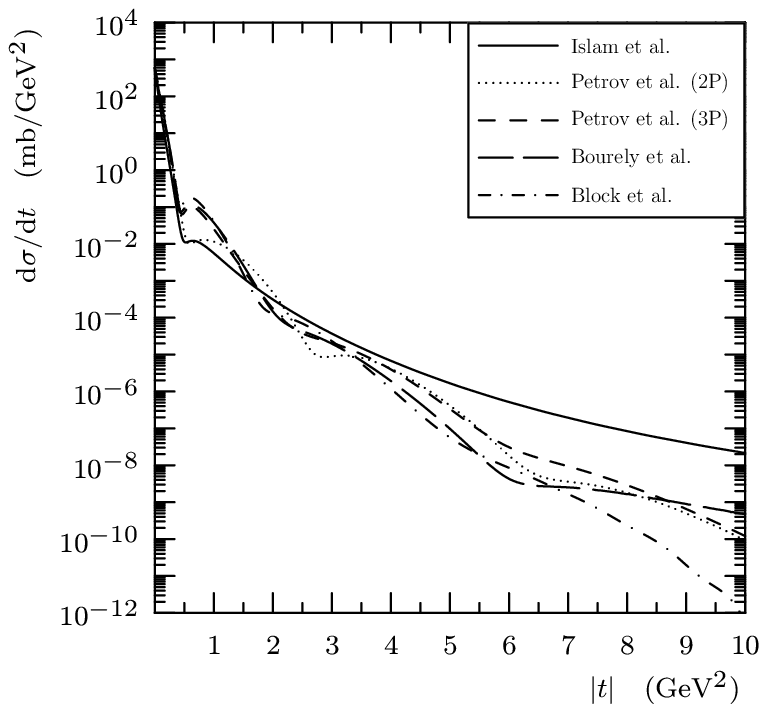}\\[-0.7cm]
\begin{minipage}[t]{0.45\textwidth}
\caption{\label{fig:vkun_rhoBconst}
Different contributions to ${{d \sigma} \over {dt}}$ for $pp$ scattering at 53 GeV}
\end{minipage} & \begin{minipage}[t]{0.45\textwidth}
\caption{\label{fig:vkun_sigma_large}${{d \sigma} \over {dt}}$ 
predictions for $pp$ scattering at 14 TeV according to different models.} 
\end{minipage}\\
&\\[-0.5cm]
\end{tabular}}
\end{figure}
\vspace*{-1.0cm}
\begin{figure}[ht!]
\centerline{
\begin{tabular}{cc}
\includegraphics{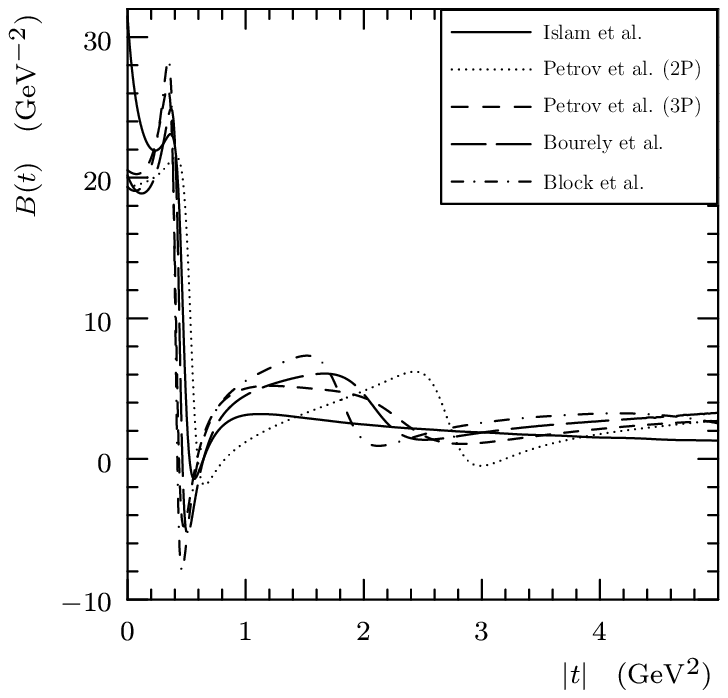} & 
\includegraphics{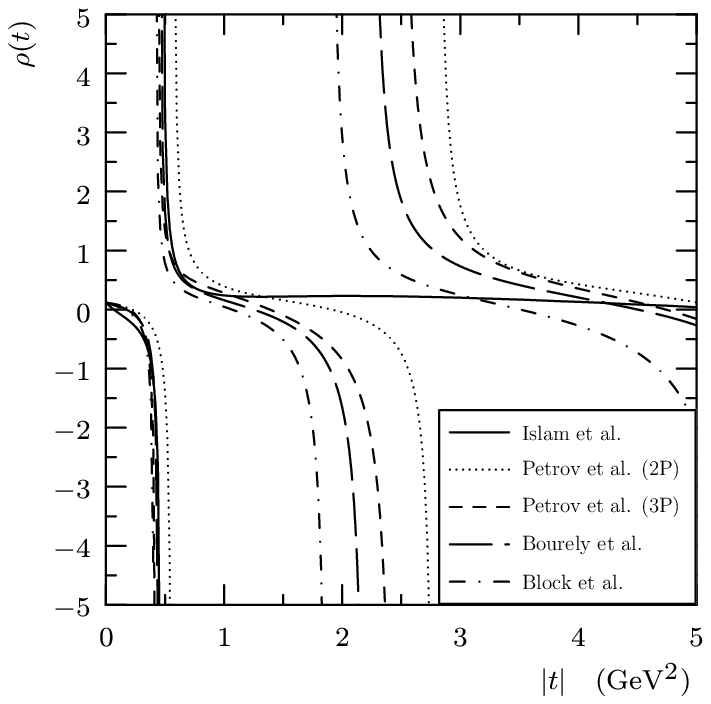}\\[-0.5cm]
\begin{minipage}[t]{0.45\textwidth}
\caption{\label{fig:vkun_slope}
The diffractive slope predictions for $pp$ scattering at 14 TeV
according to different models.} 
\end{minipage} & \begin{minipage}[t]{0.45\textwidth}
\caption{\label{fig:vkun_rho}
The $\rho (t)$ predictions for $pp$ scattering at 14 TeV
according to different models.} 
\end{minipage}\\
&\\[-0.5cm]
\end{tabular}}
\end{figure}
\vspace*{1cm}

It has been possible to determine total cross section $\sigma_{tot}(s)$,  
diffractive slope $B(s,t)$ and quantity $\rho(s,t)$ with the help of 
formulas (\ref{sl1}) and (\ref{ro1}) for each model. The integrated 
elastic hadronic cross sections have been determined by integration 
of modified Eq.~(\ref{ds1}) containing only $F^{N}(s,t)$. The values 
of $\sigma_{tot}$, $\sigma_{el}$, $B(0)$ and $\rho(0)$ are given in 
Table I. It is evident that the predictions of divers models differ 
rather significantly; the total cross section predictions range from 
95 mb to 110 mb. A higher value of $\sigma_{tot}$ has been
\begin{table}[htb]
\begin{center}
\begin{tabular}{ccccc}  
\hline  
   model $\;\;\;\;\;\;\;\;$ & $\;\;\;\;\sigma_{tot}\;\;\;\;$ & $\;\;\;\;\sigma_{el}\;\;\;\;$ & $\;\;\;\;B(0)\;\;\;\;$ & $\;\;\;\;\rho(0)$ \\ 
   & [mb] & [mb] & [GeV$^{-2}$] & \\
\hline \hline
Bourrely et al.      & 103.64 & 28.51 & 20.19 & 0.121 \\

Petrov et al. (2P)   &  94.97 & 23.94 & 19.34 & 0.097 \\

Petrov et al. (3P)   & 108.22 & 29.70 & 20.53 & 0.111 \\

Block et al.         & 106.74 & 30.66 & 19.35 & 0.114 \\

   Islam  et al.     & 109.17 & 21.99 & 31.43 & 0.123 \\

\hline \\ 
\end{tabular} 
\caption{\label{tab:models1} The values of basic parameters predicted by
different models for $pp$ elastic scattering at energy of 14 TeV} 
\end{center}
\end{table}
\vspace*{-1.2cm}
\vspace*{1.0cm}
$\;$ predicted by COMPETE collaboration \cite{cude}:
$\sigma_{tot} =  111.5\; \pm \; 1.2 ^ {\;+4.1} _{\;-2.1}$ 
mb having been determined by extrapolation of the fitted 
lower energy data with the help of dispersion relations 
technique. 
The predictions of ${d {\sigma} \over {dt}}$ values 
for higher values of $|t|$ are shown in Fig. 
\ref{fig:vkun_sigma_large}; they differ significantly 
for different models. The predictions for the $t$ dependence 
of the diffractive slopes $B(t)$ are shown in Fig. 
\ref{fig:vkun_slope}. Fig. \ref{fig:vkun_rho} 
displays the $t$ dependence of the quantity $\rho (t)$ 
that is not constant, either. 

\begin{figure}[ht!]
\centerline{
\begin{tabular}{cc}
\includegraphics{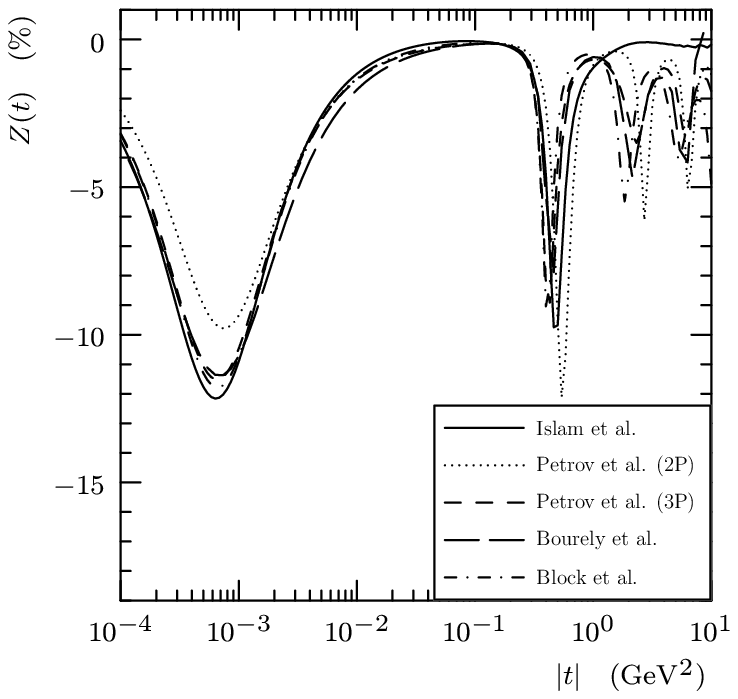} & 
\includegraphics{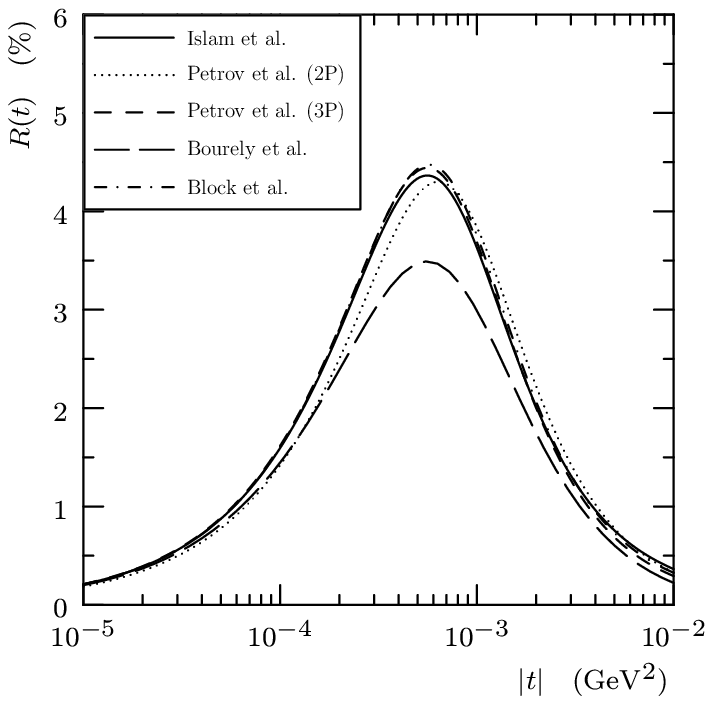}
\\[-0.7cm]
\begin{minipage}[t]{0.45\textwidth}
\caption{\label{fig:vkun_importance} The $t$ dependence of the ratio
of the interference to the hadronic contributions to the 
${{d \sigma}\over {dt}}$ for $pp$ elastic scattering at 14 TeV
according to different models.} 
\end{minipage} & \begin{minipage}[t]{0.45\textwidth}
\caption{\label{fig:vkun_R} The $R(t)$ quantity 
predictions for $pp$ scattering at 14 TeV for
different models from Sec. \ref{sec6}.} 
\end{minipage}\\
&\\[-0.5cm]
\end{tabular}}
\end{figure}

Fig. \ref{fig:vkun_importance} shows the
$t$ dependence of the ratio of interference to
hadronic contributions of the ${{d \sigma}\over {dt}}$
for all of the given models, i.e., of the quantity
\begin{equation}
Z(t) \; = \; {{|F^{C+N}(s,t)|^2 - |F^{C}(s,t)|^2 - |F^{N}(s,t)|^2 }
\over {|F^{N}(s,t)|^2}}.
\end{equation}
The graphs show clearly that the influence of the 
Coulomb scattering may not be fully neglected 
at higher values of $|t|$, either. It is interesting 
that at least for small $|t|$ the given characteristics 
are very similar for all four considered models.
The impact parameter representation of elastic hadronic
amplitude $F^N(s,t)$ allows then to establish different
root-mean-square (RMS) values of impact parameters 
that represent in principle the range of hadronic
interactions. Their values calculated with the help 
of formulas (\ref{mv2})-(\ref{mv4}) for each of 
the analyzed models are shown in Table II. The 
values of elastic RMS are in all cases lower than 
the corresponding values of the inelastic ones. 
It means that the elastic $pp$ collisions would 
be much more central then the inelastic ones 
similarly as in the case of $pp$ scattering at the ISR 
energies \cite{miet} for all these models which has 
been denoted as a 'puzzle', see Ref. \cite{giac}. 
\begin{table}[htb]
\begin{center}
\begin{tabular}{cccc}  
\hline 
     & & &     \\
 model & $\sqrt{<b^2_{tot}>}\;\;\;\;$ & $\;\;\;\;\sqrt{<b^2_{el}>}\;\;\;\;$ & $\;\;\;\;\sqrt{<b^2_{inel}>}$ \\ 
       & [fm] & [fm] & [fm] \\      
\hline \hline
Bourrely et al.     & 1.249 & 0.876 & 1.399 \\

Petrov et al. (2P)  & 1.227 & 0.875 & 1.324 \\

Petrov et al. (3P)  & 1.263 & 0.901 & 1.375 \\

Block et al.        & 1.223 & 0.883 & 1.336 \\

   Islam et al.     & 1.552 & 1.048 & 1.659 \\  
\hline \\  
\end{tabular}
\caption{\label{tab:models2} The values of root-mean-squares predicted by
different models}
\end{center}
\end{table} 
\section{Luminosity estimation and elastic scattering at LHC}
\label{sec6}
Accurate determination of elastic amplitude is very
important in the case when the luminosity of the collider 
is to be calibrated on the basis of elastic nucleon 
scattering. The luminosity $\mathcal{L}$ relates the
experimental elastic differential counting rate
${{d N_{el}}\over {dt}}(s,t)$ to the complete elastic
amplitude $F^{C+N}(s,t)$ (see Eq.~(\ref{ds1}) and Ref.
\cite{bloc}) by 
\begin{equation}
{1 \over \mathcal{L}} {{d N_{el}}\over {dt}}(s,t)  = {\pi \over {s p^2}}
 |F^{C+N} (s,t)|^2.
\label{lu1}
\end{equation}
Eq.~(\ref{lu1}) is to be valid for any admissible value of $t$. 
The value $\mathcal{L}$ might be in principle calibrated by
measuring the counting rate in the region of the smallest
$|t|$ where the Coulomb amplitude is dominant. However, 
this region may hardly be reached at the nominal LHC
energy due to technical limitations. A procedure allowing 
to avoid these difficulties may be based on Eq.~(\ref{lu1}), 
when the elastic counting rate may be, in principle, measured 
at any $t$ which can be reached, and the complete elastic
scattering amplitude $F^{C+N}(s,t)$ may be determined 
with required accuracy at any $|t|$. However,
in this case it will be very important which formula 
for the complete elastic amplitude $F^{C+N}(s,t)$ will 
be used. 
We have studied the differences between the West and Yennie 
simplified formula (see Eqs.~(\ref{wy1}) and (\ref{wy2}))
and the eikonal model (Eqs.~(\ref{kl1})-(\ref{kl3})).
The differences may be well visualized by the quantity
\begin{equation}
R(t) =  {{|F^{C+N}_{eik}(s,t)|^2 -
|F^{C+N}_{WY}(s,t)|^2}
\over {|F^{C+N}_{eik}(s,t)|^2}}, 
\label{lu2}
\end{equation}
where $F^{C+N}_{eik}(s,t)$ is the complete elastic 
eikonal model amplitude, and $F^{C+N}_{WY}(s,t)$ 
is the  West and Yennie one. The quantity $R(t)$ is plotted 
in Fig. (\ref{fig:vkun_R}) for considered models. The maximum 
deviations lie approximately in the center of interference 
region where the differences between the physically consistent 
eikonal model and the West and Yennie formula may reach 
almost systematic error of 5 $\%$ \cite{kun11,kun13}. 

\newpage
\clearpage
\setcounter{affil}{0}
\setcounter{section}{0}
\setcounter{figure}{0}
\setcounter{table}{0}
\setcounter{equation}{0}


\title{Total pp Cross Section at TOTEM}

\author{Ji\v{r}\'{\i} Proch\'{a}zka on behalf of the TOTEM Collaboration}
\affiliation{%
 CERN, Geneva, Switzerland
}%



\begin{abstract}
One of the main aims of the TOTEM experiment at CERN is to measure the pp total
cross section at LHC energies, for which current predictions based on diverse
approaches and models are quite different. Typical estimations give values in
the range 90-130~mb at a center-of-mass energy of 14 TeV.  Only a precise
measurement may give the distinction between individual models and TOTEM should
provide the value with a  precision of 1-2\%.  To achieve this goal TOTEM plans
to use the so-called ``luminosity independent method''.  An overview of this
method will be given.  It will be clearly distinguished between experimental
input, i.e.~what is necessary to measure, and model dependent calculations
which are also needed. 
\end{abstract}

\maketitle
\addcontentsline{toc}{part}{Total pp Cross Section at TOTEM - {\it  J.Proch\'{a}zka}}

\section{\label{sec:introduction} Luminosity independent method}

The TOTEM experiment at CERN \cite{1748-0221-3-08-S08007,LoI,TP,TDR} (TOTal Elastic and
diffractive cross section Measurement) plans to employ the so-called luminosity
independent method to measure the pp total cross section $\sigma_{\mathrm{tot}}$.
This method is based on the optical theorem:
\begin{equation}
  \sigma_{\mathrm{tot}} = \frac{4\pi}{p\sqrt{s}}\Im F^\mathrm{N}_{\mathrm{el}}(s,t=0)
\end{equation}

\noindent
where $F^\mathrm{N}_{\mathrm{el}}(s,t)$ is the elastic hadronic scattering
amplitude; $p$ is the momentum of the incident proton, $t$ is four-momentum
transfer squared and $s$ is the square of the center-of-mass energy.  Adding
the relation for the elastic hadronic differential cross section 
\begin{equation}
 \frac{\text{d}\sigma^\mathrm{N}_{\mathrm{el}}}{\text{d}t} = \frac{\pi}{sp^2} \left|F^\mathrm{N}_{\mathrm{el}}(t)\right|^2
\end{equation}

\noindent
and the relation between luminosity and total rate $\rate{}{tot} = \rate{N}{el}+\rate{}{inel}$
\begin{equation}
 L \sigma_{\mathrm{tot}} = \rate{N}{el}+\rate{}{inel}
\end{equation}

\noindent
where \rate{N}{el} is the elastic hadronic rate and $\rate{}{inel}$ is the inelastic
rate, we may derive 
an expression for the total cross section which is independent from the luminosity 
\begin{equation}
   \sigma_{\mathrm{tot}} = \frac{16 \pi}{1+\rho^2} \;\;
   \frac{\left.\difrate{N}{el}\right|_{t=0}}{\rate{N}{el}+\rate{}{inel}}\text{; \qquad} 
\label{eq:sigtot}
\end{equation}

\noindent
where the $\rho$ parameter is defined as 
\begin{equation}
\rho = \frac{\Re F^\mathrm{N}_{\mathrm{el}}(s,t=0)}{\Im F^\mathrm{N}_{\mathrm{el}}(s,t=0)}.
\end{equation} 

\noindent
  One may similarly derive a formula for the luminosity  which does not depend on $\sigma_{\mathrm{tot}}$: 
\begin{equation}
   L = \frac{1+\rho^2}{16 \pi}\;\;\frac{(\rate{N}{el}+\rate{}{inel})^2}{\left.\difrate{N}{el}\right|_{t=0}}. 
\label{eq:luminosity}
\end{equation}


\indent
The inelastic rate \rate{}{inel} is a quantity that can be measured ``directly'', 
while the elastic hadronic rate \rate{N}{el} (\difrate{N}{el}) can not
be measured directly. The reason is that one measures the complete elastic rate given in
general not only by the hadronic interaction but also by the Coulomb interaction,
i.e.~we measure \rate{C+N}{el} and not \rate{N}{el}. It is, therefore, necessary
to separate the hadronic scattering from the Coulomb scattering. This can be
done by employing a formula for the complete scattering amplitude \ampl{C+N} which
takes into account the interference between the Coulomb and the hadronic interaction.
One possibility is to use the simplified West and Yennie formula \cite{PhysRev.172.1413}
\begin{equation}
  \ampl[WY]{C+N}= \ampl{C} e^{i\alpha\phi} + \frac{\sigma_{tot}}{4\pi} p \sqrt{s}(\rho+i)e^{Bt/2}.
\label{eq:FCNWY}
\end{equation}
where $\phi(s,t) = \mp \left [\ln{\left(\frac{-Bt}{2}\right)}+\gamma \right]$ is the relative phase ($\gamma \approx 0.5772$ is
Euler's constant).  The upper (lower) sign in the relative phase $\phi$ in
\refp{eq:FCNWY} corresponds to the scattering of particles  with the same
(opposite) charges. This WY formula has been derived under the assumption that both the
quantities $\rho$ and $B$ are $t$-independent in the whole kinematically
allowed region of $t$. However, the assumption of a constant diffractive slope
contradicts experimental data of differential cross section at higher $\abs{t}$ where
diffractive structure has been observed.  For other limitations and deficiencies of
this formula see Ref.~\cite{Kaspar:2009nf}.

Another possibility to make the separation is to use the more general eikonal formula 
  \begin{equation}
  \ampl[eik]{C+N} = \ampl{C}+\ampl{N}[1 \mp i \alpha G(s,t)]
  \label{eq:FCNeik}
   \end{equation}

\noindent
where
\begin{align} 
  G(s,t) &= \int\limits^{0}_{t_{\mathrm{min}}} \text{d}t' \left \{\ln
  \left (\frac{t'}{t} \right ) \frac{\text{d}}{\text{d}t'}[f_1(t')f_2(t')] -
  \frac{1}{2\pi}\left[ \frac{F^\mathrm{N}(s,t')}{F^\mathrm{N}(s,t)}-1\right ]I(t,t') \right \}\\
  I(t,t') &= \int\limits^{0}_{2\pi} \text{d} \Phi
  ''\frac{f_1(t'')f_2(t'')}{t''},
\end{align}

\noindent
$t'' = t + t' + 2\sqrt{tt'}\cos{ \Phi ''}$ and $t_{\mathrm{min}}=-s+4m^2$, see
\cite{KLunpolarized1994}. The upper (lower) sign in \refp{eq:FCNeik}
corresponds again to the scattering of particles  with the same (opposite)
charges.  The important property of the eikonal formula is that it properly
combines the Coulomb amplitude \ampl{C} for all values of $t$ with an arbitrary
hadronic amplitude \ampl{N}. While the Coulomb amplitude is assumed to be well
known and is expressed generally as $\pm\frac{\alpha s }{t}f_1(t)f_2(t)$ where
$\alpha$ is the fine structure constant and $f_{1,2}(t)$ are proton
electromagnetic form factors, the hadronic amplitude is unknown; there are only
several phenomenological models of \ampl{N}. For example, in the simplified WY
formula \refp{eq:FCNWY} it is assumed that the hadronic amplitude has purely
exponential modulus and constant phase in the whole kinematically allowed
region of $t$.

To determine the elastic hadronic rate $\left.\difrate{N}{el}\right|_{t=0}$ one
has to measure first the differential rate \difrate{C+N}{el} down to small
values of $\abs{t}$, make the separation of the Coulomb and the hadronic
scattering to obtain just the hadronic rate \difrate{N}{el} and then
extrapolate this quantity to optical point $t=0$. To do the separation and
extrapolation one needs to know \ampl{N}, i.e.~a model of elastic hadronic
scattering needs to be employed. The luminosity independent method is,
therefore, a model dependent method. 

The last quantity which we need to know to calculate $\sigma_{\mathrm{tot}}$
given by \eref{eq:sigtot} is the $\rho$ parameter. It may be taken as predicted
by the COMPETE collaboration \cite{PhysRevLett.89.201801} ($\rho = 0.132$ at
$\sqrt{s} = 14~\text{ TeV}$) or it may be calculated on the basis of a model of
\ampl{N} which is needed also for the extrapolation of elastic hadronic rate
(more consistent approach). The values of the $\rho$ also at $\sqrt{s} =
14~\text{ TeV}$ taken from \cite{Kaspar:2009nf} as predicted by different
models
\cite{Islam2005115,Petrov2005182,BSW1991,BSW1992,Bourrely1993195,Bourrely2003,PhysRevD.60.054024}
are: Islam et al.~0.123, Petrov et al.~(2P)~0.0968, Petrov et al.~(3P)~0.111,
Bourrely et al.~0.121 and Block et al.~0.114.  If we take, e.g., the value of
$\rho$ predicted by the COMPETE collaboration ($\rho = 0.132$), the error
contribution from ($1 + \rho^2$) to $\sigma_{\mathrm{tot}}$ at $\sqrt{s} =
14~\text{ TeV}$ assuming the full COMPETE error band $\frac{\delta\rho}{\rho}=
0.33$ is $\pm1.2\%$.

\ccc{
\begin{table}
  \justifying
\begin{tabular}{c c}

\begin{minipage}[h!]{0.4\textwidth}%
\vspace{1.2cm}
\hspace{1cm}
\begin{tabular}{p{3cm} c}
\hline \hline
Model & $\rho$  \\ 
\hline
Islam et al.       & 0.123  \\
Petrov et al.~(2P) & 0.0968 \\
Petrov et al.~(3P) & 0.111  \\
Bourrely et al.    & 0.121  \\
Block et al.       & 0.114  \\
\hline
COMPETE            & 0.132 \\
\hline \hline
\end{tabular}
  \justifying
\justifying

\captionof{table}{The values of $\rho$ predicted by different models
and the COMPETE collaboration for $pp$ elastic scattering at $\sqrt{s} = 14~\text{
TeV}$.}
\label{tab:rhoValues}
\end{minipage}
&  
\begin{minipage}[h!]{0.50\linewidth}
\vspace{-0.6cm}
\includegraphics[width=0.95\textwidth]{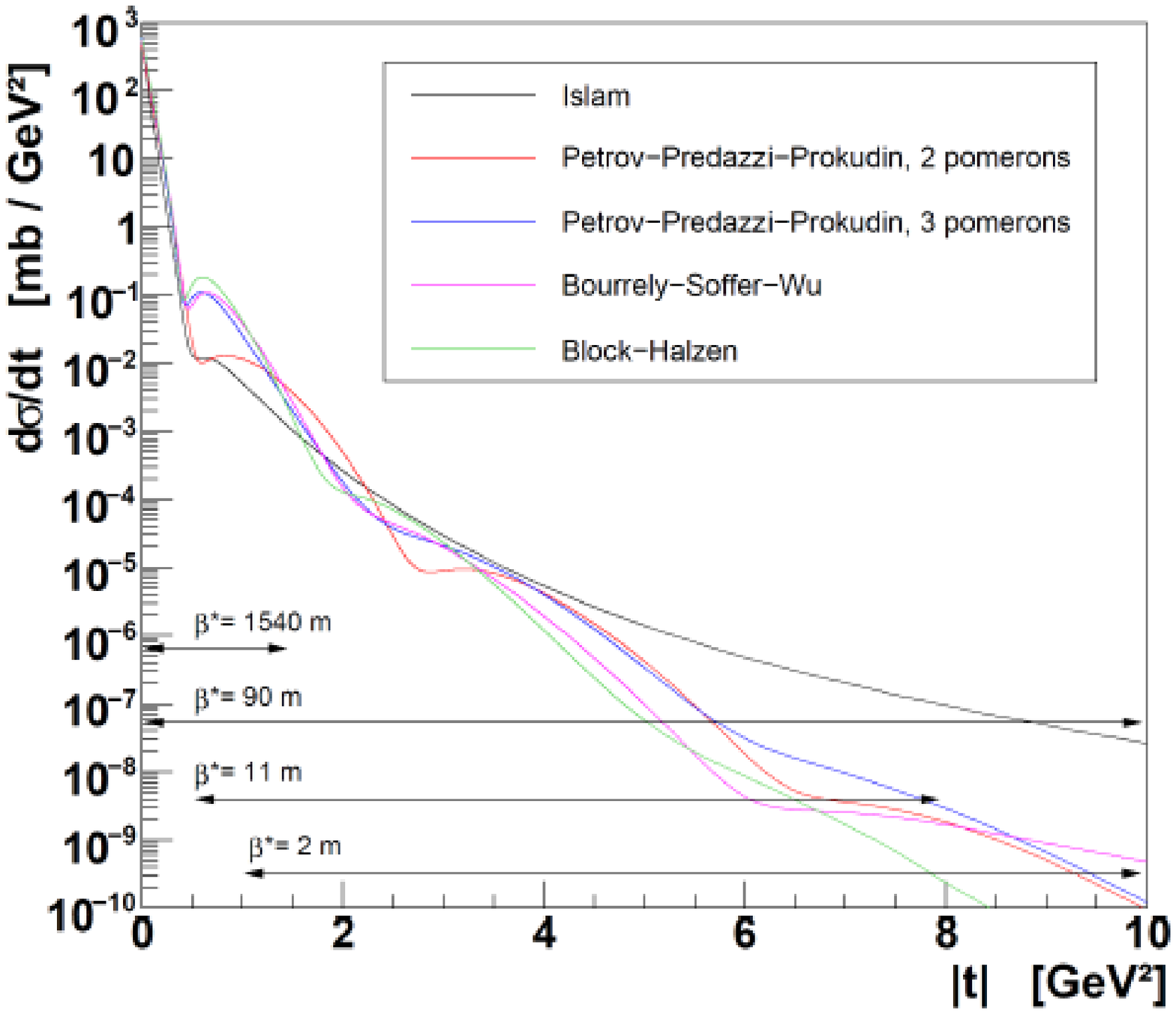}\\
  \justifying
\vspace{-0.3cm}
\captionof{figure}{Elastic differential cross section at $\sqrt{s} = 14~\text{
TeV}$ as predicted by various models together with the $t$-acceptance ranges for
different optics settings.}
\label{fig:elmodels}
\end{minipage}
\end{tabular}
\end{table}
}

\begin{figure}
\includegraphics[width=0.45\textwidth]{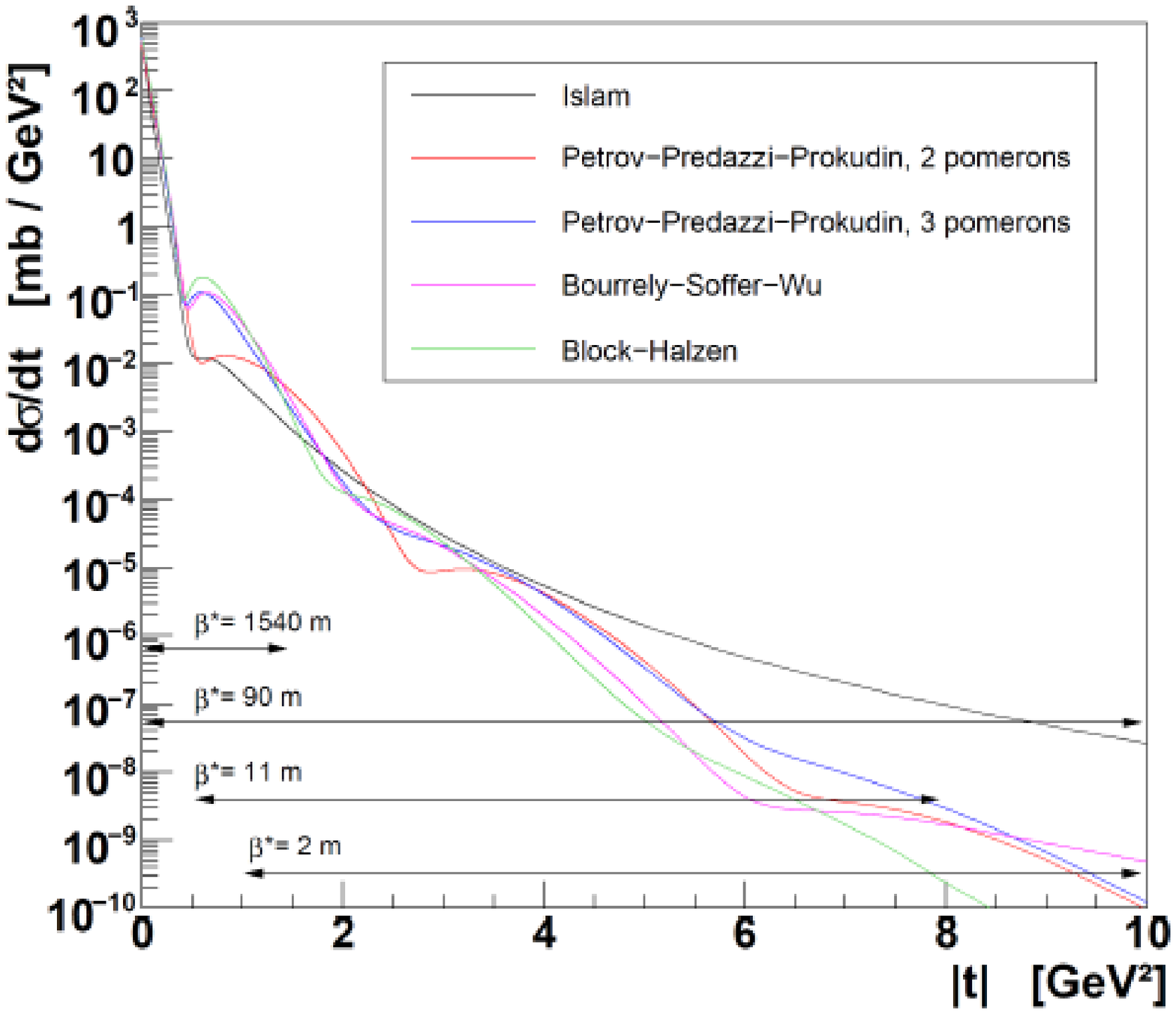}
\caption{\label{fig:elmodels}Elastic differential cross section at $\sqrt{s} = 14~\text{
TeV}$ as predicted by various models together with the $t$-acceptance ranges for
different optics settings.}
\end{figure}

Hence the quantities to be measured by TOTEM are the complete elastic rate
\difrate{C+N}{el} and the inelastic rate \rate{}{inel}. Their measurement method will
be the subject of the next section. For a study of contemporary models of elastic
nucleon scattering and their predictions for the LHC see \cite{Kaspar:2009nf}.

\FloatBarrier
\section{\label{sec:experiment} Experimental method}

\subsection{Experimental set-up}
The TOTEM apparatus is placed symmetrically on both sides of the interaction point
five (IP5) of the LHC. Schematic drawings of the ``right'' arm of the TOTEM detectors are shown in
Figs.~\ref{fig:totemApparatus} and \ref{fig:romanpots}.

To detect elastically scattered protons in the very forward direction TOTEM uses a
system of Roman Pots (RPs) -- movable beam-pipe insertions
which are equipped with edgeless silicon strip detectors designed by TOTEM with the
specific objective of reducing the insensitive area at the edge facing the beam
to only a few tens of microns ($\approx 50 \mu\text{m}$). High efficiency up to the
physical detector border is an essential feature in view of maximising the
experiment's acceptance for protons scattered elastically (or diffractively) at
polar angles down to a few microradians at the interaction point. Currently, two
RP stations at $\pm 220$~m from the interaction point are fully equipped with
detectors.  

Two tracking telescopes, T1 (Cathode Strip Chambers - CSC) and T2 (Gas Electron
Multiplier - GEM), centered at $\pm 9$~m and $\pm13.5$~m from the interaction
point detect inelastically scattered charged particles from the
interaction point. The T1 coverage in pseudo-rapidity range is $3.1 < |\eta| <
4.7$, and for T2 the range is $5.3 < |\eta| < 6.5$. 
Both telescopes have $2\pi$ (full) azimuthal coverage. Simulations show that about 99.5\%
of all non-diffractive minimum bias events and 84\% of all diffractive events
have charged particles within the acceptance of T1 or T2 and are thus triggerable
with these detectors (all TOTEM detectors are trigger capable). Thus T1 and T2
allow to measure the inelastic rate \rate{}{inel}. For more details
concerning the TOTEM apparatus see \cite{1748-0221-3-08-S08007}.

\begin{center}
\begin{figure}[h!] 
    \includegraphics[width=.75\textwidth]{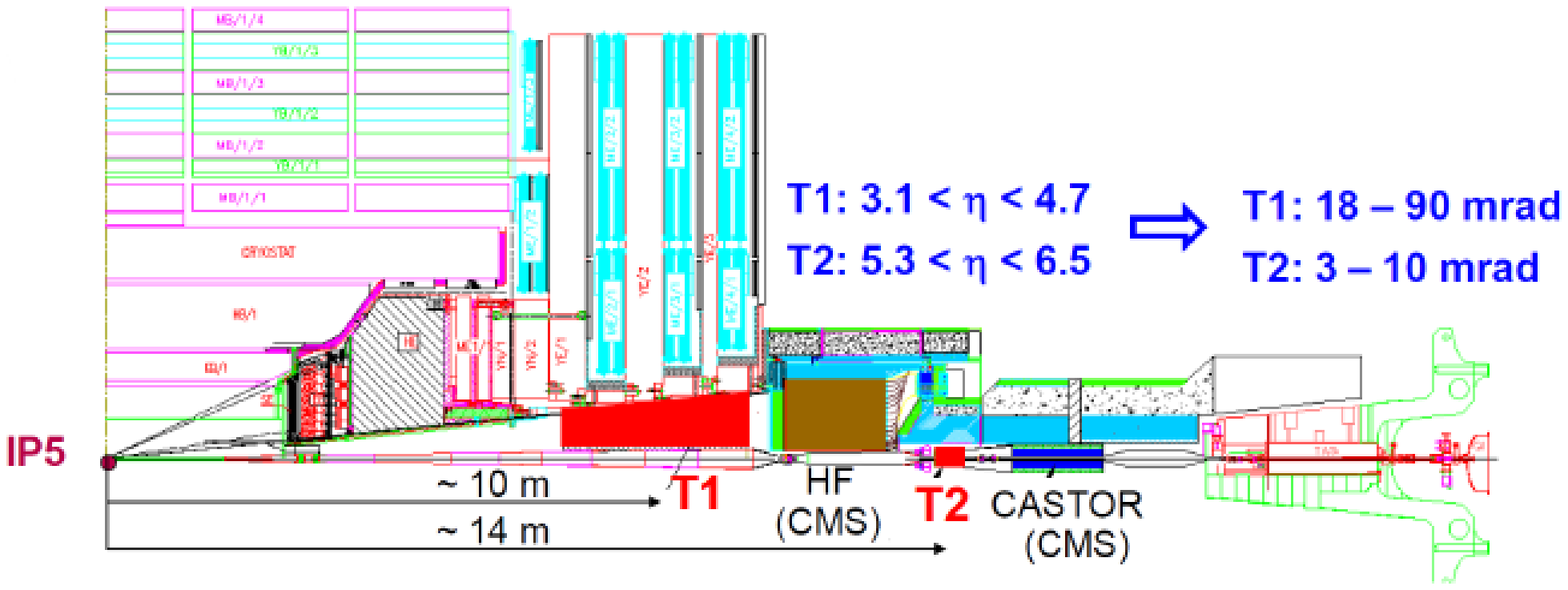}\\
\caption{The TOTEM forward trackers T1 and T2 embedded in the CMS forward region.}
\label{fig:totemApparatus}
\end{figure}
\begin{figure}[h!] 
    \includegraphics[width=.7\textwidth]{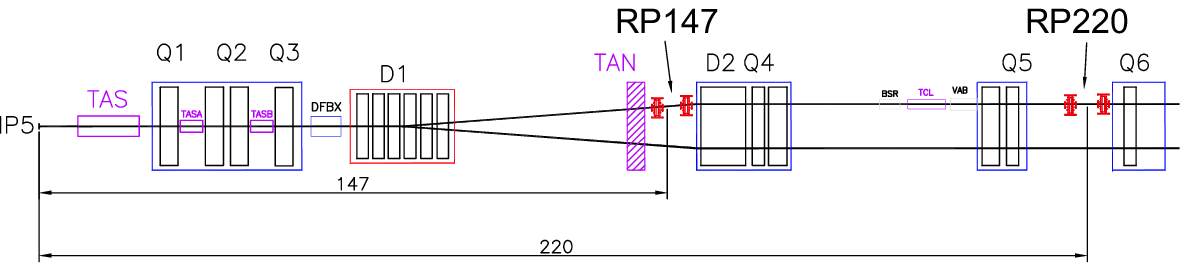}\\
\caption{The LHC beam line on the ``right'' side of the interaction point IP5 and the
TOTEM Roman Pots at 147 and 220~m (RP147 and RP220).}
\label{fig:romanpots}
\end{figure}
\end{center}

\FloatBarrier
\vspace{-2cm}
\subsection{Measurement of the elastic rate}
\begin{figure}[ht]
\setcounter{subfigure}{0}
\subfigure[\text{ }$\beta^* = 90$ m; 50\% acceptance: $0.025 \text{ GeV}^2 \lesssim |t| \lesssim 0.2 \text{ GeV}^2$]{ \includegraphics[width=0.45\textwidth]{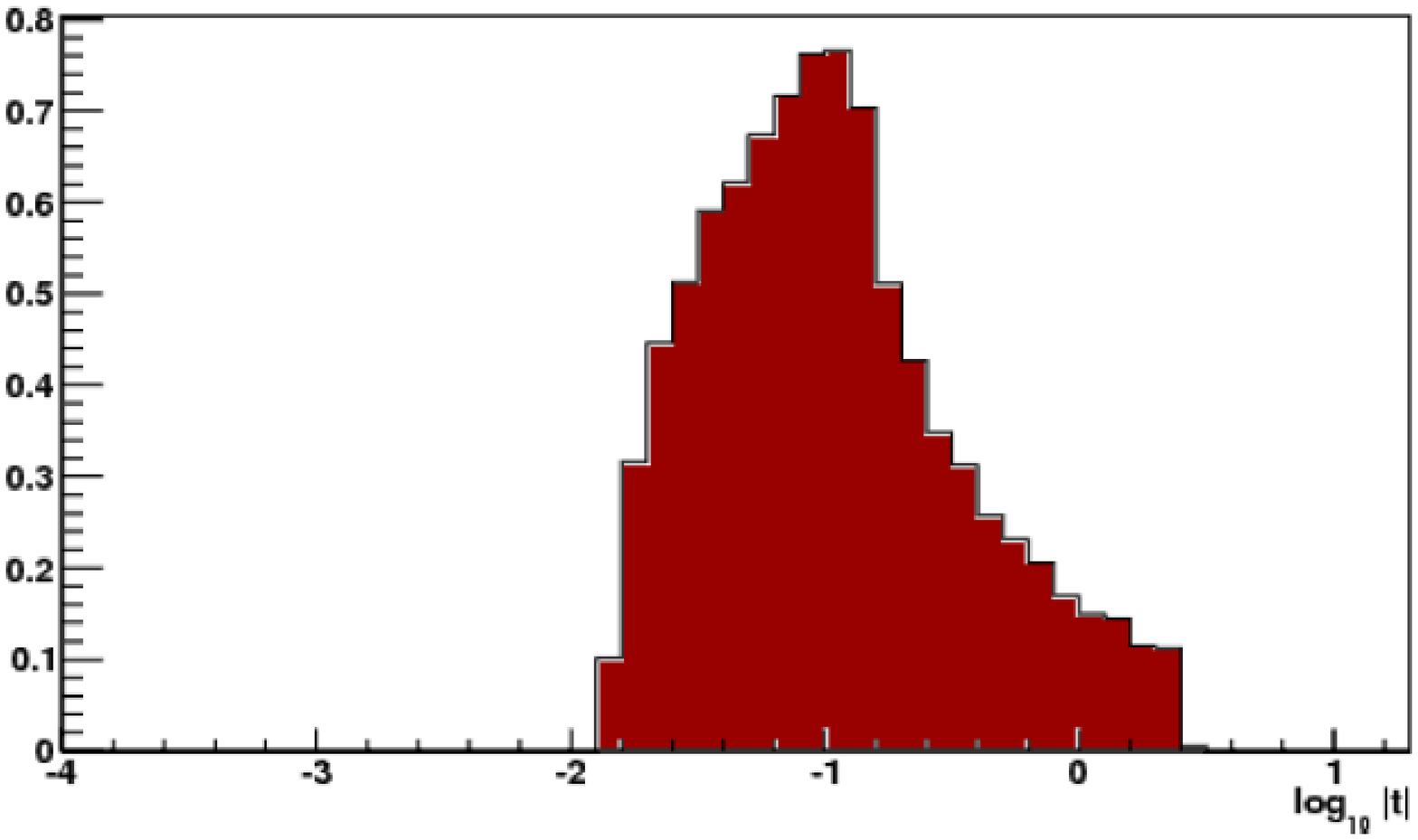} }
\subfigure[\text{ }$\beta^* = 1535$ m; 50\% acceptance: $8\times10^{-4} \text{ GeV}^2 \lesssim |t| \lesssim 0.2 \text{ GeV}^2$]{ \includegraphics[width=0.45\textwidth]{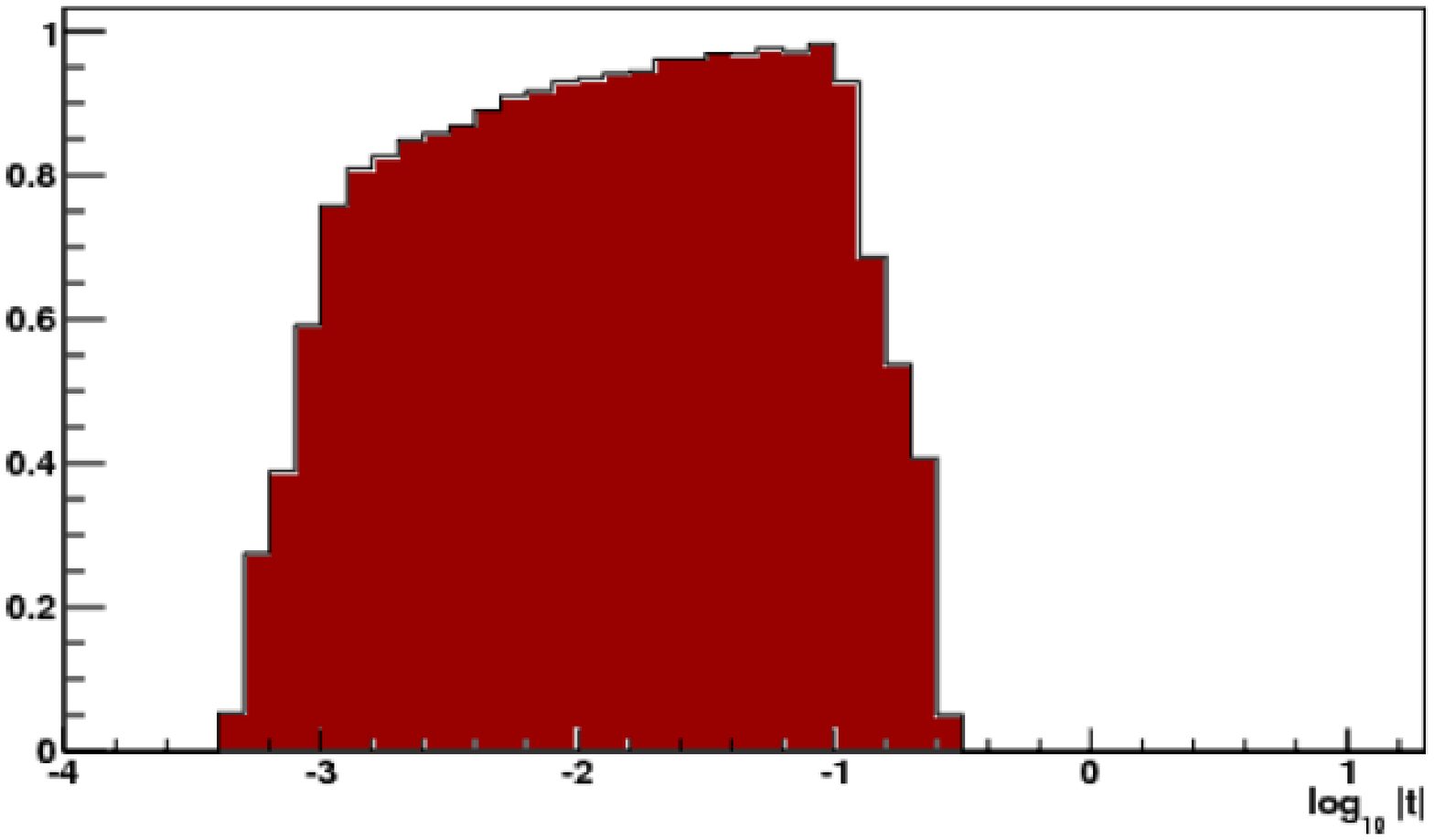} }
\caption{The acceptance of RP detectors at 220~m at $\sqrt{s} = 7$ TeV for two different beam optics.}
\label{fig:acceptance}
\end{figure}

\vspace{-.5mm}
To measure the elastic rate \difrate{C+N}{el} (from which we may separate the
elastic hadronic rate \difrate{N}{el} and extrapolate it to $t$=0, see
\sref{sec:introduction}) one needs proton acceptance at small values of
$\abs{t}$. Two optics have been proposed by TOTEM for this purpose.  One with
$\beta^* = 90 \text{ m}$ and the ultimate one with $\beta^* = 1535 \text{ m}$
foreseen at a later stage. The former one can use the standard injection optics
and thus easier to commission in early LHC operation. The acceptance of RP
detectors at 220~m at $\sqrt{s} = 7$ TeV for both optics is in
\fref{fig:acceptance}. This acceptance has been calculated on the basis of a
simulation for detectors at distance $10 \sigma + 0.5 \text{ mm}$ from the beam
($\sigma$ is size of the beam). From \fref{fig:acceptance} we may see that the
minimal achievable value of $t$ is $\abs{t_{\mathrm{min}}} \approx 0.025 \text{
GeV}^2$ for $\beta^* = 90 \text{ m}$ and $\abs{t_{\mathrm{min}}} \approx
8\times10^{-4} \text{ GeV}^2$ in the case of $\beta^* = 1535 \text{ m}$.
of $\xi = \frac{\Delta p}{p}$. For the $t$-acceptance at $\sqrt{s} = 14$ TeV see
\cite{1748-0221-3-08-S08007}. With such $t$-acceptance TOTEM will not be able
to reach the so-called Coulomb region of $t$ values where the elastic
scattering is given practically just by the Coulomb interaction. The
Coulomb-hadronic region (where neither the Coulomb nor the hadronic interaction
can be neglected) will be reached instead, see \cite{1748-0221-3-08-S08007} and
also \cite{kaspar:2010mv} for details. Once \difrate{N}{el} is known, the
elastic rate \rate{N}{el} can be obtained by integration. Because the
elastic differential cross section for small values of $\abs{t}$ is relatively
very high (see \fref{fig:elmodels}) just a few days of running are needed to measure
elastic scattering for small values of $\abs{t}$. 

To discriminate between different models of elastic scattering which give
different predictions of differential cross section (mainly for higher values of
$\abs{t}$, see \fref{fig:elmodels}), it is important to measure elastic scattering
in the widest possible region. The estimated error contribution from elastic
rate to the total cross section $\sigma_{\mathrm{tot}}$ is given in
\tref{tab:precision} (for details see \cite{kaspar:2010mv}).

\subsection{Measurement of the inelastic rate}

\begin{figure}[h!] 
    \includegraphics[width=.45\textwidth]{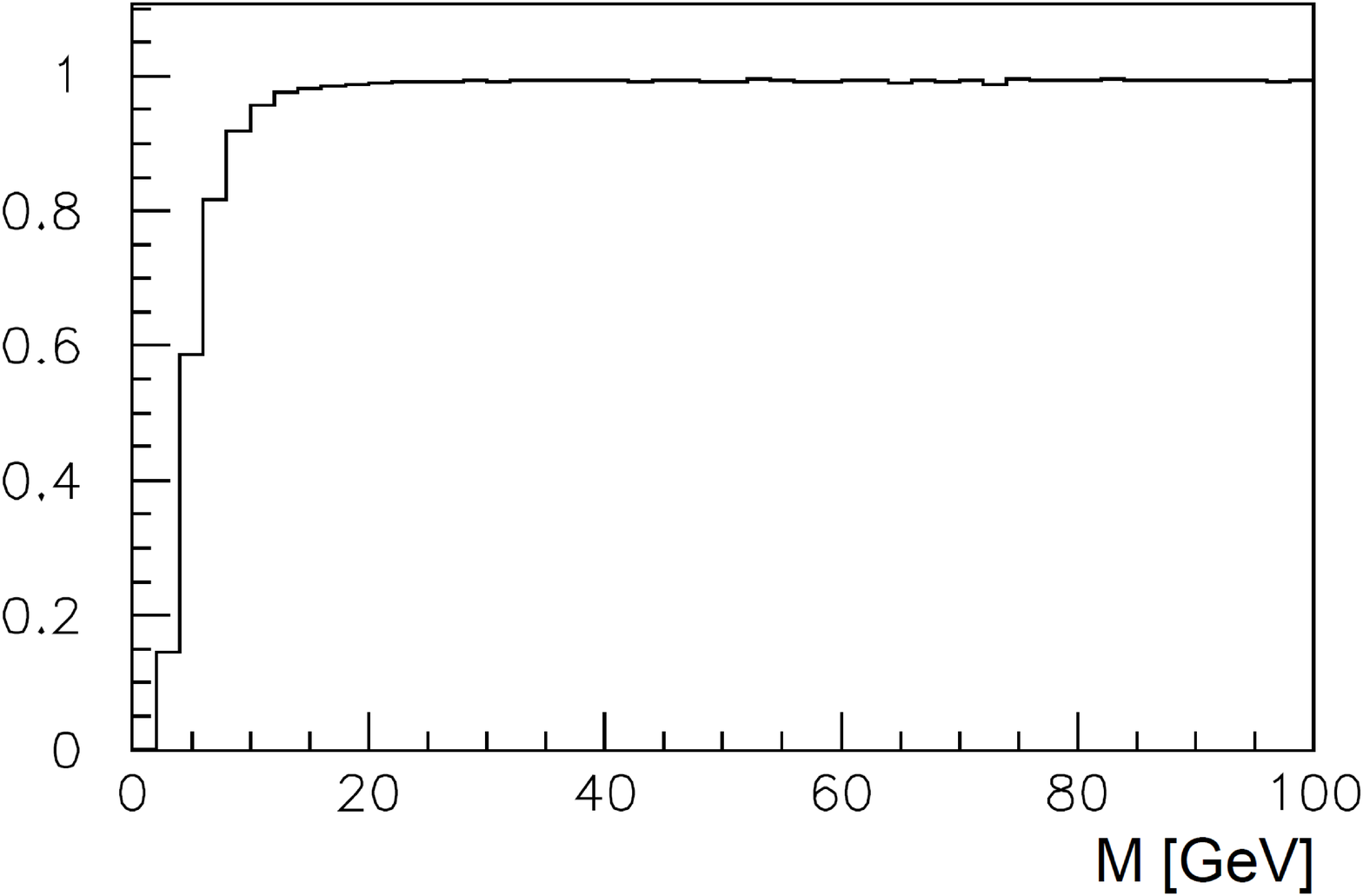}\\
\caption{The acceptance for single diffractive events as a function of the diffractive mass $M$ at  $\sqrt{s} = 14 \text{ TeV}$.}
\label{fig:sdMAcceptance}
\end{figure}

The inelastic rate \rate{}{inel} will be measured by the T1 and T2 telescopes.
The main background in the cross section measurement comes from beam-gas events
which can be largely rejected by primary vertex reconstruction. Studies
repeated in \cite{1748-0221-3-08-S08007} show that mainly single and double
diffractive events cause a major loss in the inelastic rate. The undetected
single diffractive events are mainly those with very low mass below $\approx 10
\text{ GeV}^2$, see \fref{fig:sdMAcceptance}, since all their particles are
produced at pseudo-rapidities beyond the acceptance of T1 or T2. The fraction
of events which are not detected must therefore be estimated by a model. The
error contribution to the total cross section coming from the total inelastic
rates and dominated by the inelastic trigger losses was estimated to $\pm1$\%
for $\beta^* = 90 \text{ m}$ and $\pm0.8$\% in the case of $\beta^* = 1535
\text{ m}$.

\begin{table}
\resizebox{0.8\textwidth}{!}{
\centering
\newcolumntype{V}{>{\centering\arraybackslash} m{.1\linewidth} }
\begin{tabular}{V p{0.6\textwidth} V V}
\hline \hline
 & & \multicolumn{2}{c}{$\beta^*$ optics} \\[-4mm]
 & Quantity  &  90~m & 1535~m  \\
\hline
$\left.\frac{\text{d}N_{\mathrm{el}}}{\text{d}t}\right|_{t=0}$ & {\bf  Extrapolation of elastic cross-section to $t=0$ } 

 {\footnotesize(Smearing effect due to beam divergence, statistical errors, 
 uncertainty of effective length $L_{eff}$, RP alignment, model dependent deviation)} & $\pm4\%$   &  $\pm0.2\%$   \\
$N_{\mathrm{el}}$ & {\bf Total elastic rate}

{\footnotesize(strongly correlated with extrapolation)}                            & $\pm2\%$   &  $\pm0.1\%$ \\
$N_{\mathrm{inel}}$  & {\bf Total inelastic rate }

{\footnotesize(error dominated by single diffractive losses)}                       & $\pm1\%$   &  $\pm0.8\%$ \\
$\rho$ &{\bf  Error contribution from ($1+\rho^2$) }

{\footnotesize (using full COMPETE error band $\frac{\delta \rho}{\rho} = 33\%$)}  & \multicolumn{2}{c}{$\pm1.2\%$} \\
\hline
\multicolumn{2}{r}{{\bf Total uncertainty in $\sigma_{\mathrm{tot}}$}} & $\pm5\%$ &  $\pm1-2\%$ \\
\multicolumn{2}{r}{{\bf Total uncertainty in \hspace{1.5mm}$L$\hspace{1.5mm}}   }         & $\pm7\%$ &  $\pm2\%$ \\
\hline \hline
 
\end{tabular}
}
\caption{\label{tab:precision}
Estimated error contributions to the total cross section
$\sigma_{\mathrm{tot}}$ and the luminosity $L$ at $\sqrt{s} = 14 \text{ TeV}$ from all quantities entering in
Eqs.~\refp{eq:sigtot} and \refp{eq:luminosity}.}
\end{table}

\FloatBarrier
\section{Conclusion}
The determination of the total pp cross section based on the
luminosity independent method requires from theory a model of elastic
scattering (i.e., hadronic scattering amplitude \ampl{N}), which can be used both
for extrapolation of \difrate{N}{el} to the optical point $t=0$ and also for
calculation of the $\rho$ parameter) and a method for the separation of Coulomb and
hadronic scattering.  The elastic rate \difrate{C+N}{el}
and inelastic rate \rate{}{inel} have to be measured. 
The total uncertainty in $\sigma_{\mathrm{tot}}$ at
$\sqrt{s} = 14 \text{ TeV}$ is $\pm5\%$ for $\beta^* = 90$~m and $\pm(1-2)\%$ in
the case of the final optics $\beta^* = 1535$~m. The same quantities which
determine $\sigma_{\mathrm{tot}}$ determine also the luminosity $L$ given by
\eref{eq:luminosity}. The total uncertainty in $L$ is slightly worse because
the total rate $\rate{N}{el}+\rate{}{inel}$ in \eref{eq:luminosity} is squared.

\FloatBarrier


\newpage
\clearpage
\setcounter{affil}{0}
\setcounter{section}{0}
\setcounter{figure}{0}
\setcounter{table}{0}
\setcounter{equation}{0}


\title{QCD mini-jets contribution to the total cross-section  for pions and protons and expectations at LHC} 



\author{Agnes Grau}
\email[]{igrau@ugr.es}
\affiliation{Departamento de Fisica Teorica y del Cosmos, Universidad de Granada, Spain}
\author{Rohini M. Godbole}
\email[]{rohini@cts.iisc.ernet.in}
\affiliation{
Centre for High Energy Physics, Indian Institute of Science, Bangalore, 560012, India }
\author{Giulia Pancheri}
\email[]{pancheri@lnf.infn.it}
\homepage[]{{\tt http://www.lnf.infn.it/theory/pancheri/Welcome.html}}
\author{Olga Shekhovtsova}
\email[]{olga.shekhovtsova@lnf.infn.it}\affiliation{INFN Frascati National Laboratories, Va E. Fermi 40, I00044 Frascati, Italy}
\author{Andrea Achilli}
\email[]{achilli@fisica.unipg.it}
\author{Yogendra N. Srivastava}
\email[]{yogendra.srivastava@pg.infn.it}
\affiliation{Physics Department and INFN, University of Perugia, Perugia I06123, Italy}


\begin{abstract}
We describe our \ktresummation \ model for total \x s and show its application to \pp \ and \pbarp \ scattering. The model uses mini-jets to drive the rise of the cross-section and soft gluon resummation in the infrared region to transform the violent rise of the mini-jet cross-section into a logarithmic behaviour in agreement with the Froissart bound. 
\end{abstract}

\pacs{}

\maketitle 
\addcontentsline{toc}{part}{QCD mini-jets contribution to the total cross-section  for pions and protons and expectations at LHC - {\it A.Grau, R.M.Godbole, L.Pancheri, O.Shekhovtsova, Y.N.Srivastava}}
\section{Introduction}
\label{sec:introduction}
The total \x \ is an observable dominated by very large distances: the so far
basic unsolved problem of QCD. As such, we do not yet have an understanding from first principles 
of its energy behaviour or process dependence, but models abound. Models are  based mostly on the optical theorem, analyticity, eikonal representation, Glauber theory, Reggeon field theory. One  most  popular and simple model   from Donnachie and Landshoff \cite{Donnachie:1992ny}  describes  all known total cross-sections with just  two  terms, one   with a decreasing behaviour, from  Regge exchanges, and one rising, from  Pomeron exchange, namely
\be
\sigma_{ab}=Y_{ab}s^{-\eta}+X_{ab}\ s^{\epsilon}
\ee
This model  fits reasonably well all known total  cross-sections with two universal powers
and different coefficients for different processes. Apart from the fact that there is some evidence from photon-photon scattering that the  rise is not universal \cite{withalbert}, a  major objection to this model is that  it suffers from violating the Froissart bound. The general consensus is that it may give  a good description in the intermediate energy region, but not at the extremely high energies where the Froissart bound is to be valid.

 It has been known for  quite some time that the bulk of the total \x\ is due to semi-hard QCD processes \cite{GLR}. Such  model input can be  used in  the eikonal representation, derived from an optical description of scattering. Among the eikonal models, there are the so-called mini-jet models, which use PDF calculable QCD mini-jets to describe the rise of the total \x \ \cite{DurandPi,Grau:1999em,liplus}, or use QCD inspired formulations, among them the papers by Block et al. \cite{Aspen}.  Models are all in need of some amount of data fitting, since none of them is able to obtain normalizations from first principles. Other predictions come from Regge based low energy fits 
implemented by analyticity constraints, and limitations from the Froissart bound.

An example of such strategy is the one followed by the COMPETE collaboration, which is reproduced in PDG, and has the following general expression \cite{PDG},  
\be
\sigma_{ab,{\bar a}b}
=Z^{ab}+B\ln^2(\frac{s}{s_0})+Y_1^{ab}(\frac{s_1}{s})^{\eta_1}\mp Y_2^{ab}(\frac{s_1}{s})^{\eta_2}
\ee
 where the last two terms reflect the exchange of different Regge trajectories in different processes,  the first term is sometime referred to as the "Pomeron" and the squared logarithm reflects a geometrical picture, in addition to saturating the Froissart bound. Other fits, such as  the one by Block and Halzen\cite{BH}, have an additional term linear in $\ln {s}$.
 As useful as fits  are, one still needs models if one wants to learn something fundamental from total cross-section measurements. In particular, since the total \x \ reflects the large distance behaviour of the underlying theory, we need models which, albeit phenomenologically, probe the infrared region. 
 
\begin{figure}
\begin{center}
\resizebox{0.7\textwidth}{!}{
\includegraphics{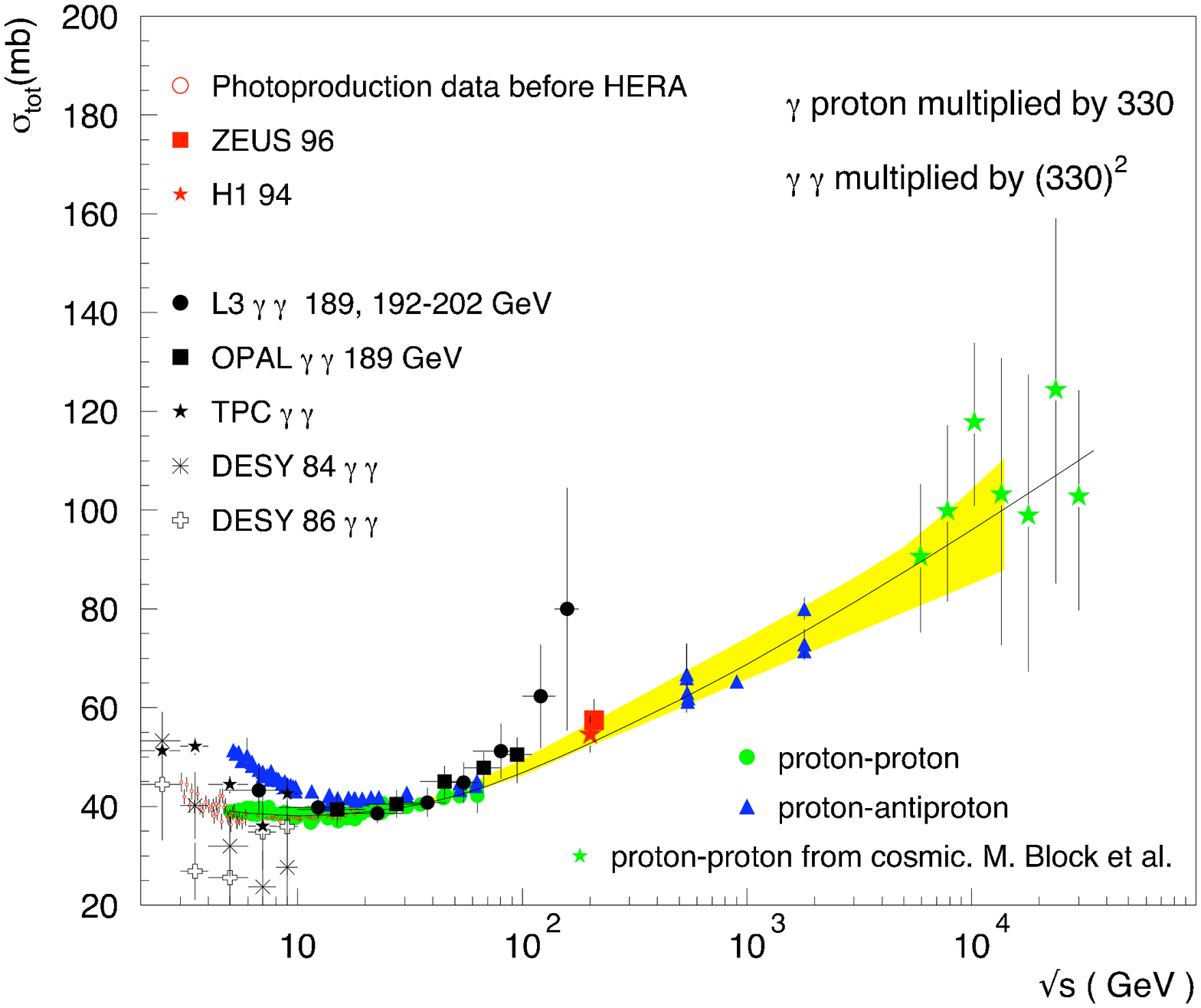}}
\caption{A compilation of proton and photon total cross-sections, as described in the text. }\label{fig:all}
\end{center}
\end{figure}

An overview of proton and photon  total cross-section data in the medium to high energy range \cite{ourEPJC} is shown in Fig.~\ref{fig:all}. For the sake of comparison, we have plotted photon and proton data, with an {\it ad hoc} normalization factor. The full line represents the result from a  model which we have developed through a number of years, and which probes the large distance behaviour through soft gluon \ktresummation \ with an ansatz for the effective coupling of gluons with quarks in the  very low momentum region \cite{Corsetti:1996wg,Grau:1999em,Godbole:2004kx}.

In the following we outline how  the model works, using the case of \pp \ and \pbarp \ scattering.  Then  we show how one can apply the model   to pion-proton scattering, to predict the $\pi p$ \x \ at energies reacheable at LHC through neutron detection in the very forward region.  

\section{Applying the \ktresummation \ model to \pp \ and \pbarp \ scattering}
\begin{figure}[h]
\centering
\begin{minipage}{.45\linewidth}
\includegraphics[width=.85\textwidth]{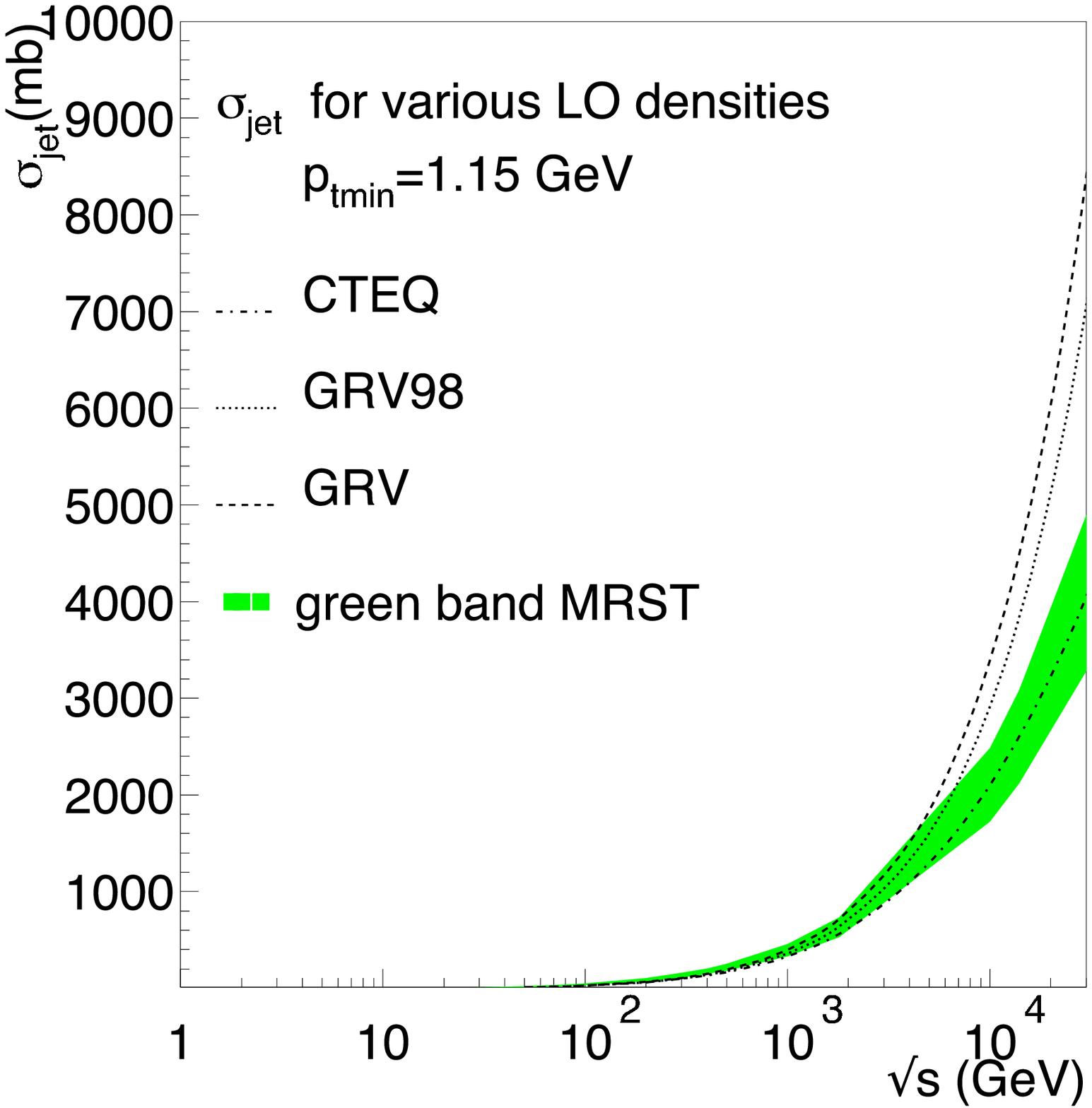}
\caption{The mini-jet cross-section for different energies and different densities.}
\label{fig:sigjetband}
\end{minipage}%
\hspace{1cm}%
\begin{minipage}{.45\linewidth}
\includegraphics[width=.85\textwidth]{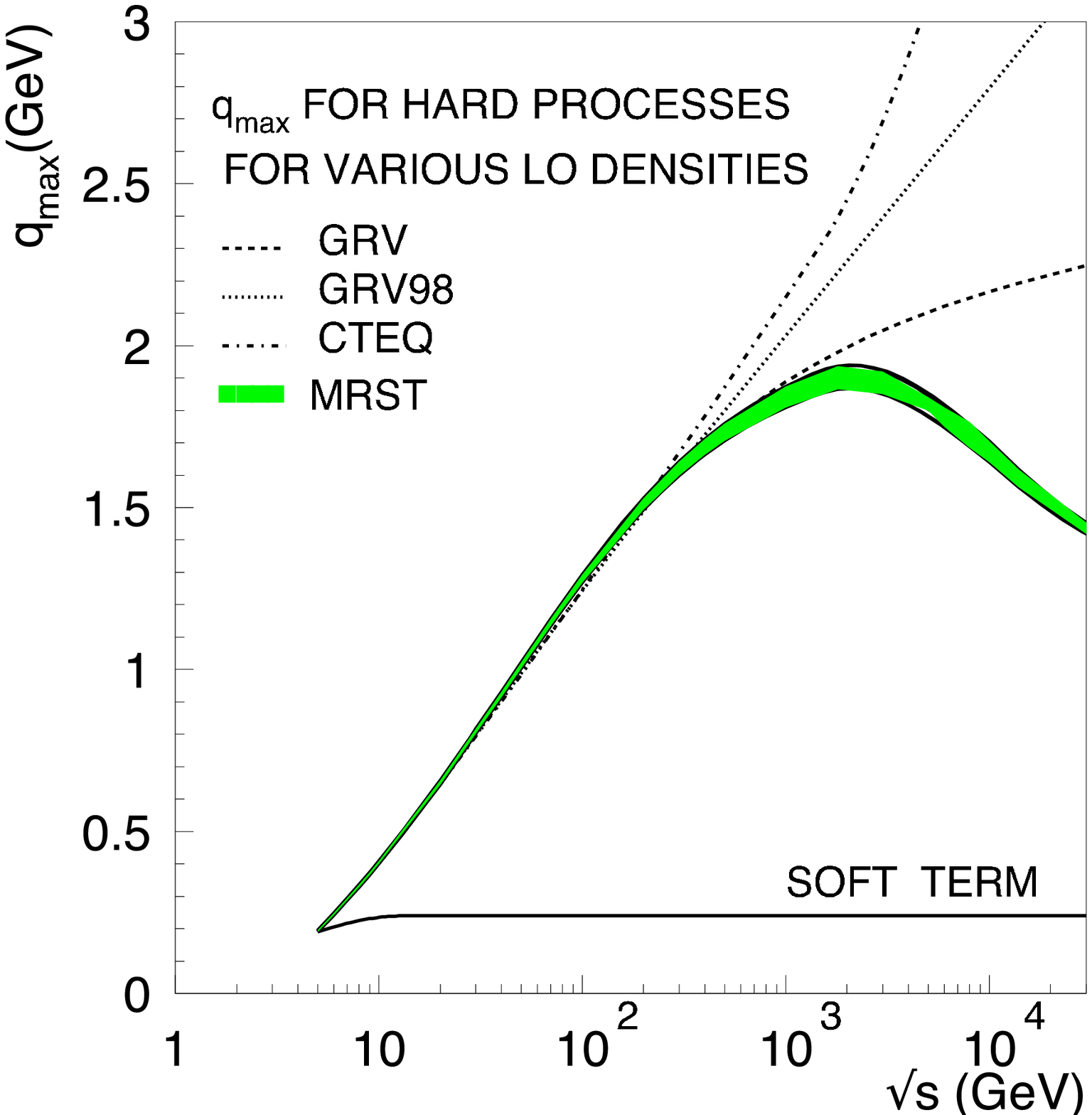}
\caption{The maximum single soft gluon transverse momentum  in \pp\  \ / \pbarp  \ scattering.
 }
\label{fig:qmaxband}
\end{minipage}
\end{figure}

Our model blends together a QCD calculated input, given by mini-jet \x s, and \ktresummation. The model is characterized by three different momentum regions, as follows:
\begin{enumerate}
\item $p_t\ge p_{tmin}$,   for parton parton collisions, 
where the  perturbative QCD description 
leading to the mini-jet \x \ 
is applied,  with $p_{tmin}\sim (1-2)\ GeV$  kept fixed and independent of energy;
\item  $\Lambda_{QCD}\le k_t\le q_{max}$  for  
single soft gluons emitted from initial state quarks before the hard parton-parton collision,
 through the usual asymptotic freedom perturbative coupling $\alpha_s(k_t)$,   
 with \cite{Corsetti:1996wg} $ q_{max}\sim p_{tmin} \ln{\sqrt{s}/p_{tmin}}$; 
\item   $k_t\le \Lambda_{QCD}$ for ultrasoft gluons in  a region which is dominated by a singular,  but integrable, coupling of the gluons with the emitting quarks, $\alpha_{eff}(k_t)\sim k_t^{-2p}$ as $k_t \rightarrow 0$.
\end{enumerate}
Neglecting the real part of the scattering amplitude, the above modeling for the interactions  is then  input   to the following expressions:
\bea
\sigma_{total}&\approx & 2\int d^2\vecb [1-e^{-\nbar (b,s)/2}]\\
\nbar ^{AB} (b,s)&=&
\nbar _{soft}^{AB}(b,s)
  +A_{BN}^{AB} (b,s) \sigma^{AB}_{\rm jet} (s,p_{tmin})\\
   A_{BN}^{AB} (b,s)&=&{\cal N}
exp
\{
-\frac{16}{3\pi}\int_0^{qmax} \frac{dk_t}{k_t}{\alpha_{eff}}(k_t)\ln 
(\frac{2q_{max}}{
k_t})
[1-J_0(bk_t)]\}
\eea
where the impact parameter distribution for the collision is obtained from the normalized Fourier transform of the resummed expression for soft gluon emission from the initial state in the collision.
 In Fig.~\ref{fig:sigjetband} we show the energy behaviour of the mini-jet \x s for different Parton Density Functions (PDFs) as indicated. The lower cut-off for the parton-parton \x s is given by $p_{tmin}=1.15\ GeV$. With such a value and current PDFs,   up to  $\sqrt{s} \approx 10 \ GeV$  the contribution from  mini-jets     to the total \x \  is quite small and the bulk of the cross-section comes from   $\nbar _{soft}^{AB}(b,s)$. This quantity can then be   parametrized through the convolution of the form factors of the colliding particles  \cite{Grau:1999em} or through other appropriate  methods  \cite{Godbole:2004kx}.
 
\begin{figure}[h]
\centering
\begin{minipage}{.45\linewidth}
\includegraphics[width=.85\textwidth]{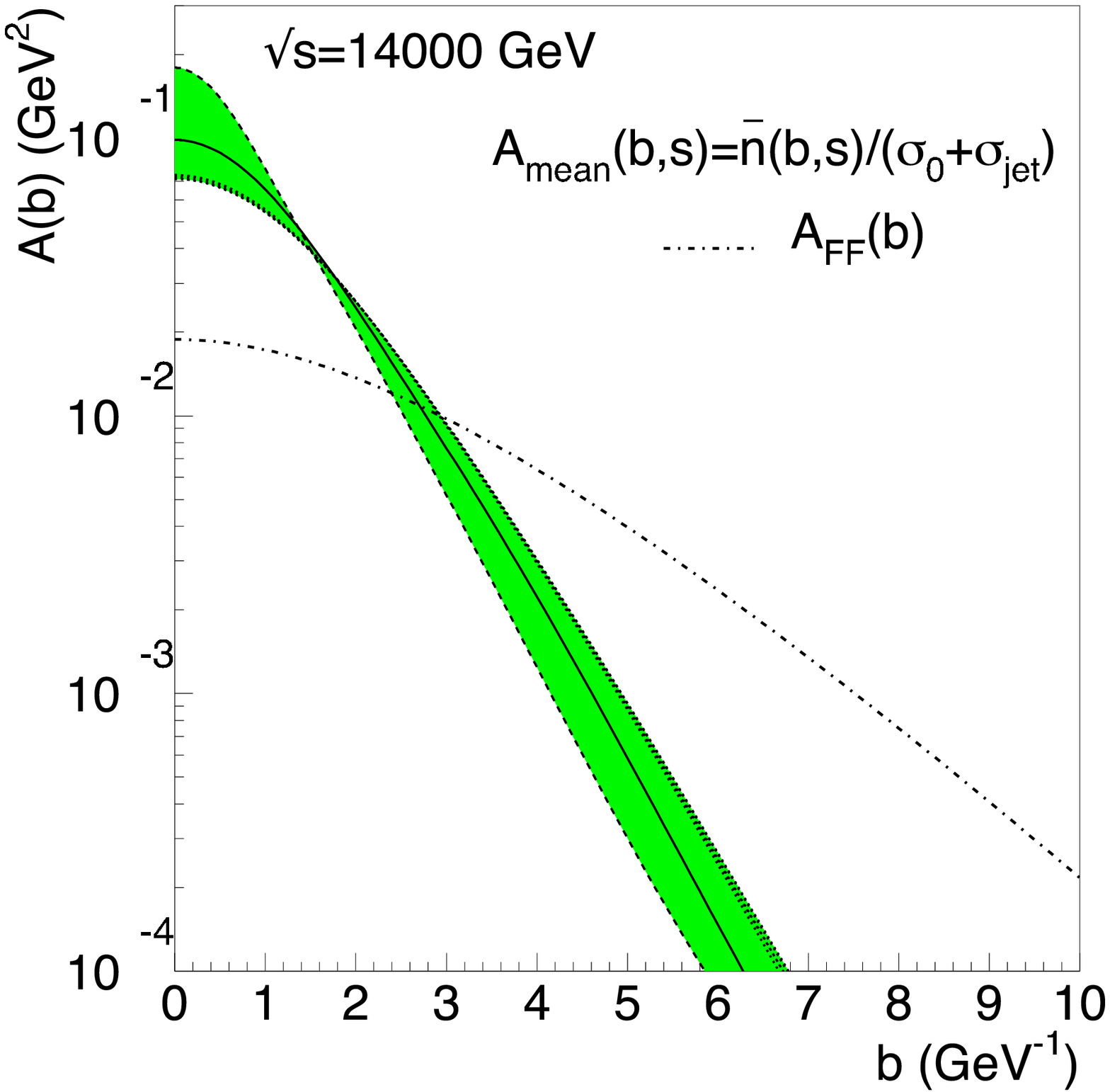}
\caption{The impact parameter distribution for different PDFs at LHC energy, compared with the Form Factor expression. Range of PDF's is the same as in the other figures.}
\label{fig:abmean14band}
\end{minipage}%
\hspace{1cm}%
\begin{minipage}{.45\linewidth}
\includegraphics[width=.85\textwidth]{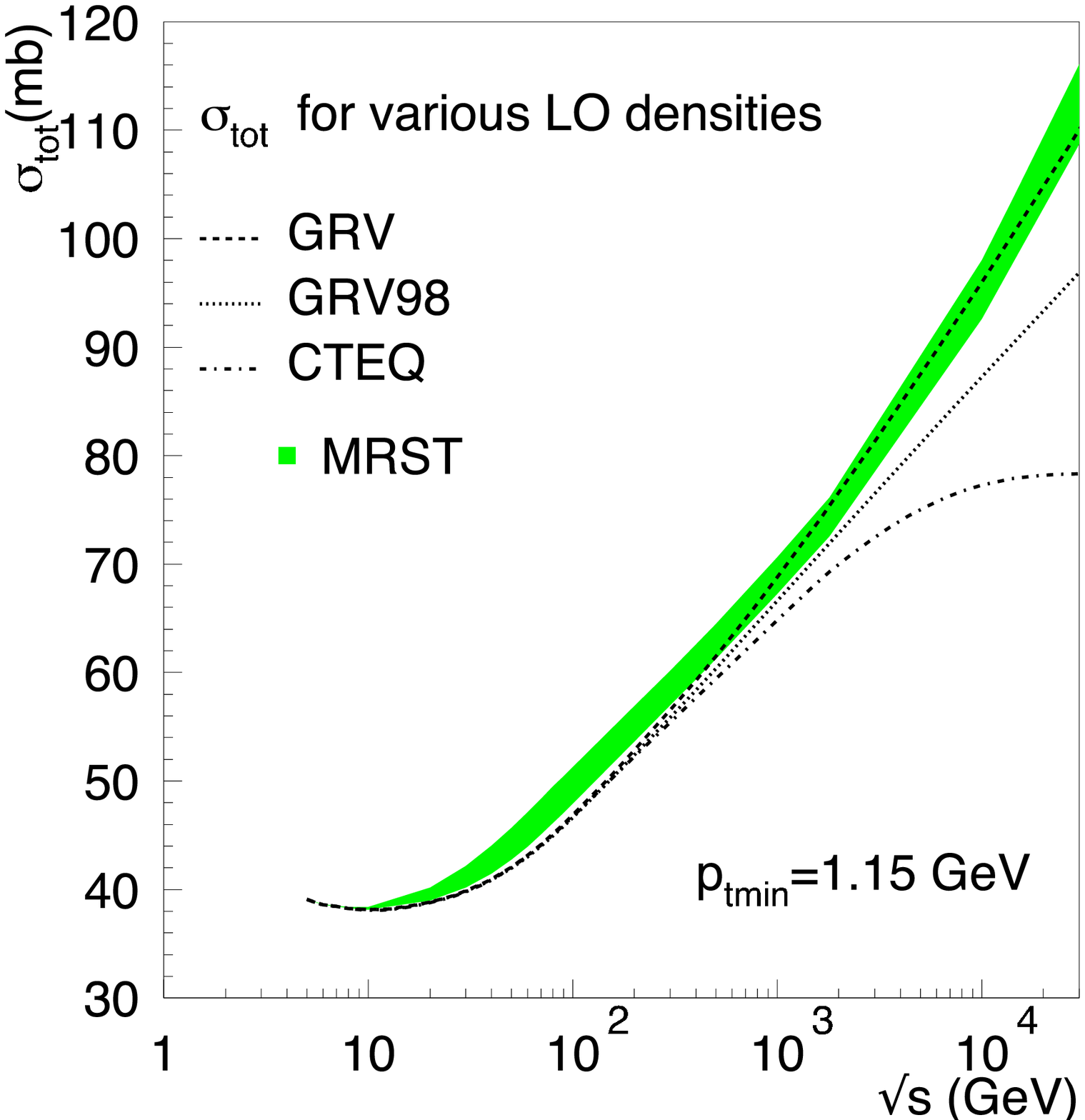}
\caption{The result for the total \x \  as is obtained by folding mini-jet \x s and impact parameter distributions  in the eikonalized expression.}
\label{fig:eiksigband}
\end{minipage}
\end{figure}

 Through these inputs, we describe the rise of the total \x \ . One can follow the various steps which lead to our final result through Figs.~\ref{fig:sigjetband}, \ref{fig:qmaxband}, \ref{fig:abmean14band} and \ref{fig:eiksigband}.

\begin{figure}
\begin{center}

\includegraphics[width=0.8\textwidth]{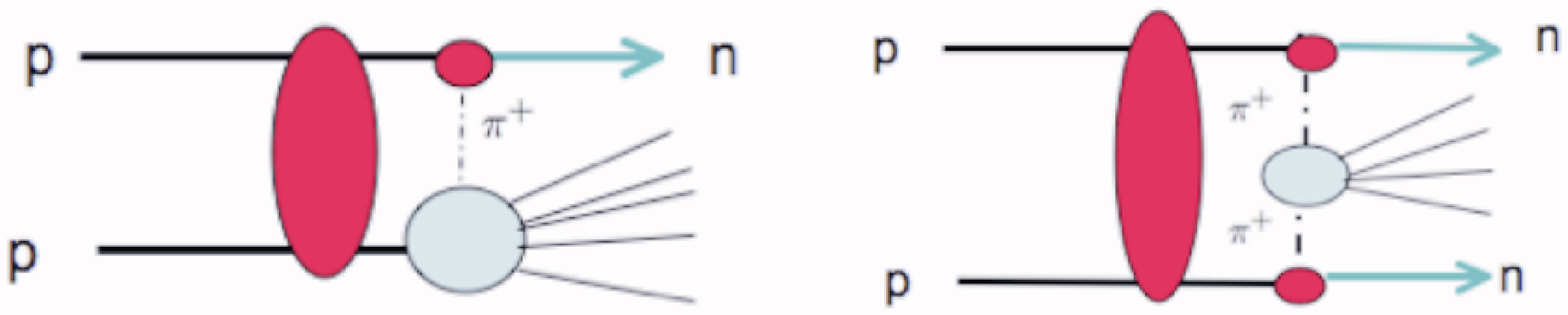}
\caption{\label{fig:pionexchange}}
\end{center}
\end{figure}

The model we just described can be easily extended to pion processes. In Fig. ~\ref{fig:pionexchange} we show the pion exchange process which might make  the measurement of $\pi p$ and $\pi \pi$ total \x \  possible at LHC, through the Zero Degree Calorimeters (ZDC) 
detecting neutrons very near the beam direction. The feasibility and interest in such processes has  continued for a long time, and has recently been discussed in \cite{soffer,PRS,murray}. 
While expectation at LHC are strongly determined by the many lower energy experiments, up to the Tevatron results at $\sqrt{s}=1800\ GeV$, the situation for pion processes is very different, as direct measurements from target experiments do not extend beyond $\sqrt{s} \sim 40\ GeV$. Thus,
predictions for $\pi p$  and $\pi \pi$ total cross-sections suffer from lack of data in the high energy region, namely the region  where the cross-section starts rising because of the onset of parton-parton processes. 
 \begin{figure}[h]
 \centering
 \includegraphics[width=0.6\textwidth]{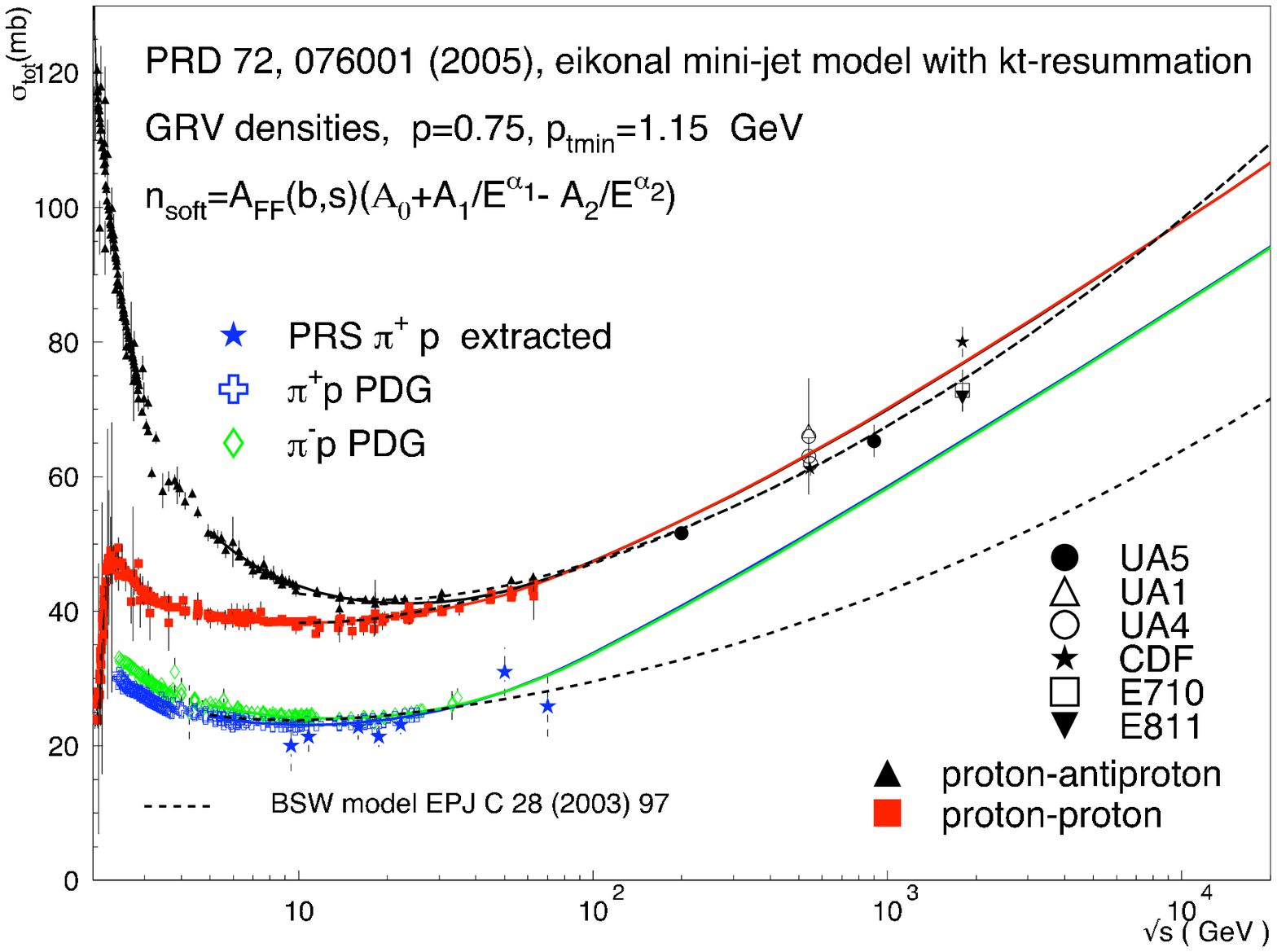}
 \caption{Predictions from our \ktresummation\  model for proton and pions total cross-sections from very low to very high energies and comparison with BSW model \cite{BSW}. }
 \label{fig:pionproton} %
 \end{figure}
 We have plotted in Fig. ~\ref{fig:pionproton} a preliminary estimate for $\pi p$ total cross-section  comparing it  with predictions  from our model for \pp \ and \pbarp \ scattering \cite{our2008}.  In Fig. ~\ref{fig:pionproton} PRS data for $\pi p$   have been extracted by Petrov, Ryutin and Sobol   \cite{PRS} from earlier measurements and obtained using   the charge exchange mechanism shown in Fig.~\ref{fig:pionexchange}. A recent discussion \cite{ourpion} indicates that our model satisfies factorization for pion and proton total \x s. For all the processes considered we have also made a comparison  with the BSW model \cite{BSW}.  The comparison with the BSW result indicates a good agreement for the pure proton processes but a different high energy behaviour for  $\pi p$ and a faster increase at low energy. Given the lack of fixed target data, this gives further arguments for   the advocated  \cite{PRS} need for future measurements in the high energy region.  More work  is in progress and will be published soon.


%
%

%

\begin{acknowledgments}
G.P. thanks the MIT Center for Theoretical Physics  and Brown University Physics Department for hospitality. Work partially supported by the Spanish MEC 
(FPA2006-05294 and FPA2008-04158-E/INFN) and by Junta de Andaluc\'\i a 
(FQM 101).
\end{acknowledgments}

\newpage
\clearpage
\setcounter{affil}{0}
\setcounter{section}{0}
\setcounter{figure}{0}
\setcounter{table}{0}
\setcounter{equation}{0}

\pagestyle{empty}

\begin{center}


\begin{LARGE} 
\textbf{The ATLAS Forward Physics Program} 
\end{LARGE}
   
\vskip 0.5 cm
\begin{large}
Sara Diglio{\small$^{^{{\,a}}}$} on behalf of the ATLAS Collaboration.\\
\vspace{0.5cm}
{\small{$^{a}$}} 
\small{{\sl{ LAL, Univ Paris-Sud, CNRS/IN2P3, Orsay, France }}}\\   
\end{large}


\end{center}

\begin{center}

 \textbf{Abstract} 
\end{center} 
{
A brief review of the ATLAS forward detector system is presented. The ATLAS forward physics program with early data and luminosity determination are introduced and the relevant analysis
strategies reviewed. A proposed high luminosity upgrade project is also discussed.
 }

\addcontentsline{toc}{part}{The ATLAS Forward Physics Program - {\it S.Diglio}}



\pagestyle{plain}

\section{The ATLAS Forward Detectors}
\label{sez:FD}
\begin{figure}[h!]
\epsfig{figure=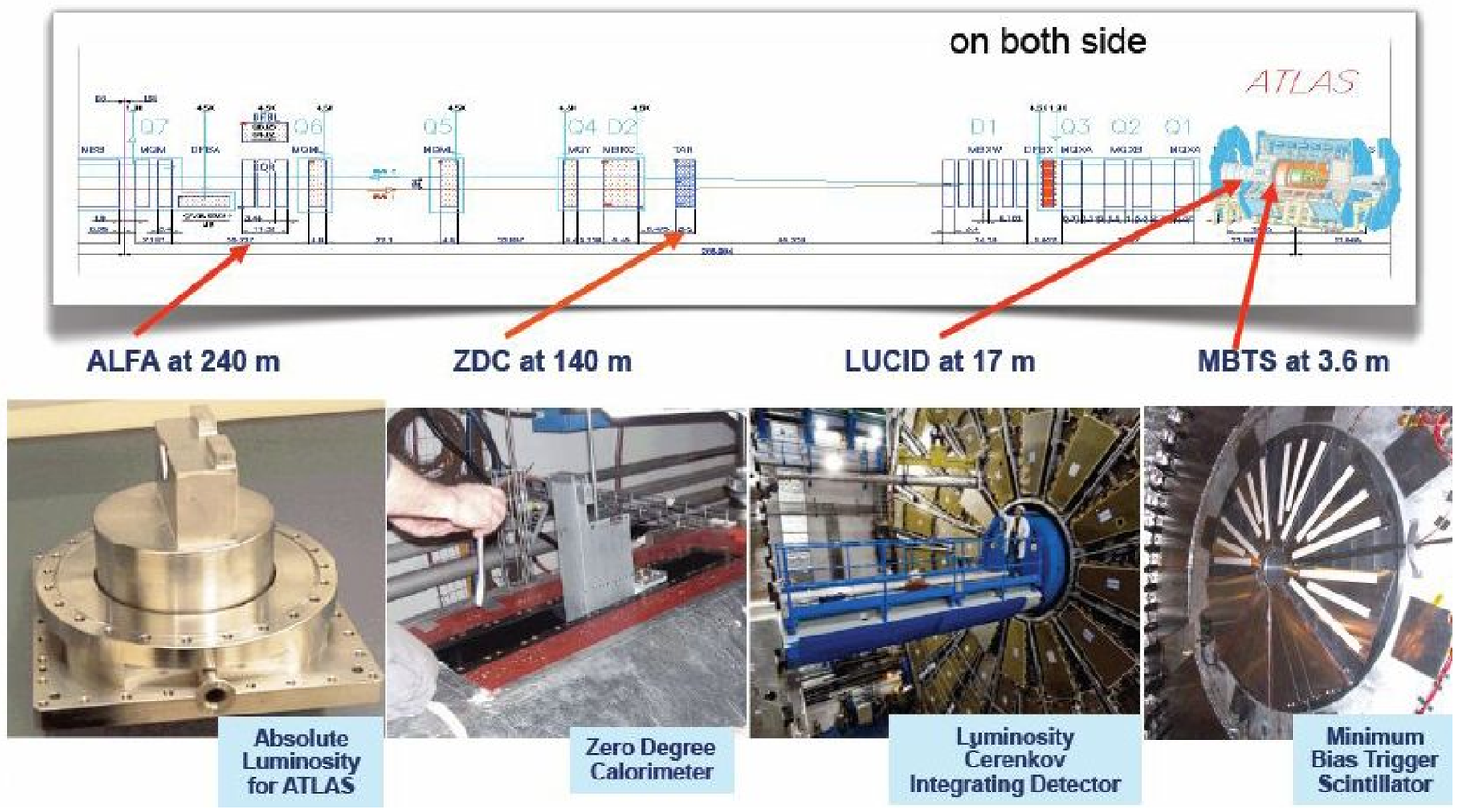,height=7cm,width=11cm} 
\caption{\it The layout of the approved ATLAS Forward detectors.
\label{fig:layout_FD}}
\end{figure}

The aim of forward physics is to study processes in which particles are produced at very small polar angle $\theta$ with respect to the
beam. In terms of psudorapidity $\eta = - \ln [\tan \frac{\theta}{2}] $ it is possible to divide the ATLAS detector into a central part
and a number of subdetectors to measure forward particle production. \\
The central detector consists of an inner detector for tracking purposes ($|\eta|<$ 2.5), electromagnetic calorimeters to measure the
energy of electrons and photons ($|\eta|<$ 3.2), hadronic calorimeters for measuring the energy of baryons and mesons ($|\eta|<$ 4.9), and a muon
spectrometer ($|\eta|<$ 2.7).\\
The forward detectors are designed to measure forward particle production. These are ALFA, LUCID, MBTS and ZDC: they will be presented in
the following sections. A layout sketch of these detectors is presented in fig. \ref{fig:layout_FD}.
\subsection{ALFA}
\label{sez:ALFA}
The ALFA (Absolute Luminosity For ATLAS)  \cite{ALFA} Roman Pots (RP) are located 240 m from the interaction point. There will be two RP
on each side, the distance between these two stations is 4 m. The system provides a pseudorapidity coverage 10.6 $<|\eta|<$ 13.5.
The RP system is not fixed relative to the beam. 
At injection the ALFA detectors are in a withdraw position far from the beam. When the
beam is stabilized the detectors are inserted to the measurement position (within 1.5 mm to the beam).\\
The ALFA detectors are primarily designed for measurement of the integrated luminosity and elastic scattering at small angles. 
This will be done by measuring the position of elastic and diffractive protons that have been deflected at very small angles with respect to the beam 
and that pass through arrays of scintillating fiber trackers ( 20 $\times$ 64 fibers in each array). These fibers provide a spatial resolution of about 30 $\mu$m.\\
In order to realize such kind of measurement ALFA will collect data in special LHC runs at low luminosity ($10^{27}$ $cm^{-2}$ $s^{-1}$), 
with high $\beta^{*}$ optics and reduced beam emittance.\\
The infrastructure and part of the mechanics have been installed in the LHC tunnel, while the
detectors will undergo a final test beam in summer 2010, and be installed along with the final
mechanics at the first available shutdown.
\subsubsection*{Elastic scattering at small angles and luminosity measurement}
\label{sez:ALFA_el}
The optical theorem relates the forward scattering amplitude to the total cross section of the scatterer.
It is usually written in the form $\sigma_{tot}= {4 \pi} \Im[f_{el}(t=0)]$ 
where it can be seen that the total cross section ($\sigma_{tot}$) is directly proportional to the imaginary part of the
nuclear forward elastic scattering amplitude extrapolated to zero momentum transfer ($f_{el}(t=0)$).\\
For small scattering angle, the four-momentum transfer can be determined by $-t = (p_i-p_o)^2 \approx (p \theta)^2$ where
$p$ states for the proton momentum of the beam ($p_i$ and $p_o$ refer respectively to the incoming and outgoing proton) and 
$\theta$ for the scattering angle at the interaction point.\\
The ALFA detector allows to determine the the total cross section and the absolute luminosity measuring the elastic scattering down to such
small $t$-values where the cross section becomes sensitive to the electromagnetic amplitude via the Coulomb interference term.
Since we know very well the electromagnetic amplitude, it will be possible to add a constraint if the Coulomb region will be reached.
Taking into account the optical theorem and the Coulomb term, the rate of elastic scattering
at small $t$-values can be written in a simplified version as :
\begin{equation}
\label{eq:el_rate} 
\centering 
\frac{d N}{d t} = L \pi \, |f_C+f_N|^2 \approx L \pi \, (- \frac{2 \alpha}{|t|} + \frac{\sigma_{tot}}{4 \pi} \, (i+\rho) \, e^{{- \frac{b |t|}{2}}})^2
\end{equation}
where the first term corresponds to the Coulomb and the second to the strong interaction
amplitude. $\alpha$ is the fine structure constant.
Fitting the $t$-distribution allows to determine the absolute luminosity ($L$), the total cross section ($\sigma_{tot}$),
the slope ($b$) and the ratio of the real over the imaginary part of the forward elastic
scattering amplitude ($\rho$) without measuring the inelastic rate. It is expected that the absolute luminosity will be
determined to $\sim$ 2-3 $\%$ accuracy.
\subsection{LUCID}
\label{sez:LUCID}
LUCID (Luminosity measurement using Cerenkov Integrating Detector) \cite{ATLAS, LUCID} is composed of two modules located at 17 m from
the interaction point on both side of ATLAS. Each module is composed of 20 Cherenkov tubes at the end of which photomultiplier tubes (PMT) are placed
to collect the light. Each Cherenkov tube is made of aluminium, 15 mm in diameter, filled with $C_{4}F_{10}$. 
It covers a pseudorapidity range 5.6 $<|\eta|<$ 5.9 for charged particles with a Cherenkov threshold of 10 MeV for electrons and 2.8 GeV for charged
pions.\\
LUCID is one of the main ATLAS on-line monitors for instantaneous and integrated luminosity measurement. The principle of the measurement is based on 
the fact that the average number of interactions in a bunch crossing is proportional to the number of particles detected in LUCID.\\
In order to provide the actual luminosity rather than the change in luminosity, LUCID must be calibrated using a known absolute luminosity.
At the beginning of LHC running the calibration procedure is based on LHC parameters using a so called  Van Der Meer or beam separation scans. 
The present accuracy from this method  is 11 $\%$ but this might be improved in the future \cite{lumi_conf_note}. Later standard candles like Z-boson production can be used. 
Here the estimated accuracy is in the 5-8 $\%$ range. Ultimately ,LUCID will be calibrated using information from ALFA measurements that should allow to reach an accuracy of
 $\sim$ 2-3 $\%$ on the absolute luminosity (see sec. \ref{sez:ALFA_el}). 
\subsection{MBTS}
\label{sez:MBTS}
MBTS (Minimum Bias Trigger Scintillators) \cite{ATLAS} are segmented scintillator paddles quite close to the beam-pipe. The system consists of 32 scintillator paddles, 
2 cm thick, organised into 2 disks, one on each side of the interaction point of ATLAS.
The system is placed between inner detector and end-cap cryostat and provides a pseudorapidity coverage 2.1 $<|\eta|<$ 3.8 .\\
The main purpose is to provide a trigger on minimum collision activity during the proton-proton collisions at low luminosities.
The apparatus is particularly well suited for the measurement of the start-up LHC luminosity. Because of heavy radiation, it is expected that the inefficiency of
the MBTS will increase after some time of higher luminosity operation. For this reason MBTS will be active only during the initial running phase where the 
average number of interactions per bunch crossing is expected to be low.
\subsection{ZDC}
\label{sez:ZDC}
The ZDC (Zero Degree Calorimeter) \cite{ZDC} is placed at 140 m on both sides of interaction point in the TAN region (target absorber for neutrals) between the 
tubes at the point where the single beam pipe splits into two.
It is a sampling calorimeter composed of four modules: one electromagnetic and three hadronic tungsten/quartz calorimeters.
The ZDC is able to measure neutral particles at pseudorapidity $|\eta|>$ 8.3 .\\
At the LHC startup phase, in the early data taking period, the electromagnetic module will not be installed and its position will be occupied by the LHCf experiment \cite {LHCf}.
After initial running LHCf will be removed and the full ZDC installed. \\
The main purpose of the ZDC is to measure the centrality of the collisions in heavy ion runs.
During proton-proton collision ZDC is used for beam halo, beam gas suppression, luminosity monitor 
and also as an additional minimum bias trigger. 
It will also be used to tag diffractive processes. 
When the luminosity of $10^{33}$  $cm^{-2}s^{-1}$ will be reached, the ZDC modules will be removed in order to minimize the radiation damage.
The ZDC will be reinstalled for heavy ion runs. 
\section{Forward Physics Measurements}
\label{sez:FPM}
Forward region instrumentation at LHC provides a new window to QCD
physics. In this section a short review of the ATLAS forward physics measurements that have just started with early data taking from December 2009 and that are planned to be done 
in the future 
is presented.
\subsection{Soft Diffraction}
\label{sez:Soft}
Single (double) diffraction is a low-t process in which a colour singlet (i.e. Pomeron) is exchanged between the two protons and one
(both) of the protons breaks up into a dissociative system. Diffractive events can be tagged by identifying a rapidity gap between
the outgoing proton and the dissociative system for the single diffraction (SD) or between the two dissociative systems in the double diffraction (DD) case.
Both single and double diffractive dissociation have large cross section of the order of 10 mb. \\
There are two approaches that will be used to measure soft SD at ATLAS. The first one will be focused on the invariant mass of the dissociated system $M_{X}$ and the fractional 
longitudinal momentum loss $\xi = \frac{M_{X}^2}{s}$ (where $s$ is the center of mass energy for the proton proton collisions).
It is clear thet events with low-$\xi$ will be contained only in the forward detectors, whereas high-$\xi$ events will have activity in many areas of the central detector as
well. The dissociative system will be identified using the inner detector, calorimeters, LUCID and the ZDC. The last two subdetectors and MBTS will be used as trigger.\\
The second approach implies the use of the ALFA subdetector. The outgoing proton in SD exchange can be tagged and measured using special LHC runs with high-$\beta^{*}$ optics at
a luminosity of $10^{27}$ $cm^{-2}$ $s^{-1}$. ALFA will be able to measure the longitudinal momentum loss $\xi$ directly using $\xi = 1- \frac{|p_{o}|}{|p_{i}|}$ ( where $p_{i}$ and $p_{o}$
are respecively the longitudinal momenta of the incoming and outgoing  protons).
The resolution of the $\xi$ measurement is approximately 8$\%$ for $\xi =0.01$, falling to $\approx$ 2$\%$
for  $\xi =0.1$. 1.2-1.8 million of events are expected in 100 hours at a luminosity of $\approx$ $10^{27}$  $cm^{-2}s^{-1}$ .
\subsection{Hard Diffraction}
\label{sez:Hard}
An interesting measurement is to look for hard scattering events with gap on one side of the detector.
The aim is to study diffractive parton density functions (dPDF), the ratio of single diffractive (SD) di-jets to non
diffractive (ND) di-jets and the ratio of double pomeron (DPE) to single diffractive di-jets.\\
A few thousand SD di-jet events in 100 $pb^{-1}$ with $E_T>$ 20 GeV (after trigger prescale in Level 1 trigger for low transverse energy jets and gap requirement) are expected.
\subsection{Central Exclusive Di-jet Production}
\label{sez:CEP}
Central exclusive di-jet production (CEP) is defined as the process $pp \rightarrow p \, + \, jj \, + \, p$ where '+' states for a large rapidity gap from the outgoing protons.
In these events all of the energy lost by the protons goes into the production of a hard central system without any other activity. One of the possibility is that this system will 
be constitued from the
Higgs boson: this is why  recently such kind of process has received a great deal of attention.\\
A few hundred di-jet CEP events after trigger and analysis cuts with $E_T>$ 20 GeV in 20 $pb^{-1}$ of data are expected. The main background comes from double diffractive di-jet
production. To overcome this background, the idea is to measure the di-jet mass fraction defined as $R_{jj} = \frac{M_{jj}}{M_{calo}}$ where $M_{jj}$ is the invariant mass of the
di-jets and $M_{calo}$ is the mass of all energy deposit in the calorimeter. Typically an exclusive event will have $R_{jj} \sim$ 1 while inclusive/diffractive events will have
$R_{jj} \ll$ 1.
\subsection{Gaps between Jets}
\label{sez:Gaps}
Gaps between jets arises from a 2 $\rightarrow$ 2 scatter via a colour singlet exchange. The typical signature is constitued by two high $p_t$ jets separated 
in the detector by a large pseudo-rapidity gap $\Delta \eta >$ 3. This process has been previously measured at HERA \cite{gap_hera} and the Tevatron \cite{gap_tevatron} but due to the increase of center of
mass energy, an improved measurement should be possible at the LHC. ATLAS should be able to reach a gap fraction up to$\Delta \eta \sim$ 9
$\rightarrow$ 9.5.	
We aim to first rediscover this process before studying the dependence on the gap size and jet energies in the kinematic regime of the ATLAS
detector. Colour singlet exchange also provides a useful early opportunity to study jet reconstruction
and triggering in the forward calorimeter.
Measurements should be possible with 10 $pb^{-1}$ of data.
\section{Possible Future: the AFP}
\label{sez:AFP}
The AFP (ATLAS Forward Protons) \cite{AFP} is a project of installing silicon and fast timing forward detectors at 220 m (AFP220) and 420 m (AFP420) from the ATLAS interaction point 
for measurements at high luminosity. AFP420 will approach the beam down to 5 mm while AFP220 should reach 2-3 mm. 
Being so close to the beam at luminosities in the $10^{33}-10^{34}$  $cm^{-2}$ $s^{-1}$ range requires very radiation hard detectors. 
Another important requirement is the very good position resolution in order to obtain a high mass resolution of the order of 1-2 $\%$ for 420+420 and 4-5 $\%$ for 420+220 configurations.
The position and angular resolutions are required to be respectively of the order of 10 $\mu$m and 1 $\mu$rad for both AFP220 and AFP420. In order to satisfy those requirements, a 3D silicon
detector has been chosen. A precision time of flight (ToF) system composed of Cherenkov photon detectors with a resolution of 10-20 ps has been chosen 
to identify the primary vertex. This allows
to obtain a large reduction in overlap backgrounds. \\
The capability to detect both outgoing protons in diffractive and photoproduction processes opens up the possibility  
for a rich QCD, electroweak, Higgs and beyond the Standard Model experimental program.
A complete summary of the forward physics program at LHC using the proton taggers is described in \cite{AFPPhys}.
\section{Summary and conclusions}
\label{sez:summary}
ATLAS has a variety of Forward Detectors (MBTS, LUCID, ZDC, ALFA) that will allow to exploit several Forward physics topics and to determine the absolute luminosity with high precision.
We expect to be able to study soft single, double diffraction and gaps between jets when an integrated luminosity of the order of 10 $pb^{-1}$ will be reached. 
Exclusive di-jet and single diffractive di-jet production measurements will be possible increasing the value of integrated luminosity.
After 2010 ATLAS expect to use the ALFA subdetector to study single diffraction and elastic scattering. The elastic scattering
measurement is interesting in itself but will also be used for a
measurement of the absolute luminosity of the experiment. 
Finally the AFP project will allow to extend the ATLAS physics program at high luminosities.


\newpage
\clearpage
\setcounter{affil}{0}
\setcounter{section}{0}
\setcounter{figure}{0}
\setcounter{table}{0}
\setcounter{equation}{0}

\vspace*{4cm}
\title{Diffraction at TOTEM}

\author{K. \"Ostergberg \\
(on behalf of the TOTEM Collaboration)}

\affiliation{Helsinki Institute of Physics and Department of Physics, University of Helsinki, P.O. Box 64, FI-00014 Helsinki, Finland }
\begin{abstract}
The TOTEM experiment at the LHC is devoted to a deeper understanding of the proton structure with the main goal being a precise measurement of the
total proton-proton cross-section at LHC energies. In addition, TOTEM will measure
elastic scattering over a wide $|t|$-range and perform comprehensive studies of diffractive processes and forward event topologies in inelastic proton-proton collisions. This article reviews the status of the experiment as well as the expected physics, especially in terms of diffraction, from the data taken in 2010 and to be taken in 2011, both in standard LHC
and special dedicated TOTEM runs.
\end{abstract}
\maketitle
\addcontentsline{toc}{part}{Diffraction at TOTEM - {\it K.\"Osterberg}}
\section{Introduction}\label{sec:intro}

TOTEM~\cite{Totem_TDR} is the only LHC experiment exploring the forward region by charge particle measurement at pseudorapidities $\eta$ larger than three and thus TOTEM physics programme, aiming at an deeper understanding of the proton ($p$) structure, is very different compared to the other LHC experiments. The main goal of TOTEM is a precise measurement of the total $pp$ cross section $\sigma_{tot}$ at 1-2 \% level using the luminosity-independent method described in detail
elsewhere~\cite{Totem_TDR, Totem_JNST, Jiri_sigmatot}.

Elastic and diffractive protons scattered at very small angles are measured by TOTEM with proton detectors embedded in "Roman Pots" (RP), placed along the LHC beam line on both sides of interaction point 5 (IP5). Charged particles produced by inelastic interactions in the $\eta$-range of 3.1 $\le |\eta| \le$ 6.5 are measured on both sides of IP5 with full azimuthal coverage by the inelastic telescopes T1 and T2 of TOTEM embedded in the CMS end-caps. This will allow TOTEM to determine, in addition to $\sigma_{tot}$, the elastic scattering cross section over a wide $|t|$-range ($\sim$10$^{-3} \le |t| \le$ 10 GeV$^2$) and study soft diffractive processes and the charged particle event topology in the forward region. These studies will be complemented at a later stage by studies of a wide range of diffractive and forward physics topics~\cite{CMS_TOTEM_TDR}, taking advantage of an unprecedented tracking and calorimetry coverage in $\eta$, in common data taking with the CMS~\cite{CMS} experiment.

\section{Experimental apparatus and status}\label{sec:detec}

\begin{figure}
\includegraphics[scale=1.15]{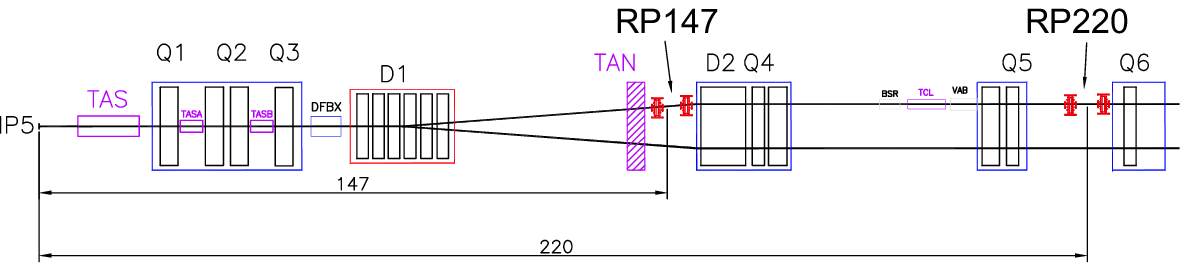} \caption{The LHC beam line on one side of interaction point 5 with the positions of the TOTEM Roman Pots at 147 and 220 m (RP147 and RP220) indicated.}
\label{fig:totem_rp}
\end{figure}

To measure elastically and diffractively scattered protons requires the reconstruction of the protons tracks by detectors moved as close as $\sim$1 mm from the center of the outgoing beam. This is obtained with two RP stations installed, symmetrically measuring protons circulating both
clockwise ("sector 56") and anticlockwise ("sector 45"), at distances of
$\sim$ 147 m (RP147) and $\sim$ 220 m (RP220) from IP5 (Fig.~\ref{fig:totem_rp}). Each RP station is composed of two units at a distance of several meters allowing a local track reconstruction. Each unit consists of three
pots, two approaching the beam vertically from the top and the bottom and one horizontally to complete the acceptance for diffractively scattered protons. Each pot contains a stack of 10 planes of silicon strip "edgeless" detectors with half with their strips oriented at an angle of $+$45$^o$
and half at an angle of $-$45$^o$ with respect to the edge facing the beam. These detectors~\cite{RP_Silicon}, designed by TOTEM with the objective of reducing the insensitive area at the edge facing the beam to only a few tens
of microns, have a spatial resolution of $\sim$20 $\mu$m. High efficiency up to the physical detector border is essential in view of maximizing the elastic and diffractive proton acceptances. All detectors for the 220 m stations were installed in 2009 and succesfully commissioned in 2010, while those for the 147 m stations have been tested and will be installed during
the 2010/11 winter shutdown.

\begin{figure}
\includegraphics[scale=0.325]{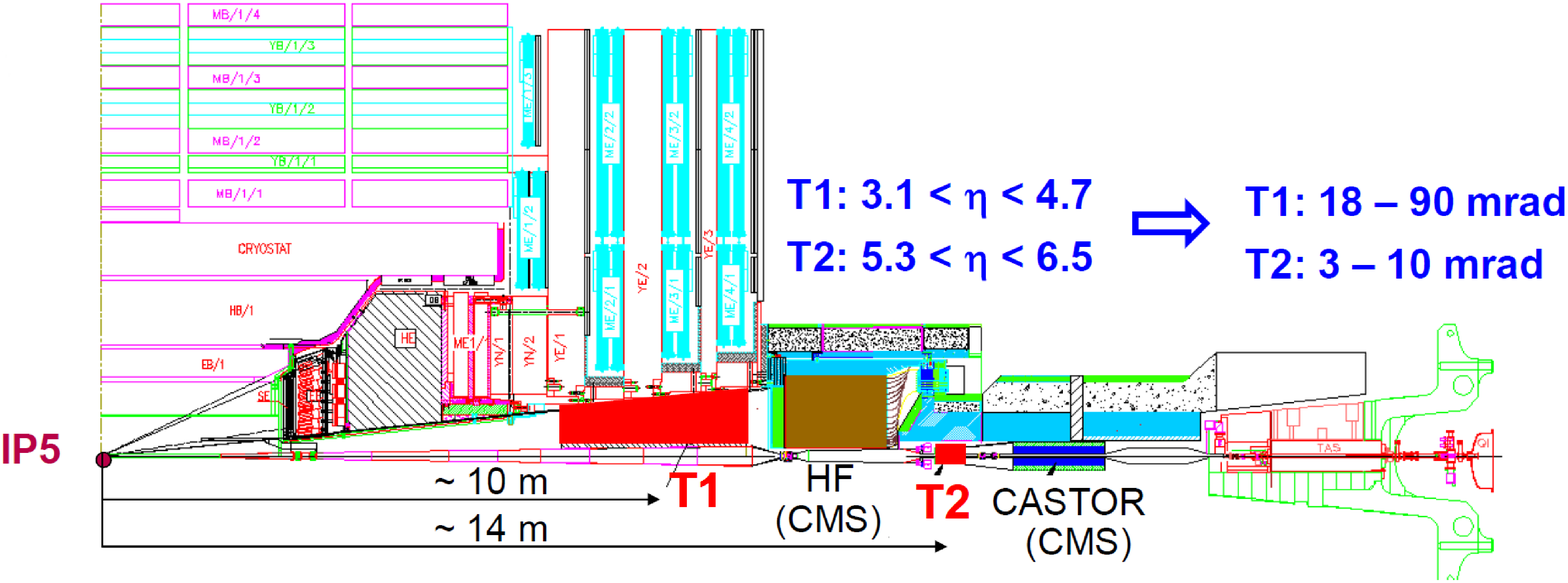}
\caption{The positions of the TOTEM T1 and T2 telescopes in the CMS forward region. The pseudorapidity and angular coverage of T1 and T2 are given.}
\label{fig:totem_T1T2}
\end{figure}

To measure charged particles produced in inelastic proton-proton collisions with a very good efficiency, TOTEM has the T1 and T2 telescopes embedded in the forward region of CMS (Fig.~\ref{fig:totem_T1T2}). The T1 telescope consists of Cathode Strip Chambers (CSC) and T2 telescope of Gas Electron Multipliers (GEM)~\cite{GEM}. The $\eta$ coverage of T1 and T2 is 3.1
$\le |\eta| \le$ 4.7 and 5.3 $\le |\eta| \le$ 6.5, respectively. Each T1 telescope
arm consists of five planes made up of six trapezoidal formed CSC’s with a spatial resolution of $\sim$1 mm.
Each T2 telescope arm consists of 20 semicircular shaped triple-GEM detectors with a spatial resolution
of  $\sim$100 $\mu$m in the radial direction and a inner radius that matches the beam-pipe. Ten aligned detectors
mounted back-to-back are combined to form one T2 half arm on each side of the beam-pipe. The full T2 telescope was installed in 2009 and successfully commissioned in 2010. The full T1 telescope has been commissioned with particles beams at the CERN SPS H8 test line and will be installed during the 2010/11 winter shutdown. TOTEM has been taking data with all
installed detectors (both T2 arms and
both RP220 stations) since the start of LHC operations in 2009. Data
taking settings like the timing and thresholds of both detectors has been carefully tuned in order to optimize the running conditions.

\begin{figure}
\includegraphics[scale=0.65]{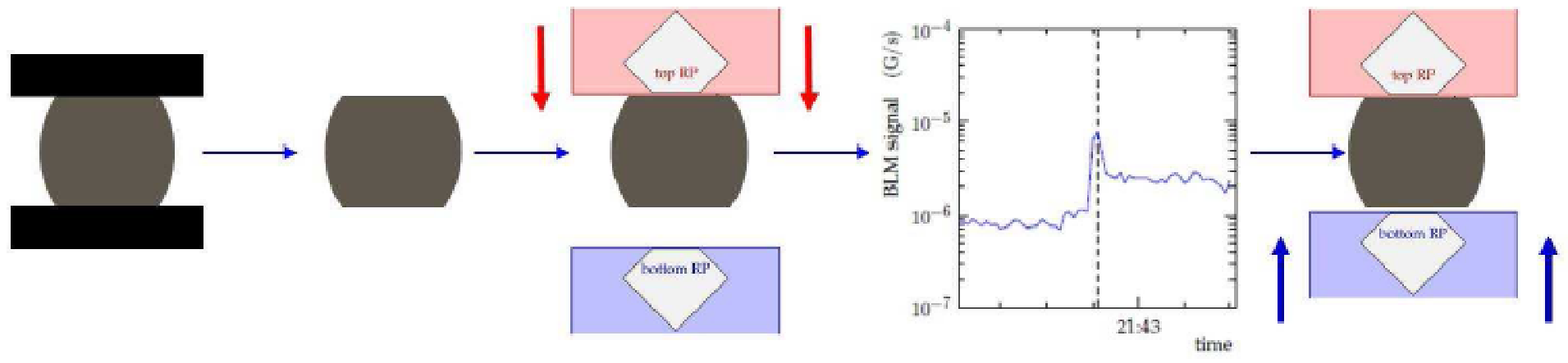}
\caption{The principle of the collimator-based RP alignment. The
collimators scrape the beam (left) creating a beam with sharp edges
(second left). The top RP approaches the beam in small steps
(middle) until large signal seen in beam loss monitor (second right).
The procedure is repeated for the bottom RP (right).}
\label{fig:totem_align}
\end{figure}

The knowledge of the positions of the RPs w.r.t. beam center is a
key issue. Therefore a collimator-based RP alignment procedure (see Fig.~\ref{fig:totem_align}) has been developed
in collaboration with the LHC operators taking advantage of the sharp beam edges produced by the LHC collimation system. Each RP is approaching the beam in small steps until a large signal increase is seen in the Beam Loss monitors directly downstream indicating that the RP is out of the shadow of the collimators and thus at the same number of $\sigma_{beam}$ from the beam center as the collimators. The beam center position obtained from the
collimator-based alignment is verified and corrected vertically and
horizontally using the vertical position distribution from all particles
in the horizontal pot and using the horizontal position distribution from
a pure sample of elastic candidates, respectively.
As a result of these collimator-based alignment exercises, the LHC operators
has gained confidence on the absolute positions of the RPs to allow the vertical (horizontal) RPs to approach closer to the beam in steps,
taking data at 30 (30), 25 (30), 20 (25) and 18 (20) times the beam size,
$\sigma_{beam}$, in standard LHC runs at $\sqrt{s}$ = 3.5 TeV and
$\beta^*$ = 3.5 m with nominal bunch intensity. Most ($\sim$ 3.9 pb$^{-1}$)
of the integrated luminosity was taken with the vertical and horizontal RPs
at 18$\sigma_{beam}$ and 20$\sigma_{beam}$, respectively. In addition, the RPs have taken data with vertical and horizontal RPs at
7$\sigma_{beam}$ and 16$\sigma_{beam}$, respectively, during a
collimator-based alignment exercise and during a dedicated few hours TOTEM run
at the end of the data taking of 2010 corresponding to a total integrated
luminosity of $9.5$ nb$^{-1}$. In the latter, a combined RP and T2 data taking was performed on a specially introduced bunch with a reduced number of
protons ($\sim 10^{10}$) for creating collisions without pileup for
precise studies of forward charged particle multiplicity and diffractive 
processes.

\section{Physics of the 2010 runs and potential of the 2011 runs}\label{sec:phys}

The diffractive studies in TOTEM covers several diffractive processes:
elastic scattering $pp \rightarrow pp$, single
$pp \rightarrow p + X$, double $pp \rightarrow Y + X$ and
central diffraction $pp \rightarrow p + X + p$, where $+$ indicates a
rapidity gap, an $\eta$-range without final state particles,
and $X$ and $Y$ hadronic systems.

\begin{figure}
\includegraphics[scale=0.715]{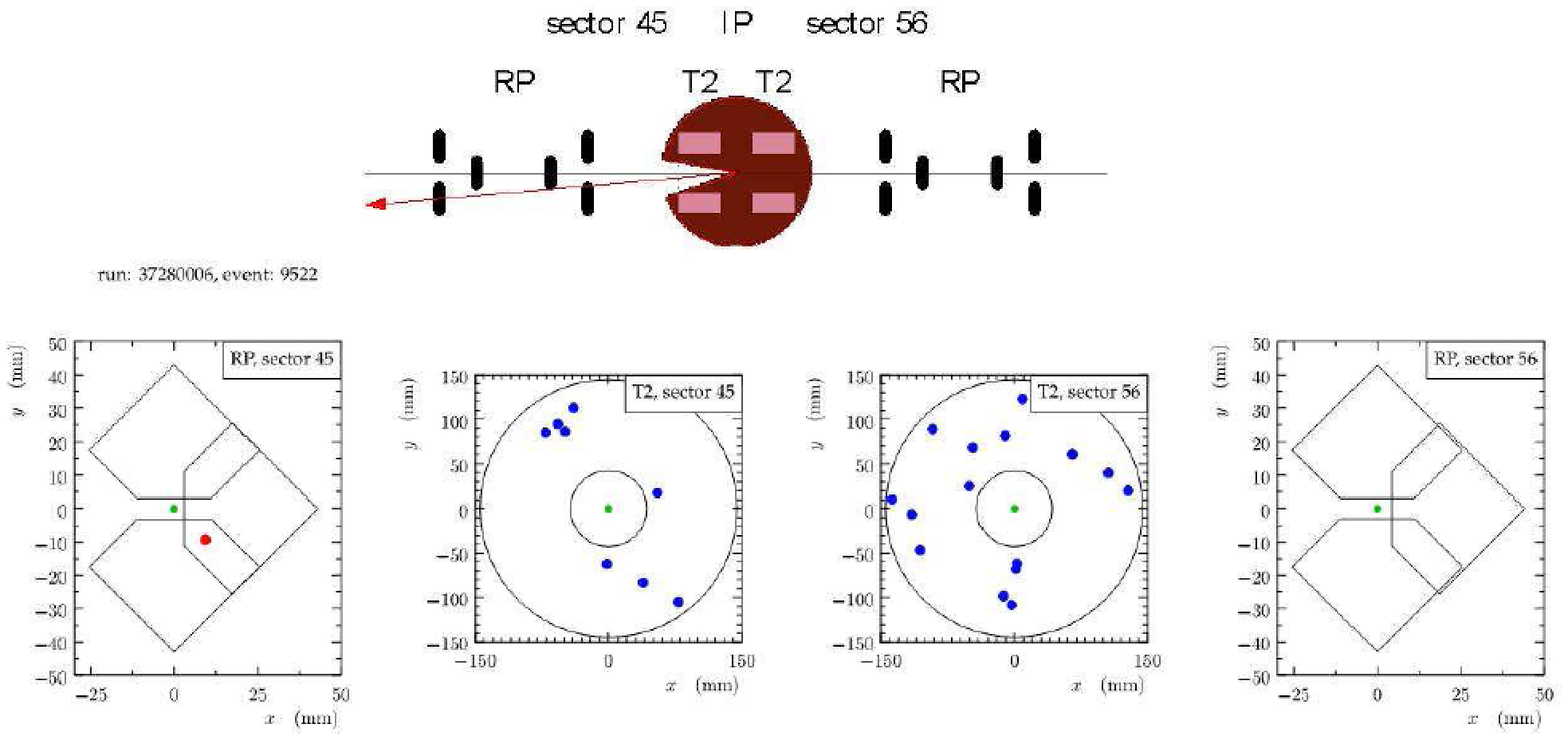}
\includegraphics[scale=0.715]{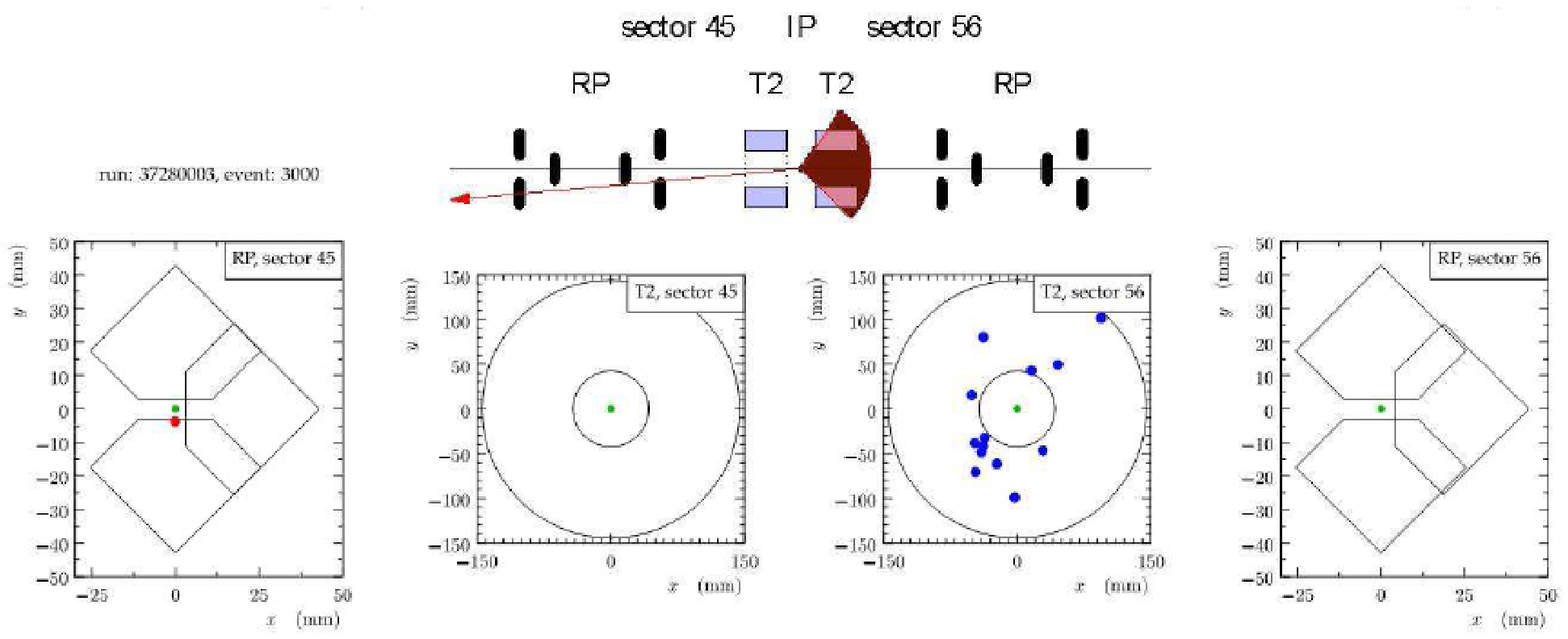}
\caption{Top: A high mass single diffraction candidate from data taken in collisions of bunches with reduced number of protons October 2010. 
Bottom: A low mass single diffraction candidate from the
same run.}
\label{fig:totem_SD}
\end{figure}

The 2010 LHC optics conditions of $\sqrt{s}$ = 3.5 TeV and
$\beta^*$ = 3.5 m allow both protons in an elastic $pp$ collision be
detected by sector 45 top(bottom) and sector 56 bottom(top) RP
combinations if $|t_y|$ $\ge$ $\sim$0.4
and $|t_y|$ $\ge$ $\sim$2.5 GeV$^2$ in runs with vertical RPs at 
7$\sigma_{beam}$ and 18$\sigma_{beam}$, respectively. Diffractive 
protons can be seen in
the 2010 LHC optics conditions either in the horizontal RPs (and
possibly also in either top or bottom RPs if $|t_y|$ is sufficiently 
large) if their
$\xi$ $\ge$ $\sim$2 \% or only the vertical RPs with similar $|t_y|$
requirements as elastic protons (if $\xi \le$ $\sim$2 \%). By
selecting protons reconstructed in the horizontal RPs or only in the
vertical RPs, one can enforce certain kinematics on the diffractive final
state. As a demonstration, high and low mass single diffractive (SD)
events can be selected by requiring the proton to be reconstructed in the
horizontal RP or only in the vertical RP, respectively, as shown in
Fig.~\ref{fig:totem_SD}. This kinematical selection can then be confirmed by the charged particles reconstructed in the T2 measuring charged particles
5.3 $\le |\eta| \le$ 6.5 on both sides of IP5 with high mass SD having charged particles in both sides and low mass SD having charged particles only on the
opposite side with respect to the proton (see Fig.~\ref{fig:totem_SD}).
The expected physics results of the full analysis of the
data taken by TOTEM in 2010 at $\sqrt{s}$ = 7 TeV are:                                   \begin{itemize}
\item a determination of the elastic scattering cross-section
for $\sim$0.4 $\le |t| \le$ $\sim$5 GeV$^2$;
\item a study of the kinematics of central diffractive events in the
kinematically accessible range i.e. both $p$'s with large $\xi$'s,
one $p$ with large $|t|$ and other with large $\xi$ or
both $p$'s with large $|t|$'s and a determination of the mass distribution of high mass central diffractive events;
\item a study of the $\xi$ and rapidity gap correlation for single
diffractive events and a determination of the mass distribution for high mass single diffractive events combining RP and T2 data;
\item a determination of the forward charged particle multiplicity in
the 5.3 $\le |\eta| \le$ 6.5 region;
\item a study of the charged particle multiplicity correlation over a
large rapidity range i.e. the forward-backward charged particle
multiplicity correlations in inelastic $pp$ collisions.
\end{itemize}
The analysis of the data and the estimation of the systematic errors
affecting these measurements is still in progress. The RP measurements
on one hand are especially affected by uncertainties related to the LHC
optics, the RP alignment and the trigger efficiencies and the T2 measurements on the other hand
by uncertainties related to production of secondaries from interactions
with material in front of the T2. Of the physics topics listed above,
only the determination of the elastic cross-section in the high end of the
accessible $|t|$-spectrum will be statistics limited. Therefore, in early 2011, the collimator-based RP alignment should be repeated with the new optics
in nominal beam conditions to enable constant data taking with the RPs at 18$\sigma_{beam}$ or closer in normal LHC runs to improve significantly
the statistics
of elastic scattering at the highest accessible $|t|$-values.

The installation of the T1 and the RP147 in the winter shutdown of 2010/11
and the efforts of the LHC machine to prepare for high $\beta^*$ running
in 2011 enables TOTEM to start pursuing its full physics programme during
2011. Having the T1 increases the charged particle detection range considerably from 5.3 $\le |\eta| \le$ 6.5 to 3.1 $\le |\eta| \le$ 6.5,
enabling the inelastic events to be classified into non-diffractive minimum
bias, single and double diffractive events with high purity and efficiency.
A low $\beta^*$ run with collisions of bunches with reduced number of protons
will allow the determination of the inelastic rate to a few \% and
individual process rates at 5-10 \%, and to extend the forward charged
particle studies to the range 3.1 $\le |\eta| \le$ 6.5. At $\beta^*$ = 90 m, all protons with $|t| \ge$
$\sim$0.1 GeV$^2$ will be detected by the RP220s irrespective of their
$\xi$ if the vertical RPs are approached to 7$\sigma_{beam}$.
Short runs at $\beta^*$ = 90 m opens up the possibility for following
physics:
\begin{itemize}
\item a determination of the total $pp$ cross-section $\sigma_{tot}$
and luminosity at
5 \% and 7 \% level, respectively, using the luminosity-independent
method. In addition, in conjunction with the $\sigma_{tot}$ measurement,
the elastic and inelastic cross sections will be determined at 5 \% level;
\item a comprehensive study of soft singel and central diffraction at any diffractive mass value.
\end{itemize}

\section{Summary and Conclusions}
The TOTEM experiment has successfully commissioned its Roman Pot (RP)
detectors at 220 m and its forward inelastic telescope T2. In 2010,
TOTEM has also taken data with the vertical (horizontal) RP detectors
at a distance of 7$\sigma_{beam}$ (16$\sigma_{beam}$) in special and
at a distance of 18$\sigma_{beam}$ (20$\sigma_{beam}$) in nominal LHC
runs allowing TOTEM to complete in its first physics
measurement, the measurement of the elastic $pp$ cross-section at
$\sqrt{s}$ = 7 TeV for the 0.4 $\le |t| \le$ 5 GeV$^2$ range, in the
upcoming months. In addition,
TOTEM will in the near future produce results on high mass single and
central diffraction as well as the forward charged particle multiplicity
and forward-backward charged particle multiplicity correlation for
in the 5.3 $\le |\eta| \le$ 6.5 range.

With the installation of the T1 telescope in the 2010/2011 winter shutdown
and the preparation of the $\beta^*$ = 90 m optics from the LHC machine side, TOTEM will be allowed to start pursuing its full physics programme in 2011
by taking data in a  combination of nominal LHC runs, dedicated high
$\beta^*$ TOTEM runs and special low $\beta^*$ runs with bunches with reduced number of protons.
$\beta^*$ = 90 m runs will allow a first measurement of the total $pp$
cross-section at 5 \% level and in conjunction measurements of the luminosity,
elastic and inelastic cross-sections at 5-7 \% level. In addition, 
$\beta^*$ = 90 m runs will allow a
comprehensive study of single and central diffractive processes at any mass
as well as a determination of the cross-section of non-diffractive minimum
bias, single and double diffraction at the 5-10 \% level. The forward
charged particle multiplicity in the 3.1$ \le |\eta| \le$6.5 region and
corresponding forward-backward multiplicity correlation will be studied either in a
$\beta^*$ = 90 m run or in a special low $\beta^*$ run with bunches with reduced number of protons.

\newpage
\clearpage
\setcounter{affil}{0}
\setcounter{section}{0}
\setcounter{figure}{0}
\setcounter{table}{0}
\setcounter{equation}{0}

\preprint{}

\title{LHCf and the connection with Cosmic Ray Physics }

\author{O. Adriani, on behalf of the LHCf Collaboration}
\email{adriani@fi.infn.it}
\affiliation{%
University of Florence and INFN Sezione di Firenze
}%


\begin{abstract}
The LHCf experiment has been designed to precisely measure the $\gamma$ and $n$ energy spectra in the very forward region at LHC; these measurements are important to calibrate the Monte Carlo models widely used in the High Energy Cosmic Ray induced air showers analysis, allowing a better understanding of the experimental systematics. 
LHCf has started the data taking at the end of 2009, at 900 GeV center of mass energy, and has later on continued the data taking during the first part of 2010, when LHC center of mass energy was increased up to 7 TeV. Preliminary results of 900 GeV and 7 TeV data taking periods are reported in this paper.
\end{abstract}

\maketitle
\addcontentsline{toc}{part}{LHCf and the connection with Cosmic Ray Physics - {\it O.Adriani}}

\section{\label{sec:intro} LHCf physics case}
The LHCf experiment is devoted to the precise measurements of the energy and transverse momentum spectra of $\gamma$ and neutrons produced in the very forward region proton-proton interactions at the LHC collider, in the $|\eta|>8.3$ 
pseudo-rapidity region. The detector consists of a double arm (Arm1 and Arm2) -- double tower  
sampling and imaging calorimeter, placed at $\pm$ 140 m from ATLAS interaction point (IP1) 
inside the zero-degree neutral absorbers (Target Neutral Absorber, 
TAN). Charged particles from the IP are swept away by the inner beam separation dipole 
before reaching the TAN, so that only photons mainly from $\pi^0$ decays, neutrons and 
neutral kaons reach the LHCf calorimeters. \\
Each calorimeter tower is made 
of 16 layers of plastic scintillators interleaved by tungsten layers as converter, complemented by 
a set of four X-Y position sensitive layers which provide incident shower positions, in order to 
obtain the transverse momentum of the incident primary and to 
correct for the effect of leakage from the edges of the calorimeters.  \\
 While the two calorimeters are 
identical for the calorimetric structure, they slightly differ for the geometrical 
arrangement of the two towers and for the position sensitive layers made by 1 mm$^2$ scintillating fibers 
in one calorimeter and silicon micro-strip layers in the other.
A detailed description of the LHCf detector and it's performances can be found in~\cite{lhcf.jinst}.

The LHCf experiment differs from the other LHC experiment not only in dimensions but also 
for the main physics motivation, which for LHCf is strictly connected to astroparticle physics.
The goal of the experiment is indeed to measure neutral particle spectra to 
calibrate Monte Carlo codes used in High Energy Cosmic Ray (HECR) physics. A good knowledge of 
nuclear interaction model of primary cosmic rays with earth's atmosphere is mandatory to 
better understand many properties of primary cosmic rays, like the 
energy spectrum and the composition, whose knowledge is finally strictly related to our 
capability to understand the origin of high energy phenomena in the Universe.
 Dedicated extensive air shower experiments are taking data since many years and 
have strongly contributed to our understanding of High and Ultra High Energy 
Cosmic Ray (UHECR) Physics. However, the results 
of these experiments are in some cases not fully in agreement and, in addition, 
the interpretation of their data in terms of primary cosmic ray properties 
is strongly affected by the knowledge of the nuclear interactions   
in the earth's atmosphere. This is true, for instance, for the interpretation  
of the behaviour of the energy spectrum in the UHE region, in particular the 
existence of events above the so called GZK cut-off, and the chemical composition 
of cosmic rays.  Indeed, evidence of UHECR, above the GZK cut-off, has been reported
for the first time by the AGASA experiment~\cite{Takeda}.
On the contrary, the results of the HiRes~\cite{Abbasi:2007sv} experiment and, more recently, the ones of 
the Pierre Auger Collaboration~\cite{Yamamoto:2007xj} are consistent with the 
existence of the cut-off. 
The disagreement among data would be 
reduced by adjusting the energy scales of the different experiments to account for 
systematic effects in the determination of 
the particle energy, that might be due to different detecting 
techniques. Similar considerations hold for the interpretation of 
cosmic ray composition since it is directly related to their 
primary sources. Accelerator experiments validating the interaction model chosen are hence essential. 
As a matter of fact air shower development is dominated by the
forward products of the interaction between the
primary particle and the atmosphere. The only available data on the 
production cross-section of neutral pions emitted in the
very forward region have been obtained more than twenty years ago by the 
UA7 Collaboration~\cite{UA7} at the CERN Sp${\mathrm{\overline p}}$S up to an energy of 
10$^{14}$ eV and in a very narrow pseudo-rapidity range. 
The LHCf experiment at LHC has the unique opportunity to take data at 
energies ranging from $\sqrt{s} = 0.9$ TeV up to 14 TeV, thus extending significantly 
the energy range up to a region of great interest for high energy cosmic rays, the region 
between the ``knee'' and the GZK cut-off.

\begin{figure}[h!]
\begin{center}
\includegraphics[width=0.8\linewidth,clip=]{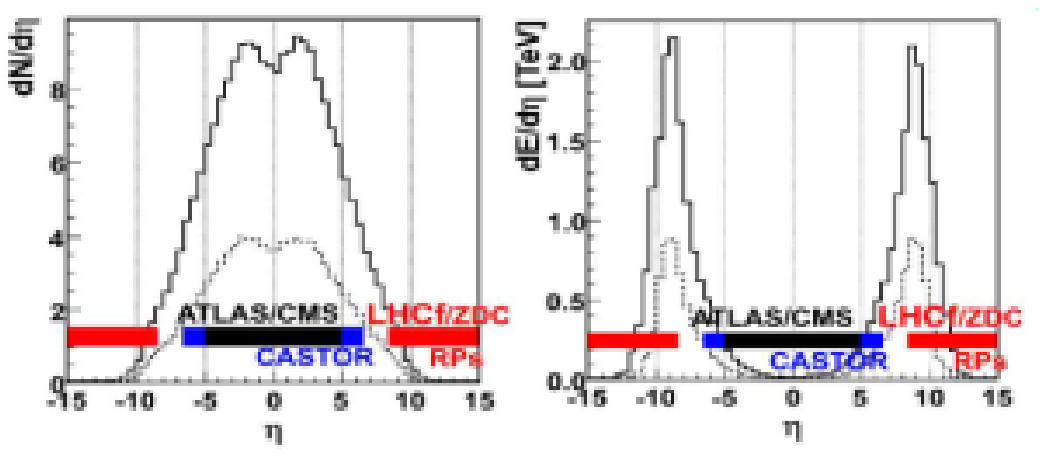}
\end{center}
\caption{Distributions of the number of particles (left side) and of the energy flux (right side) as function of pseudo-rapidity in the p-p collisions at 14 TeV center of mass energy. Typical coverage of the various central and forward detectors are also shown.}
\label{fig.energyflow}
\end{figure}

Additionally, the possibility to install the detector in the region where the single LHC beam pipe splits in two allows us to cover the pseudo-rapidity range $|\eta|>8.3$, where most of the energy flux produced in the p-p collisions is concentrated. Fig.~\ref{fig.energyflow} shows the distribution of the number of particles (left side) and of the energy flux (right side) as function of the pseudo-rapidity; most of the final state particles in the p-p reaction are produced in the central region, but most of the energy flux is concentrated in the very forward region, $6<|\eta|<12$, clearly demonstrating the advantage of a very forward detector in the energy flux measurement. 

\section{\label{sec:experimental} The experimental aspects}
\begin{figure}[h!]
\begin{center}
\includegraphics[width=0.5\linewidth,clip=]{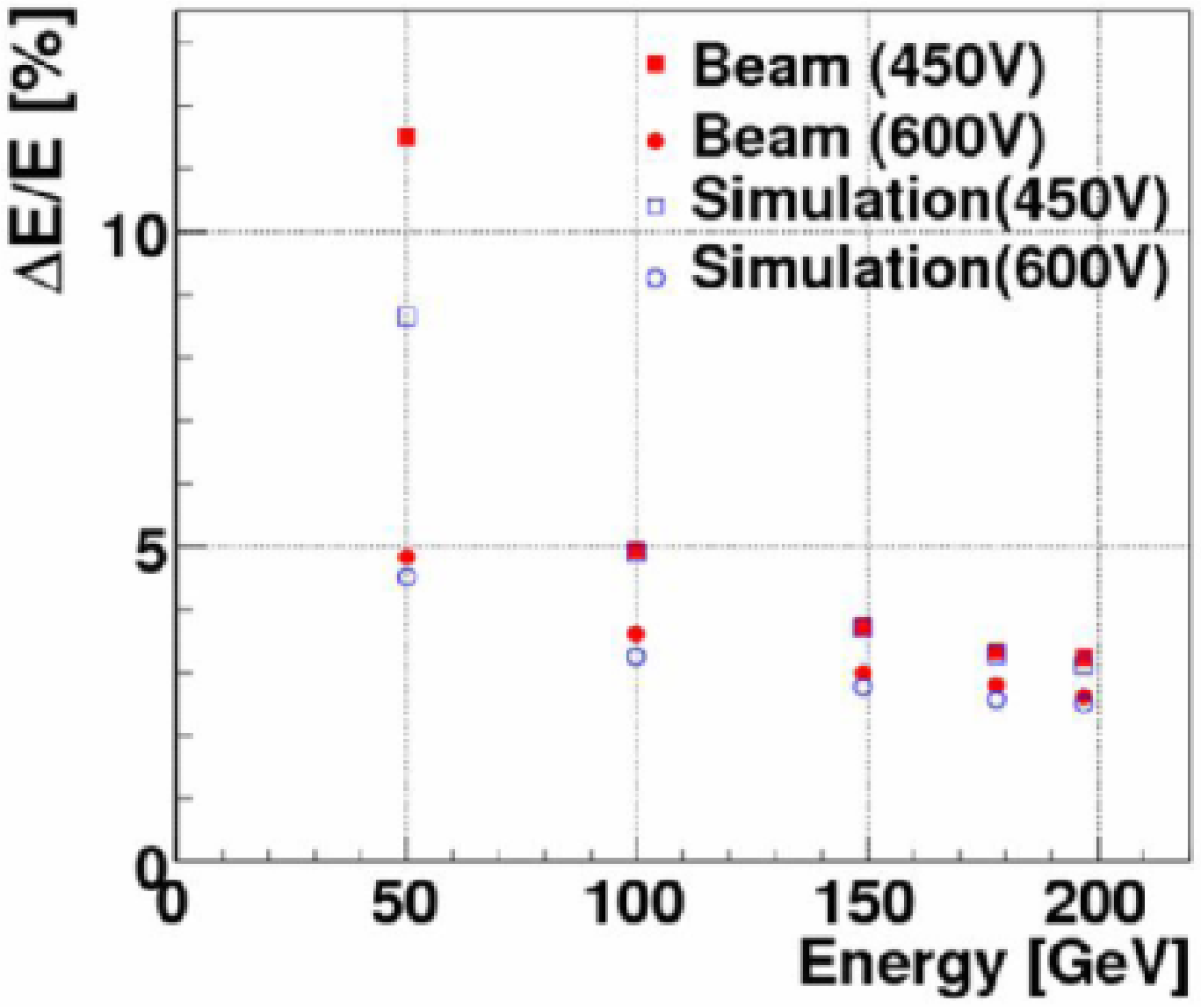}
\end{center}
\caption{Energy resolution for electromagnetic particles measured on beam test (red markers) and expected from the Monte Carlo simulation (empty markers). The results are shown for two different setting of the photomultipliers high voltage, corresponding to Low Gain and High Gain mode.}
\label{fig.energyresolution}
\end{figure}

In order to calibrate the Monte Carlo codes used in HECR physics, the LHCf experiment 
should be able to have a detailed knowledge of the absolute energy scale. For this reason LHCf relies on 
a very precise reconstruction of the $\pi^0$ mass, by reconstructing in the two towers 
the showers from the 2 $\gamma$ from $\pi^0$ decays. \\
The performances of the detector have been careful measured using beam test data and 
well satisfy the design requirements~\cite{lhcf:tdr}. The measured position resolution in locating the shower center for particles above 100 GeV (which is the region of interest for LHCf) is about 200 $\mu$m for 
scintillating fibre layers (ARM1) and 
about 50 $\mu$m for the silicon micro-strip layers (ARM2). \\
The energy resolution for the calorimeters is better than 4\% at 
200 GeV, as can be seen from Fig.~\ref{fig.energyresolution}, that shows the energy resolution for electromagnetic particles
measured on beam test and expected from the Monte Carlo simulation, for two different photomultipliers high voltage setting (Low Gain and High Gain mode). This figure demonstrate the excellent performances of LHCf in the high energy electromagnetic particles reconstruction, despite the small transverse size of the towers (20~x~20 mm$^2$ and 40~x~40 mm$^2$).

\begin{figure}[htb!]
\vfill \begin{minipage}[t]{.5\linewidth}
\begin{center}
{\includegraphics[height=0.85\linewidth,angle=0,clip=]{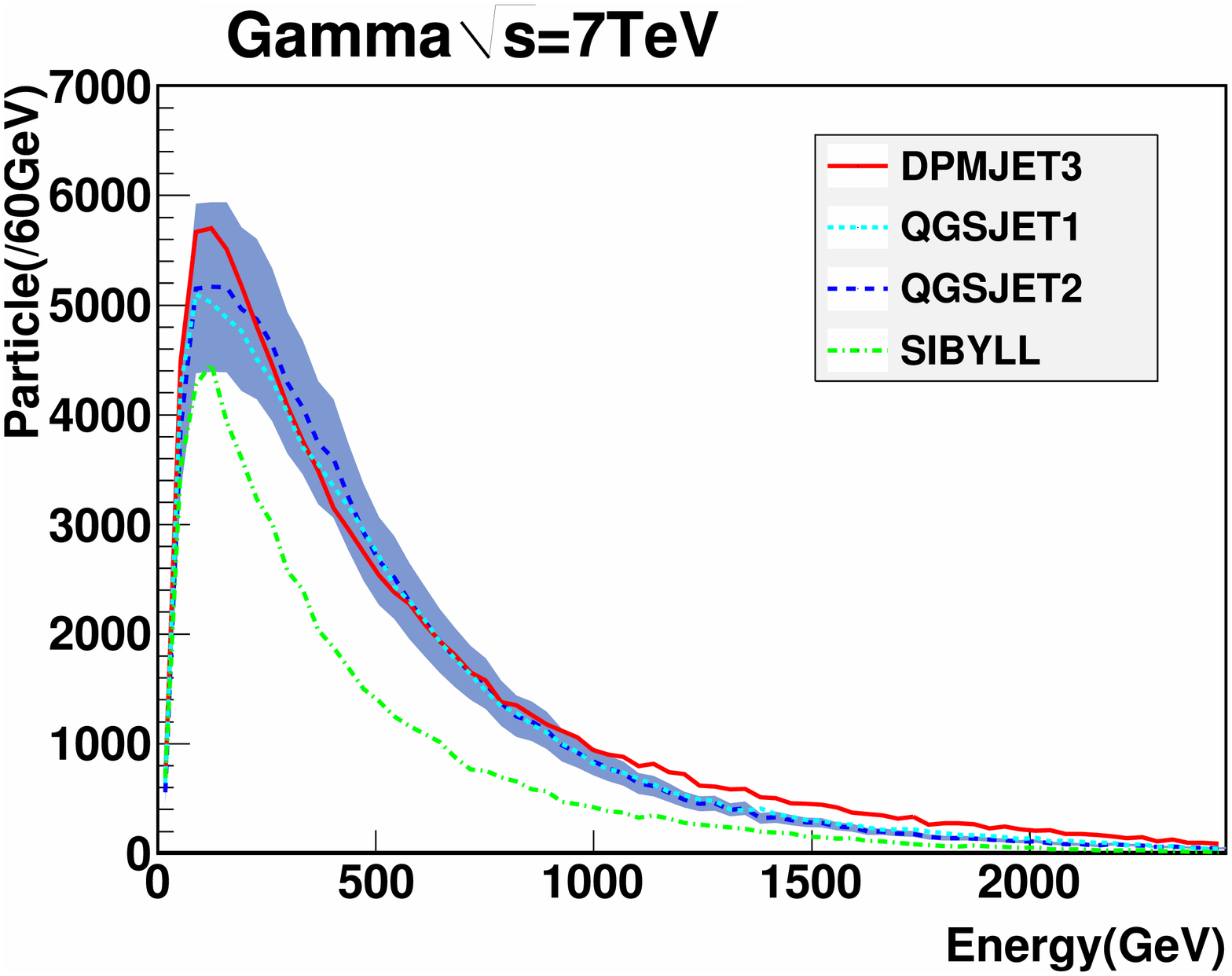}}
\end{center}
\end{minipage}\hfill
\begin{minipage}[t]{.5\linewidth}
\begin{center}
\includegraphics[height=0.85\linewidth,angle=0,clip=]{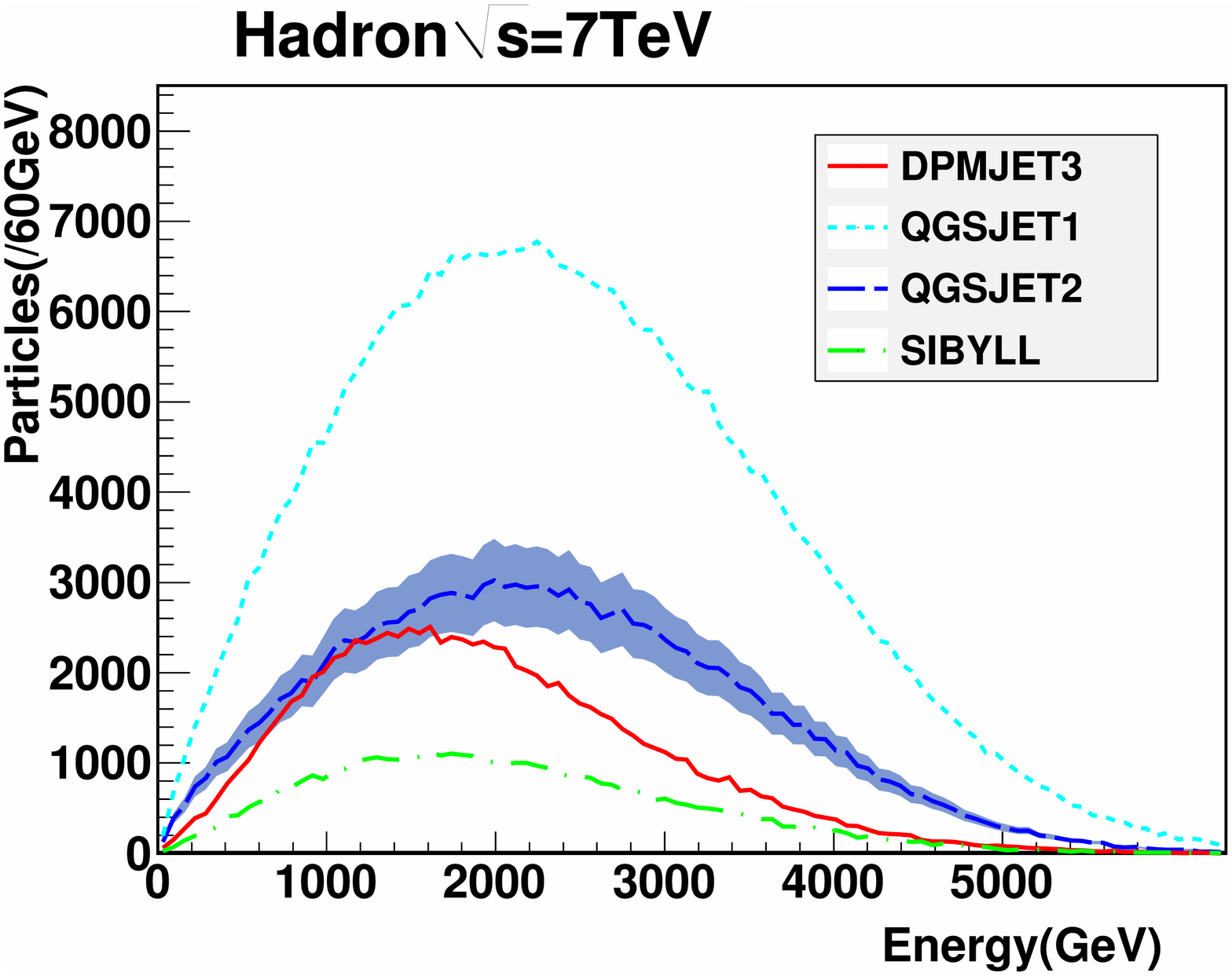}
\end{center}
\end{minipage}
\caption{Expected energy spectrum for photons (left) and neutral hadrons (right) according to 
different interaction models at 3.5+3.5 TeV center of mass energy p-p collisions.}
\label{fig.spectra}
\end{figure}

Fig.~\ref{fig.spectra} shows LHCf expectations for the $\gamma$ and neutron energy spectra 
for few minutes exposure at 10$^{29}$ cm$^{-2}$s$^{-1}$ with 3.5+3.5 TeV center of mass energy p-p collisions.  
Depending on the nuclear interaction
model used, the energy spectra change more or less significantly.  
As can be seen from this plot, the LHCf experiment will be 
able to disentangle 
different interaction models already at lower energy and with very low statistics, thus ensuring a calibration 
of cosmic ray Monte Carlo in an energy range wider than the one expected at the beginning of the project. 

\section{\label{sec:900GeV} Data taking at 900 GeV}
\begin{figure}[htb!]
\begin{center}
\includegraphics[height=8 truecm,clip=]{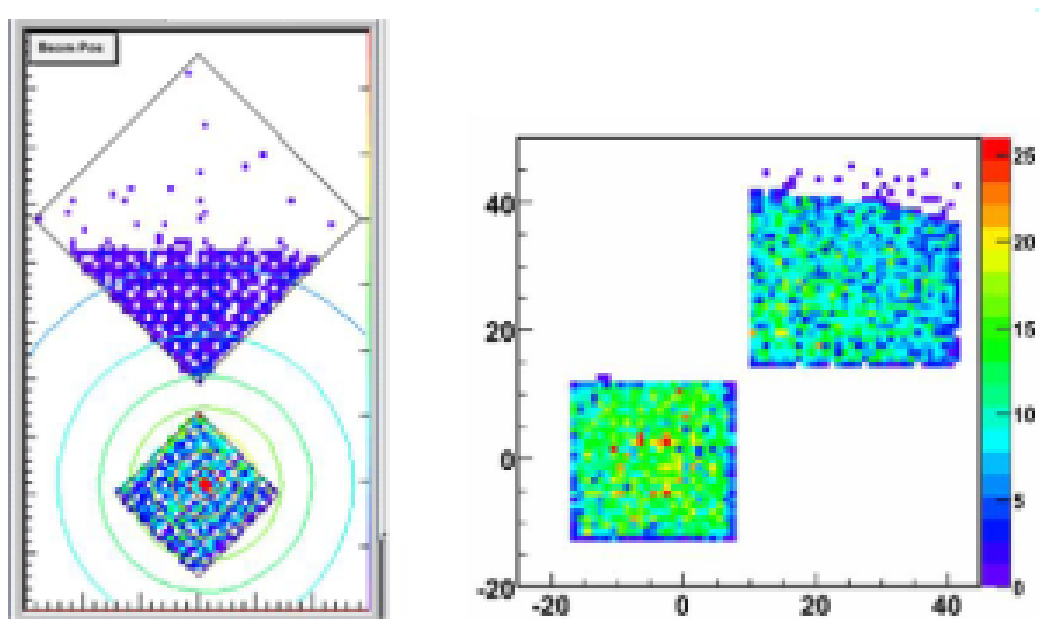}
\end{center}
\caption{Hit maps obtained in a typical LHCf run for the Arm1 (left side) and Arm2 (right side) detectors. The acceptance reduction due to the elliptical beam pipe located close to the bending dipole magnet D1 in the region between interaction point and the LHCf location is clearly visible. }
\label{fig.hitmap}
\end{figure}

\begin{figure}[htbp!]
\begin{center}
\includegraphics[height=8 truecm,clip=]{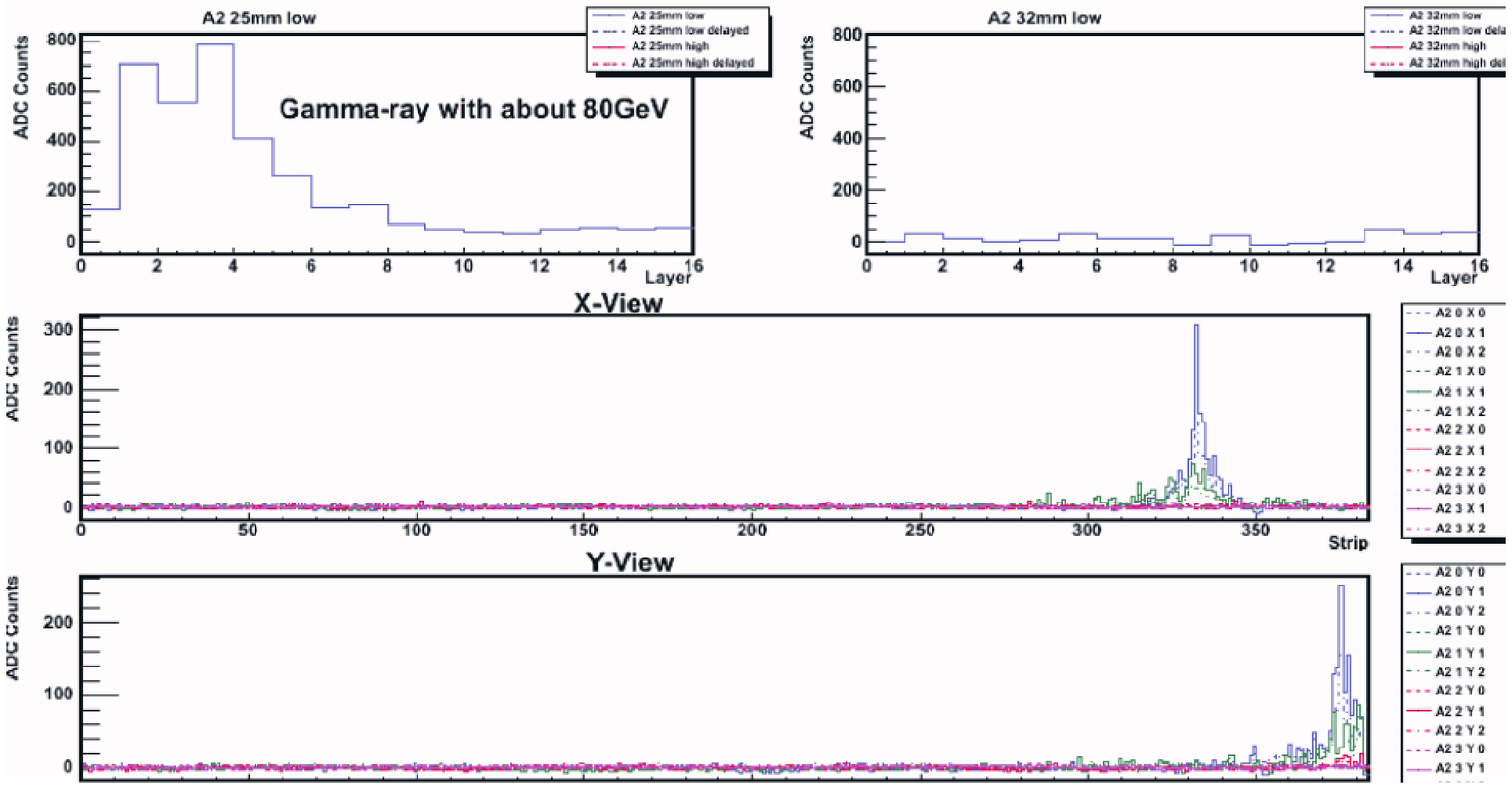}
\end{center}
\caption{A typical $\gamma$ event registered on ARM2 detector in 900 GeV collision data. The two upper panels 
show the longitudinal energy profile deposited on each tower of the calorimeter, while the two lower panels 
show the transverse energy X and Y profile deposited on each of the four silicon layers. }
\label{fig.event900GeV}
\end{figure}

At the end of November 2009 LHC has started to provide collisions to the experiments at 
900 GeV center of mass energy. The LHCf experiment has taken data from December 6 till 
December 15, accumulating about 6500 shower triggers in total on both arm of 
the calorimeter. The hit maps obtained in a typical run for Arm1 and Arm2 detectors are shown in Fig~\ref{fig.hitmap}, clearly showing the effect on the detector acceptance of the elliptical beam pipe close to the bending dipole magnet D1 in the region between interaction point and the LHCf location. 
A typical $\gamma$ event registered on ARM2 detector is shown in 
Fig.~\ref{fig.event900GeV}.

\begin{figure}[htb!]
\begin{center}
\includegraphics[width=0.8\linewidth,clip=]{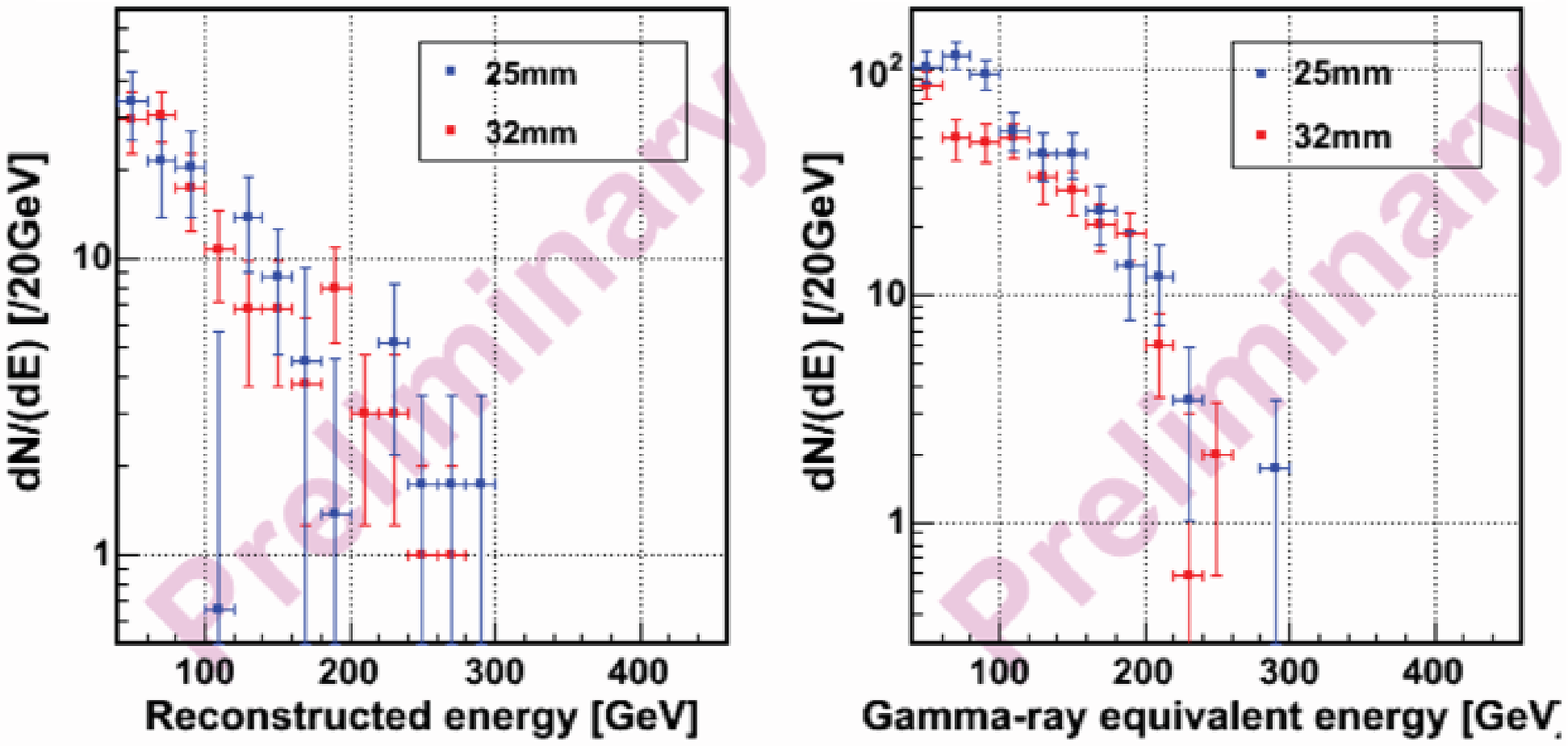}
\end{center}
\begin{center}
\includegraphics[width=0.8\linewidth,clip=]{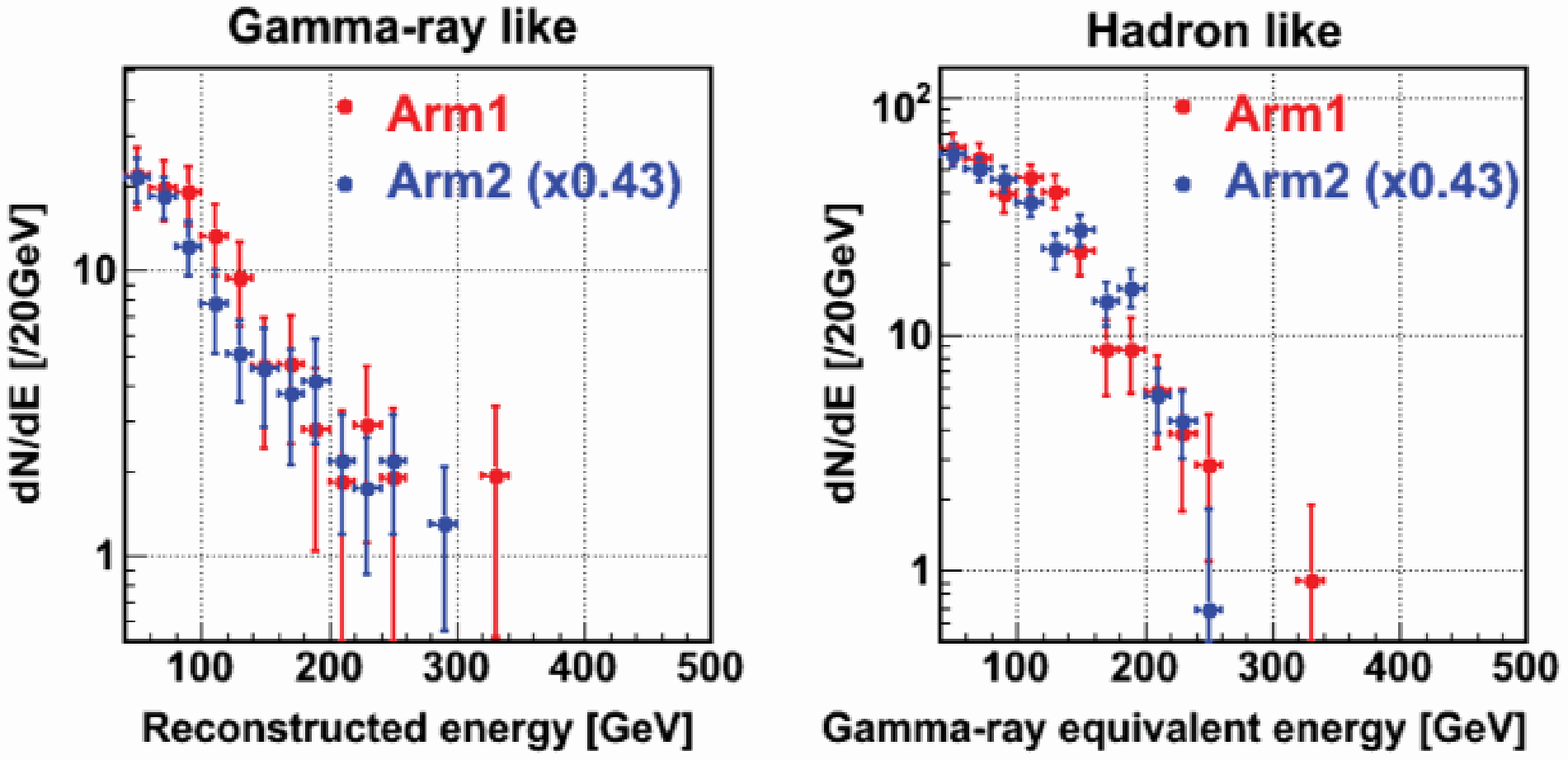}
\end{center}
\caption{Comparison of reconstructed $\gamma$ and hadron spectra in small and large tower of ARM2 
detector (upper plot) and in ARM1 and ARM2 detector (bottom plot) for 900 GeV p-p data. 
Spectra are normalized to take into account different geometrical acceptance.}
\label{fig.900GeVdata}
\end{figure}

A preliminary analysis has been carried on to reconstruct $\gamma$ and hadron spectra. The 
particle identification has been achieved through the use of transition curve information.
Results obtained in the two towers as well as in the two arms are consistent each other both 
for photons and hadrons as can be seen from Fig.~\ref{fig.900GeVdata}. 

900 GeV data have been accumulated also in a second run during spring 2010 and whole analysis 
of the events collected in 2009 and 2010 is ongoing and almost ready for publication.

\section{\label{sec:7TeV} Data taking at 7 TeV}
\begin{figure}[htbp!]
\begin{center}
\includegraphics[height=8 truecm,clip=]{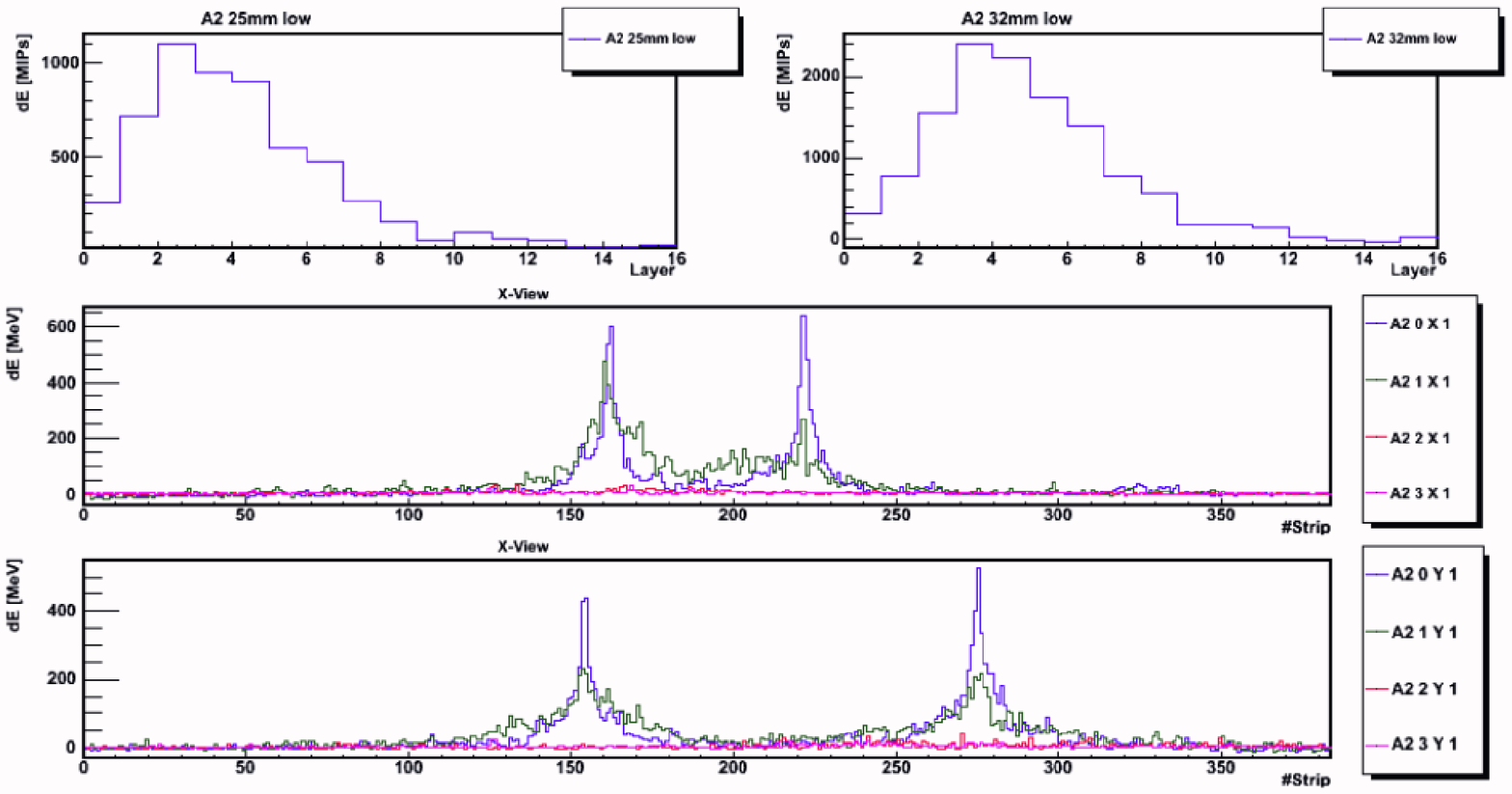}
\end{center}
\caption{A $\pi^0$ candidate event registered on ARM2 detector in 7 TeV collision data.
 The plots follow the same convention as in Fig.~\protect{\ref{fig.event900GeV}}.}
\label{fig.pi0event}
\end{figure}

At the beginning of 2010 the LHC beam energy was increased up to 3.5 TeV, allowing LHCf to take data at 7 TeV center of mass energy. The geometrical configuration of the detectors and the kinematic of the decays allow LHCf to reconstruct $\pi^0$ events, by measuring the two $\gamma$ of the decay in the two separate towers. In this way the absolute energy scale calibration can be cross checked by looking at the kinematically reconstructed $\pi^0$ invariant mass. A typical $\pi^0$ candidate event collected in ARM2 detector is shown in Fig.~\ref{fig.pi0event}. 

\begin{figure}[htbp!]
\begin{center}
\includegraphics[height=10 truecm,clip=]{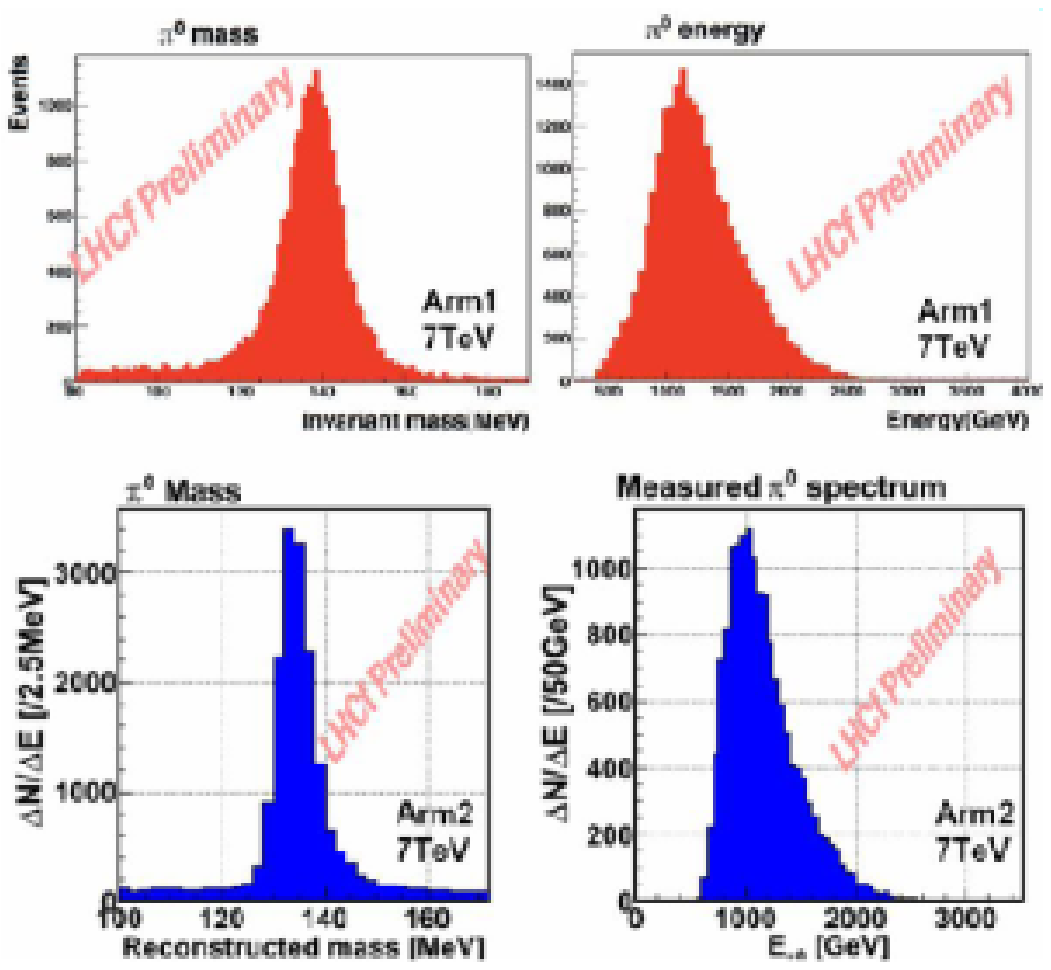}
\end{center}
\caption{Measured invariant mass and energy spectra for $\pi^0$ at 7 TeV center of mass energy, for Arm1 (top plots) and Arm2 (bottom plots). The invariant mass resolution is of the order of 5\% and 2.3\% respectively. }
\label{fig.pi0spectra}
\end{figure}

Data analysis for the 7 TeV running is still ongoing. As preliminary result, Fig.~\ref{fig.pi0spectra} shows the measured $\pi^0$ invariant mass and energy spectra obtained for Arm1 and Arm2. The invariant mass distributions demonstrate the excellent performances in the $\pi^0$ reconstruction ($\Delta m/m\simeq 5\%$ for Arm1 and $\simeq 2.3\%$ for Arm2), even for the extremely high energy $\pi^0$ (up to 3.5 TeV). 

\section{\label{sec:future} Future activities}
\begin{figure}[htb!]
\begin{center}
\includegraphics[width=16 truecm,clip=]{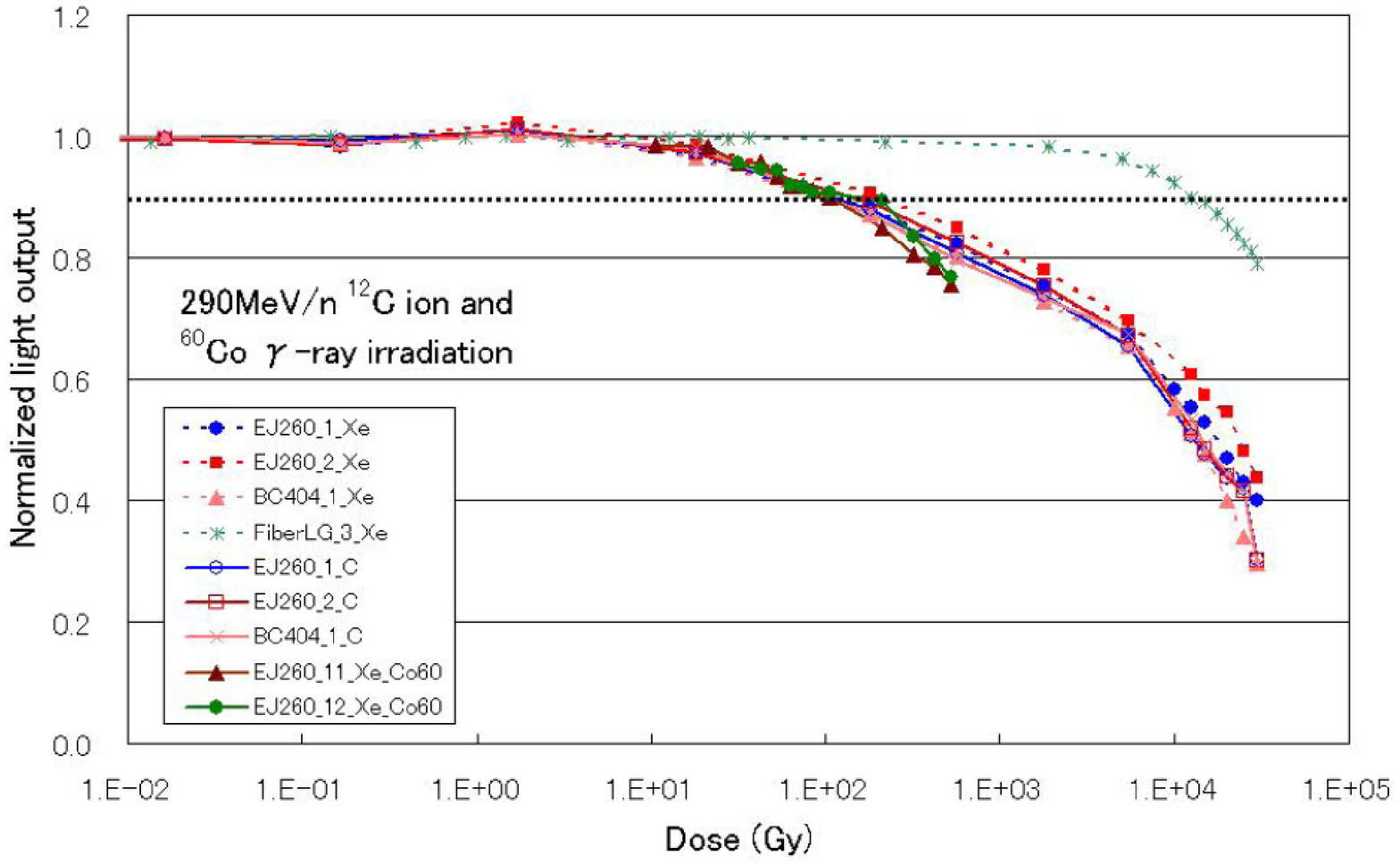}
\end{center}
\caption{The variation of the scintillator light outputs for various types of plastic scintillators, including EJ260, as function of the dose integrated by using heavy ions and $\gamma$ rays. }
\label{fig.radiationdamage}
\end{figure}

LHCf has been originally designed to take data at high energy (14 TeV) and low luminosity (L$\simeq 10^{29}$ cm$^{-2}$s$^{-1}$), in the early LHC operation phase. However, the 2008 LHC incident significantly changed the running plans, with an extended initial running period at reduced energy (7 TeV) and quite high luminosity ($> 10^{31}$ cm$^{-2}$s$^{-1}$). 
The plastic scintillators used for the energy measurement are intrinsically radiation weak, and can not sustain for long periods the luminosity foreseen during the 2010 running. 
Fig.~\ref{fig.radiationdamage} shows the results of the measurements that have been done on the plastic scintillators with the heavy ion beam from a synchrotron at HIMAC of NIRS (National Institute of Radiological Science, Japan) and with $\gamma$ rays
at the $^{60}$Co Radiation Facility of Nagoya University. The light output reduction for the EJ260 plastic scintillators was measured as function of the integrated dose, showing a significant reduction in the light yield for doses greater than few tens of Grays. In the normal LHC running conditions at 7 TeV, a 10 Gy dose is gathered for an integrated luminosity of the order of 20 nb$^{-1}$, corresponding to few days of data taking at $> 10^{29}$ cm$^{-2}$s$^{-1}$ luminosity. 

The detectors were hence removed at the end of July 2010, during a 3 days LHC technical stop, after a $\simeq$\ 10 Gy accumulated dose. 
They will be tested at the H4 SPS beam test facility during Autumn 2010, to confirm the absolute energy scale. Later on, during 2011, the detectors will be upgraded, by replacing plastic scintillators with radiation hard GSO crystals, able to sustain doses 
of $\simeq 10^6-10^7$ Gy without significant damages. In this way, LHCf will be able to be installed again in the tunnel in 2013, when the LHC energy will be increased up to 14 TeV.

\section{Conclusions}

The LHCf experiment has taken data at LHC both at 900 GeV as well at 
7 TeV center of mass energy, before being removed at the end of July 2010. It will be upgraded with radiation hard scintillators, and re-installed in the LHC tunnel in 2013, when the machine will provide 7+7 TeV p-p collisions. \\
Thanks to the excellent detector performance for 
the reconstruction of photon, neutral meson and 
neutron spectra at different energies from 900 GeV up to 14 TeV p-p runs, LHCf will allow to 
calibrate air shower Monte Carlo codes covering the most interesting energy range for HECR Physics, 
thus providing invaluable input to our understanding of high energy phenomena in the Universe.



\newpage
\clearpage
\setcounter{affil}{0}
\setcounter{section}{0}
\setcounter{figure}{0}
\setcounter{table}{0}
\setcounter{equation}{0}

\title{Mueller Navelet jets, jet gap jets and anomalous $WW\gamma \gamma$
couplings in $\gamma$-induced processes at the LHC}
\author{C. Royon}\email{christophe.royon@cea.fr}
\affiliation{IRFU/Service de physique des particules, CEA/Saclay, 91191 Gif-sur-Yvette cedex, France}

\begin{abstract}

We describe two different important measurements to be performed at the LHC.
The Mueller Navelet jet and jet gap jet cross section represent a test of BFKL
dynamics and we perform a NLL calculation of these processes and compare it
with recent Tevatron measurements. The study of the $WW\gamma \gamma$ couplings
at the LHC using the forward detectors proposed in the ATLAS Forward Physics
project as an example allows to probe higgsless and extradimension
models via anomalous quartic
couplings since the reach is improved by four orders of magnitude with respect 
to the LEP results..
\end{abstract}

\maketitle
\addcontentsline{toc}{part}{Mueller Navelet jets, jet gap jets and anomalous $WW\gamma \gamma$
couplings in $\gamma$-induced processes at the LHC - {\it C.Royon}}

\section{Mueller Navelet jets at the LHC}
In this section, we give the BFKL NLL cross section calculation for Mueller
Navelet processes at the Tevatron and the LHC. Since the starting point of this
study was the description of forward jet production at HERA, we start by
describing briefly these processes.

\subsection{Forward jets at HERA}

\begin{figure}
\centerline{\includegraphics[width=0.45\columnwidth]{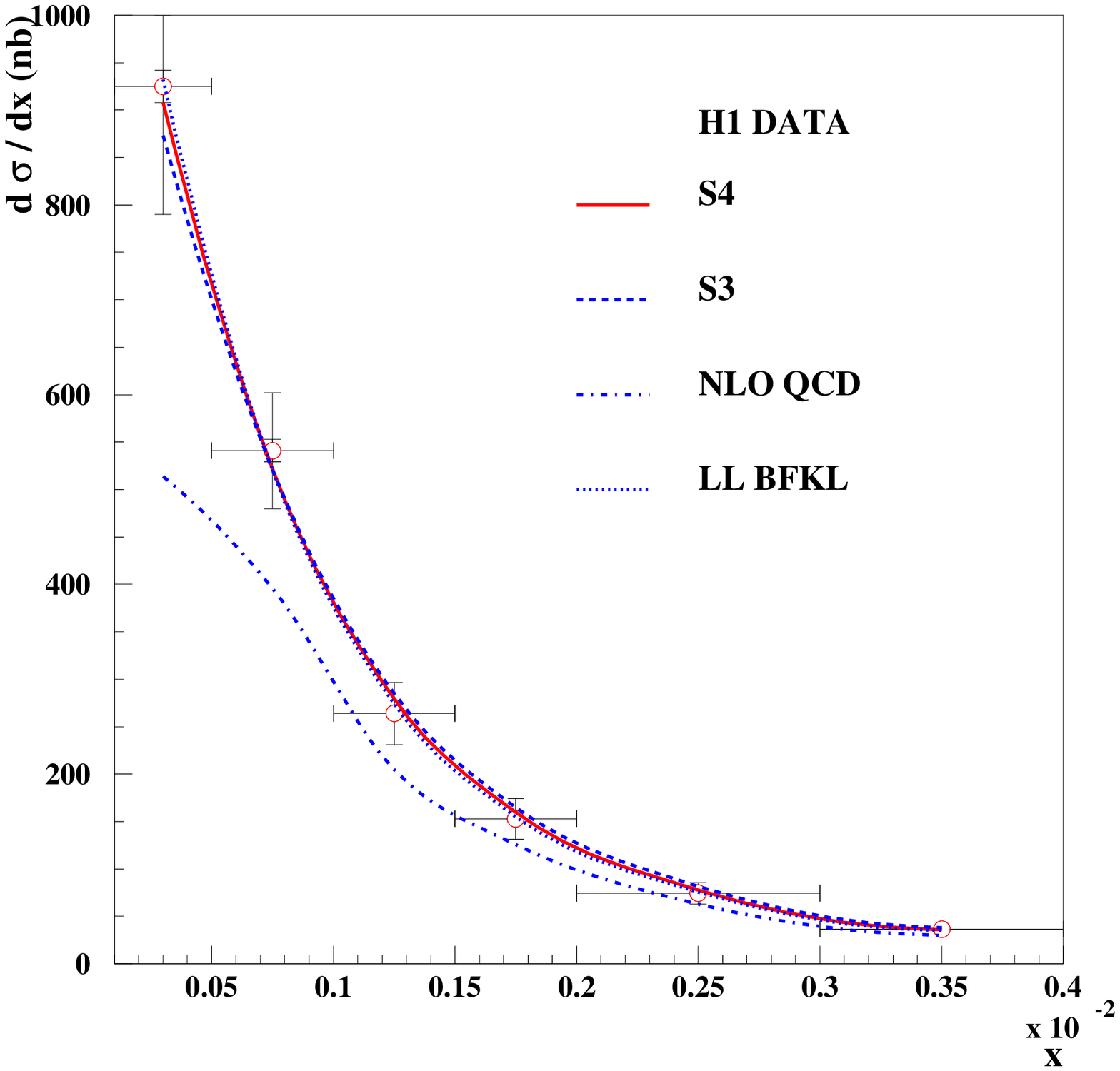}}
\caption{Comparison between the H1 $d \sigma /dx$ measurement 
with predictions for BFKL-LL, BFKL-NLL (S3 and S4 schemes) and DGLAP NLO calculations
(see text). S4, S3 and LL BFKL cannot be distinguished on that figure.}\label{Fig1a}
\end{figure}

Following the successful BFKL \cite{bfkllo} parametrisation of the forward-jet 
cross-section $d\sigma/dx$ at Leading Order (LO)
at HERA \cite{mr,mrb}, it is possible to perform a similar study using Next-to-leading (NLL)
resummed BFKL kernels.
Forward jets at HERA are an ideal observable to look for BFKL resummation
effects. The interval in rapidity between the scattered lepton and the jet in
the forward region is large, and when the photon virtuality $Q^2$ is close to
the transverse jet momentum $k_T$, the DDLAP cross section is small because of
the $k_T$ ordering of the emitted gluons.
In this short report, we will only discuss the
phenomelogical aspects and all detailed calculations can be found in 
Ref.~\cite{fwdjet} for forward jets at HERA and in Ref.~\cite{mnjet} for Mueller Navelet jets
at the Tevatron and the LHC.

\subsection{BFKL NLL formalism}
The BFKL NLL \cite{bfkl} longitudinal transverse cross section reads:
\begin{eqnarray}
\frac{d\sigma^{\gamma*p\!\rightarrow\!JX}_{T,L}}{dx_Jdk_T^2}=
\frac{\alpha_s(k_T^2)\alpha_s(Q^2)}{k_T^2Q^2}\ f_{eff}(x_J,k_T^2)
\int d\gamma \left(\frac{Q^2}{k_T^2}\right)^{\gamma} \phi^{\gamma}_{T,L}(\gamma)\ 
e^{\bar\alpha(k_T Q)\chi_{eff}[\gamma,\bar\alpha(k_T Q)]Y}
\label{nll}
\end{eqnarray}
where $x_J$ is the proton momentum fraction carried by the forward jet,
$\chi_{eff}$ is the effective BFKL NLL kernel and the $\phi$s are 
the transverse and longitunal impact factors taken at LL. The effective kernel
$\chi_{eff}(\gamma,\bar\alpha)$ is
defined from the NLL kernel $\chi_{NLL}(\gamma,\omega)$ by solving the implicit 
equation numerically
\begin{eqnarray}
\chi_{eff}(\gamma,\bar\alpha)=\chi_{NLL}\left[\gamma,\bar\alpha\ 
\chi_{eff}(\gamma,\bar\alpha)\right]\ ,
\label{eff}
\end{eqnarray}

The integration over $\gamma$ in Eq.~\ref{nll} is performed numerically.
It is possible to fit directly $d \sigma/dx$ measured by the H1 collaboration
using this formalism with one single parameter, the normalisation.
The values of $\chi_{NLL}$ are
taken at NLL~\cite{bfkl} using different resummation schemes to remove spurious
singularities defined as S3 and S4~\cite{resum}. Contrary to LL BFKL, it is
worth noticing that the coupling constant $\alpha_S$ is taken using the
renormalisation group equations, the only free parameter in the fit being the
normalisation.

\begin{figure}
\centerline{\includegraphics[width=0.75\columnwidth]{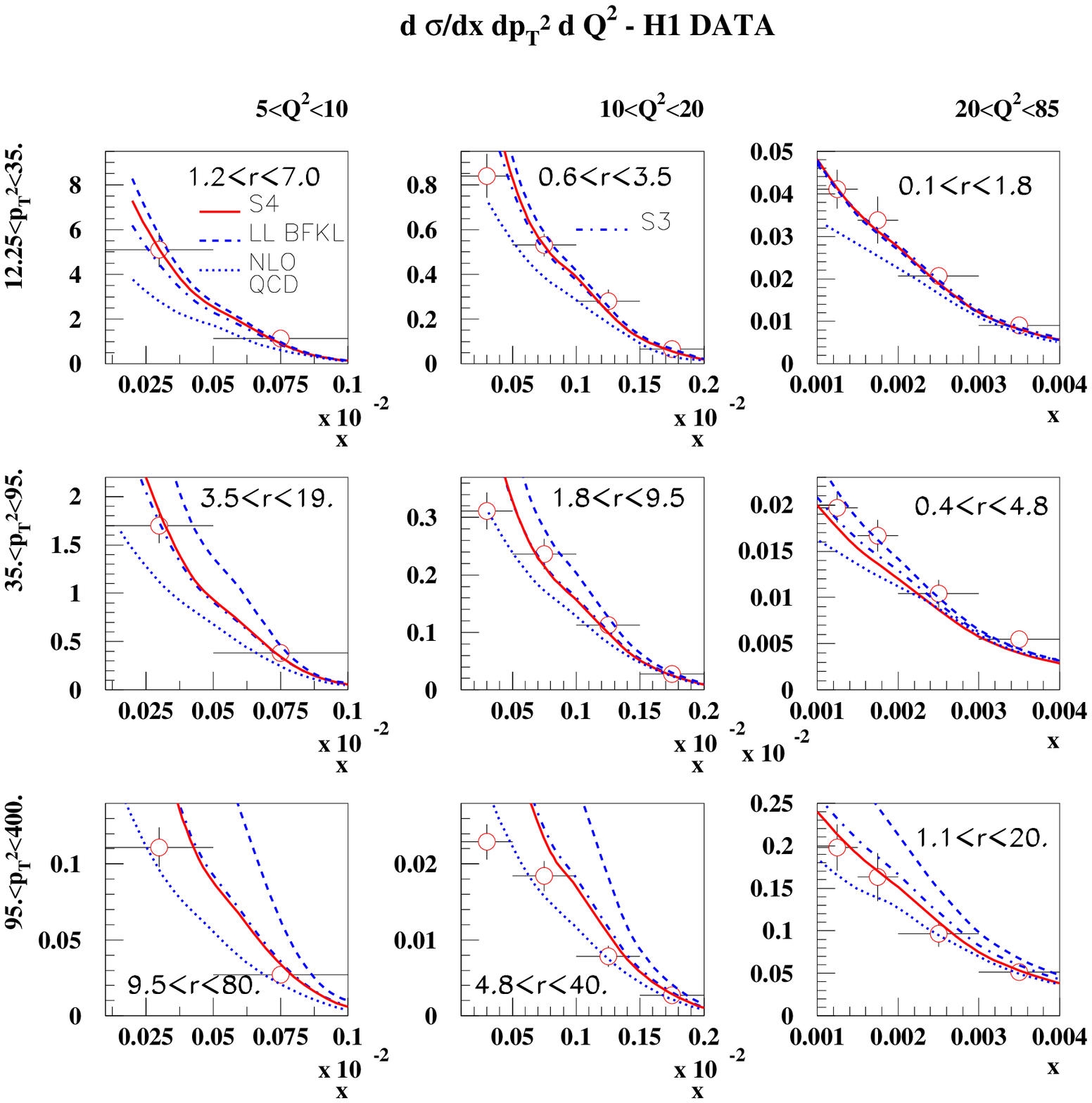}}
\caption{Comparison between the H1 measurement of the triple differential cross
section with predictions for BFKL-LL, BFKL-NLL and DGLAP NLO calculations
(see text).}\label{Fig2}
\end{figure}

To compute $d \sigma/dx$ in the experimental bins, we need to integrate
the differential cross section on the bin size in $Q^2$, $x_{J}$ (the momentum
fraction of the proton carried by the forward jet), $k_T$ ,
while taking into account the experimental cuts. To simplify the numerical
calculation, we perform the integration on the bin using the
variables where the cross section does not change rapidly, namely $k_T^2/Q^2$,
$\log 1/x_{J}$, and $1/Q^2$. Experimental cuts are treated directly at the
integral level (the cut on $0.5<k_T^2/Q^2<5$ for instance) or using a toy Monte
Carlo. 
More detail can be found about the fitting procedure 
in Appendix A of Ref.~\cite{mrb}.

The NLL fits~\cite{fwdjet} can nicely
describe the H1 data~\cite{h1} for the S4  and S3 schemes~\cite{mr,mrb,fwdjet} ($\chi^2=0.48/5$
and  $\chi^2=1.15/5$ respectively
per degree
of freedom with statistical and systematic errors added in quadrature).
The curve using a LL fit is indistinguishable in Fig.~\ref{Fig1a} from the result of the
BFKL-NLL fit.
The DGLAP NLO calculation fails to
describe the H1 data at lowest $x$ (see Fig.~\ref{Fig1a}). We also checked the effect
of changing the scale in the exponential of Eq.~\ref{nll} from $k_TQ$ to
$2k_TQ$ or $k_TQ/2$ which leads to a difference of 20\% on the cross section while
changing the scale to $k_T^2$ or $Q^2$ modifies the result by less than 5\%
which is due to the cut on $0.5 < K_T^2/Q^2<5$. Implementing the higher-order
corrections in the impact factor due to exact gluon dynamics in the $\gamma^*
\rightarrow q \bar{q}$ transition~\cite{robi} changes the result by less than
3\%.

The H1 collaboration also measured the forward jet triple differential cross
section~\cite{h1} and the results are given
in Fig.~\ref{Fig2}. We keep the same normalisation coming from the fit to $d \sigma/dx$
to predict the triple differential cross section.
The BFKL LL formalism leads to a good description 
of the data when $r=k_T^2/Q^2$
is close to 1 and deviates from the data when $r$ is further away from 1. This
effect is expected since DGLAP radiation effects are supposed to occur when
the ratio between
the jet $k_T$ and the virtual photon $Q^2$ are further away from 1. The BFKL 
NLL calculation
including the $Q^2$ evolution via the renormalisation group equation leads to a
good description of the H1 data on the full range. We note that the higher order
corrections are small when $r \sim 1$, when the BFKL effects are
supposed to dominate. By contrast, they are significant as expected when $r$ is different from
one, ie when DGLAP evolution becomes relevant. We notice that the DGLAP NLO calculation
fails to describe the data when $r \sim 1$, or in the region where
BFKL resummation effects are expected to appear. 

In addition, we checked the dependence of our results on the scale taken in the
exponential of Eq.~\ref{nll}. The effect is a change of the cross section of
about 20\% at low $p_T$ increasing to 70\% at highest $p_T$. Taking the correct
gluon kinematics in the impact factor lead as expected to a better description
of the data at high $p_T$~\cite{fwdjet}. 

\subsection{Mueller Navelet jets at the Tevatron and the LHC}

\begin{figure}
\centerline{\includegraphics[width=0.45\columnwidth]{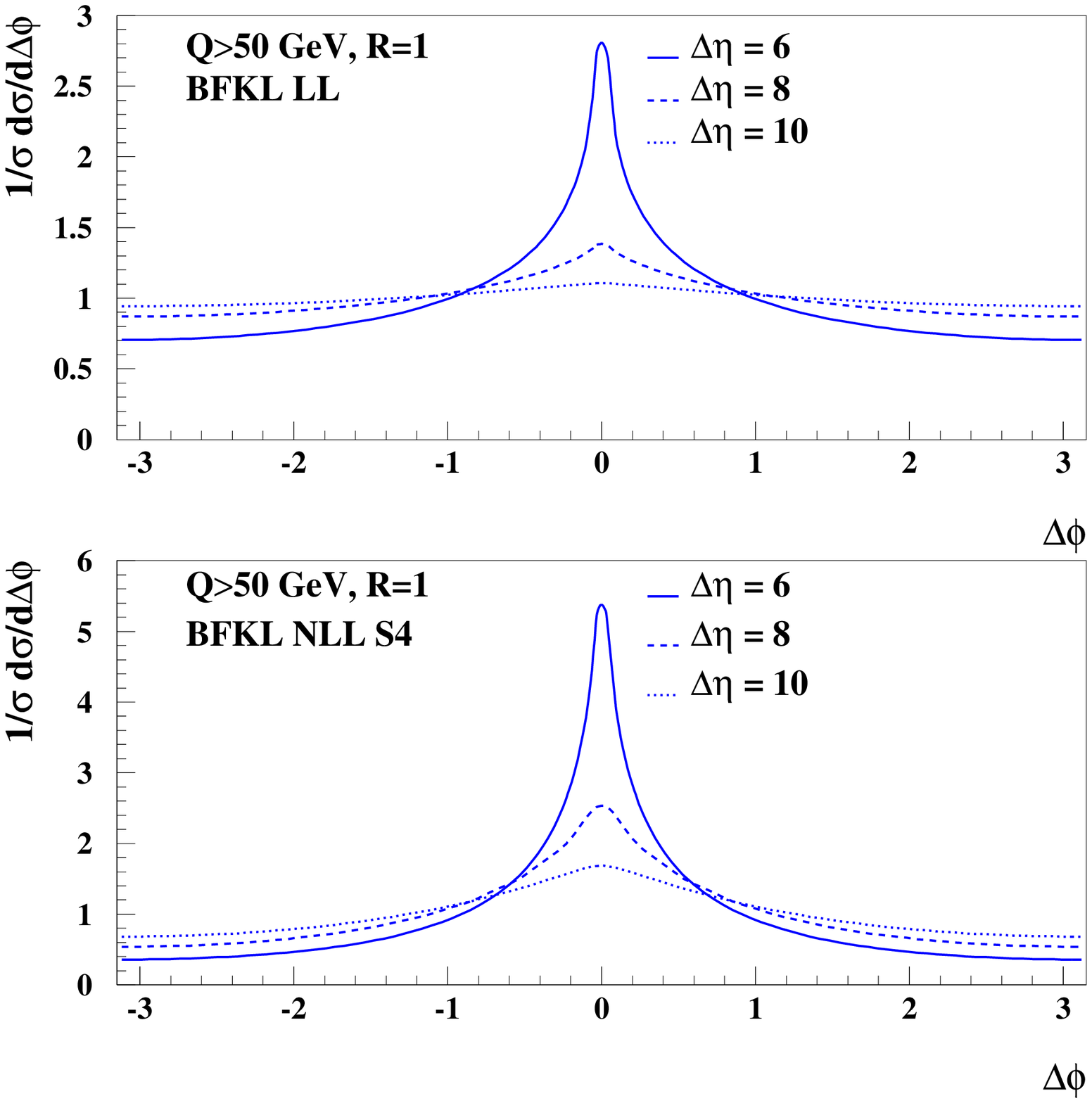}}
\caption{The Mueller-Navelet jet $\Delta\Phi$ distribution for LHC kinematics in the BFKL framework at 
LL (upper plots) and NLL-S4 (lower plots) accuracy for $\Delta\eta=6,\ 8,\ 10.$}\label{Figlhc}
\end{figure}

\begin{figure}
\centerline{\includegraphics[width=0.45\columnwidth]{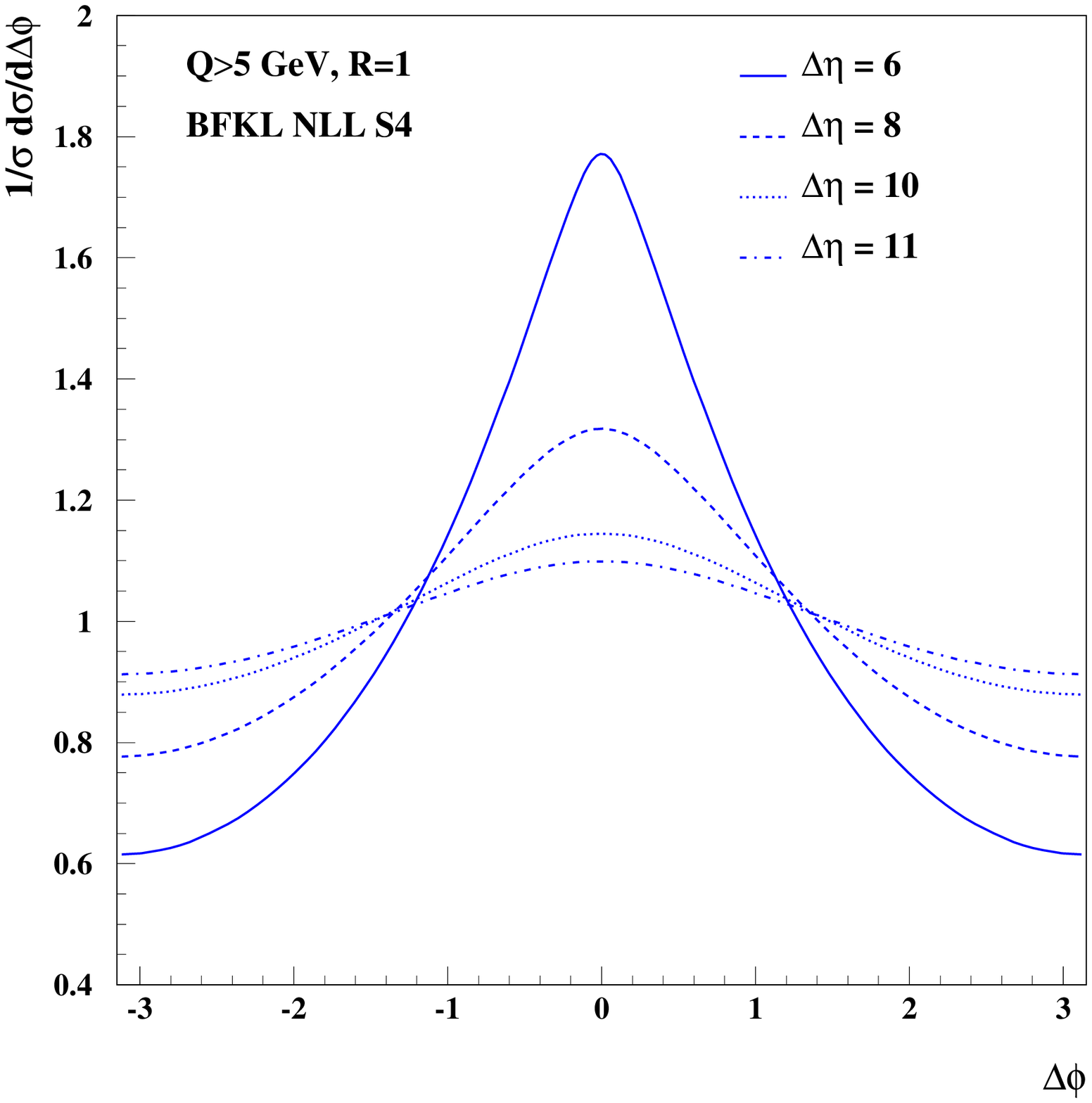}}
\caption{Azimuthal correlations between jets with $\Delta \eta=$6, 8, 10
and 11 and $p_T>5$ GeV in the CDF acceptance. This measurement will represent a
clear test of the BFKL regime.}\label{Fig3}
\end{figure}

Mueller Navelet jets are ideal processes to study BFKL resummation effects~\cite{mn}.
Two jets with a large interval in rapidity and with similar
tranverse momenta are considered. A typical observable to look for BFKL effects
is the measurement of the azimuthal correlations between both jets. The DGLAP
prediction is that this distribution should peak towards $\pi$ - ie jets
are back-to-back- whereas
multi-gluon emission via the BFKL mechanism leads to a smoother distribution.
The relevant variables to look for azimuthal correlations are the following:
\begin{eqnarray}
\Delta \eta &=& y_1 - y_2  \nonumber \\
y &=& (y_1 + y_2)/2 \nonumber \\
Q &=& \sqrt{k_1 k_2} \nonumber \\
R &=& k_2/k_1  \nonumber 
\end{eqnarray}
where $y_{1,2}$ and $k_{1,2}$ are respectively the jet rapidities and transverse
momenta.
The azimuthal correlation for BFKL reads:
\begin{eqnarray}
2\pi\left.\frac{d\sigma}{d\Delta\eta dR d\Delta\Phi}
\right/\frac{d\sigma}{d\Delta\eta dR}=
1+ \nonumber 
\frac{2}{\sigma_0(\Delta\eta,R)}\sum_{p=1}^\infty \sigma_p(\Delta\eta,R) \cos(p\Delta\Phi)
\nonumber 
\end{eqnarray}
where in the NLL BFKL framework,
\begin{eqnarray}
\sigma_p&=& \int_{E_T}^\infty \frac{dQ}{Q^3}
\alpha_s(Q^2/R)\alpha_s(Q^2R) \nonumber 
\left( \int_{y_<}^{y_>} dy x_1 f_{eff}(x_1,Q^2/R)x_2f_{eff}(x_2,Q^2R) 
\right) \nonumber
\\
&~& \int_{1/2-\infty}^{1/2+\infty}\frac{d\gamma}{2i\pi}R^{-2\gamma}
\ e^{\bar\alpha(Q^2)\chi_{eff}(p, \gamma, \bar{\alpha})\Delta\eta}  \nonumber
\end{eqnarray}
and $\chi_{eff}$ is the effective resummed kernel.
Computing the different $\sigma_p$ at NLL for the resummation schemes S3 and S4
allowed us to compute the azimuthal correlations at NLL. As expected, the
$\Delta \Phi$ dependence is less flat than for BFKL LL and is closer to the
DGLAP behaviour \cite{mnjet}. In Fig.~\ref{Figlhc}, we display the observable $1/\sigma d \sigma/d\Delta \Phi$ as a function
of $\Delta\Phi$, for LHC kinematics. The results are displayed for different values of 
$\Delta\eta$ and at both LL and NLL accuracy using the S4 resummation scheme. In general, the 
$\Delta\Phi$ spectra are peaked around $\Delta\Phi\!=\!0,$ which is indicative of jet emissions occuring back-to-back. 
In addition the $\Delta\Phi$ distribution flattens with increasing 
$\Delta\eta\!=\!y_1\!-\!y_2$. Note the change of scale on the vertical axis 
which indicates the magnitude of the NLL corrections with respect to the LL-BFKL results. The NLL corrections 
slow down the azimuthal angle decorrelations for both increasing $\Delta\eta$ and $R$ deviating from $1.$ We also
studied the $R$ dependence of our prediction which is quite weak~\cite{mnjet}
and the scale dependence
of our results by modifying the scale $Q^2$ to either $Q^2/2$ or $2Q^2$ and the effect on the azimuthal distribution 
is of the order of 20\%. The effect of the energy conservation in the BFKL equation~\cite{mnjet} is large when $R$ goes
away from 1. The effect is to reduce the effective value of $\Delta \eta$ between the jets and thus the decorrelation 
effect. However, it is worth noticing that this effect is negligible when $R$ is close to 1 where this measurement 
will be performed.

A measurement of the cross-section 
$d\sigma^{hh\!\to\!JXJ}/d\Delta\eta dR d\Delta\Phi$ at the Tevatron (Run 2) or the LHC will allow for a 
detailed study of the BFKL QCD dynamics since the DGLAP evolution leads to much less jet angular
decorrelation (jets are back-to-back when $R$ is close to 1). In particular, measurements with 
values of $\Delta\eta$ reaching 8 or 10 will be of great interest, as these could allow to distinguish between 
BFKL and DGLAP resummation effects and would provide important tests for the relevance of the BFKL formalism. 

To illustrate this result, we give in Fig.~\ref{Fig3} the azimuthal
correlation in the CDF acceptance. The CDF collaboration installed the
mini-Plugs calorimeters aiming for rapidity gap selections in the very forward
regions and these detectors can be used to tag very forward jets. A measurement
of jet $p_T$ with these detectors would not be possible but their azimuthal
segmentation allows a $\phi$ measurement. In Fig.~\ref{Fig3}, we display the jet
azimuthal correlations for jets with a $p_T>5$ GeV and $\Delta \eta=$6, 8, 10
and 11. For $\Delta \eta=$11, we notice that the distribution is quite flat,
which would be a clear test of the BFKL prediction.

\section{Jet gap jets at the Tevatron and the LHC}

In this section, we describe another possible measurement which can probe BFKL
resummation effects and we compare our predictions with existing D0 and CDF
measurements~\cite{usb}.

\subsection{BFKL NLL formalism}

The production cross section of two jets with a gap in rapidity between them reads
\begin{equation}
\frac{d \sigma^{pp\to XJJY}}{dx_1 dx_2 dE_T^2} = {\cal S}f_{eff}(x_1,E_T^2)f_{eff}(x_2,E_T^2)
\frac{d \sigma^{gg\rightarrow gg}}{dE_T^2},
\label{jgj}\end{equation}
where $\sqrt{s}$ is the total energy of the collision,
$E_T$ the transverse momentum of the two jets, $x_1$ and $x_2$ their longitudinal
fraction of momentum with respect to the incident hadrons, $S$ the survival probability,
and $f$ the effective parton density functions~\cite{usb}. The rapidity gap
between the two jets is $\Delta\eta\!=\!\ln(x_1x_2s/p_T^2).$ 

The cross section is given by
\begin{equation}
\frac{d \sigma^{gg\rightarrow gg}}{dE_T^2}=\frac{1}{16\pi}\left|A(\Delta\eta,E_T^2)\right|^2
\end{equation}
in terms of the $gg\to gg$ scattering amplitude $A(\Delta\eta,p_T^2).$ 

In the following, we consider the high energy limit in which the rapidity gap $\Delta\eta$ is assumed to be very large.
The BFKL framework allows to compute the $gg\to gg$ amplitude in this regime, and the result is 
known up to NLL accuracy
\begin{equation}
A(\Delta\eta,E_T^2)=\frac{16N_c\pi\alpha_s^2}{C_FE_T^2}\sum_{p=-\infty}^\infty\intc{\g}
\frac{[p^2-(\g-1/2)^2]\exp\left\{\bar\alpha(E_T^2)\chi_{eff}[2p,\g,\bar\alpha(E_T^2)] \Delta \eta\right\}}
{[(\g-1/2)^2-(p-1/2)^2][(\g-1/2)^2-(p+1/2)^2]} 
\label{jgjnll}\end{equation}
with the complex integral running along the imaginary axis from $1/2\!-\!i\infty$ 
to $1/2\!+\!i\infty,$ and with only even conformal spins contributing to the sum, and 
$\bar{\alpha}=\alpha_S N_C/\pi$ the running coupling.

Let us give some more details on formula \ref{jgjnll}. The NLL-BFKL effects are 
phenomenologically taken into account by the effective kernels $\chi_{eff}(p,\g,\bar\alpha)$.
The NLL 
kernels obey a {\it consistency condition} which allows to reformulate the 
problem in terms of $\chi_{eff}(\g,\bar\alpha).$ The effective kernel
$\chi_{eff}(\g,\bar\alpha)$ is obtained from the NLL kernel $\chi_{NLL}\lr{\g,\omega}$ by 
solving the implicit equation
$\chi_{eff}=\chi_{NLL}\lr{\g,\bar\alpha\ \chi_{eff}}$ as a solution of the 
consistency condition as it was also performed for forward jets.

In this study, we performed a parametrised distribution of $d \sigma^{gg\rightarrow gg}/dE_T^2$
so that it can be easily implemented in the Herwig Monte Carlo~\cite{herwig} since performing the integral over
$\gamma$ in particular would be too much time consuming in a Monte Carlo. The implementation of the
BFKL cross section in a Monte Carlo is absolutely necessary to make a direct comparison with data.
Namely, the measurements are sensititive to the jet size (for instance, experimentally the gap size
is different from the rapidity interval between the jets which is not the case by definition in the
analytic calculation).

\begin{figure}
\begin{center}
\epsfig{file=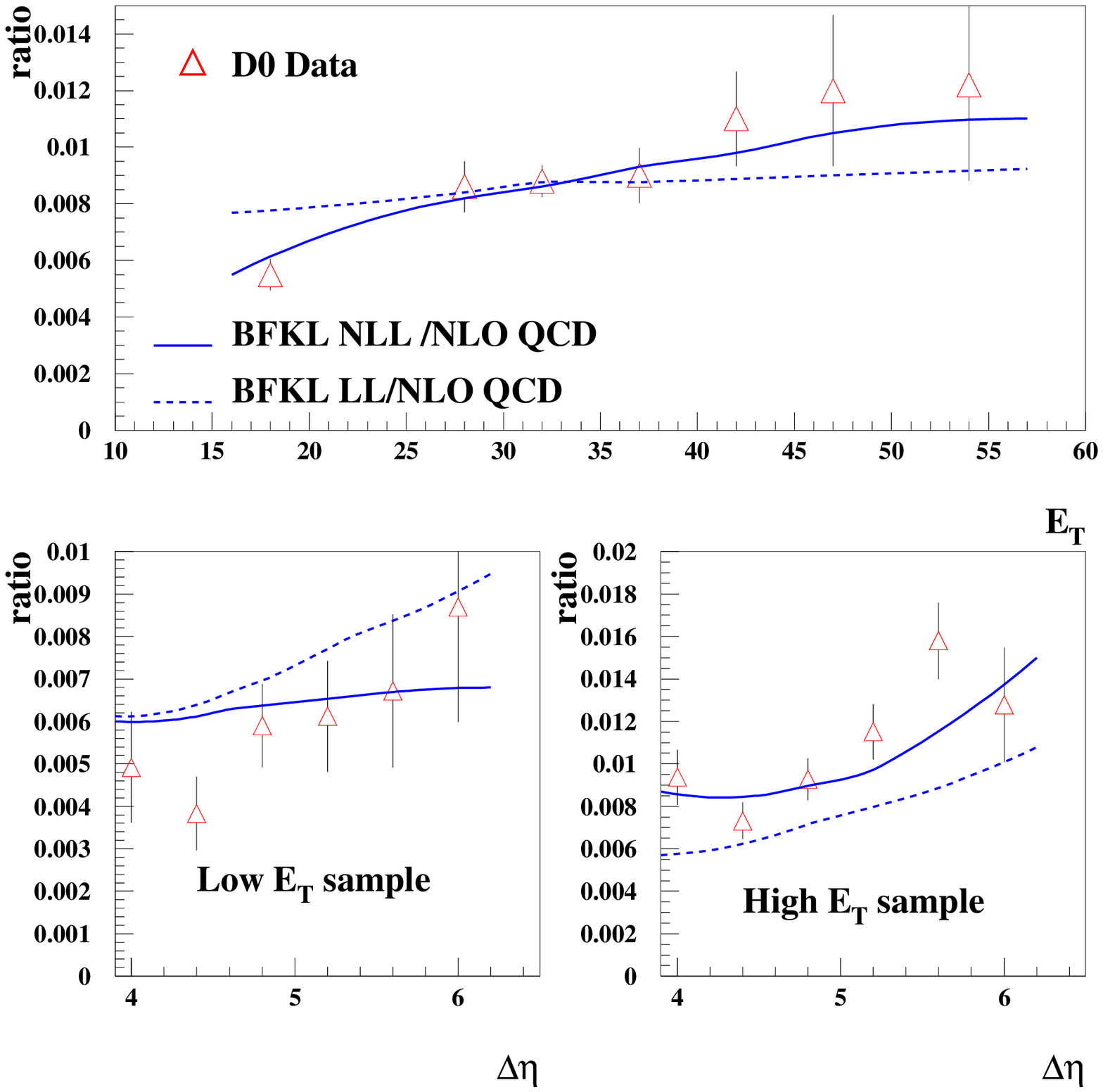,width=8.5cm}
\caption{Comparisons between the D0 measurements of the jet-gap-jet event ratio with the NLL- 
and LL-BFKL calculations. The NLL calculation is in fair agreement with the data. 
The LL calculation leads to a worse description of the data.}
\label{d0}
\end{center}
\end{figure}

\begin{figure}
\begin{center}
\epsfig{file=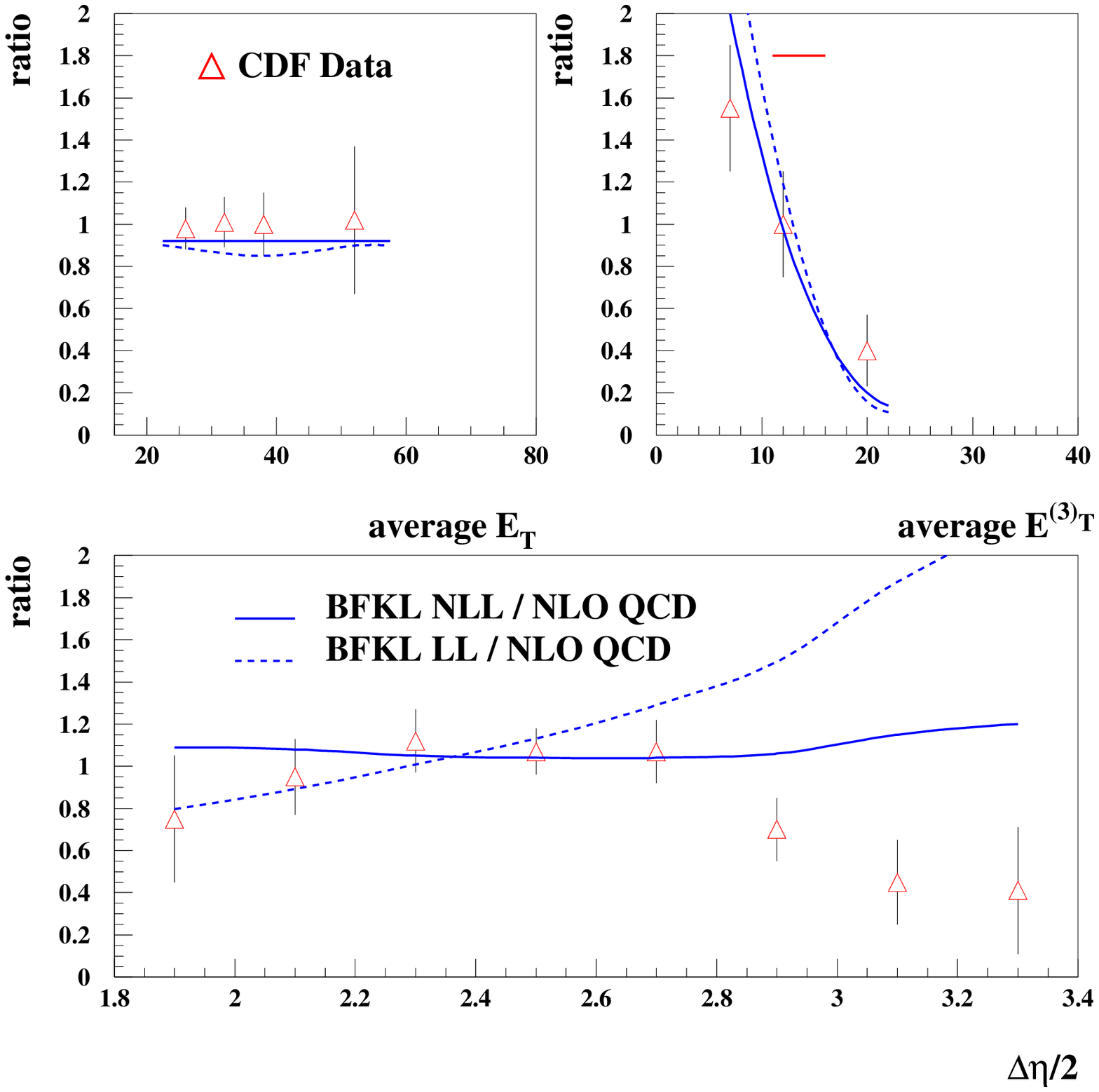,width=8.5cm}
\caption{Comparisons between the CDF measurements of the jet-gap-jet event ratio with the NLL- 
and LL-BFKL calculations. The NLL calculation is in fair agreement with the data. 
The LL calculation leads to a worse description of the data.}
\label{cdf}
\end{center}
\end{figure}

\subsection{Comparison with D0 and CDF measurements}
Let us first notice that the sum over all conformal spins is absolutely necessary. Considering
only $p=0$ in the sum of Equation~\ref{jgjnll} leads to a wrong normalisation and a wrong jet $E_T$
dependence, and the effect is more pronounced as $\Delta \eta$ diminishes.

The D0 collaboration measured the jet gap jet cross section ratio with respect to the total dijet
cross section, requesting for a gap between -1 and 1 in rapidity, as a function of the second
leading jet $E_T$,
and $\Delta \eta$ between the two leading jets for two different low and high $E_T$ samples
(15$<E_T<$20 GeV and $E_T>$30 GeV). To compare with theory, we compute the following quantity
\begin{eqnarray}
Ratio = \frac{BFKL~ NLL~HERWIG}{Dijet~Herwig} \times \frac{LO~QCD}{NLO~QCD} 
\end{eqnarray}
in order to take into account the NLO corrections on the dijet cross
sections, where $BFKL~ NLL$ $HERWIG$ and $Dijet~Herwig$ denote the BFKL NLL and the dijet cross section
implemented in HERWIG. The NLO QCD cross section was computed using the NLOJet++ program~\cite{nlojet}.

The comparison with D0 data~\cite{d0jgj} is shown in Fig. 5. We find a good agreement between the data
and the BFKL calculation. It is worth noticing that the BFKL NLL calculation leads to a better result
than the BFKL LL one (note that the best description of data is given by the BFKL LL formalism 
for $p=0$ but it does not make sense theoretically to neglect the higher spin components and this
comparison is only made to compare with previous LL BFKL calculations).

The comparison with the CDF data~\cite{d0jgj} as a function of the average jet $E_T$ and the
difference in rapidity between the two jets is shown in Fig. 6, and the conclusion remains the same:
the BFKL NLL formalism leads to a better description than the BFKL LL one.

\subsection{Predictions for the LHC}
Using the same formalism, and assuming a survival probability of 0.03 at the LHC, it is possible to
predict the jet gap jet cross section at the LHC. While both LL and NLL BFKL formalisms lead to a
weak jet $E_T$ or $\Delta \eta$ dependence, the normalisation is found to be
quite different (see Fig. 7)
leading to higher cross section for the BFKL NLL formalism.

\begin{figure}
\begin{center}
\epsfig{file=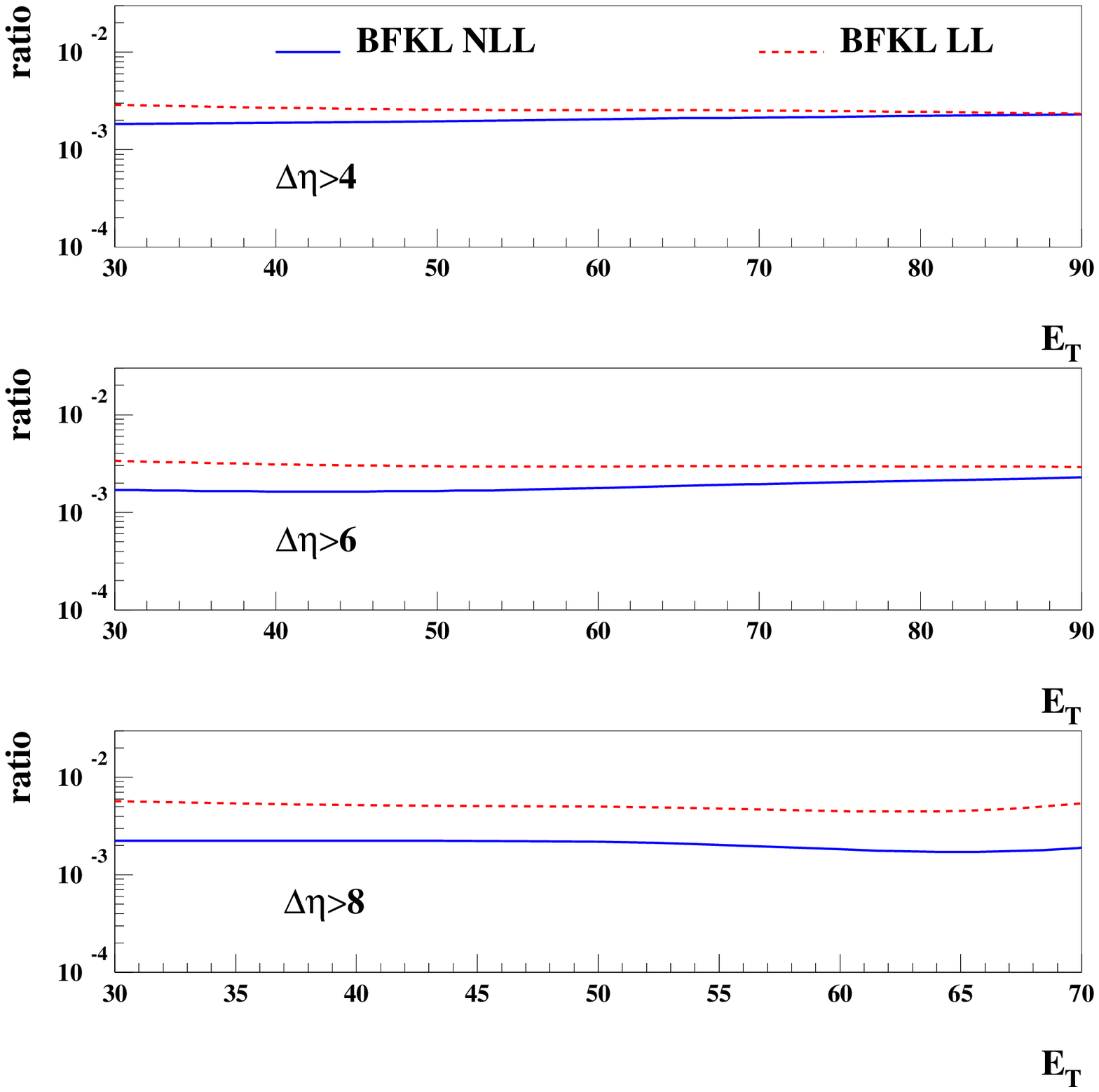,width=8.5cm}
\hfill
\epsfig{file=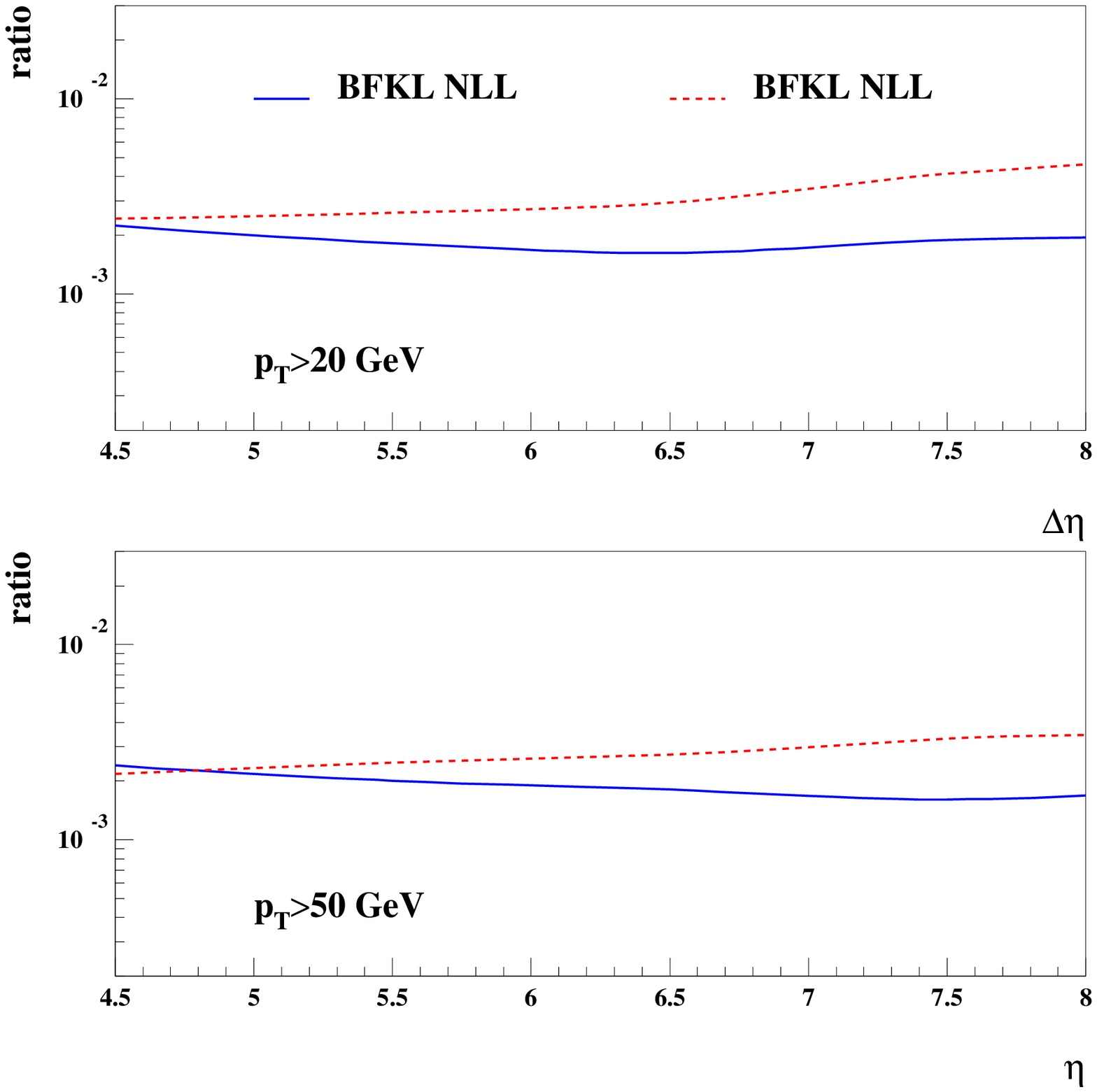,width=8.5cm}
\caption{Ratio of the jet gap jet to the inclusive jet cross sections at the LHC as a function of jet $p_T$ and $\Delta \eta$.}
\label{fig6}
\end{center}
\end{figure}

\section{Quartic anomalous couplings at the LHC}

In the third part of this report, we discuss a completely different topic,
namely the possibility to probe anomalous quartic couplings between photons and
$W$ or $Z$ bosons at the LHC with an unprecedent precision using forward
detectors to be installed in CMS and ATLAS experiments~\cite{us}.
In the Standard Model (SM) of particle physics, the couplings of fermions and 
gauge bosons are constrained by the gauge symmetries of the Lagrangian.
The measurement of $W$ and $Z$ boson pair productions via the exchange of
two photons allows to provide directly stringent tests
of one of the most important and least understood
mechanism in particle physics, namely the
electroweak symmetry breaking~\cite{stirling}. The non-abelian gauge nature of the SM
predicts the existence of quartic couplings
$WW\gamma \gamma$
between the $W$ bosons and the photons which can be probed directly at the 
Large Hadron Collider (LHC) at CERN.
The quartic coupling to the $Z$ boson $ZZ\gamma \gamma$ is not present in the
SM. Quartic anomalous couplings between the photon and the $Z$ or $W$ bosons
are specially expected to occur in higgsless or extradimension
models~\cite{higgsless}.

\subsection{Photon exchange processes in the SM}
The process that we intend to study is the $W$ pair production shown in Fig.~8
induced by the 
exchange of two photons~\cite{piotr,us}. It is a pure QED process
in which the decay products of the $W$ bosons are measured in the central 
detector and the scattered protons leave intact in
the beam pipe at very small angles, contrary to inelastic collisions. Since 
there is no proton remnant the process is purely exclusive; only $W$ decay products 
populate the central detector, and the intact protons can be detected in
dedicated detectors located along the beam line far away from the interaction
point.

The cross section of the $pp\rightarrow p WW p$ process which proceeds through 
two-photon exchange is calculated as a convolution of the 
two-photon luminosity and the total cross section $\gamma\gamma\rightarrow WW$.
The total two-photon cross section is 95.6 fb. 

All considered processes (signal and background) were produced using the Forward
Physics Monte Carlo~\cite{fpmc} (FPMC) generator. The aim of FPMC is to produce different
kinds of processes such as inclusive and exclusive diffraction, photon-exchange
processes. FPMC was interfaced to as fast simulation of the ATLAS
detector~\cite{atlfast}. To reduce the amount of considered background, we only
use leptonic (electrons and muons) decays of $Z$ and $W$ bosons.  The following
backgrounds were considered:
$\gamma\gamma\rightarrow l\bar{l}$ --- two-photon dilepton production,
DPE$\rightarrow l\bar{l}$ ---- dilepton production through double pomeron 
exchange,
DPE$\rightarrow W^+W^-\rightarrow l\bar{l}\nu\bar{\nu}$ --- diboson 
production through double pomeron exchange.

After simple cuts to select exclusive $W$ pairs decaying into leptons, such
as a cut on the proton momentum loss of the proton ($0.0015<\xi<0.15$) --- we
assume the protons to be tagged in the ATLAS Forward Physics
detectors~\cite{afp} ---,
on the transverse momentum of the leading and second leading leptons at 25 and
10 GeV respectively, on $\met>20$ GeV, $\Delta \phi>2.7$ between leading
leptons, and $160<W<500$ GeV, the diffractive mass reconstructed using the
forward detectors, the background is found to be less than 1.7 event for 30
fb$^{-1}$ for a SM signal of 51 events. In this channel, a 5 $\sigma$ discovery
of the Standard Model $pp\rightarrow pWWp$ process is possible after 5 fb$^{-1}$.  

\begin{figure}
\begin{center}
\epsfig{file=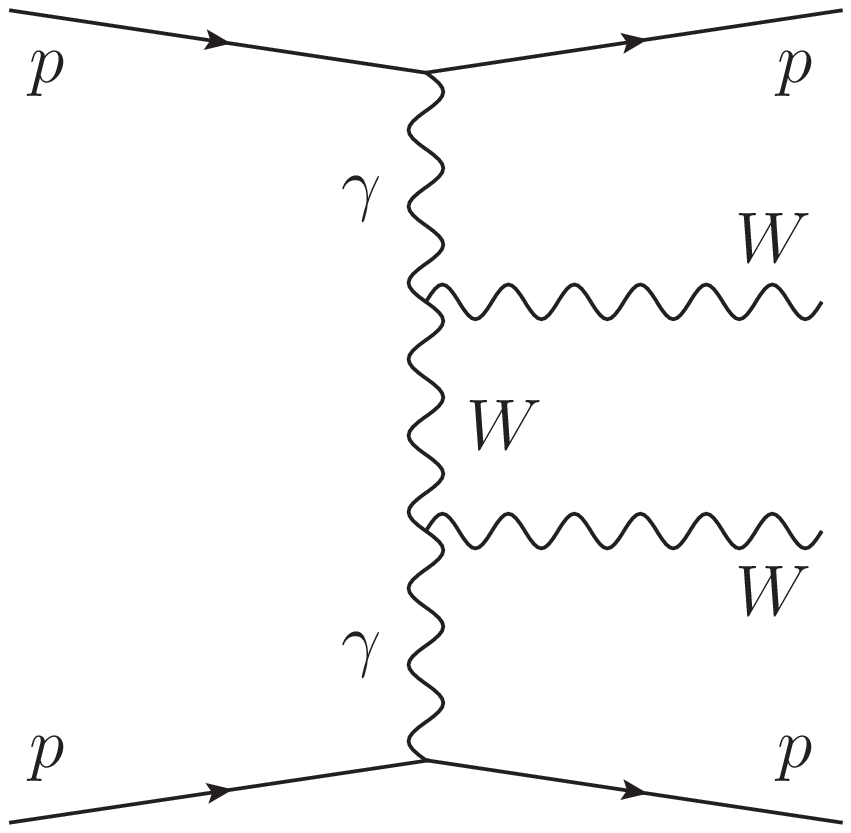,width=6.cm} 
\end{center}
\caption{Sketch diagram showing the two-photon production of a central system.}
\end{figure}

\begin{figure}
\begin{center}
\epsfig{file=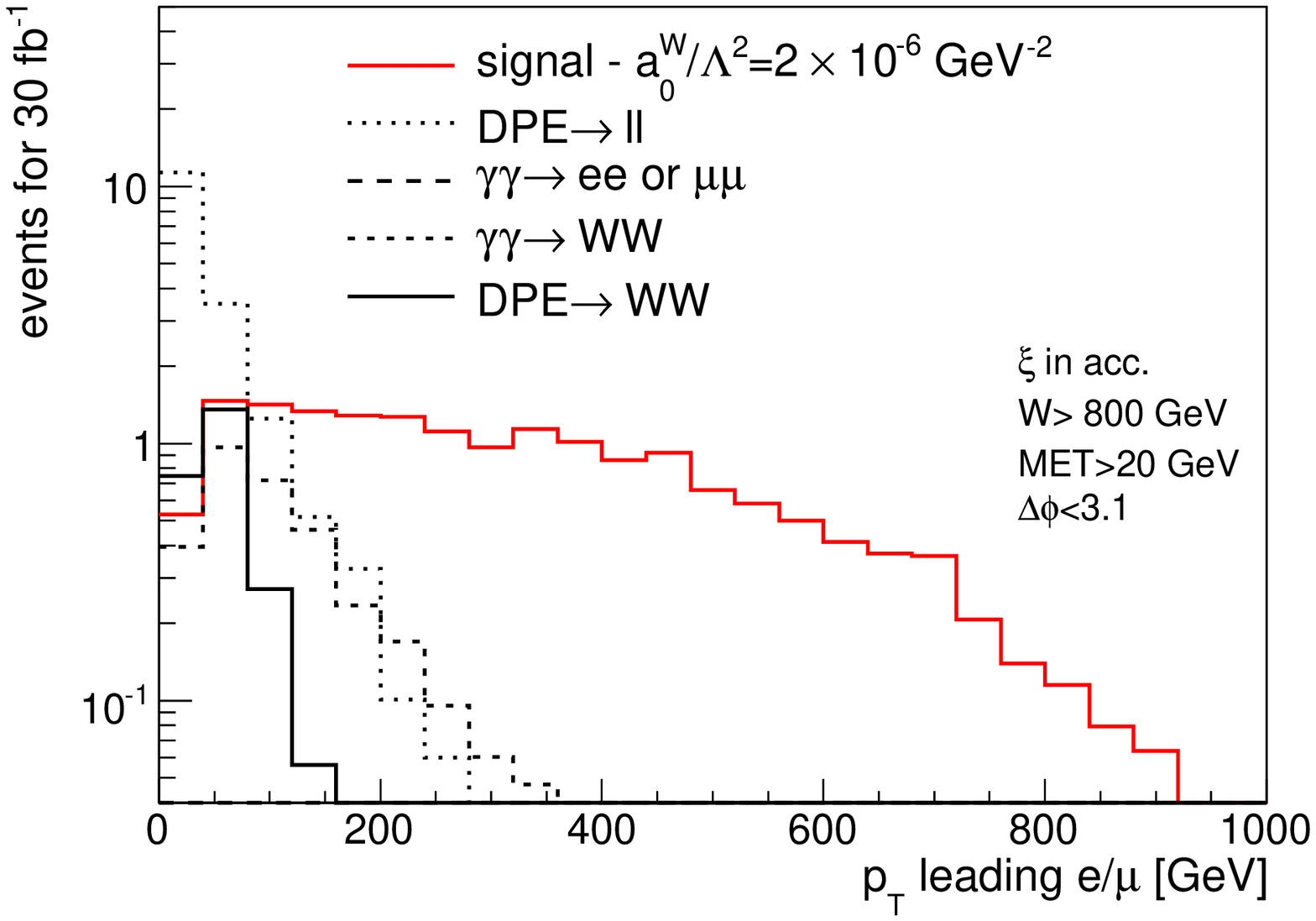,width=8.cm} 
\end{center}
\caption{Distribution of the transverse momentum of the leading lepton for
signal and background after the cut on $W$, $\met$, and $\Delta \phi$ between
the two leptons.}
\end{figure}

\subsection{Quartic anomalous couplings}
The parameterization of the quartic couplings
based on \cite{Belanger:1992qh} is adopted. We concentrate on the lowest order 
dimension operators which have
the correct Lorentz invariant structure and obey the $SU(2)_C$ 
custodial symmetry in order to fulfill the stringent experimental bound on the 
$\rho$ parameter. 
The lowest order interaction
Lagrangians which involve two photons are dim-6 operators. 
The following expression for the effective quartic Lagrangian is used
 \begin{eqnarray}
     \mathcal{L}_6^0 &=& \frac{-e^2}{8} \frac{\aZerow}{\Lambda^2} F_{\mu\nu} 
     F^{\mu\nu} W^{+\alpha} W^-_\alpha 
     - \frac{e^2}{16\cos^2 \theta_W} \frac{a^Z_0}{\Lambda^2} F_{\mu\nu} 
     F^{\mu\nu} Z^\alpha
Z_\alpha\nonumber \\
     \mathcal{L}_6^C & = & \frac{-e^2}{16} \frac{\aCw}{\Lambda^2} 
     F_{\mu\alpha} F^{\mu\beta} (W^{+\alpha} W^-_\beta + W^{-\alpha} 
     W^+_\beta) 
	- \frac{e^2}{16\cos^2 \theta_W} \frac{a^Z_C}{\Lambda^2} 
	F_{\mu\alpha} F^{\mu\beta} Z^\alpha Z_\beta
\label{eq:anom:lagrqgc}
\end{eqnarray}
where $a_0$, $a_C$ are the parametrized new coupling constants and the new scale $\Lambda$ is introduced
so that the Lagrangian density has the correct dimension four and is 
interpreted as the typical mass scale of  new
physics.
In the above formula, we allowed the $W$ and $Z$ parts of the Lagrangian to 
have specific couplings, i.e.
 $a_0\rightarrow (\aZerow$, $a^Z_0$) and similarly $a_C\rightarrow(\aCw$, 
 $\aCz$).

The $WW$ and $ZZ$ two-photon cross sections rise quickly at high energies when 
any of the anomalous parameters are non-zero. The cross section rise has to be 
regulated by a form factor which vanishes in the high energy limit to 
construct a realistic physical model of the BSM theory. We therefore 
modify the couplings by form factors 
that have the desired behavior, i.e. they modify the coupling at small 
energies only slightly but suppress it  when the center-of-mass energy $W_{\gamma\gamma}$ 
increases. The form of the form factor that we consider is the following
\begin{eqnarray}
a\rightarrow \frac{a}{(1+W^2_{\gamma\gamma}/\Lambda^2)^n}
\label{eq:anom:formfactor}
\end{eqnarray}
where $n$=2, and $\Lambda \sim$2 TeV.

The cuts to select quartic anomalous gauge coupling $WW$ events are similar as the
ones we mentioned in the previous section, namely $0.0015<\xi<0.15$ for the
tagged protons, $\met>$ 20 GeV, $\Delta \phi<3.13$ between the two leptons. In
addition, a cut on the $p_T$ of the leading lepton $p_T>160$ GeV and on the
diffractive mass $W>800$ GeV are requested since anomalous coupling events
appear at high mass. Fig~9 displays the $p_T$ distribution of the leading lepton
for signal and the different considered backgrounds. 
After these requirements, we expect about 0.7 background
events for an expected signal of 17 events if the anomalous coupling is about
four order of magnitude lower than the present LEP limit ($|a_0^W / \Lambda^2| =
5.4$ 10$^{-6}$) for a luminosity of 30 fb$^{-1}$. The strategy to select anomalous coupling $ZZ$ events is
analogous and the presence of three leptons or two like sign leptons are 
requested. 
Table 1 gives the reach on anomalous couplings at the LHC for a
luminosity of 30 and 200 fb$^{-1}$ compared to the present OPAL limits~\cite{opal}. We note that we can gain almost
four orders of magnitude in the sensitivity to anomalous quartic gauge couplings
compared to LEP experiments, and it is possible to reach the values expected in Higgsless
or extra-dimension models which are of the order of 5 10$^{-6}$. 
The tagging of the protons using the ATLAS Forward
Physics detectors is the only method at present to test such small values of
quartic anomalous couplings and thus to probe the higgsless models in a clean
way. The reach on anomalous triple gauge couplings is much less improved at the
LHC compared to LEP experiments~\cite{kepka}.

To conclude, the ATLAS Forward Physics program (and the CMS one) will allow to
study Higgsless models with an unprecedent precision as well as to probe the
Higgs boson by allowing its mass and spin measurements~\cite{higgs} using the
forward detectors proposed for installation at 220 and 420 m in ATLAS and CMS.

\begin{table}
\begin{center}
   \begin{tabular}{|c||c|c|c|}
    \hline
    Couplings & 
    OPAL limits & 
    \multicolumn{2}{c|}{Sensitivity @ $\mathcal{L} = 30$ (200) fb$^{-1}$} \\
    &  \small[GeV$^{-2}$] & 5$\sigma$ & 95\% CL \\ 
    \hline
    $a_0^W/\Lambda^2$ & [-0.020, 0.020] & 5.4 10$^{-6}$ & 2.6 10$^{-6}$\\
                      &                 & (2.7 10$^{-6}$) & (1.4 10$^{-6}$)\\ \hline               
    $a_C^W/\Lambda^2$ & [-0.052, 0.037] & 2.0 10$^{-5}$ & 9.4 10$^{-6}$\\
                      &                 & (9.6 10$^{-6}$) & (5.2 10$^{-6}$)\\ \hline               
    $a_0^Z/\Lambda^2$ & [-0.007, 0.023] & 1.4 10$^{-5}$ & 6.4 10$^{-6}$\\
                      &                 & (5.5 10$^{-6}$) & (2.5 10$^{-6}$)\\ \hline               
    $a_C^Z/\Lambda^2$ & [-0.029, 0.029] & 5.2 10$^{-5}$ & 2.4 10$^{-5}$\\
                      &                 & (2.0 10$^{-5}$) & (9.2 10$^{-6}$)\\ \hline               
    \hline
   \end{tabular}
\end{center}
\caption{Reach on anomalous couplings obtained in $\gamma$ induced processes
after tagging the protons in the final state in the ATLAS Forward Physics
detectors compared to the present OPAL limits. The $5\sigma$ discovery and 95\%
C.L. limits are given for a luminosity of 30 and 200 fb$^{-1}$} 
\end{table}

\newpage
\clearpage

\setcounter{affil}{0}
\setcounter{section}{0}
\setcounter{figure}{0}
\setcounter{table}{0}
\setcounter{equation}{0}

\title{
Experimental results on diffraction at CDF}

\author{Michele Gallinaro\footnote{On behalf of the CDF Collaboration}
}
\affiliation{The Rockefeller University (USA) and LIP Lisbon, Portugal}

\begin{abstract}
Diffractive events are studied by means of identification of one or more rapidity gaps and/or a leading antiproton.
Measurements of soft and hard diffractive processes have been performed at the Tevatron $p\bar p$ collider and presented.
We report on the diffractive structure function obtained from dijet production 
in the range $0<Q^2<10,000$~GeV$^2$, and on the
$|t|$ distribution in the region $0<|t|<1$~GeV$^2$ for both soft and hard diffractive events up to $Q^2\approx 4,500$~GeV$^2$.
Results on single diffractive W/Z production, forward jets, 
and central exclusive production of both dijets and Z-bosons are also presented.
\end{abstract}
\maketitle

\addcontentsline{toc}{part}{Experimental results on diffraction at CDF - {\it M.Gallinaro}}

\section{Introduction}

Diffraction can be described as an exchange of a combination of quarks and gluons carrying the quantum numbers of the vacuum~\cite{dino}.
As no radiation is expected from such an exchange, diffractive processes are characterized by the presence of large rapidity regions 
not filled with particles (``rapidity gaps'').

%

%


At the Fermilab Tevatron collider, proton-antiproton collisions have been used to study diffractive interactions 
in Run~I (1992-1996) at an energy of $\sqrt{s}=1.8$~TeV and continued in Run~II (2003-present) with new and upgraded detectors at $\sqrt{s}=1.96$~TeV.
The goal of the CDF experimental program at the Tevatron is to provide results help decipher
the QCD nature of hadronic diffractive interactions, and to measure exclusive production rates which could be used to 
establish the benchmark for exclusive Higgs production at the Large Hadron Collider (LHC).
The study of diffraction has been performed by tagging events either with a rapidity gap or with a leading hadron.
The experimental apparatus includes a set of forward detectors\cite{fd} that 
extend the rapidity~\cite{rapidity} coverage to the forward region. 
The Miniplug (MP) calorimeters cover the region $3.5<|\eta|<5.1$; the Beam Shower Counters (BSC) surround the beam-pipe 
at various locations and detect particles in the region $5.4<|\eta|<7.4$;
the Roman Pot spectrometer (RPS) tags the leading hadron scattered from the interaction point after 
losing a fractional momentum approximately in the range $0.03<\xi<0.10$.

\section{Diffractive dijet production}

The gluon and quark content of the interacting partons can be investigated by comparing 
single diffractive (SD) and non diffractive (ND) events.
SD events are triggered on a leading anti-proton in the RPS
and at least one jet, while the ND trigger requires at least one jet in the calorimeters.
The ratio of SD to ND dijet production rates ($N_{jj}$) is proportional to the ratio 
of the corresponding structure functions ($F_{jj}$),
$R_{\frac{SD}{ND}}(x, \xi, t)= \frac{N_{jj}^{SD}(x, Q^2, \xi, t)}{N_{jj}(x, Q^2)}
\approx \frac{F_{jj}^{SD}(x, Q^2, \xi, t)}{F_{jj}(x, Q^2)}$,
and can be measured as a function of the Bjorken scaling variable $x\equiv x_{Bj}$\cite{xbj}.
In the ratio, jet energy corrections approximately cancel out, thus avoiding dependence on Monte Carlo (MC) simulation.
Diffractive dijet rates are suppressed by a factor of {\cal O(10)} with respect to expectations based on the proton PDF obtained 
from diffractive deep inelastic scattering at the HERA $ep$ collider~\cite{dino}.
The SD/ND ratios (i.e. gap fractions) of dijets, W, b-quark, $J/\psi$ production are all approximately 1\%, indicating that 
the suppression factor is the same for all processes and it is related to the gap formation.

\begin{figure}[htbp]
\begin{center}
\includegraphics[width=0.49\textwidth]{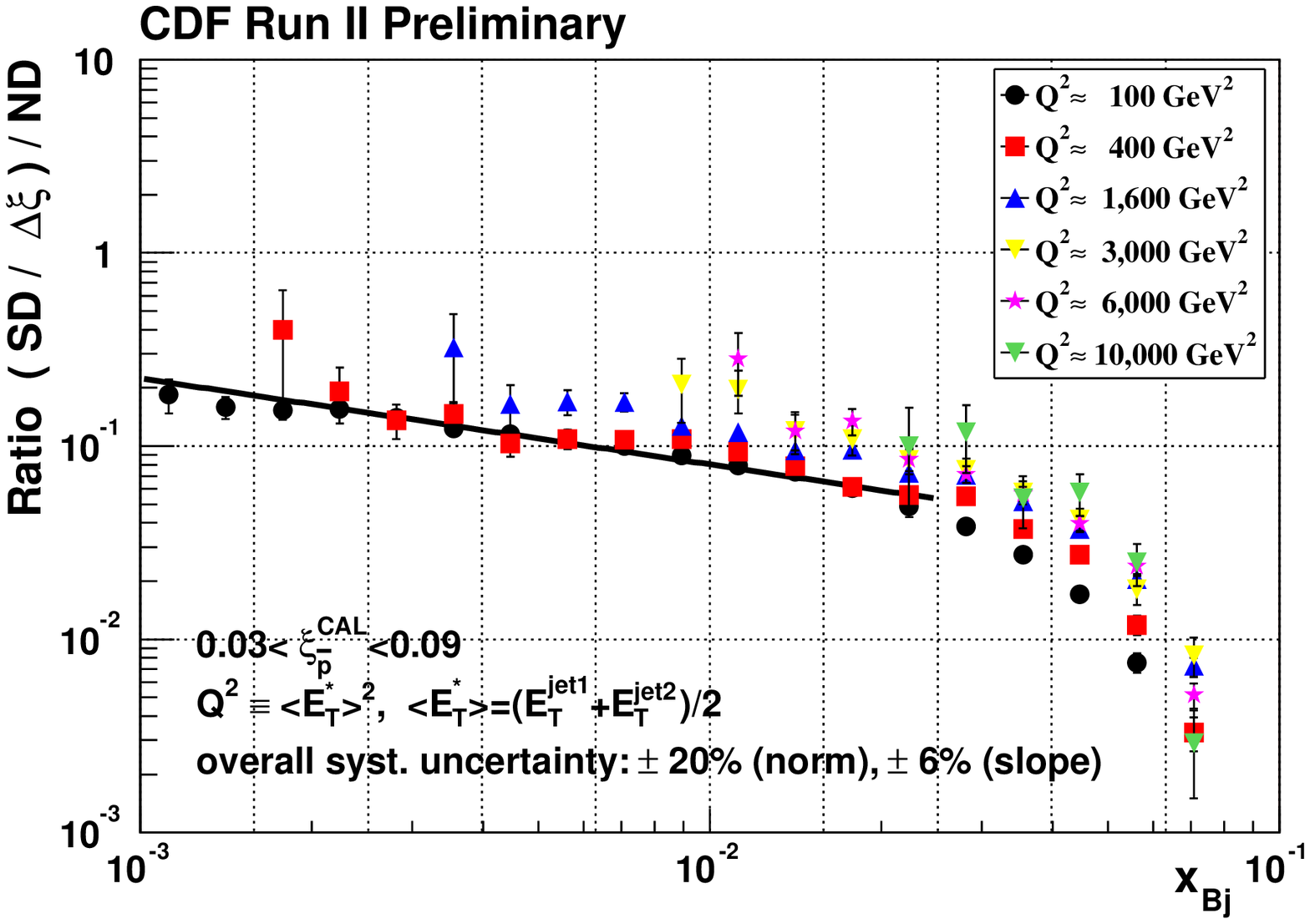}
\includegraphics[width=0.49\textwidth]{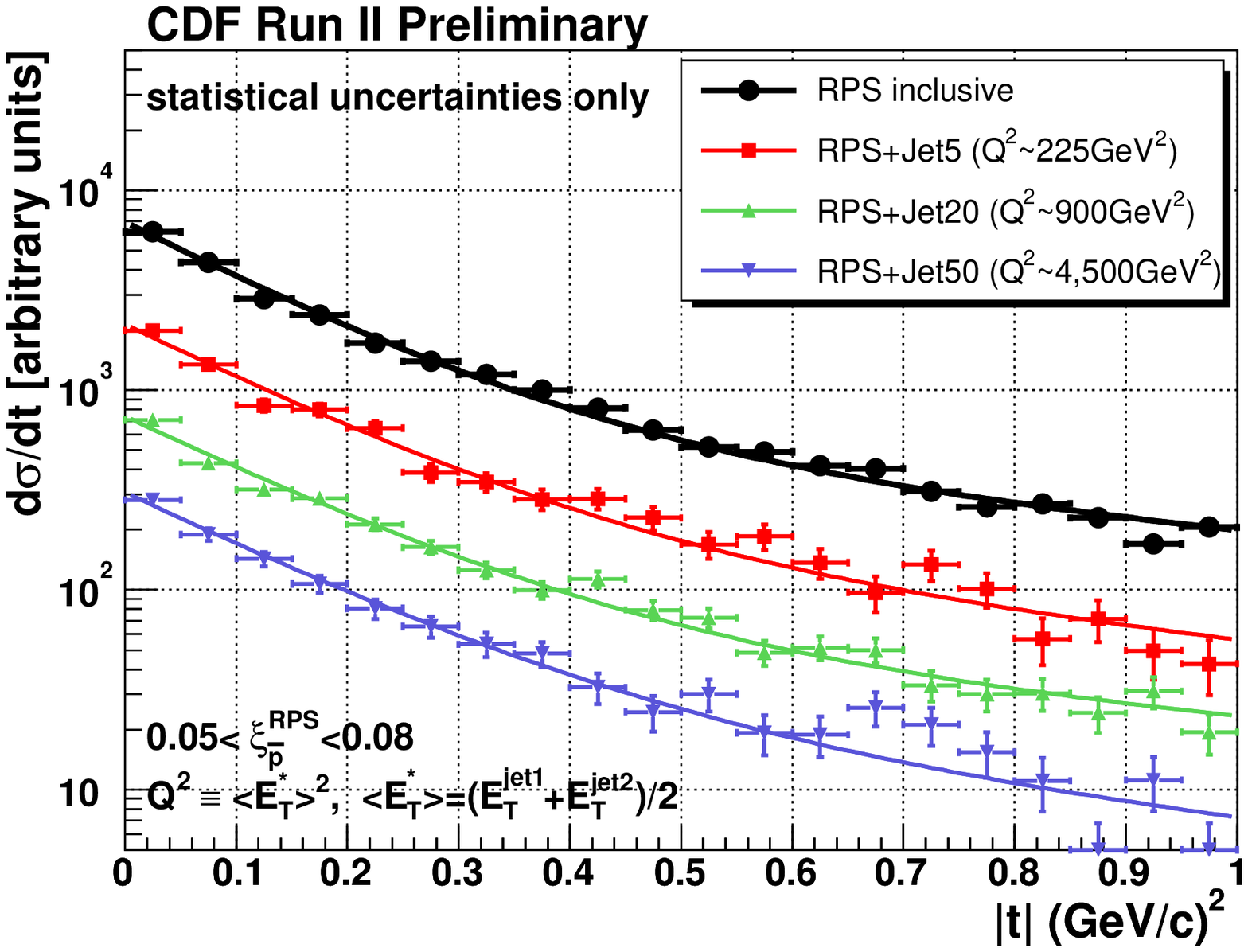}
\caption{
{\em Left}: Ratio of diffractive to non-diffractive dijet event rates
as a function of $x_{Bj}$ (momentum fraction of struck parton in the anti-proton) for different values of $E{_T}^2 \equiv Q^2 $;
{\em Right}: Measured $|t|$-distributions for soft and hard diffractive events.
}
\label{fig:dsf}
\end{center}
\end{figure}

In Run~II, the jet $E_T$ spectrum extends to $E_T^{\rm jet}\approx 100$~GeV, and 
results are consistent with those of Run~I\cite{run1_dsf}, hence confirming a breakdown of factorization. 
Preliminary results indicate that the ratio does not strongly depend on $E_T^2\equiv Q^2$ 
in the range $100<Q^2<10,000$~GeV$^2$ (Fig.~\ref{fig:dsf}, left).
The relative normalization uncertainty cancels out in the ratio, and the results indicate that the $Q^2$ evolution, 
mostly sensitive to the gluon density, is similar for the proton and the Pomeron.
A novel technique~\cite{dis06} to align the RPS is used to measure 
the diffractive dijet cross section
as a function of the $t$-slope in the range up to $Q^2\simeq 4,500$~GeV/c$^2$ (Fig.~\ref{fig:dsf}, right).
The shape of the $t$ distribution does not depend on the $Q^2$ value, in the region $0\le |t| \le 1$~GeV$^2$.
Moreover, the $|t|$ distributions do not show diffractive minima, which could be caused by the interference 
of imaginary and real parts of the interacting partons.

\section{Diffractive W/Z production}

Studies of diffractive production of the W/Z bosons are an additional handle to the understanding of diffractive interactions.
At leading order (LO) diffractive W/Z bosons are produced by a quark interaction in the Pomeron. 
Production through a gluon can take place at NLO, which is suppressed by a factor $\alpha_s$ 
and can be distinguished by the presence of one additional jet.

In Run~I, the CDF experiment measured a diffractive $W$ boson event rate $R_W=1.15\pm0.51$~(stat)$\pm 0.20$~(syst)\%. 
Combining the $R_W$ measurement with the dijet production event rate (which takes place both through quarks and gluons) and with the b-production rate
allows the determination of the gluon fraction carried by the Pomeron which can be estimated to be $54^{+16}_{-14}$\%~\cite{gluonfraction}.

\begin{figure}[htbp]
\begin{center}
\includegraphics[width=0.44\textwidth]{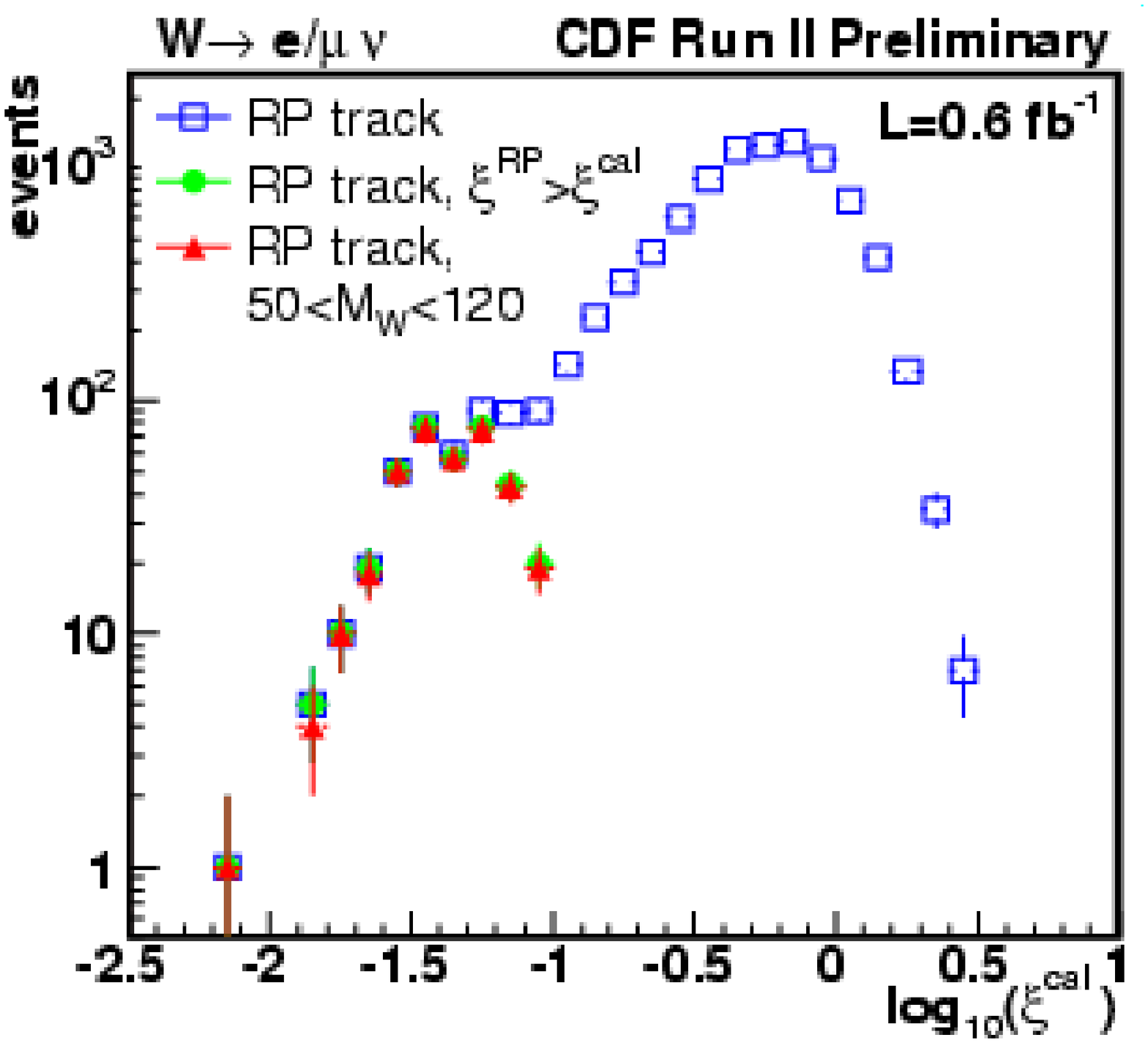}
\includegraphics[width=0.49\textwidth]{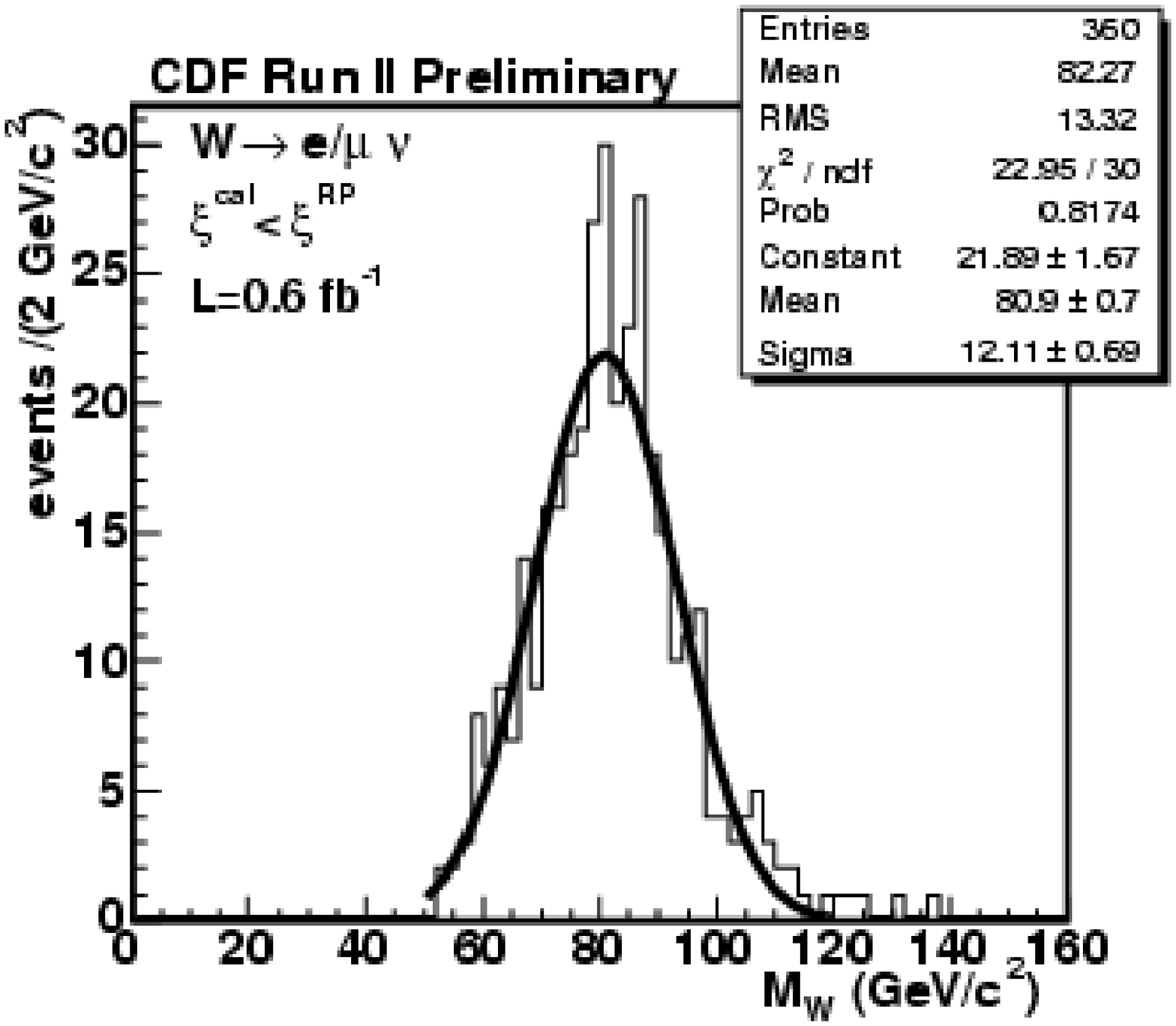}
\caption{Calorimeter $\xi^{cal}$ distribution in $W$ events with a reconstructed Roman Pot track ({\it left}).
Due to the neutrino, $\xi^{cal}<\xi^{RPS}$ is expected. The difference $\xi^{RPS}-\xi^{cal}$ is used to determine the $W$ boson mass ({\it right}).}
\label{fig:diffW}
\end{center}
\end{figure}

In Run~II, the RPS provides an accurate measurement of the 
fractional energy loss ($\xi$) of the leading hadron (Fig.~\ref{fig:diffW}, left), removing the ambiguity of the gap survival probability.
The innovative approach of the analysis takes advantage of the full
$W\rightarrow l\nu$ event kinematics including the neutrino.
The missing transverse energy ($\met$) is calculated as usual from all calorimeter towers, and the neutrino direction (i.e. $\eta_\nu$) 
is obtained from the comparison between the fractional energy loss measured in the Roman Pot spectrometer ($\xi^{RPS}$) 
and the same value estimated from the calorimeters ($\xi^{cal}$): 
$\xi^{RPS}-\xi^{cal}={\frac{\met}{\sqrt{s}}}\cdot e^{-\eta_\nu}$. 
The reconstructed $W$ mass (Fig.~\ref{fig:diffW}, right) yields $M_W=80.9\pm 0.7$~GeV/c$^2$, in good agreement with the world average value of
$M_W=80.398\pm 0.025$~GeV/c$^2$\cite{pdg}.
After applying the corrections due to the RPS acceptance, trigger and track reconstruction efficiencies, and taking into account the effect of multiple interactions,
both $W$ and $Z$ diffractive event rates are calculated:
$R_W=0.97\pm 0.05 {\rm (stat)}\pm 0.10{\rm (syst)}\%$, and $R_Z$=0.85$\pm$0.20~(stat)$\pm$0.08~(syst)\%~\cite{mary}.

Search for exclusive Z candidate events, i.e. $p\bar p\rightarrow p + Z + \bar p$, has been explored with a null result.
In the SM, the process takes places through photo-production. 
The search requires that nothing else is found in the detector, except the two leptons from $Z\rightarrow ll$.
The method consists in comparing the total energy in the calorimeter ($M_X$) to the dilepton invariant mass ($M_{ll}$). 
Exclusive events are expected to be found on the diagonal $M_X$=$M_{ll}$. Some effects that may artificially change the ``exclusive'' behavior.
By increasing the calorimeter thresholds the value of $M_X$ moves closer to the diagonal $M_X$=$M_{ll}$. 
Because of charge conservation, $W$ bosons cannot be produced exclusively and are used as the control sample. 
Additional control of the background is performed by looking at ``empty crossings'', where no tracks are reconstructed and calorimeter noise is the dominant effect.
No exclusive candidate are found in the data.

\section{Forward jets}

\begin{figure}[htbp]
\begin{center}
\includegraphics[width=0.49\textwidth]{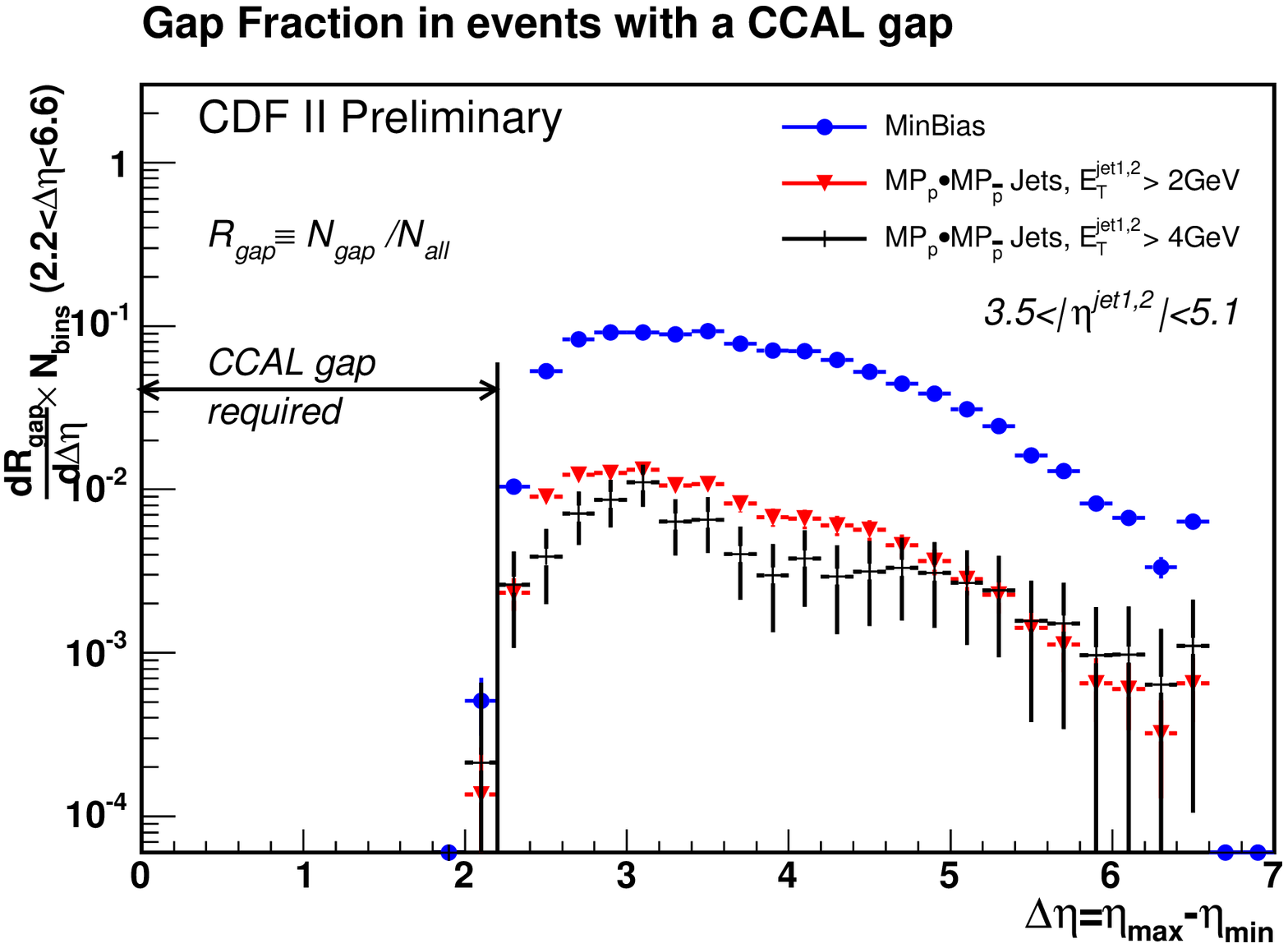}
\includegraphics[width=0.49\textwidth]{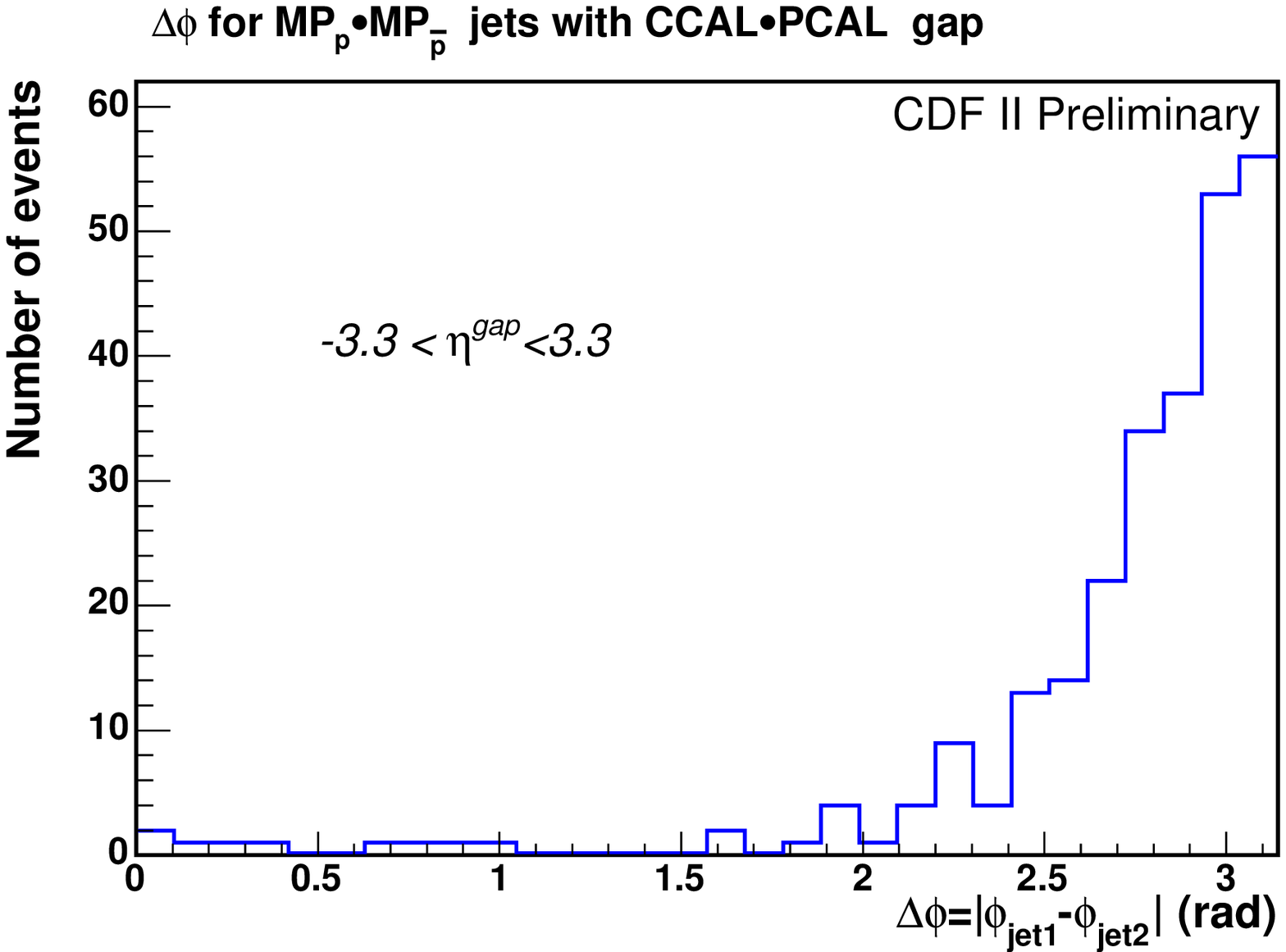}
\caption{
{\it Left:} Event rate gap fraction defined as $R_{gap}=N_{gap}/N_{all}$, for minimum bias (MinBias) and MP jet events with $E_T>2 (4)$~GeV;
{\it Right:} Azimuthal angle difference $\Delta\phi$ distribution of the two leading jets in a DD event with a central rapidity gap ($|\eta^{gap}|<3.3$).
}
\label{fig:forwardjets}
\end{center}
\end{figure}

An interesting process is dijet production in double diffractive (DD) dissociation.
DD events are characterized by the presence of a large central rapidity gap and are presumed to be due to 
the exchange of a color singlet state with vacuum quantum numbers.
A study of the dependence of the event rate on the width of the gap was performed using Run~I data with small statistics.
In Run~II larger samples are available.
Typical luminosities (${\cal L}\approx 1\div 10 \times 10^{31}$cm$^{-2}$sec$^{-1}$) 
during normal Run~II run conditions hamper the study of gap ``formation''
due to multiple interactions which effectively ``kill'' the gap signature.
Central rapidity gap production was studied in soft and hard diffractive events collected during a special 
low luminosity run (${\cal L}\approx 10^{29}cm^{-2}sec^{-1}$).
Figure~\ref{fig:forwardjets} (left) shows a comparison of the gap fraction rates, as function of the gap width (i.e. $\Delta\eta$) 
for minimum bias (MinBias), and MP jet events.
Event rate fraction is calculated as the ratio of the number of events in a given rapidity gap region divided by all events: $R_{gap}=N_{gap}/N_{all}$. 
The fraction is approximately 10\% in soft diffractive events, and approximately 1\% in jet events.
Shapes are similar for both soft and hard processes, and gap fraction rates decrease with increasing $\Delta\eta$.
The MP jets of gap events are produced back-to-back (Fig.~\ref{fig:forwardjets}, right).

\section{Exclusive production}

\begin{figure}[htp]
\begin{center}
\includegraphics[width=0.30\textwidth]{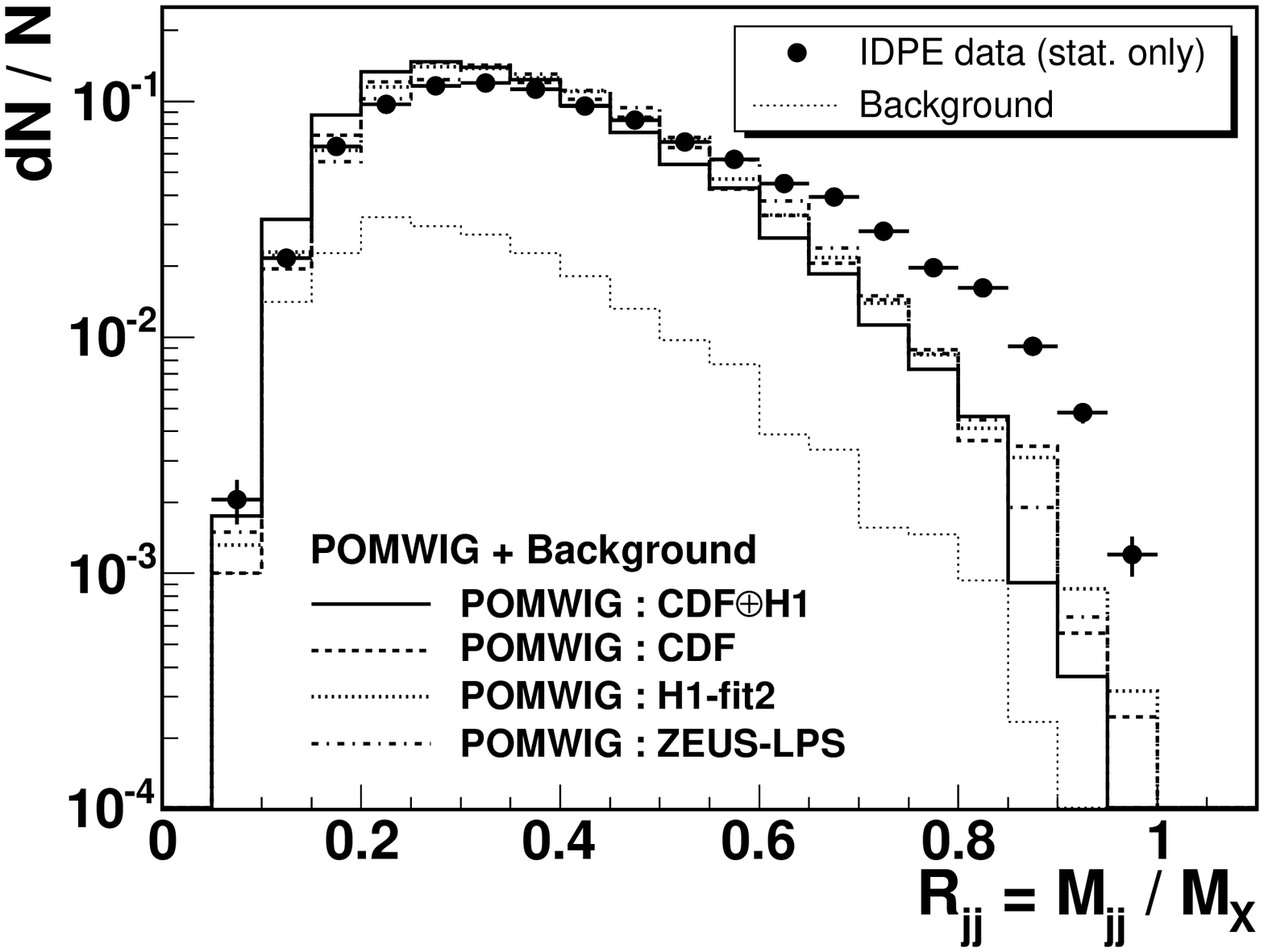}
\includegraphics[width=0.35\textwidth]{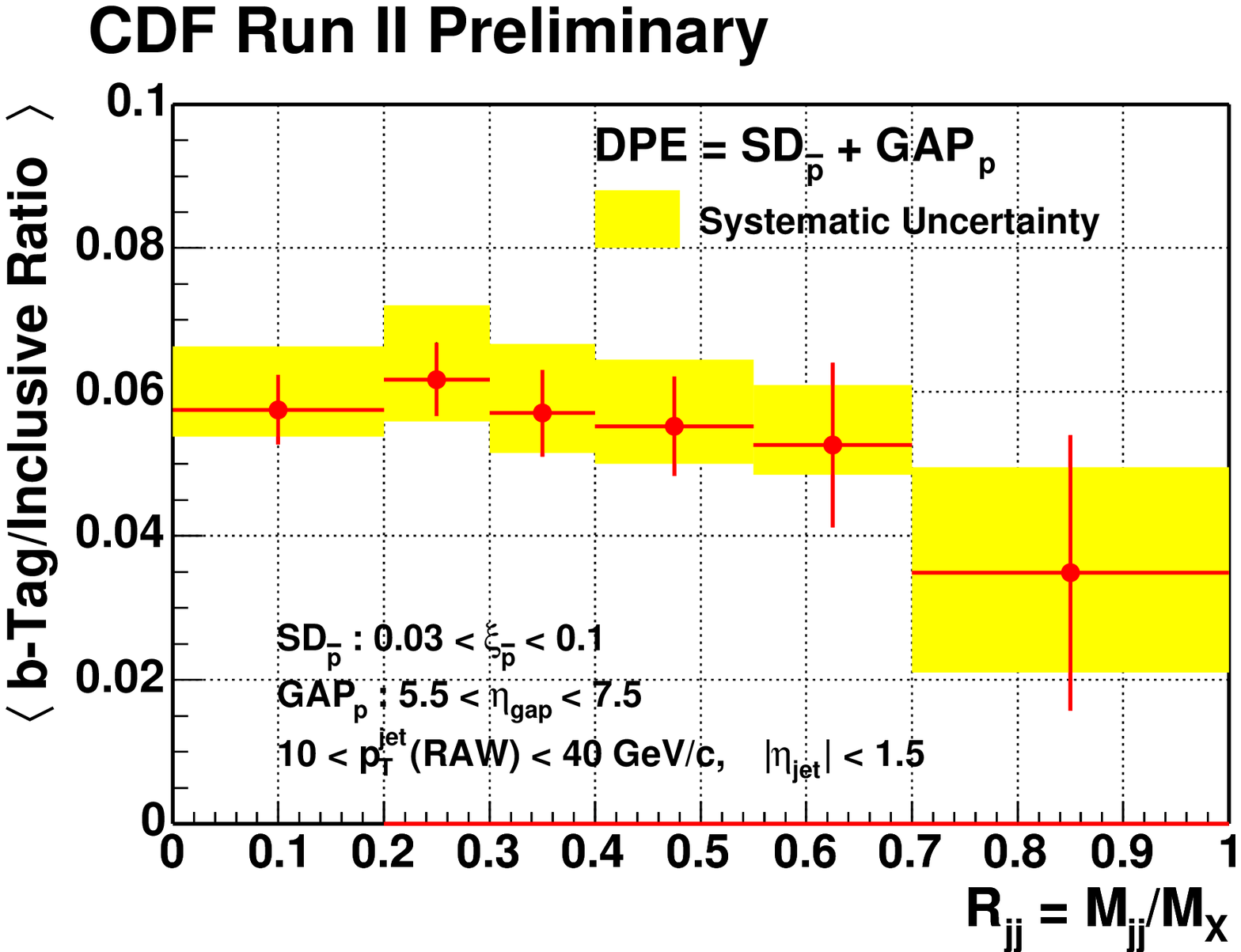}
\includegraphics[width=0.33\textwidth]{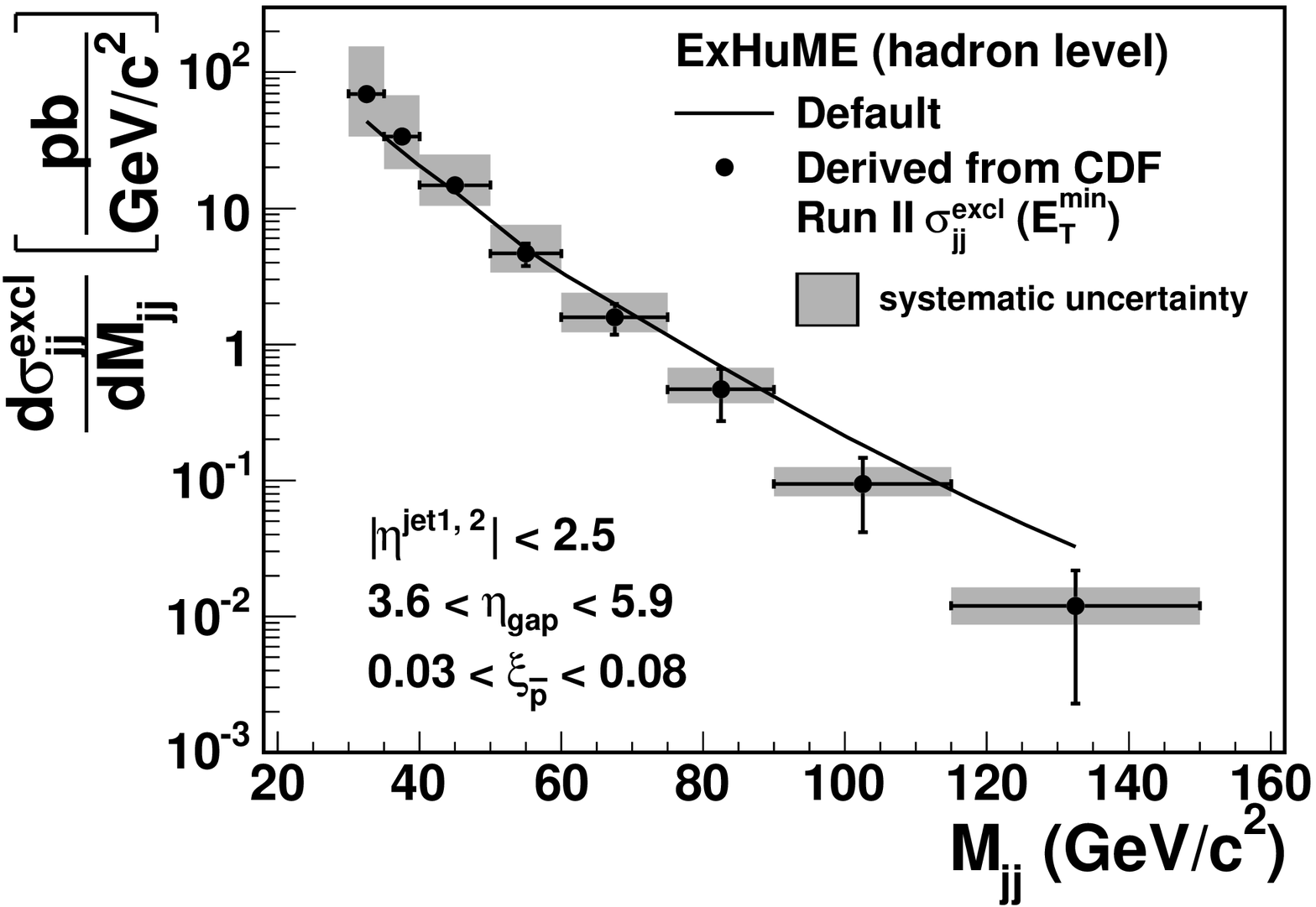}
\caption{{\it Left:} Dijet mass fraction $R_{jj}$ in inclusive DPE dijet data. 
An excess over predictions at large $R_{jj}$ is observed as a signal of exclusive dijet production;
{\it Center:} Ratio of b-tagged jets to all inclusive jets as a function of the mass fraction $R_{jj}$. 
The error band corresponds to the overall systematic uncertainty;
{\it Right:} The cross section for events with $R_{jj}>0.8$ is compared to predictions.}
\label{fig:exclusive}
\end{center}
\end{figure}

The first observation of the process of exclusive dijet production can be used as a benchmark to establish predictions 
on exclusive diffractive Higgs production, a process with a much smaller cross section\cite{kmr}.
A wide range of predictions was attempted to estimate the cross section for exclusive dijet and Higgs production. 
In Run~I, the CDF experiment set a limit on exclusive jet production~\cite{runI_excldijet}.
First observation of this process was made in Run~II.
The search strategy is based on measuring the dijet mass fraction ($R_{jj}$), 
defined as the ratio of the two leading jet invariant mass divided by the total mass calculated using all calorimeter towers. 
An exclusive signal is expected to appear at large $R_{jj}$ values (Fig.~\ref{fig:exclusive}, left).
The method used to extract the exclusive signal from the $R_{jj}$ distribution is based on fitting the data to MC simulations.
The quark/gluon composition of dijet final states can be exploited to provide additional hints on exclusive dijet production.
The $R_{jj}$ distribution can be constructed using inclusive or b-tagged dijet events. 
In the latter case, as the $gg\rightarrow q\bar{q}$ is strongly suppressed for $m_q/M^2\rightarrow 0$ ($J_z=0$ selection rule),
only gluon jets will be produced exclusively and heavy flavor jet production is suppressed. 
Figure~\ref{fig:exclusive} (center) illustrates the method that was used to determine the heavy-flavor composition of the final sample.
The falling distribution at large values of $R_{jj}$ ($R_{jj}>0.7$) indicates the suppression of the exclusive b-jet events.
The CDF result favors the model in Ref.~\cite{kmr2} (Fig.~\ref{fig:exclusive}, right).
Details can be found in Ref.~\cite{runII_excldijet}.

Exclusive $e^+e^-$ and di-photon production were studied using a trigger that requires forward gaps 
on both sides of the interaction point and at least two energy clusters in the electromagnetic calorimeters with transverse energy $E_T>5$~GeV. 
All other calorimeter towers are required to be below threshold.
In the di-electron event selection, the two tracks pointing at the energy clusters are allowed.
The CDF experiment reported the first observation of exclusive $e^+e^-$ production~\cite{excl_ee}.
A total of 16 $\gamma\gamma\rightarrow e^+e^-$ candidate events are observed, consistent with QED expectations.
Exclusive di-photon events can be produced through the process $gg\rightarrow \gamma\gamma$. Three candidate events were selected,
where one is expected from background sources (i.e. $\pi^0\pi^0$). 
A 95\%C.L. cross section limit of 410~pb can be set~\cite{excl_gg}, about ten times larger than expectations~\cite{kmr3}.

\section{Conclusions}

The results obtained  during the past two decades have led the way to the identification of
striking characteristics in diffraction. Moreover, they have significantly contributed to an understanding of
diffraction in terms of the underlying inclusive parton distribution functions.
The regularities found in the Tevatron data and the interpretations of
the measurements can be extrapolated to the LHC era.
At the LHC, the diffractive Higgs can be studied but not without challenges, as triggering and event acceptance
will be difficult.
Still, future research at the Tevatron and at the LHC holds much promise for further understanding of diffractive processes.

\section{Acknowledgments}
My warmest thanks to the organizers for the kind invitation and for a warm atmosphere, and in particular to S. Lami for providing the financial support.

\newpage
\clearpage
\setcounter{affil}{0}
\setcounter{section}{0}
\setcounter{figure}{0}
\setcounter{table}{0}
\setcounter{equation}{0}

\title{Forward Physics at LHCb\\
{\small -- Prospects for the Study of Diffractive Interactions--}}
\author{Michael Schmelling}
\thanks{on behalf of the LHCb Collaboration}
\affiliation{MPI for Nuclear Physics, Saupfercheckweg 1, D-69117 Heidelberg, Germany}
\email{Michael.Schmelling@mpi-hd.mpg.de}

\begin{abstract}
LHCb, the smallest of the large LHC experiments, is a forward
spectrometer covering the angular range $2<\eta<5$\/ with 
tracking, calorimetry and particle identification. Partial
coverage of the backward hemisphere is also provided by the vertex
detector (Vertex Locator, VeLo), a silicon strip detector surrounding 
the interaction region. Generator level Monte Carlo studies suggest 
that using the VeLo to ask for a rapidity gap of $\Delta\eta=2.5$\/ 
in the backward region allows to select event samples dominated by 
diffractive processes. Making use of the excellent tracking, vertexing
and particle identification capabilities of the LHCb detector, 
the characteristics of diffractive particle production thus 
can be studied in detail in the forward acceptance covered
by the experiment. 
\end{abstract}

\maketitle
\addcontentsline{toc}{part}{Forward Physics at LHCb: Prospects for the Study of Diffractive Interactions - {\it M.Schmelling}}
\section{Introduction}
With the startup of LHC an energy regime has become experimentally
accessible which will allow to probe fundamental physics with 
unprecedented sensitivity. Although built with the focus on finding
the Higgs boson and doing searches for physics beyond the standard model,
also basic questions of particle production in high energy collisions
have to be addressed. Minimum bias physics at the LHC differs from 
previous studies at hadron machines in that the center-of-mass 
energy has reached a level, where even collisions between very soft
partons can contribute to final state particle production. The typical 
scale is given by the requirement $x_1\cdot x_2\cdot  s > 4m_{\pi}^2$,
i.e. $x\sim 2m_{\pi}/\sqrt{s}$. Since the parton densities at small
$x$\/ and small momentum transfer rise faster than $1/x$, it is 
expected that multi-parton interactions become important and that 
a new holistic picture including diffractive processes for describing
such interactions is required. In this paper some thoughts addressing 
this issue are discussed together with first ideas how the LHCb 
experiment can contribute in this area.

\section{The LHCb Experiment}
\begin{figure}[htb]
\includegraphics[width=0.8\textwidth]{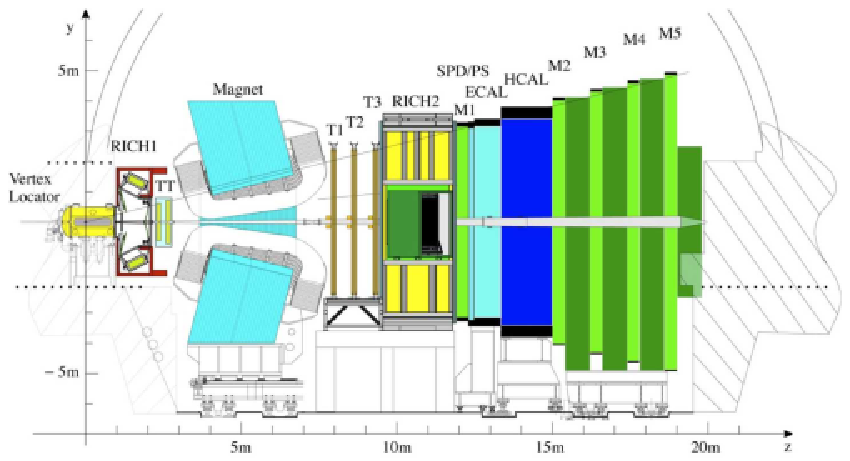}
\caption{
Schematic view of the LHCb single arm forward spectrometer. 
The interaction region is located on the left inside the Vertex 
Locator. The tracking system and the RICH detectors for particle 
identification are installed both before and after the dipole 
magnet, calorimetry and the muon system are located downstream
of the magnet.}
\label{fig:lhcb}
\end{figure}

The LHCb detector \cite{Alves:2008zz} is constructed as a forward 
spectrometer, covering the angular range of $15 < \theta < 300$\,mrad with 
respect to the beam axis. A schematic view of the experiment is shown in 
Fig.\,\ref{fig:lhcb}. Momentum measurement is performed with a dipole
magnet with a field integral of 4\,Tm. In front of the magnet the 
Vertex Locator (VeLo) surrounds the interaction region. Going downstream,
a first RICH detector and the so-called TT tracking station are still 
located in front of the magnet. Immediately behind the magnet follows the 
second part of the tracking system, consisting of a high granularity 
Inner Tracker (IT) in the region of large particle densities close to the 
beam pipe and the Outer Tracker system at larger transverse distances.
VeLo, TT and IT are silicon strip detectors, the OT consists of straw
tubes. Following the tracking system is a second RICH detector, a
pre-shower and scintillating pad detector (SPD/PS), electromagnetic
calorimeter (ECAL), hadron calorimeter(HCAL) and muon system for the
identification of electrons and photons, neutral hadrons and muons,
respectively. The RICH detectors allow pion, kaon, proton separation
in the momentum range between $2 < p < 100$\,GeV/$c$. The detector 
is constructed such that it offers tracking, calorimetry and
particle identification over most of its forward acceptance. 

\begin{figure}[htb]
\includegraphics[width=0.8\textwidth]{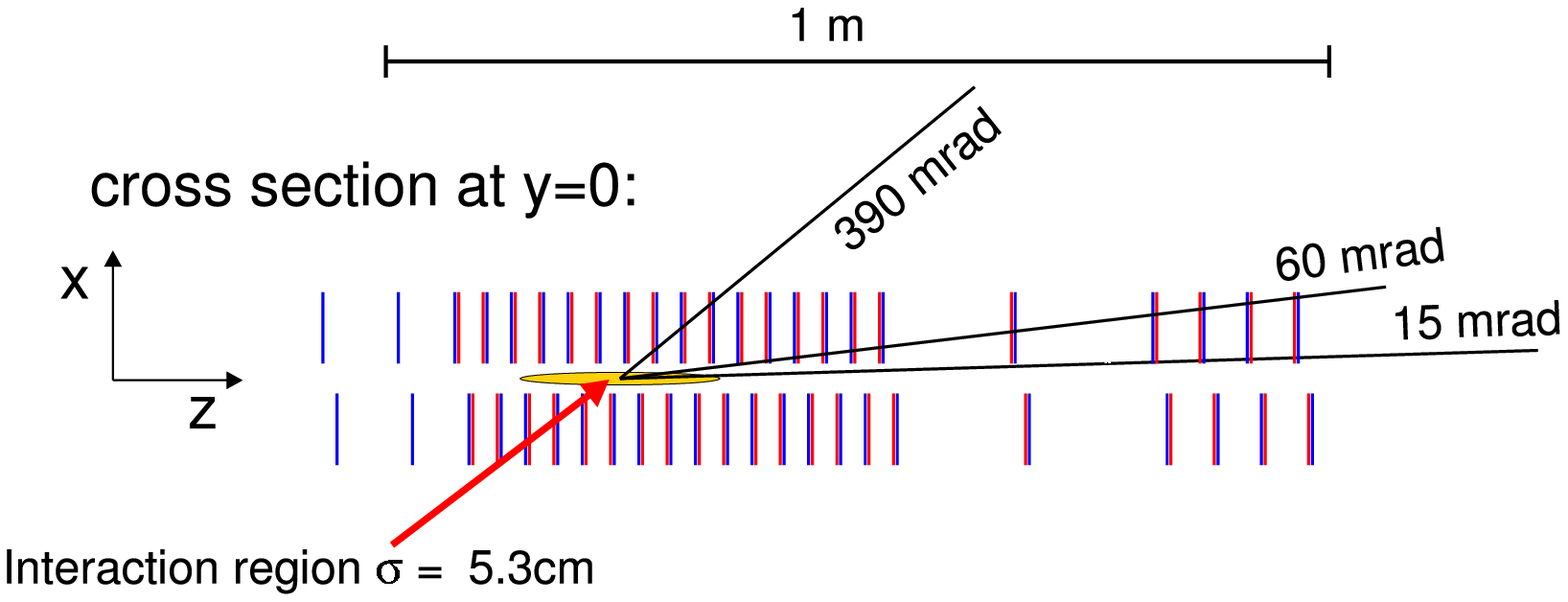}
\caption{
Layout of the LHCb Vertex Locator (VeLo) in the horizontal 
$(x,z)$-plane, with the $z$-axis along the direction of the 
proton beams. 21 sensor planes measure space points, 
the two most backward ($-z$) layers provide only radial 
coordinates of charged particle tracks.}
\label{fig:velo}
\end{figure}

The Vertex Locator has 21 double-layer sensor planes around the 
interaction region for measuring space points, plus two additional 
planes providing only radial track coordinates. The layout of the VeLo
is shown in Fig.\,\ref{fig:velo}. It has a larger angular acceptance 
than the rest of the tracking system and covers even part of the 
backward hemisphere. However, being located outside of the magnetic 
field, VeLo track segments do not have momentum information. 
Furthermore, since at least three planes are required to reconstruct
a track segment, the VeLo is blind in the central region. Charged 
particle tracks are reconstructed in the rapidity ranges $-4<\eta<-1.5$\/
and $1.5<\eta<5$. As will be shown below, the large angular coverage of 
the VeLo is vital for the study of diffractive processes.

\section{Inelastic Proton-Proton Interactions}
\begin{figure}[t]
\includegraphics[width=0.9\textwidth]{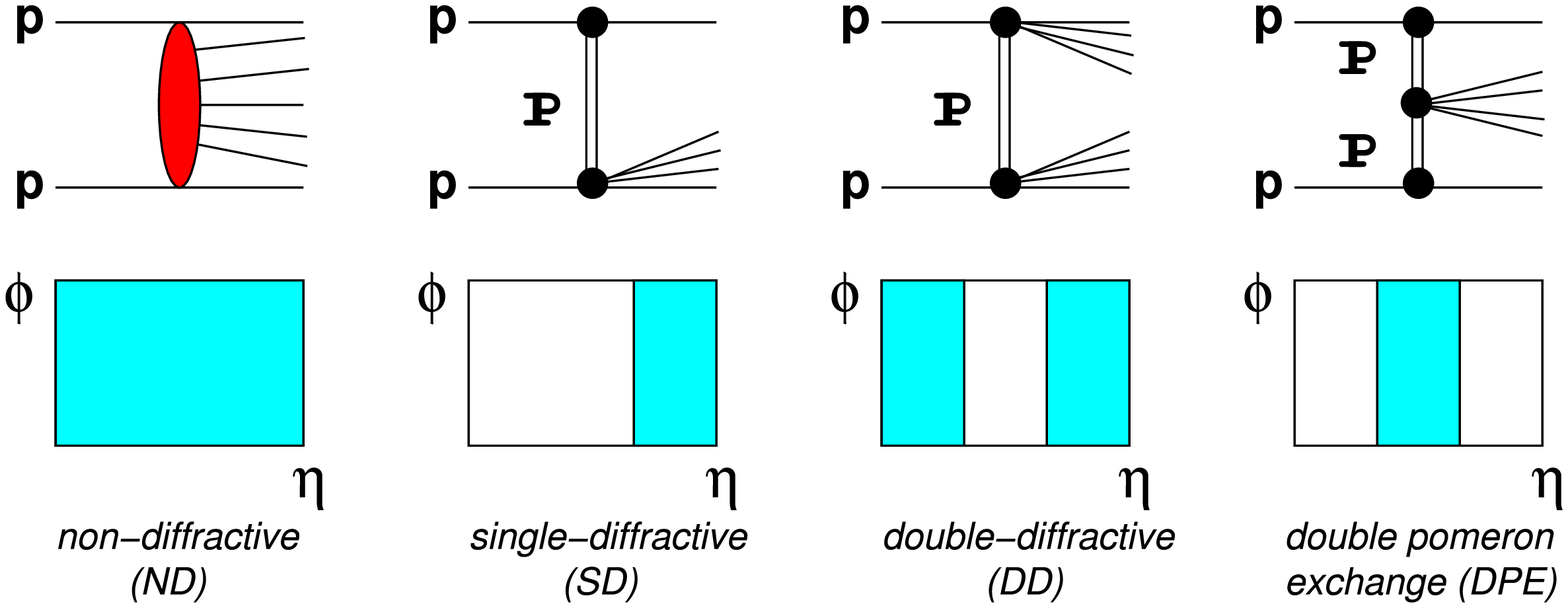}
\caption{
Schematic classification of inelastic proton-proton collisions.
The upper row shows some born-level type diagrams for different 
classes of interactions, the lower row illustrates the angular
range into which particles produced in the collision are emitted.
Note that in the upper row rapidity runs from top to bottom,
while it goes from left to right in the lower row. Also, while 
the born-level type diagrams represent scattering amplitudes,
the particle flow corresponds to cross sections, i.e. the 
squares of the respective amplitudes.}
\label{fig:ppint}
\end{figure}

A schematic view of different types of inelastic $pp$-interactions 
is presented in Fig.\,\ref{fig:ppint}. Here the basic distinction 
is colour(-octet)- and colour-singlet exchange, respectively, between 
the colliding protons. Colour-exchange implies that the structure of 
both protons is resolved with the consequence that the colour fields 
stretched between the partons lead to particle production in the full 
rapidity range. In contrast, colour-singlet exchange is 
phenomenologically described by pomerons coupling to the protons as 
a whole. No colour is transferred and the protons can either scatter 
elastically or be excited into a high mass state which then decays 
to produce multi-particle final states. Depending on whether only 
one or both protons are excited these processes are referred to as 
single- or double-diffractive scattering. An example for a higher 
order process involving pomerons is double pomeron exchange, where 
both protons stay intact and the two pomerons interact to form 
a massive central system.

The above classification of $pp$-interactions is most adequate 
at small momentum transfers. In the language of QCD then colour 
exchange means gluon exchange and the pomeron can be understood 
as a colour-singlet two gluon state. With increasing momentum 
transfer, however, the simple picture breaks down. The two
protons are resolved into an increasing number of partons 
and the interaction is described by ladder-diagrams of all 
possible topologies. Diffractive and non-diffractive scattering
is no longer an unambiguous classification, and even the 
notion of colour singlet exchange becomes frame 
dependent \cite{Gustavson:2010zz}. It follows that eventually
a unified description of hadron-hadron collisions is 
required which covers diffractive and non-diffractive physics
within a coherent framework \cite{Jung:2010zz}.

Figure\,\ref{fig:ppint} illustrates another conceptual problem with 
respect to diffractive and non-diffractive interactions. While
the upper row represents Born-level type diagrams of the different
processes, i.e. amplitudes contributing to the inelastic 
proton-proton scattering, the lower one is a pictorial 
representation of the cross-section. In many Monte Carlo models,
such as e.g. {\tt PYTHIA} \cite{Sjostrand:2006za}, the different
components making up the total cross-section are generated 
independently. Interference terms between the different amplitudes 
are ignored. Physics-wise, however, these terms certainly contribute, 
as it is experimentally not possible to distinguish the fast forward 
proton from a single-diffractive scattering from a non-diffractive 
interaction where a fast proton is generated in the fragmentation
process \cite{Skands:2010zz}. An unambiguous classification of an 
event into one of several types of inelastic processes thus is 
impossible. Furthermore, any such separation within the context 
of a specific Monte Carlo simulation is to some extent arbitrary
since parameter tuning generally allows to trade e.g. a larger 
diffractive cross-sections against a smaller non-diffractive part 
by choosing a different setting of the hadronization parameters. 

In the past experimental measurements often were corrected for e.g.
single diffractive contributions to the cross section. From the
previous discussion it is clear, that such attempts to focus on
the non-diffractive cross section are always model dependent, even
though they work to a certain degree since, in the language of 
Fig.\,\ref{fig:ppint}, some amplitudes are dominant in specific
regions of phase space. Nevertheless, a better approach would 
be to avoid any such artificial classifications, and instead
perform measurements subject to experimental cuts which enhance 
or suppress certain contributions to the cross section. That way 
the results of the measurements do not rely on a particular
model, and measurements and their interpretation are cleanly 
separated.

\section{Monte Carlo Studies}
To study the prospects for experimentally studying the properties
of events with dominantly diffractive contributions, a simple 
generator level study has been performed. The study is based
on {\tt PYTHTIA 8.135} available from \cite{Sjostrand:2010zz}.
Single proton-proton collisions with a center-of-mass energy
$\sqrt{s}=7$\,TeV were generated with process selection
{\tt pythia.readString("SoftQCD:all=on")}. 

The study focuses on a measurement of the inclusive charged particle 
transverse momentum spectrum for all events and for events with an 
enhanced fraction of diffractive contributions. For this tracking 
based study the VeLo was simulated with its nominal geometry. A track 
was assumed to be measured by the VeLo if at least three stations 
were hit. The event selection was based on the VeLo track segments  
only. For accepted events the transverse momentum spectrum then 
was determined using all tracks with a VeLo-segment and within
the acceptance of the tracking system behind the magnet. The 
latter was approximated by the requirement $p>2$\,GeV/$c$\/ 
the pseudo-rapidity range $2<\eta<5$.  

A diffraction-enriched event sample was selected by exploiting 
the fact that diffractive events are characterized by enhanced 
probabilities for large rapidity gaps in the final state particle 
distribution. To obtain a quantitative measure for the level of 
enrichment which can be achieved, the {\tt PYTHIA} process type 
was analyzed for all events. While evidently giving model dependent
estimates for the fractions of different events, the qualitative 
picture is expected to be generic.

Denoting by $n_B$\/ and $n_F$\/ the number of VeLo track segments 
in the backward ($\eta<0$) and forward ($\eta>0$) hemispheres, two
selection criteria were studied: (a) $n_B+n_F>0$\/ and 
(b) $n_B==0 \;\&\&\; n_F>0$. Criterion (a) corresponds to the
so called micro-bias trigger of LHCb, which is close to 100\% 
efficient for non-diffractive $pp$-collisions. The second
criterion asks for no charged tracks in the backwards acceptance
of the VeLo, i.e. it corresponds to a requirement of a rapidity 
gap $\Delta\eta=2.5$\/ for charged tracks. 

\begin{figure}[t]
\includegraphics[width=0.9\textwidth]{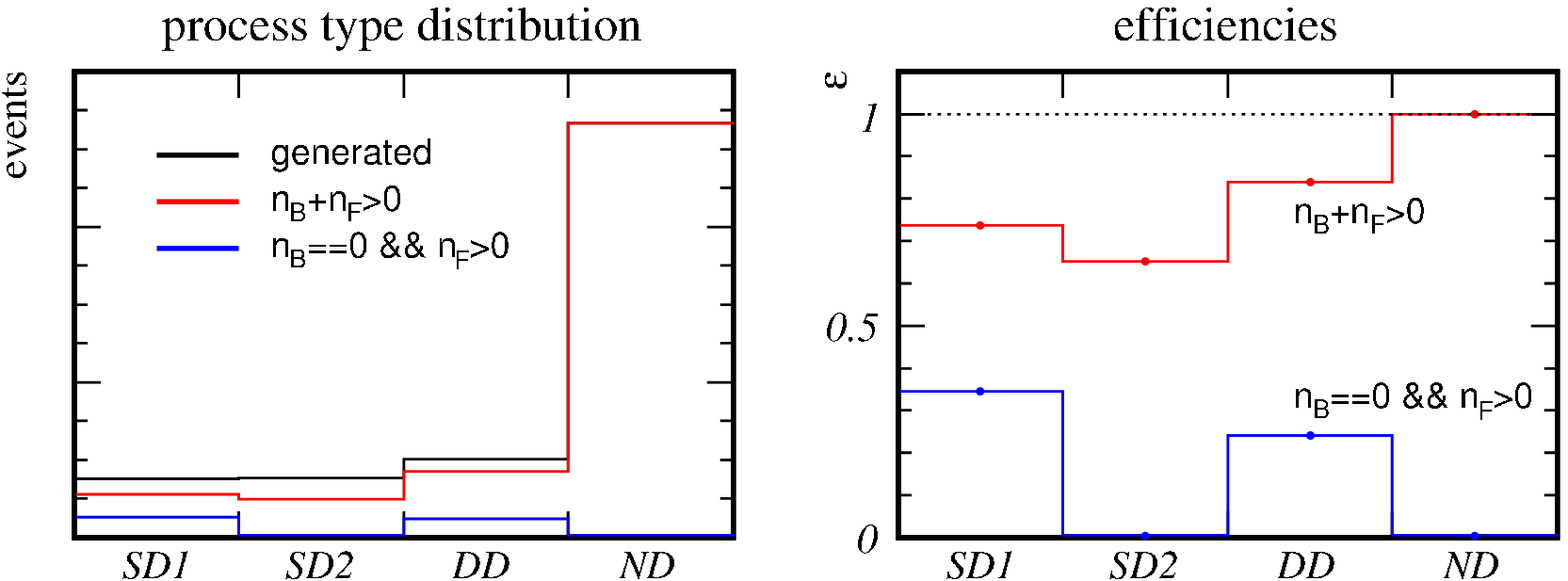}
\caption{
Generator level MC study: Mix of single diffractive, double diffractive
and non-diffractive events generated by {\tt PYTHIA 8.135}. Here {\tt SD1}
refers to single diffractive scattering where the excited proton travels 
in the direction of the LHCb detector, in {\tt SD2} the decaying heavy 
mass moves into the opposite direction. The left hand plot shows the mix 
of events generated (black), passing the micro-bias trigger(red) and the 
diffraction selection (blue). The  right hand plot shows the selection 
efficiencies for the two cases.}
\label{fig:compeff}
\end{figure}

\begin{figure}[t]
\includegraphics[width=0.975\textwidth]{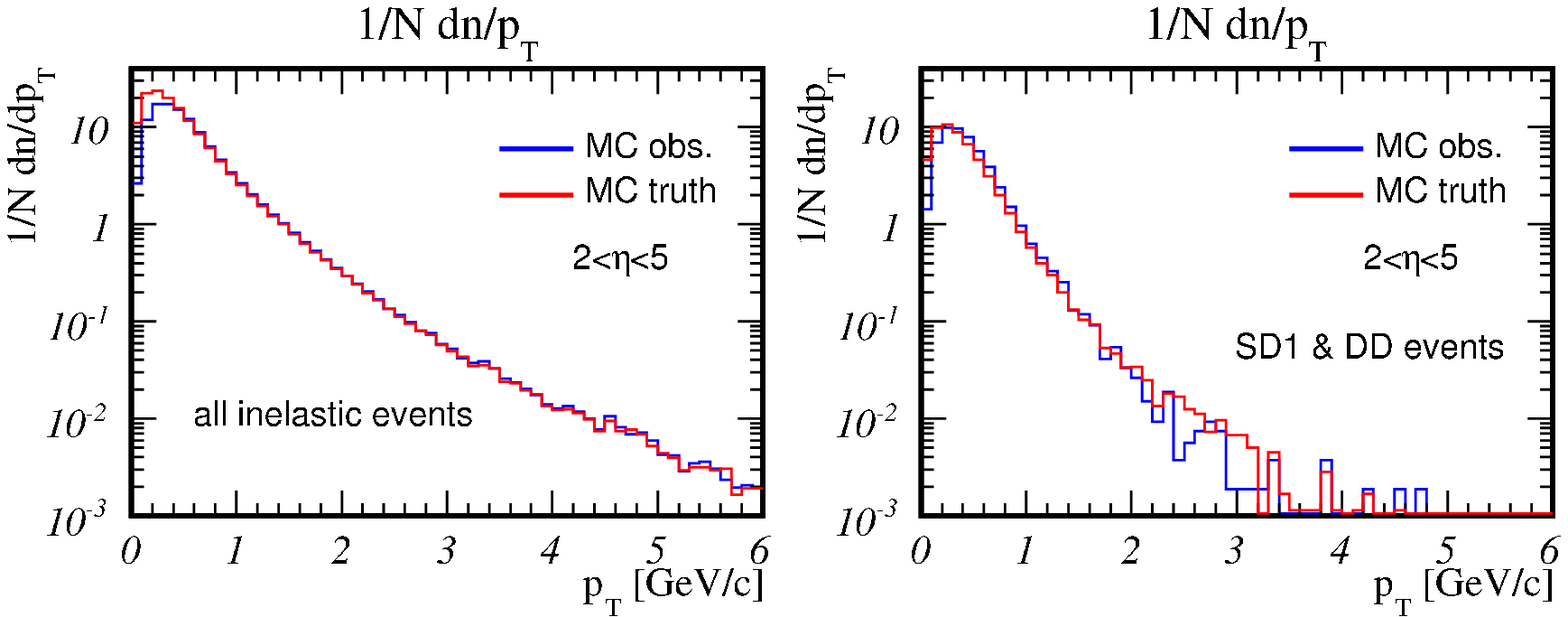}
\caption{
Generator level MC study: comparison of generated and observable
transverse momentum spectra of charged particles in the 
pseudo rapidity range $2<\eta<5$\/ when asking for a micro-bias 
trigger (left) or a rapidity gap (right). The spectra are 
normalized to the number of accepted events. Finite resolution 
or imperfect tracking efficiency has not been modeled. For the
diffraction-enhanced event sample the MC-truth level has been
defined to include only events of type ${\tt SD1}$\/ and 
${\tt DD}$. The losses at small transverse momenta are due to 
incomplete geometric coverage at low $p_T$\/ and low $\eta$.}
\label{fig:genrec}
\end{figure}

Results are shown in Figs.\,\ref{fig:compeff} and \ref{fig:genrec}.
The left hand plot of Fig.\,\ref{fig:compeff} shows the mix
of single diffractive, double diffractive and non-diffractive events
generated by {\tt PYTHIA 8.135}. One clearly sees that the requirement
of a rapidity gap in the backwards region almost completely suppresses
non-diffractive events while keeping between $20\%$\/ and $30\%$\/ of 
diffractive interactions. The comparison of the transverse momentum 
spectra in Fig.\,\ref{fig:genrec} shows very good agreement between 
generated and observed distribution, i.e. a robust measurement 
comparing the fully inclusive transverse momentum spectra and the
spectra in events dominated by diffraction seems feasible. Other
observables like charge ratios, the production cross-sections 
for identified particles or particle ratios are a natural extension 
of these studies.

\section{Summary and Outlook}
A model independent approach toward the study of diffractive particle
production in minimum bias events has been presented. Generator 
level Monte Carlo studies suggest that asking for a rapidity gap 
of $\Delta\eta=2.5$\/ in the backward region of LHCb VeLO allows
to select event samples dominated by diffractive processes. Making 
use of the excellent tracking, vertexing and particle identification 
capabilities of the LHCb detector, the characteristics of particle 
production in those events can be studied in detail in the 
pseudo-rapidity range $2<\eta<5$.


\begin{thebibliography}{99}
\bibitem{rapidity}
Rapidity, $y=\frac{1}{2}\ln\frac{E+p_L}{E-p_L}$, and pseudorapidity, $\eta=-\ln\tan\frac{\theta}{2}$, are used interchangeably, since in the kinematic regions of interest they are approximately equal.
\bibitem{missingET} Transverse energy is defined as $E_T=E\sin\theta$, and missing $E_T$ as $\met = |\overrightarrow{\met}|$ with  $\overrightarrow{\met}= -\sum_iE_T^i\hat{n_i}$, where $\hat{n_i}$ is a unit vector perpendicular to the beam axis and pointing at the $i^{th}$ calorimeter tower. The sum $E_T$ is defined by $\sum E_T=\sum_iE_T^i$. Both sums are over all calorimeter towers above the set thresholds. The missing $E_T$ significance is defined as $S_{\not E}\equiv{\not\!E_T}/\sqrt{\Sigma{E_T^2}}$
\end{thebibliography}
\begin{thebibliography}{0}
\bibitem{rapidity}
The terms rapidity and pseudorapidity are used interchangeably.
The pseudorapidity $\eta$ is defined as $-\ln \tan(\theta/2)$, where $\theta$ is
 the polar angle with respect to the proton beam direction.
The azimuthal angle is denoted by $\phi$, and the transverse energy of a jet, $E_T$, is defined as $E_T\equiv E\cdot \sin\theta$.
\bibitem{xbj}
For each event, $x_{Bj}$ is evaluated from the $E_T$ and $\eta$ of the jets using the equation
$x_{Bj}=\frac{1}{\sqrt{s}}\sum_{i=1}^3E_T^ie^{-\eta^i}$.
\end{thebibliography}

\newpage
\clearpage
\setcounter{affil}{0}
\setcounter{section}{0}
\setcounter{figure}{0}
\setcounter{table}{0}
\setcounter{equation}{0}

\preprint{}

\title{Forward Physics with the CMS experiment at the Large Hadron Collider}

\thanks{Presented at Forward Physics at LHC Workshop (May 27-29, 2010), Elba Island, Italy}

\author{Dmytro Volyanskyy\footnote{on behalf of the CMS collaboration}}
 \email{Dmytro.Volyanskyy@cern.ch}
\affiliation{%
Deutsche Elektronen-Synchrotron DESY \\
Notkestrasse 85, 22607 Hamburg, Germany
}%

\begin{abstract}
The forward physics program of the CMS experiment at the LHC spans a broad range
of diverse physics topics including studies of low-$x$ QCD and diffractive scattering,
multi-parton interactions and underlying event structure,
$\gamma$-mediated processes and luminosity determination,
Monte Carlo tuning and even MSSM Higgs discovery in central exclusive production.
In this article, the forward detector instrumentation around the CMS interaction point
is described and the prospects for diffractive and forward physics using
the CMS forward detectors are summarized. In addition, first observation of forward jets
as well as early measurements of the forward energy flow in the pseudorapidity
range $3.15<|\eta|<4.9$ at $\sqrt{s}=0.9$~TeV, $2.36$~TeV and $7$~TeV are presented.
\begin{description}
\item[PACS numbers:] 
\item[Keywords:] forward physics, diffraction, energy flow
\end{description}

\end{abstract}


\maketitle
\addcontentsline{toc}{part}{Forward Physics with the CMS experiment at the Large Hadron Collider - {\it D.Volyanskyy}}
\section{The CMS experiment at the LHC}

The Compact Muon Solenoid~(CMS)~[1] is one of two general-purpose particle physics detectors
built at the Large Hadron Collider~(LHC) at CERN. The detector has been designed to study
various aspects of proton-proton ($pp$) collisions at $\sqrt{s}=$14~TeV and heavy-ion(Pb-Pb) collisions
at $\sqrt{s}=$5.5~TeV, that will be provided by the LHC at a design luminosity of $10^{34}~ \rm cm^{-2}s^{-1}$
and of $10^{27}~ \rm cm^{-2}s^{-1}$, correspondingly.
To enhance the physics reach of the experiment the CMS subcomponents must provide high-precision
measurements of the momentum and the energy of collision-products. The CMS detector comprises
the tracking system covering the pseudorapidyty range $-2.5<\eta<2.5$ and the calorimetry system covering
the pseudorapity range $-5<\eta<5$. In addition to that, CMS includes several very forward calorimeters,
whose design and physics potential will be described later in this article.
It should be emphasized that the CMS detector is one of the largest scientific instruments ever built.
It comprises about $76.5$ millions of readout channels in total. 
The detector has been designed, constructed and currently operated by the collaboration consisting
of more than $3500$ scientists from $38$ countries.

First collision data taking at CMS took place in November 2009.
Since then and by the end of May 2010, CMS has collected around $10$~$\rm nb^{-1}$ of collision data.
It should be noted that the quality of collected data is rather good:
more than $99\%$ of CMS readout channels are operational and the CMS data taking efficiency is above $90\%$.
Several tens of $\rm pb^{-1}$ of the $pp$ collision data are expected to be collected by the end of 2010.

\section{Forward detectors around the CMS interaction point}

The maximum possible rapidity at the LHC in $pp$ collisions at $\sqrt{s}=14$~TeV
is $y_{max}=ln~(\sqrt{s}/m_{\pi})\approx 11.5$ and one of the great features of the CMS experiment is that it includes several subdetectors
covering the kinematic region at very small polar angles and so, large values of rapidity.
A schematic view of CMS forward detectors is shown in Figure~1. As can be seen, the CMS forward instrumentation
consists of the Hadronic Forward calorimeter~(HF), the CASTOR and ZDC calorimeters.
All of them are sampling calorimeters. That is, they are made of repeating layers
of a dense absorber and tiles of scintillator. A separate experiment TOTEM as well as proton detectors FP420
are additional forward detectors around the CMS interaction point~(IP5).
They further extend the forward reach available around IP5.
\begin{figure}
\begin{center}
\includegraphics[width=0.8\textwidth]{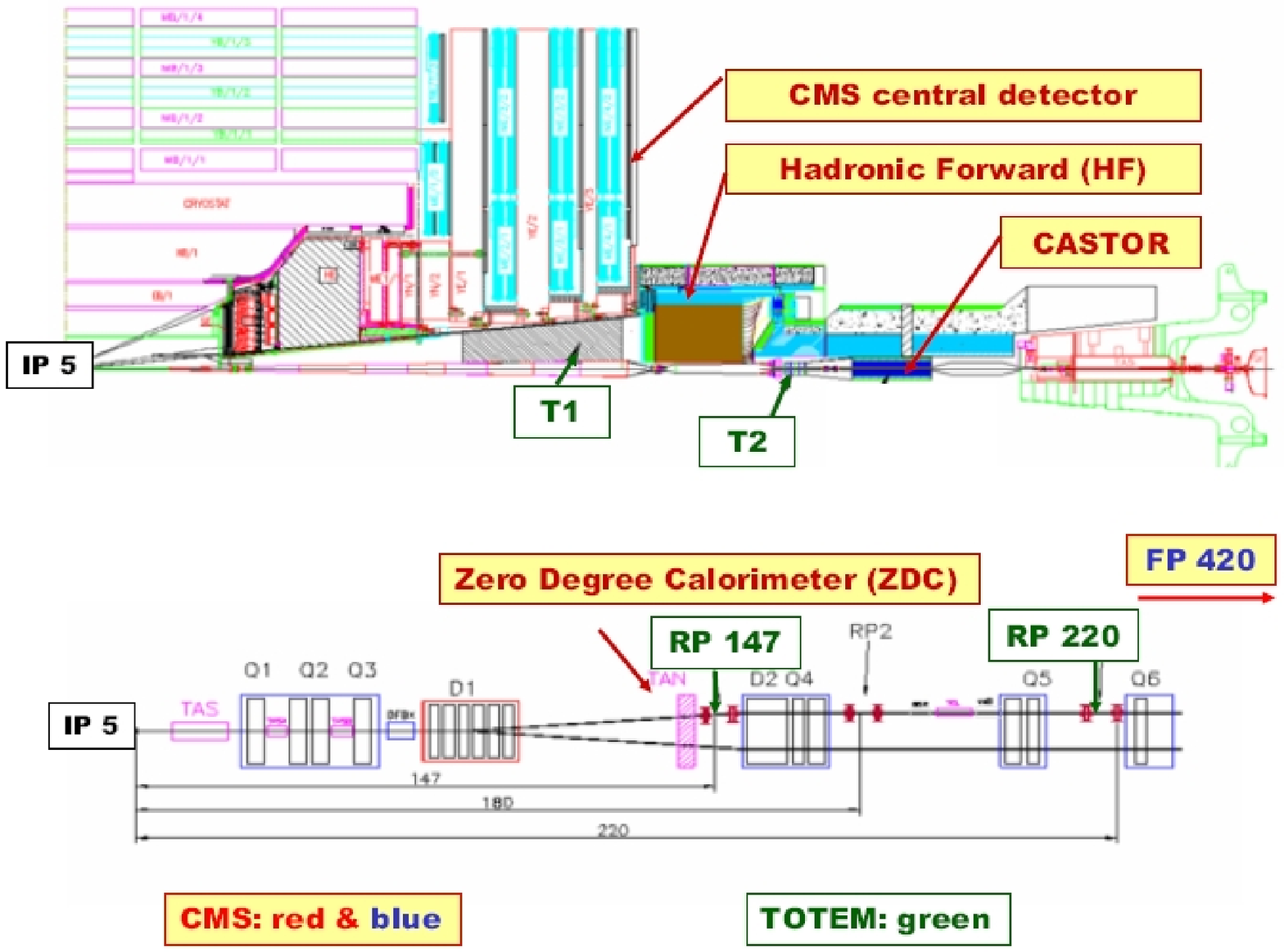}
\caption{Layout of the forward detectors around the CMS interaction point.}
\end{center}
\end{figure}
%

\subsection{HF}

The CMS HF detector~[2] includes two calorimeters HF+ and HF--,
which are located at a distance of $11.2$~m on the both sides from the IP5
covering the pseudorapidity range $3<|\eta|<5$. The detector is designed
to carry out the measurements of the forward energy flow and forward jets.
The HF is a Cerenkov sampling calorimeter which uses radiation hard
quartz fibers as the active material and steel plates as the absorber.
The signal in the HF is produced when charged shower particles
pass through the quartz fibers with the energy above the Cerenkov threshold ($190$~keV for electrons).
The generated Cerenkov light is then collected by air-code light guides,
which are connected to photo-multipliers tubes PMTs. The detector fibers run parallel to the beamline
and are bundled to form $0.175 \times 1.175$ ($\Delta\eta \times \Delta\phi$) towers.
Half of the fibers run over the full depth of the absorber,
whereas the other half starts at a depth of $22$~cm from the front of the
detector. These two sets of fibers are read out separately.
Such a structure allows to distinguish showers
generated by electrons and photons, which deposit a large
fraction of their energy in the first $22$~cm, from those generated
by hadrons, which produce signals in both calorimeter
segments. The detector is embedded into a cylindrical steel structure
with the outer radius of $131$~cm and the inner radius of $12.5$~cm
to accommodate the beam pipe. It is azimuthally subdivided into
$20^{0}$ modular wedges, each of which consists of two azimuthal sectors
of $10^{0}$. The detector extends over $10$ interaction lengths and includes
$1200$ towers in total.

\subsection{CASTOR}

The CASTOR (Centauro And STrange Object Reseacrh) detector~[3] is a
quartz-tungsten Cerenkov sampling calorimeter, which is located at a distance of $14.4$~m
from the IP5 and covering the pseudorapidity range $-6.6<\eta<-5.2$.
The tungsten metal has been chosen as the absorber medium in CASTOR,
since it provides the smallest possible shower size. In this detector,
the radiation hard quartz plates used as the active medium are tilted at $45^{0}$
to efficiently capture the Cerenkov light produced by relativistic particles
passing the detector. As in the case of the HF, the produced Cerenkov light is collected
by air-code light guides that are connected to PMTs, which produce signals
proportional to the amount of light collected.
The CASTOR detector is a compact calorimeter with the physical size of about $\rm 65~cm \times 36~cm \times 150~cm$ 
and having no segmentation in $\eta$.  It is embedded into a skeleton, which is made of stainless steel.
The detector consists of $14$ longitudinal modules, each of which comprises $16$ azimuthal sectors
that are mechanically organized in two half calorimeters.
First $2$ longitudinal modules form the electromagnetic section,
while the other $12$ modules form the hadronic section.
In the electromagnetic section, the thicknesses of the tungsten and quartz plates are $5.0$ and $2.0$~mm respectively,
whereas in the hadronic section the corresponding thicknesses are $10.0$ and $4.0$~mm.
With this design, the diameter of the showers of electrons and positrons produced by hadrons
is about one cm, which is an order of magnitude smaller than in other types of calorimeters.  
The detector has a total depth of $10.3$ interaction lengths and includes $224$ readout channels.

\subsection{ZDC}

The CMS ZDC~(Zero Degree Calorimeter) detector~[4] consists of two calorimeters that are located
inside the TAN absorbers at the ends of the straight section of the LHC beampipe at a distance
of 140~m on both sides from the IP5. These are Cerenkov sampling calorimeters with quarz fibers
as the active material and tungsten plates as the absorber material.
The ZDC detector is designed to measure neutrons and very forward photons
providing detection coverage in the pseudorapidity region $|\eta|>8.4$.  
Each ZDC is made up of separate electromagnetic and hadronic sections.
The electromagnetic section consists of $33$ layers of $2$~mm thick tungsten plates
and $33$ layers of $0.7$~mm thick quartz fibers. The hadronic section is made of $24$ layers
of $15.5$~mm thick tungsten plates and $24$ layers of $0.7$~mm thick quartz fibers.
The electromagnetic section is segmented into $5$ horizontal individual readout towers,
whereas the hadronic section is longitudinally segmented into $4$ readout segments.
The tungsten plates are oriented vertically in the electromagnetic section
whereas they are tilted by $45^{0}$ in the hadronic section.
The detector is read out via aircore light guides and PMTs.
It has a total depth of $6.5$ interaction length.
 
\subsection{TOTEM and FP420}

TOTEM~[5] is an independent experiment at the CMS interaction point whose main objectives
are the precise measurement of the total $pp$ cross-section and a study of
elastic and diffractive scattering at the LHC. To achieve optimum forward coverage
for charged particles, TOTEM comprises two tracking telescopes, T1 and T2, that are installed on both sides
from the IP5 in the pseudorapidity region $3.1<|\eta|<6.5$, and Roman Pot stations
that are located at distances of $\pm147$~m and $\pm220$~m from the IP5.
The T1 telescope is located in front of HF and consists of $5$ planes of cathode strip chambers,
while the T2 telescope is located in front of CASTOR and comprises $10$ planes of gas electron multipliers.
For efficient reconstruction of very forward protons, silicon strip detectors are housed in the Roman Pot stations.

FP420~[6] is a proposed detector system, which is supposed to provide
proton detection at a distance of $\pm420$~m from the IP5.
The FP420 detector comprises a silicon tracking system that can be moved transversely
and measure the spatial position of protons, which have been bent out
by the LHC magnets due to the loss of a small fraction of their initial momentum.
The potential physics topics that can be studied
with this detector system include Higgs central exclusive production as well as
a rich QCD and electroweak program.

\section{Physics program}

Extending the physics reach of CMS, the program for forward physics
includes studies of low-$x$ QCD and diffractive scattering,
multi-parton interactions and underlying event structure,
$\gamma$--mediated processes and luminosity determination.
It is also supposed to contribute to the discovery physics
via searches of MSSM Higgs in central exclusive production.

\subsection{Low-x QCD}

A study of QCD processes at a very low parton momentum fraction $x=p_{parton}/p_{hadron}$
is a key to understand the structure of the proton, whose gluon density
is poorly known at very low values of $x$.
Low-$x$ QCD dynamics can be studied in $pp$ collisions
if the parton momentum fraction of one of the colliding protons $x_{1}$
is significantly larger than the parton momentum fraction of the other colliding proton $x_{2}$ ($x_{1}>>x_{2}$).
The result of such a collision is a creation of either jets, prompt-$\gamma$ or Drell-Yan electron pairs
at very low polar angles in the very forward region of the detector.
Low-$x$ QCD studies at CMS will be a continuation of studies of
deep inelastic scattering in electron-proton collisions at HERA,
where low-$x$ QCD dynamics has been explored down to values of $10^{-5}$.
Measurements at HERA have shown that the gluon density
in the proton rises rapidly with decreasing values of $x$.
As long as the densities are not too high this rise can either be
described by the DGLAP model~[7] that assumes strong ordering
in the transverse momentum $k_{T}$ or by the BFKL model~[8] that assumes
strong ordering in $x$ and random walk in $k_{T}$.
Eventually at low enough $x$, the gluon-gluon fusion effects
become important saturating the growth of the parton densities.

At the LHC the minimum accessible $x$ in $pp$ collisions
decreases by a factor of about $10$ for each $2$ units of rapidity.
This implies that a process with a hard scale of $Q \sim 10$~GeV
and within the CASTOR/T2 detector acceptance
can probe quark densities down $x \sim 10^{-6}$.
Such processes include the production of forward jets
and Drell-Yan electron pairs.

\subsubsection{Forward Jets}

A low-$x$ parton distribution function (PDF) of the proton
can be constrained by measuring single inclusive jet cross-section in HF.
Figure~2 illustrates the $log(x_{1,2})$ distribution for parton-parton scattering
in $pp$ collisions at $\sqrt{s}=14$~TeV requiring at least one
jet with the transverse energy above $20$~GeV in the HF acceptance.
As can be seen, by measuring forward jets in HF one can probe $x$ values as low as $10^{-5}$.
\begin{figure}
\begin{center}
\resizebox{3.2in}{!}{
\rotatebox{0}{
\includegraphics{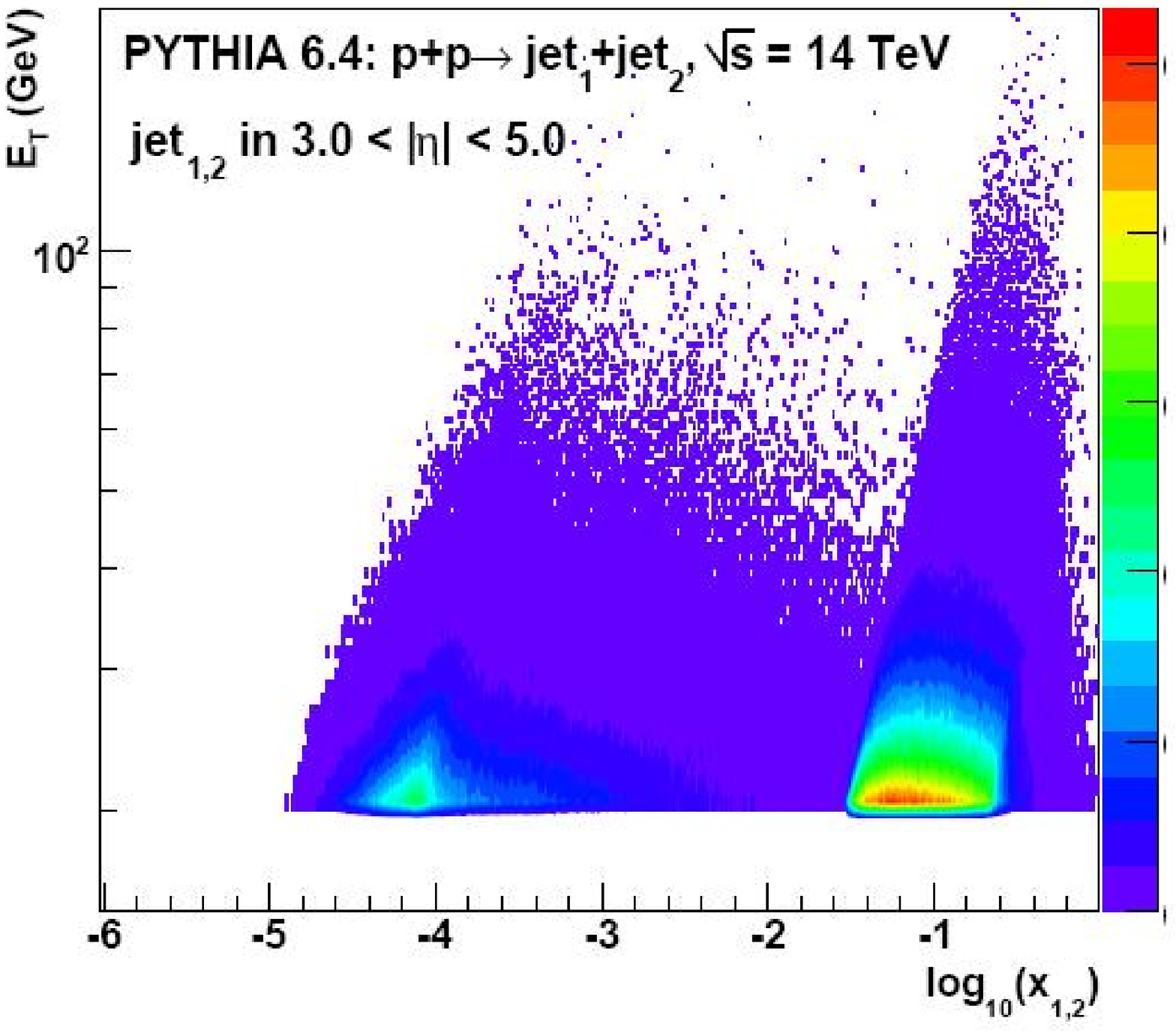}}}
\caption{$log(x_{1,2})$ distribution of two partons producing at least one jet with $E_{T}>20$~GeV in the HF acceptance.}
\end{center}
\end{figure}
A detailed analysis of fully simulated and reconstructed QCD jet events
generated with PYTHIA in the range $p_{T}=20$~GeV/c--$200$~GeV/c in $pp$ collisions
at $\sqrt{s}=14$~TeV  for an integrated luminosity of $1$~$\rm pb^{-1}$ shows that the momentum resolution 
for forward jets in HF is about $18\%$ at $p_{T}=20$~GeV/c and is gradually decreasing to $12\%$ at $p_{T}\geq 100$~GeV/c~[9].

A possibility to gain information on the full QCD evolution to study high order QCD reactions
can be provided by measuring forward jets in the CASTOR calorimeter, that will allow to probe
the parton densities down $10^{-6}$. Apart from that, it has been found that 
a BFKL like simulation
predicts more hard jets in the CASTOR acceptance than the DGLAP model.
Therefore, measurements of forward jets in CASTOR can be used as a good tool
to distinguish between DGLAP and non-DGLAP type of QCD evolution.

Further studies of low-$x$ QCD can be made with Mueller-Navalet dijet events,
which are characterized by two jets with similar $p_{T}$
but large rapidity separation. By measuring Mueller-Navalet dijets
in CASTOR and HF one can probe BFKL-like dynamics and small-$x$ evolution.
\subsubsection{Drell-Yan}
Low-$x$ proton PDFs can also be constructed by measuring electron pairs produced
via the Drell-Yan process $qq\rightarrow\gamma^{*} \rightarrow e^{+}e^{-}$ within the acceptance of CASTOR and TOTEM-T2 station,
whose usage is essential for detecting these events. 
Figure~3 illustrates the distribution of the invariant mass $M$ of the $ee$ system against
the parton momentum fraction $x_{2}$ of one of the quarks, where $x_{2}$ is chosen such that $x_{1} >> x_{2}$.
In this figure, the solid line indicates the kinematic limit,
whereas the region between the dotted lines is the acceptance window for both electrons to be detectable in CASTOR/T2.
The green points show the events with at least one electron lying in CASTOR/T2 acceptance
and the blue points indicate the events with both electrons present within the CASTOR/T2 acceptance,
while the black points correspond to any of the Drell-Yan events generated with PYTHIA.
As can be seen, by measuring two electrons in the CASTOR/T2 acceptance 
one can access $x$ values down to $10^{-6}$ for $M>10$~GeV~[10].
Futhermore, measurements of Drell-Yan events in the CASTOR/T2 acceptance
can be used to study QCD saturation effects. It has been found that
the Drell-Yan production cross section is suppressed roughly by a factor of $2$
when using a PDF with saturation effects compared to one without.
\begin{figure}
\begin{center}
\resizebox{3.2in}{!}{
\rotatebox{0}{
\includegraphics{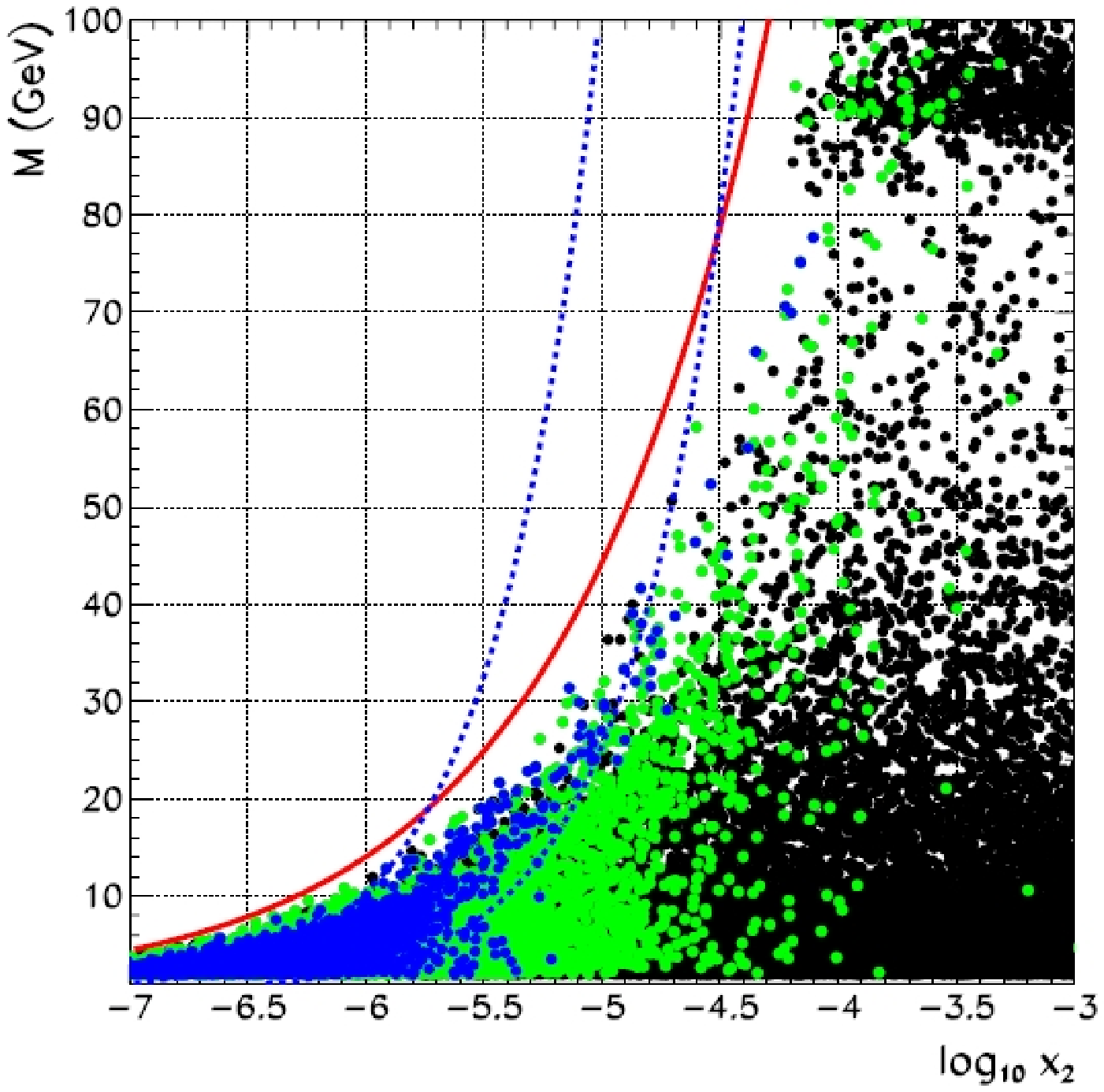}}}
\caption{Acceptance of the CASTOR/T2 detectors for Drell-Yan electrons. See text for details.}
\end{center}
\end{figure}
%

\subsection{Diffraction}
In $pp$ collisions a diffractive process is a reaction $pp \rightarrow X Y$,
where $X$ and $Y$ can either be protons or low-mass systems which may be
a resonance or a continuum state. In all cases, the final states $X$ and $Y$
acquire the energy approximately equal to that of the incoming protons and
carry the quantum numbers of the proton as well as are separated by a Large Rapidity Gap~(LRG).
Diffraction in the presence of a hard scale can be described with perturbative QCD
by the exchange of a colourless state of quarks or gluons, whereas soft diffraction
at high energies is described in the Regge Theory~[11] as a colourless exchange mediated
by the Pomeron having the quantum numbers of the vacuum.
The cross section of hard diffractive processes can be factorized into generalized parton
distributions and diffractive parton distributions functions~(dPDF), which contain
a valuable information about low-$x$ partons. However, the factorization becomes broken
when scattering between spectator partons takes place. This effect is quantified by
the so-called rapidity gap survival probability that can be probed by measuring
the ratio of diffractive to inclusive processes with the same hard scale.
At the Tevatron, the ratio is found to be $O(1\%)$, whereas theoretical expectations
at the LHC vary from a fraction of a percent to up to $30\%$~[12].

The two main types of diffractive processes occurring in $pp$ collisions
are the single diffractive dissociation~(SD) where one of the protons dissociates
and the double diffractive dissociation~(DD) where both protons are scattered into a low-mass system.
The single-diffractive productions of $W$ and dijets are in particular
very interesting processes to study, since they are sensitive
to the quark and gluon content of the PDFs, correspondingly.
They both are hard diffractive processes that can provide information on the rapidity gap survival probability.
A selection of such events can be performed using the multiplicity distributions of tracks
in the central tracker and calorimeter towers in HF plus CASTOR exploiting the fact that diffractive events
on average have lower multiplicity in the central region and in the $"$gap side$"$ than non-difractive ones.
Feasibility studies to detect the SD productions of $W$~[13] and dijets~[14]
have shown that the diffractive events peak in the regions of no activity in HF and CASTOR.

\subsection{Exclusive dilepton production}

Another interesting topic that is going to be studied at CMS is the exclusive dilepton production
$pp \rightarrow ppl^{+}l^{-}$, which can either occur via $\Upsilon$ photoproduction $\gamma p \rightarrow \Upsilon \rightarrow l^{+}l^{-}$
or via the pure QED process $\gamma \gamma \rightarrow l^{+}l^{-}$ that has been observed
by the CDF experiment at the Tevatron~[15]. The latter is an elastic process whose production cross section
is precisely known. As a result, it can potentially serve as an ideal calibration channel
and is going to be used for measuring the luminosity at the LHC.
Using this process an absolute luminosity calibration with the accuracy of $4\%$ is feasible with $100$~$\rm pb^{-1}$ of data~[16].
The dominant background source for this mode is inelastic processes, where one of the proton in the process
does not stay intact but dissociates. It can be significantly suppressed by applying
a veto condition on activity in CASTOR and ZDC.
Exclusive dilepton production occurring via $\Upsilon$ photoproduction is also a mode of interest,
since the cross section of the $\Upsilon$ photoproduction process is sensitive to the generaliszed PDF
for gluons in the proton. Finally, it should be noted that exclusive dimuon production is an ideal alignment channel
for the proposed FP420 proton detectors.
 
\subsection{Multi-parton interactions and forward energy flow}

Multi-parton interactions~(MPI) arise in the region of small-$x$ where parton densities
are large so that the likelihood of more than one parton interaction per event is high.
According to all QCD models, the larger the collision energy the greater the contribution
from multiple parton interactions to the hard scattering process. However, the dependence
of the MPI cross section on the collision energy is not well known and needs to be studied.
A good way to study multiple parton interactions is provided by the energy flow in the forward region,
which is directly sensitive to the amount of parton radiation and MPI.
Measurements of the forward energy flow will allow to discriminate between different MPI models,
which vary quite a lot, and provide additional input to the determination
of the parameters of the existing MPI models. Furthermore, measurements of forward particle production
in $pp$ and Pb-Pb collisions at LHC energies should help to significantly improve
the existing constraints on ultra-high energy cosmic ray models.
The primary energy and composition of the ultra-high energy cosmic rays
are currently determined from Monte Carlo simulations using Regge-Gribov-based approaches~[17] (where the primary particle production
is dominated by forward and soft QCD interactions) with parameters constrained by the existing collider data at the $E_{lab}<10^{15}$~eV,
whereas the measured energies of the ultra-high energy cosmic rays extend up to $10^{20}$~eV and even beyond. 
At the LHC energy of $E_{lab}=10^{17}$~eV, a more reliable determination of the cosmic ray energy and composition becomes possible.
Finally, it should be emphasized that the forward energy flow has never previously been
measured at a hadron collider.

\section{First results from CMS}

\subsection{Observation of forward jets}

A search for forward jets in the pseudorapidity range $3<|\eta|<5$ has been made
as soon as the CMS detector has started to take collision data~[18]. One of the first candidates of a forward dijet
event recorded by CMS at $\sqrt{s}=0.9$~TeV  is shown in Figure~4. The displayed event includes one forward jet and
one backward jet both with a corrected $p_{T}$ above $10$~GeV/c.
\begin{figure}
\begin{center}
\includegraphics[width=0.7\textwidth]{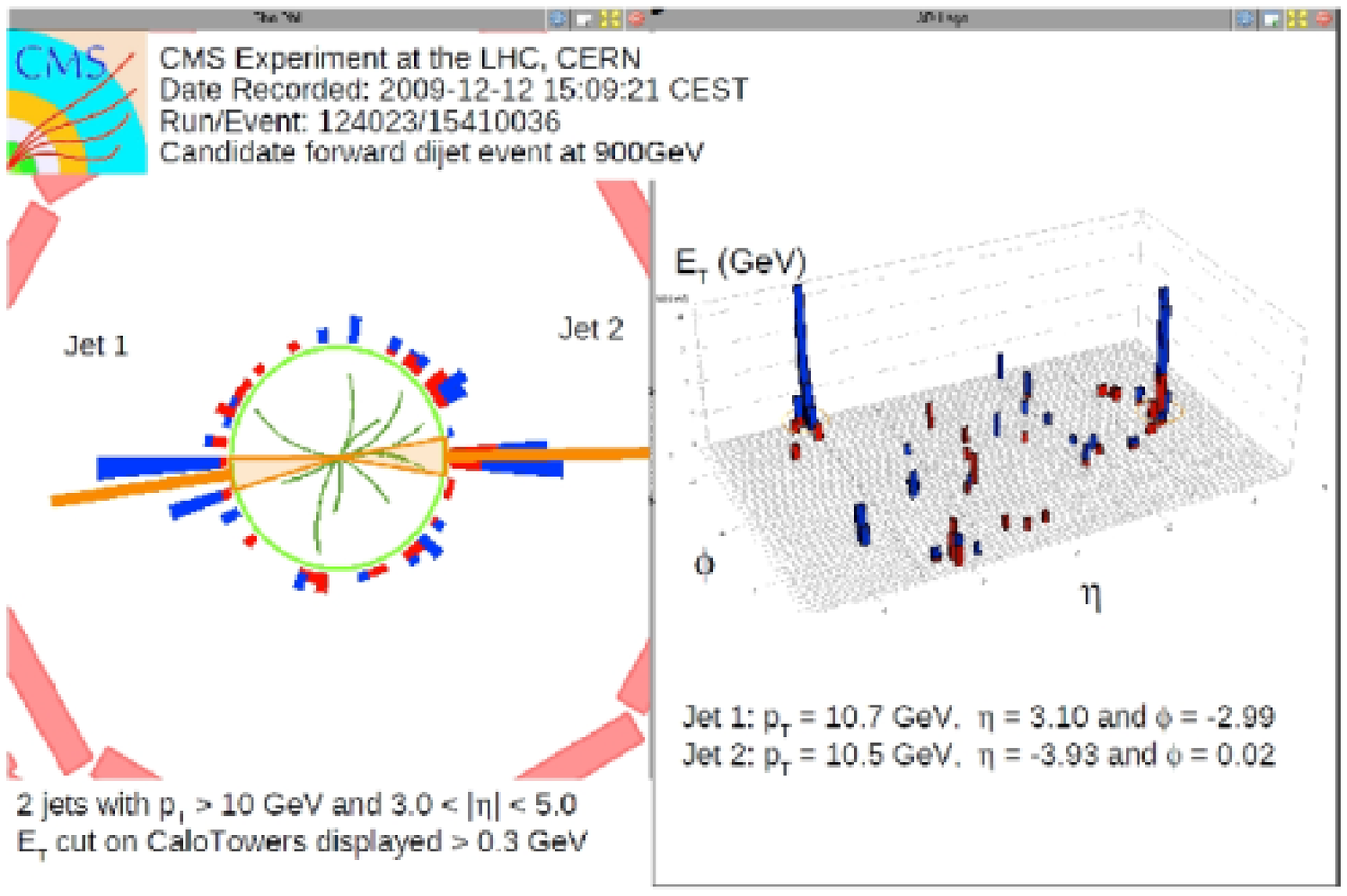}
\caption{Display of an event with two forward jets.}
\end{center}
\end{figure}
%

\subsection{Measurement of the forward energy flow}

Early measurements of the energy flow in the forward region of the CMS detector have been made
with minimum bias events using the $pp$ collision data sets collected at $\sqrt{s}=0.9$~TeV and $2.36$~TeV in the fall of 2009
and at $\sqrt{s}=7$~TeV in March 2010~[19]. To select the events of interest the following conditions were imposed.

First, the Beam Scintillator Counters~(BSC) and the Beam Pick-up Timing for the eXperiments~(BPTX),
both are elements of the CMS detector monitoring system, were used to trigger the detector readout.
The BSC devices are located at a distance of $10.86$~m on both sides from the interaction point
covering the pseudorapidity range $3.23<|\eta|<4.65$ and providing hit and coincidence signals
with a time resolution of about $3$~ns. Each BSC comprises $16$ scintillator tiles.
The two BPTX elements are located around the beam pipe at a distance of $\pm175$~m from
the interaction point providing precise information on the bunch structure and timing
of the incoming beam with a time resolution better than $0.2$~ns.
To select the minimum bias events with activity in the forward regions, the coincidence between
a trigger signal in the BSC scintillators and BPTX signals was required for both beams.

Next, to ensure that the selected event is a collision candidate, the events were required
to have at least one primary vertex reconstructed from at least $3$ tracks with a $z$ distance
to the interaction point below $15$~cm and a transverse distance from the $z$-axis smaller than $2$~cm.
Further cuts were applied to reject beam-scrapping and beam-halo events.  Finally, the energy threshold
of $4$~GeV has been imposed to suppress electronic noise in HF.

In this study, the measurement of energy flow has been made at detector level
in the pseudorapidity range $3.15<|\eta|<4.9$ covered by the HF calorimeters.
The energy flow ratio, estimated in this analysis, is defined as
\begin{equation}
 R_{E flow}^{\sqrt{s_{1}}\sqrt{s_{2}}}=\dfrac{ \frac{1}{N_{\sqrt{s_{1}}}} \frac{dE_{\sqrt{s_{1}}}}{d\eta}  } {\frac{1}{N_{\sqrt{s_{2}}}}  \frac{dE_{\sqrt{s_{2}}}}{d\eta}} \;\;\;,
\label{eq:eq}
\end{equation}
where $N_{\sqrt{s}}$ is the number of selected events, $dE_{\sqrt{s}}$  is the energy deposition integrated over $\phi$ in the region $d\eta$,
$\sqrt{s_{1}}$ refers to either $2.36$~TeV or $7$~TeV, whereas $\sqrt{s_{2}}$ refers to $0.9$~TeV.
The pseudorapidity range is divided into five bins with a size of $0.35$  in units of $\eta$
following the transverse segmentation of the HF calorimeters.
In Figures~5 and 6, the energy flow ratio is shown for different collision energies
as the average of the HF(+) and HF(--) responses.
\begin{figure}
\begin{center}
\resizebox{3.2in}{!}{
\rotatebox{0}{
\includegraphics{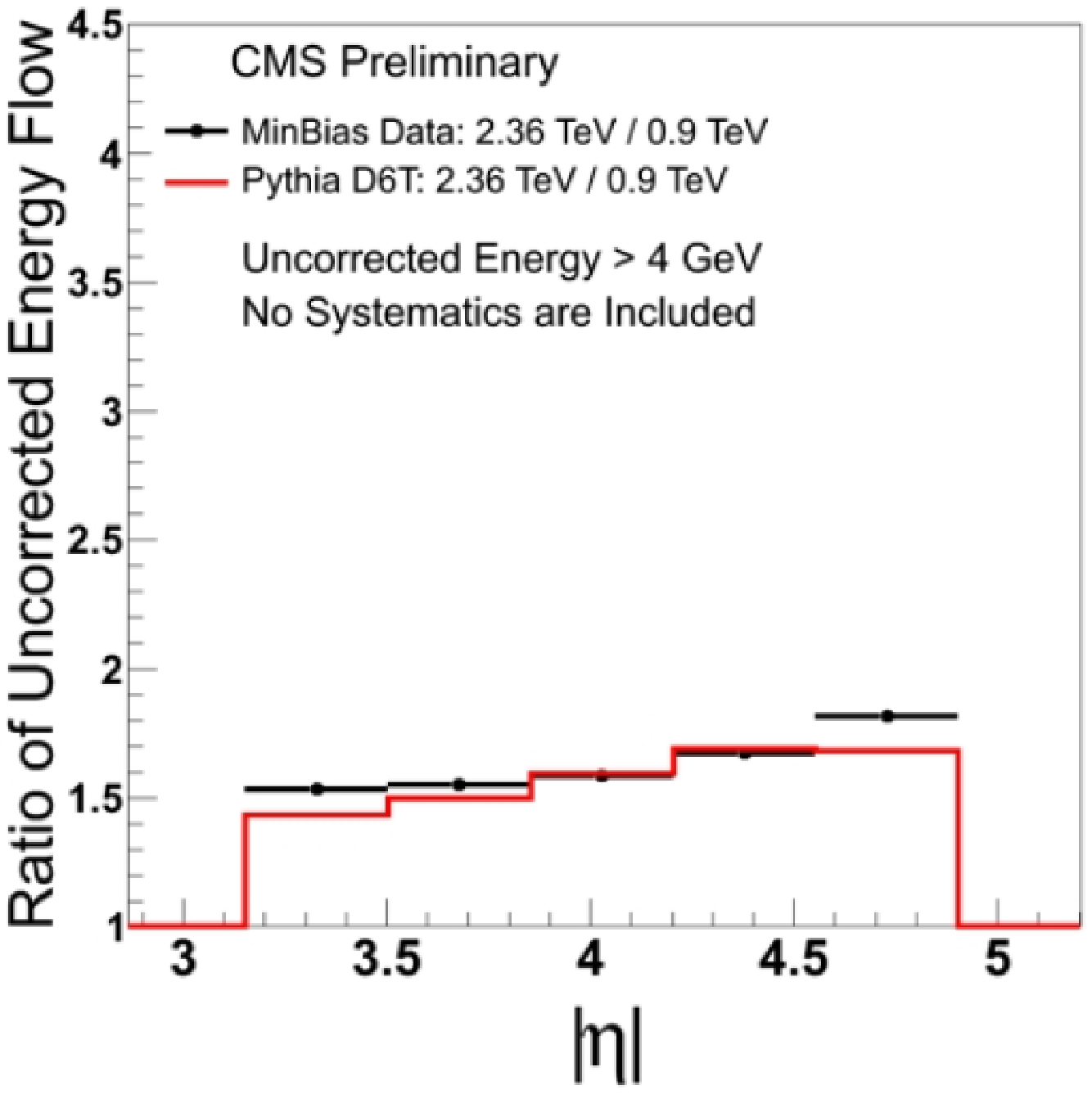}}}
\caption{Energy flow ratio for $\sqrt{s_{1}}=2.36$~TeV to $\sqrt{s_{2}}=0.9$~TeV as a function of $\eta$. See text for details.}
\end{center}
\end{figure}
\begin{figure}
\begin{center}
\resizebox{3.2in}{!}{
\rotatebox{0}{
\includegraphics{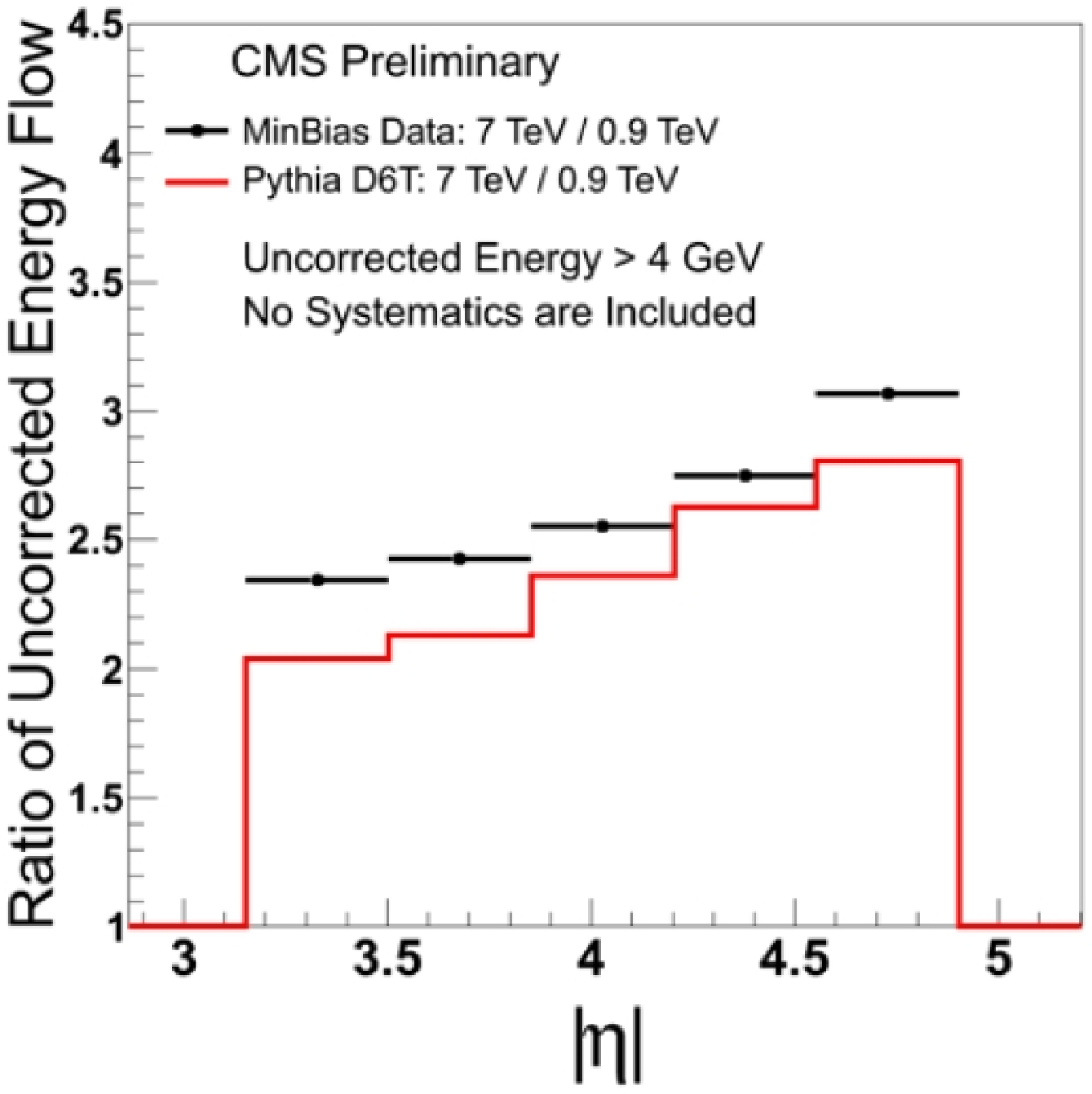}}}
\caption{Energy flow ratio for $\sqrt{s_{1}}=7$~TeV to $\sqrt{s_{2}}=0.9$~TeV as a function of $\eta$. See text for details.}
\end{center}
\end{figure}
In these plots, uncorrected data without systematic uncertainties
are compared to simulated events obtained from PYTHIA tune D6T.
As can be clearly seen, the energy flow gets larger at forward rapidities and with increasing centre-of-mass energy.
Apart from that, it should be noted that the obtained results do approximately agree with the Monte Carlo predictions.
However, no conclusions on the quality of the description can be drawn in this early study due to the missing systematic effects.

\section{Conclusions}

A very rich forward physics program can be made with the CMS detector at the LHC
due to the unprecedented kinematic coverage of the forward region.
All the CMS forward detectors have been successfully commissioned in 2009
and currently take collision data. The first measurement of the forward energy flow
has been performed and forward jets at $|\eta|>3$ have been observed
for the first time at hadron colliders.

\section{Acknowledgments}

I am very grateful to Hannes Jung, Kerstin Borras and many other colleagues working in
the CMS forward physics community for fruitful discussions and kind suggestions.



\end{document}